\documentclass[pdflatex,sn-basic]{sn-jnl}

\usepackage{graphicx}%
\usepackage{multirow}%
\usepackage{amsmath,amssymb,amsfonts}%
\usepackage{amsthm}%
\usepackage{mathrsfs}%
\usepackage[title]{appendix}%
\usepackage{xcolor}%
\usepackage{textcomp}%
\usepackage{manyfoot}%
\usepackage{booktabs}%
\usepackage{algorithm}%
\usepackage{algorithmicx}%
\usepackage{algpseudocode}%
\usepackage{listings}%

\begin{document}

\title[A review of galaxy quenching --- Part~II]{A review of galaxy quenching --- Part~II: Theoretical solutions and direct observational tests}


\author*[1]{\fnm{Asa F. L.} \sur{Bluck}}\email{abluck@fiu.edu}

\affil*[1]{\orgdiv{Stocker AstroScience Center, Dept. of Physics}, \orgname{Florida International University}, \orgaddress{\street{11200 SW 8th St.}, \city{Miami}, \postcode{33199}, \state{FL}, \country{USA}}}


\abstract{The goal of this review article series is to provide a comprehensive overview of galactic star formation and quenching from both an observational and theoretical perspective. Drawing on a vast quantity of literature, we attempt to answer a deceptively simple question: \textit{why do galaxies cease forming stars?} In Part~II, we concentrate primarily on results from theory and simulations, in addition to direct observational tests of theoretically proposed quenching routes. Over the past two decades, N-body simulations, semi-analytic models, idealized hydrodynamical simulations, cosmological hydrodynamical simulations, and zoom-in hydrodynamical simulations have all provided crucial insights into the fundamental causes and specific mechanisms of galaxy quenching. Throughout this part of the review, we discuss intrinsic routes to massive galaxy quenching from strong baryonic feedback --- including supernovae and active galactic nuclei (in both the ejective and preventative modes). Additionally, we discuss dynamical stabilization and the role of mergers. We go on to consider environmental routes to quenching via both ram pressure and dynamical stripping, and as a consequence of the location of satellite galaxies within the cosmic web. We review observational tests of the fundamental causes, specific mechanisms, and triggering of quenching from various techniques. Finally, we conclude this review with tentative answers to the major outstanding issues in this field. }




\maketitle


\setcounter{tocdepth}{3} 
\tableofcontents


\section{Introduction}\label{s1}

This review series is intended to provide an overview of the vibrant field of galaxy quenching, from both an observational and theoretical perspective. In Part~I of this review, we set up the fundamental problem of galaxy quenching and consider observational evidence for this process occurring in galaxies across a vast array of physical scales and cosmological times. In this second part of this review series, we explore theoretical routes to quenching in simulations and models. Additionally, we go on to review specific observational tests of the ideas from theory. Finally, we attempt tentative answers to how quenching operates in nature by assessing the relative success of various quenching paradigms against the most stringent observational constraints.

At the most basic level, the phenomenon of galaxy quenching appears to separate into two distinct, though not mutually exclusive, regimes. High-mass galaxies are observed to be predominantly quenched, suggesting that internal processes linked to mass (e.g., AGN feedback) play a dominant role (e.g., \citealt{Baldry2006, Peng2010}). In contrast, low-mass galaxies are typically star forming unless they reside as satellites within more massive haloes, indicating that environmental processes (e.g., stripping or strangulation) are required to suppress star formation in these systems (e.g., \citealt{Peng2012, Wetzel2012}). Additionally, it is crucial to distinguish between the regulation of star formation, where feedback processes modulate but do not halt star formation altogether, and quenching proper, where star formation is suppressed to very low levels for long cosmological time periods (e.g., \citealt{Torrey2014, Crain2015, Weinberger2017}). Any successful theoretical framework must therefore not only identify mechanisms capable of quenching individual galaxies, but also reproduce the observed demographics of galaxy populations across mass and environment.

\subsection{Motivation}\label{s11}

In the absence of strong baryonic feedback from galaxy evolution, one anticipates essentially all baryons to reside in stars by the present epoch, which is manifestly not the case (see Sect. 1.2.1 in Part~I; e.g., \citealt{White1991, Kauffmann1993, Cole2000, Fukugita2004, Bower2006, Bower2008, Shull2012, Somerville2015}). Furthermore, in the absence of a powerful heating source, one expects the hot gaseous halo surrounding massive galaxies, groups, and clusters to rapidly cool, which is not observed in most systems (see Sect. 1.2.2 in Part~I; e.g., \citealt{McNamara2005, Fabian2006, Croton2006, Bower2006, Fabian2012, HlavacekLarrondo2012}). 

Additionally, observations of the galaxy population at low and intermediate redshifts find strong evidence for bimodality in specific star formation rate (sSFR), rest-frame/ dust corrected optical colors, and stellar population ages (see Sect. 1.2.3 in Part~I; e.g., \citealt{Strateva2001, Brinchmann2004, Baldry2004, Baldry2006, Peng2010}). This bimodality evidences the existence of two fundamental types of galaxies --- actively star forming and quiescent (or `quenched') systems. A complete theory of galaxy formation and evolution must explain all of these (clearly inter-connected) problems simultaneously.

\begin{figure}[t!]  
\centering
\includegraphics[width=0.9\textwidth]{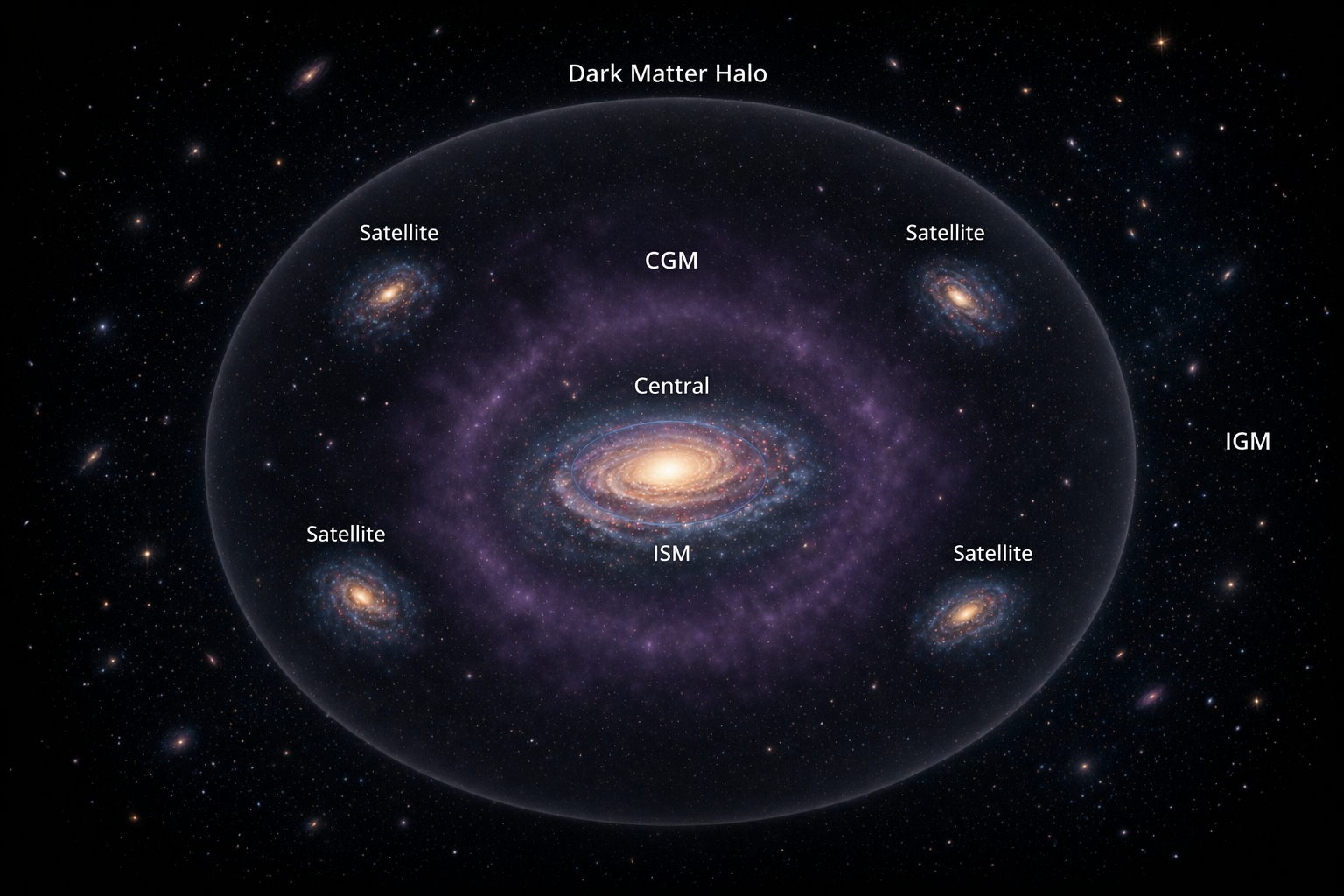}
\caption{A schematic illustration of several terms used to describe galaxies and the regions surrounding them. For definitions of acronyms see Table 1. Note that if this cartoon is taken as depicting a group, the CGM may be understood as being effectively synonymous with the IgrM. Similarly, if this figure is taken as describing a cluster, the CGM may be understood as being effectively synonymous with the ICM. This graphic was generated by ChatGPT-5 based on prompts provided by the author.}\label{f1a}
\end{figure}

These deep problems may all be resolved with a single concept: {\it galaxy quenching}. Ultimately, quenching is the process (or set of processes) which prevent baryons from forming stars within certain types of galaxies across long cosmological time periods. This naturally gives rise to bimodality in the galaxy population, and accounts for the low cosmological efficiency of star formation. Moreover, since the vast majority of baryons within high-mass haloes reside within an ionized plasma, a successful quenching mechanism must also account for how this system remains thermodynamically stable. Without this stabilization, cooling will ensue and baryons will condense into the galaxy, ultimately rejuvenating star formation.

Therefore, a complete description of quenching must explain at least two important aspects of galaxy evolution simultaneously: (i) why quenched galaxies initially cease forming stars; and (ii) why they do not excessively rejuvenate thereafter as a consequence of replenishment of gas content through cold gas accretion, halo cooling flows, and gas return from stellar evolution. The most pertinent constraint on the latter condition is the need for any rejuvenation to be sufficiently mild so as not to lead to massive galaxies being identified as star forming systems in high abundance by late cosmic times (which would manifestly contradict observational results; e.g., \citealt{Baldry2006, Peng2010}). Furthermore, throughout this work we distinguish between star formation regulation (which occurs in all galaxies) and quenching, i.e. near-complete cessation of star formation (which only occurs in some galaxies).

\begin{table*}
\centering
\small
\begin{tabular}{lll}
\hline
\textbf{Name} & \textbf{Abbreviation} & \textbf{Explanation} \\
\hline \\

Stellar Mass & $M_*$ & Total mass in stars within a galaxy [$M_\odot$]. \\

Star Formation Rate & SFR & Rate of formation of stars [$M_\odot\,{\rm yr}^{-1}$]. \\

Specific Star Formation Rate & sSFR & Star formation rate per unit stellar mass [${\rm yr}^{-1}$]. \\

Star Forming main Sequence & SFMS & SFR - $M_*$ [$z$] relationship for star forming galaxies. \\\\

Dark Matter Halo & Halo &  Gravitationally bound dark matter structure. \\

Dark Matter Subhalo & Subhalo &  Bound dark matter structure within a larger halo. \\

Halo Mass & $M_{\rm Halo}$ & Total mass of a dark matter halo [$M_\odot$]. \\

Virial Radius & $R_{\rm vir}$ & Radius within which a halo is virialized [kpc]. \\

Halo Mass (200) & $M_{200c}$ & Mass enclosed within radius $R_{200c}$ [$M_\odot$]. \\

Radius (200) & $R_{200c}$ & Radius within which $\langle \rho(<R_{200c}) \rangle = 200\, \rho_{\rm crit}$ [kpc].  \\

Central Galaxy & Cen. & The most massive galaxy in a dark matter halo. \\

Satellite Galaxy & Sat. & Any other galaxy in a dark matter halo. \\\\

Supermassive Black Hole & SMBH & Typical mass range: $M_{\rm BH} \sim 10^5 - 10^{10}\,M_\odot$.\\

Active Galactic Nucleus & AGN & Accreting supermassive black hole. \\

Supernova Type II & SN II & Core-collapse supernovae. \\

Supernova Type Ia & SN Ia & Thermonuclear supernovae. \\\\

Interstellar Medium & ISM & Gas and dust within a galaxy. \\

Circumgalactic Medium & CGM & Diffuse gas within the virial radius. \\

Intracluster Medium & ICM & Hot, X-ray emitting gas permeating clusters. \\

Intragroup Medium & IgrM & Hot, low-density gas permeating groups.\\

Intergalactic Medium & IGM & Diffuse gas filling space between galaxies.\\\\

Baryons & --- & All matter particles of the standard model. \\

Dark Matter & --- & Matter particles beyond the standard model.\\\\

Gas Accretion & --- & Inflow of gas from the CGM/IGM into the ISM. \\

Baryon Cycle & --- & Exchange of gas between the ISM, CGM and IGM. \\

Outflows (Galactic Winds) & --- & Gas driven out of the ISM into the CGM/IGM. \\

Recycling (Fountain Flows) & --- & Ejected gas that later re-accretes into the ISM. \\\\

\hline
\end{tabular}
\caption{General terminology used in galaxy formation and evolution studies.}\label{t1a}
\end{table*}

\subsection{Defining Terms}\label{s12}

The field of galaxy formation and evolution is full of bespoke terminology, which can prove quite an impediment to understanding when beginning as a researcher. In Part~I of this review series we introduce terms individually, as needed. However, in an effort to make this part of the review more accessible as a stand-alone work, we outline the key definitions of terms at the outset. In Table~\ref{t1a} we present a glossary of terms in general use throughout the galaxy formation and evolution literature. We also include in this table our chosen symbol, or abbreviation, as used throughout this part of the review.

In Table~\ref{t2a} we present a second glossary defining terms used specifically in the quenching literature. We separate this table into general quenching terminology, and terminology related to intrinsic and environmental quenching routes (respectively). Most of the definitions in Tables 1 \& 2 are fairly standard in the literature. However, some caution is warranted in taking any of these definitions as being absolute.

\begin{table*}
\centering
\small
\begin{tabular}{ll}
\hline
\textbf{Name} & \textbf{Explanation} \\
\hline \\

{\bf General:} & \\

Quenching & Cessation, or significant reduction, of star formation in a galaxy. \\

Starburst Galaxy & Galaxy which forms stars above the SFMS.\\

Star Forming Galaxy & Galaxy which forms stars on the SFMS. \\

Green Valley Galaxy & Galaxy with intermediate SFR (between star forming \& quenched). \\

Quenched Galaxy & Galaxy which forms stars well below the SFMS. \\\\

{\bf Intrinsic Quenching:} & \\

Mass Quenching & Internally driven quenching correlated with mass. \\

Halo Mass Quenching & Quenching driven by virial shock heating. \\

Shock Heating & Heating of infalling gas to the virial temperature. \\

Morphological Quenching & Suppression of star formation via dynamical stabilization. \\

AGN Feedback & Energy injection from an AGN into the surrounding galaxy \& halo. \\

Quasar-Mode AGN Feedback & Radiative/thermal AGN feedback at high accretion rates. \\

Radio-Mode AGN Feedback & Mechanical AGN feedback via jets at low accretion rates. \\

Kinetic AGN Feedback & General term for all mechanical AGN feedback. \\

Stellar Feedback & Energy input from stars (winds, radiation, supernovae). \\

Supernova Feedback & Energy injected into ISM \& CGM from SN Ia \& II\\

Ejective Feedback & Feedback that expels gas from galaxies and/or haloes. \\

Preventative Feedback & Feedback that inhibits gas accretion or cooling. \\

Maintenance Mode Feedback & Feedback that prevents rejuvenation of star formation. \\\\

{\bf Extrinsic Quenching:} & \\

Environmental Quenching & Externally driven quenching due to environment. \\

Ram Pressure Stripping & Gas removal by pressure in a dense medium. \\

Tidal Stripping & Gas or stellar removal via gravitational interactions. \\

Strangulation & Quenching via halted gas supply. \\

Harassment & Cumulative impact of repeated high-speed encounters. \\\\

\hline
\end{tabular}
\caption{Quenching-specific terminology.}\label{t2a}
\end{table*}

In Fig.~\ref{f1a} we show a schematic rendering of a group of galaxies, highlighting several aspects of the terminology used in this review. The central galaxy is taken to be the most massive system within a dark matter halo, with satellite galaxies being any other group members. The interstellar medium (ISM) refers to the gas and dust present within the galaxy itself. Conversely, the circumgalactic medium (CGM) refers to the diffuse (often ionized) gas permeating the dark matter halo beyond the galaxy. If the galaxy in question is the central of a group or cluster, further terminology is frequently used. In particular, the medium permeating the group environment is often referred to as the intragroup medium (IgrM), and the region permeating the cluster is often referred to as the intracluster medium (ICM). In both of these cases, the media is expected to be filled with hot, diffuse gas, which may be emitting in X-rays. Outside of the halo is the intergalactic medium (IGM), which refers to the gas present between galaxies in deep space.

In addition to radiation from photons, galaxy formation models generally distinguish between two types of matter: (i) `baryons' and (ii) `dark matter'. In astronomy, the former term is usually used to refer to all matter particles of the standard model\footnote{Note that this is different to in physics, where the term specifically refers to composite subatomic particles made of three quarks (or three anti-quarks), such as protons and neutrons, that participates in the strong nuclear force. Hence, for practical purposes, the main difference is that electrons are also termed baryons in astronomy.}. Alternatively, dark matter particles are taken to be matter particles beyond the standard model, which do not interact with the electromagnetic field. For pedagogical treatments of these issues, see, e.g., \cite{Peacock1999, Dodelson2003, Mo2010}.

Finally, we note that throughout this review, halo masses, radii, and velocities are denoted by $M_{200c}$, $R_{200c}$, and $V_{200c}$ (respectively), which correspond to quantities measured within a radius enclosing an average density 200 times the {\it critical density} of the Universe, $\rho_{\rm crit}(z)$. We note that some authors instead adopt these parameters with a subscript-$m$, which is defined relative to the mean matter density, $\rho_m(z)~=~\Omega_m(z)~\rho_{\rm crit}(z)$.

\subsection{Observational Constraints}\label{s13}

Observationally, the quenching of central galaxies is found to be connected to morphology (e.g., \citealt{Driver2006, Cameron2009a, Cameron2009b, Wuyts2011, Omand2014}), stellar mass (e.g., \citealt{Baldry2006, Peng2010}), central mass density (e.g., \citealt{Cheung2012, Fang2013, Barro2017, Barro2013}), bulge mass (e.g., \citealt{Lang2014, Bluck2014}), and central velocity dispersion (e.g., \citealt{Bell2008, Wake2012, Bluck2016, Bluck2022}). See Sect.~4 in Part~I for more details on these observational results. 

Additionally, the quenching of (particularly low-mass) satellites is found to be connected to local galaxy density (e.g., \citealt{Butcher1978, Dressler1980, Baldry2006, Peng2012}), the mass of their parent group/ cluster halo (e.g., \citealt{Woo2013, Bluck2020a, Goubert2024, Goubert2025}), and the location at which they reside within their parent halo (e.g., \citealt{Dressler1980, Woo2013, Goubert2024, Goubert2025}). See Sect.~5 in Part~I of this review for further details. 

Ultimately, these correlations provide clues as to the underlying causal origin of galaxy quenching, but without detailed modeling and careful control for nuisance parameters it is challenging to bridge the epistemic divide between correlation and causation. For instance, population level galaxy statistics do not directly constrain the causes or mechanisms of galaxy quenching, as various modes could lead to similar stellar mass functions or quenched fraction correlations. Alternatively, specific observations of certain mechanisms in action (e.g., ejective feedback from quasar-mode AGN, or CGM heating from AGN radio jets) may establish a process is feasible, but not that it can explain the demographics of the quenched population as a whole. As such, careful comparison between observations on various scales and simulations must be made in order to achieve meaningful progress.

\begin{figure}[t!]  
\centering
\includegraphics[width=1\textwidth]{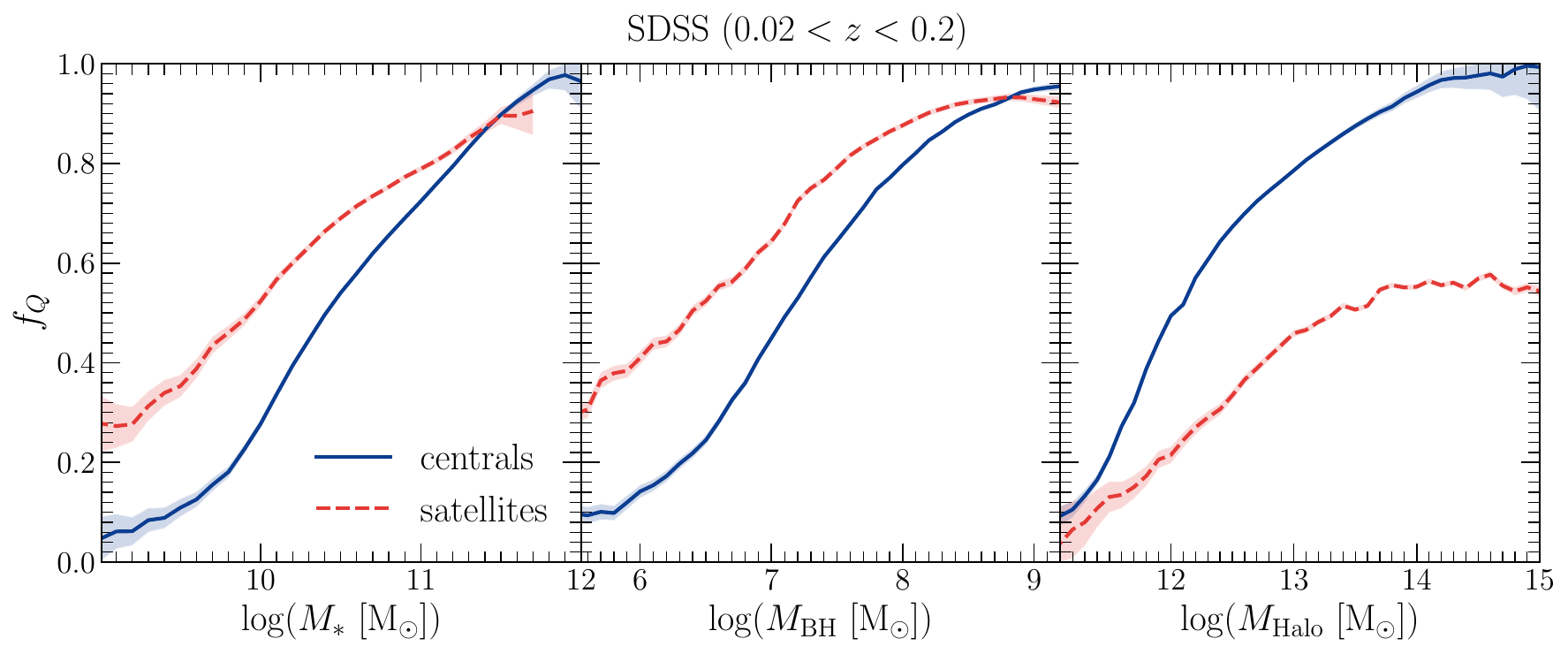}
\caption{Mass quenching vs. environment quenching. From left to right, the quenched fraction of SDSS galaxies ($0.02 < z < 0.2$) is plot as a function of stellar mass, SMBH mass, and dark matter halo mass. For centrals (shown in blue), there is a strong monotonic trend whereby systems with high masses are highly likely to be quenched, but systems with low masses are highly likely to be star forming. On the other hand, satellites are more likely to be quenched than centrals at fixed stellar and SMBH mass, but are more likely to be star forming than their centrals at a fixed halo mass. This figure highlights the observational need for at least two quenching pathways: one connected to the intrinsic properties of the galaxy (i.e., mass of various types) and one connected to the environment in which galaxies reside. This figure was produced by J. Piotrowska (private communication) based on data from \cite{Piotrowska2022}. }\label{f2a}
\end{figure}

As an illustrative example, in Fig.~\ref{f2a} we present the quenched fraction of centrals (shown in blue) and satellites (shown in red) from the SDSS (\citealt{York2000, Abazajian2009}) as a function of stellar mass (left panel), SMBH mass (center panel), and halo mass (right panel). This figure is constructed based upon data collated in \cite{Piotrowska2022}. Stellar masses are inferred from SED fitting performed in \cite{Mendel2013}; halo masses are inferred from abundance matching performed in \cite{Yang2007, Yang2009}; and SMBH masses are estimated via the $M_{\rm BH} - \sigma$ relation for all galaxy types presented in \cite{Saglia2016}.

For centrals, there is a strong positive correlation between quenched fraction and each of these inter-correlated masses (see also, \citealt{Baldry2006, Peng2012, Woo2013, Bluck2014, Bluck2016}). At high masses, central galaxies are invariably quenched, but at low masses, they are invariably star forming. This highlights the `mass-quenching' channel (\citealt{Peng2010, Peng2012}). However, it remains to be seen which (if any) of these masses is fundamentally connected to quenching.

On the other hand, satellite galaxies are much more frequently quenched than centals at low stellar and SMBH masses (see also, \citealt{Peng2012, Wetzel2012, Bluck2016, Bluck2020b, Goubert2024}). This clearly indicates the need for an additional quenching channel for satellites, likely a consequence of the dense environments in which they reside. Interestingly, satellites are less likely to be quenched within a fixed halo than their centrals. This is explained by centrals being (by definition) the most massive galaxies with their dark matter haloes. As such, this is ultimately a result of mass quenching operating within different environments. Put another way, by controlling for environment (in this case via halo mass), mass quenching is clearly still seen to operate: the most massive galaxy in a given environment is more likely to be quenched than less massive systems.

An obvious, but underappreciated, point is that not all galaxies need to quench. Indeed, the vast majority of low mass ($M_* \leq 10^{10.5} \, M_\odot$) galaxies are actively star forming at $z \sim 0$. High mass galaxies universally quench, with low mass systems only quenching if they are satellites within dense environments. On the other hand, the rate at which normal star forming systems form their stars is still likely controlled via feedback. This is a subtle point which connects to the fundamental distinction between star formation regulation and quenching. In low mass centrals, star formation may well be regulated (in practice slowed) by stellar and supernova feedback, but it cannot be halted altogether. Conversely, in high mass centrals (and many satellites across all masses), star formation must be halted (near) completely, and, moreover, for long cosmological timescales. These are very different requirements, which necessitate different theoretical causes and mechanisms.

In Part~I of this review, we also explore how quenching operates within galaxies (see Sect.~6), utilizing observations from spatially resolved spectroscopy. Central (and high-mass) galaxy quenching is found to operate `inside-out'. That is, the cores of galaxies reduce their star formation before the outskirts (e.g., \citealt{Tacchella2015, GonzalezDelgado2014, GonzalezDelgado2016, Belfiore2017, Belfiore2018, Ellison2018, Bluck2020b}). Conversely, satellite galaxy quenching (particularly at low masses) is found to operate `outside-in'. That is, the outskirts quench before the centers of these systems (e.g., \citealt{Schaefer2017, Belfiore2018, Bluck2020b}). 

This immediately points towards a quenching mechanism originating at the center of galaxies for centrals, but from the outskirts (or beyond) for satellites. These results consistently map onto the observational findings outlined above, i.e., that centrals quench as a consequence of internal (mass-correlating) conditions, but satellites quench as a function of the environments in which they reside.

Finally in Part~I, we discuss the evidence for massive quenched galaxies in the very early Universe from JWST observations, contrary to most theoretical expectations (e.g., \citealt{Carnall2023a, Carnall2023b, Carnall2024, Baker2025a, Zhang2025, Weibel2025}). Interestingly, evidence for both ejective and preventative AGN feedback is also found in the very early Universe, potentially offering a solution to this problem (e.g., \citealt{Marshall2023, Davies2024, D'Eugenio2024, Ubler2024, Roy2025, Wu2025}).

To make further progress one must tackle theory directly, in order to ascertain precisely which causes and mechanisms for quenching are viable. Moreover, one must also extract the key testable predictions from each theoretical quenching avenue to enable decisive constraints. This may then be combined with bespoke observational tests to finally reveal the origin of galaxy quenching as a function of mass and environment. Ultimately, this is the goal of this part of the review series.

\subsection{A brief overview of theoretical methods}\label{s14}

In this subsection we briefly summarize the principal theoretical and computational frameworks used to study galaxy formation and evolution, and ultimately quenching: N-body simulations, semi-analytic models (SAMs), and hydrodynamical simulations (in idealized, cosmological, and zoom-in modes). We begin with a very brief description of the background cosmology and the linear theory of structure formation, which underpin all contemporary approaches.

\subsubsection{Background cosmology \& linear structure formation}

In the standard cosmological paradigm, the large-scale Universe is assumed to be homogeneous and isotropic, leading to the Friedmann–Lemaitre-Robertson-Walker (FLRW) metric and the Friedmann solution to the Einstein field equations. In turn, this leads to a description of cosmic expansion, in which the evolution of the scale factor ($a \equiv 1/(1+z)$) is governed by the relative energy densities of matter, radiation, and dark energy. Within the currently favored $\Lambda$CDM framework, the Universe evolves from an early radiation-dominated phase, through a matter-dominated era in which gravitational instability drives the growth of density perturbations, and finally to a dark-energy-dominated phase of accelerated expansion in the late Universe. For pedagogical treatments see, e.g., \cite{Peacock1999, Dodelson2003, Hobson2006, Longair2006, Mo2010}.

In the linear regime, small overdensities grow approximately in proportion to the cosmological growth factor, while the density contrasts remain much smaller than unity ($\delta \equiv (\rho - \bar{\rho})/\bar{\rho} \ll 1 $). This phase of evolution is amenable to a direct analytical solution (see, e.g., \citealt{Peebles1980, Efstathiou1985, Frenk1988}). However, as perturbations become nonlinear, the simple analytical treatment breaks down. Thereafter, the collapse of matter proceeds hierarchically into virialized dark matter haloes (e.g., \citealt{White1978, White1991, Navarro1996, Navarro1997}), which later host galaxies, groups, and clusters. In practice, essentially all cosmological simulations impose the background expansion analytically, while numerically evolving the nonlinear growth of structure within this expanding spacetime.

\subsubsection{N-body dark matter simulations}

N-body simulations follow the collisionless evolution of cold dark matter via the Boltzmann–Poisson system (e.g., \citealt{Peebles1980, Efstathiou1985, Frenk1988, Springel2005a}). In comoving coordinates the equation of motion and field equation are given, respectively, by\footnote{If you would like to derive the equation of motion for yourself, start by taking the double temporal derivative of the physical position in comoving coordinates; set the physical acceleration equal to the Newtonian gravitational acceleration; then subtract the homogeneous background contribution. Similarly, for the Poisson equation, convert the Laplacian to comoving coordinates and subtract the background contribution from both sides.}:

\begin{equation}
\ddot{\mathbf{x}} + 2H\dot{\mathbf{x}} = -\frac{1}{a^2}\nabla \phi,
\end{equation}

\begin{equation}
\nabla^2 \phi = 4\pi G a^2 \bar{\rho}\,\delta,
\end{equation}

\noindent \noindent where $\mathbf{x}$ is the comoving position, $H$ is the Hubble parameter (defined as: $H \equiv \dot{a}/a$), $\nabla$ is taken with respect to comoving coordinates, and $\phi$ is the peculiar gravitational potential sourced by density fluctuations.

Initial conditions are drawn from the primordial power spectrum (e.g., \citealt{Zeldovich1970, Peebles1980, Efstathiou1985}). These simulations reproduce large-scale structure and halo properties (e.g., \citealt{Frenk1988, Navarro1996, Moore1999a, Navarro2004, Springel2005a, Boylan2009}) but do not include baryonic physics, which is crucial for understanding star formation and quenching.

\subsubsection{Semi-analytic models (SAMs)}

SAMs populate halo merger trees from N-body simulations with baryonic processes parameterized via simplified prescriptions (e.g., \citealt{White1991, Croton2006, Bower2006, Bower2008, Henriques2015, Somerville2015}). These include gas cooling, star formation, stellar feedback, and supermassive black hole growth and feedback. This enables connection between the fundamental theory of structure formation and observations of galaxies.

Model parameters are calibrated against key observables, including stellar mass functions and the cosmic star formation rate density evolution. While this introduces tuning, SAMs remain powerful tools for identifying missing physics. In particular, they demonstrated the necessity of AGN feedback for quenching massive galaxies from a theoretical perspective (see Sect. 2.3.3; \citealt{Croton2006, Bower2006, Bower2008}).

\subsubsection{Hydrodynamical simulations}

Hydrodynamical simulations model the gravitational interaction of dark matter and baryons, in addition to the hydrodynamics of baryons. In comoving form the fluid dynamics equation set is given by (e.g., \citealt{Dodelson2003, Longair2006, Mo2010, Vogelsberger2020}):\\

\noindent \textbf{Mass continuity:}
\begin{equation}\label{emass}
\frac{\partial \rho}{\partial t} + \frac{1}{a}\nabla \cdot (\rho \mathbf{v}) + 3H\rho = 0
\end{equation}\\

\noindent \textbf{Momentum (Euler equation):}
\begin{equation}\label{emom}
\frac{\partial \mathbf{v}}{\partial t} + \frac{1}{a}(\mathbf{v}\cdot\nabla)\mathbf{v} + H\mathbf{v}
= -\frac{1}{a\rho}\nabla P - \frac{1}{a}\nabla \phi
\end{equation}\\

\noindent \textbf{Energy equation:}
\begin{equation}\label{eenergy}
\frac{\partial u}{\partial t}
+ \frac{1}{a}\mathbf{v}\cdot\nabla u
+ \frac{P}{\rho}\left(\frac{1}{a}\nabla\cdot\mathbf{v} + 3H\right)
= \mathcal{H}_{\rm ext} - \mathcal{C}_{\rm rad}.
\end{equation}\\

\noindent \textbf{Poisson equation:}
\begin{equation}\label{eforce}
\nabla^2 \phi = 4\pi G a^2 \left(\rho_{\rm tot} - \bar{\rho}_{\rm tot}\right),
\end{equation}\\

\noindent\textbf{Equation of State:}
\begin{equation}\label{eos}
P = (\gamma - 1)\rho u, \,\,\,\, {\rm where,} \,\, \gamma = 5/3 \,\, {\rm (monatomic \,\, gas)}
\end{equation}

\noindent where $\nabla$ is taken with respect to comoving coordinates, $\rho$ is the physical gas density, $\rho_{\rm tot}$ is the total physical matter density including dark matter and baryons, $\mathbf{v}$ is the physical peculiar velocity, $P$ is the pressure, $u$ is the internal energy per unit mass, $\phi$ is the peculiar gravitational potential, $\mathcal{H}_{\rm ext}$ and $\mathcal{C}_{\rm rad}$ are the specific heating and cooling rates, respectively (i.e., energy per unit mass per unit time).

It is worth taking a moment to appreciate what these equations really mean. Briefly, eq.~\ref{emass} ensures mass conservation; eq.~\ref{emom} expresses momentum conservation for a fluid element subject to pressure gradients, gravity, and Hubble expansion; eq.~\ref{eenergy} evolves the specific internal energy of the gas, including adiabatic compression/expansion and source/sink terms from heating and cooling; eq.~\ref{eforce} is the standard gravitational field equation for a perturbation in an expanding background; and, finally, eq.~\ref{eos} closes the partial differential equation system by linking pressure to density and internal energy (in this case by assuming the fluid is an ideal gas of monatomic particles). 

These equations are solved numerically using particle-based, grid-based, or moving-mesh methods (see, e.g., \citealt{Teyssier2002, Springel2005b, Springel2010, Hopkins2018, Vogelsberger2020}). 

Throughout this review, the term `hydrodynamical simulation' is used in the sense commonly adopted within the galaxy formation community, referring specifically to cosmological simulations that evolve both dark matter and baryons, while also incorporating the key physical processes required for galaxy formation. More generally, the term `hydrodynamical simulation' may also refer to approaches which solve the equations governing fluid dynamics alone, i.e., not including any of the additional baryonic processes required to trace galaxy formation and evolution.

In a galaxy-forming hydrodynamical simulations, sub-grid models are still required for unresolved processes such as star formation, stellar evolution, SMBH formation and accretion, SN- and AGN-feedback, chemical enrichment, and radiative cooling (e.g., \citealt{Torrey2014, Weinberger2017, Nelson2019}). These remain qualitatively similar to their analogous prescriptions in SAMs, but with one key difference: the outputs of these prescriptions directly impact the broader simulation, which in turn impacts the conditions for the sub-grid recipes.

There are three types of galaxy-forming hydrodynamical simulations that are relevant to the field of galaxy quenching, which we briefly discuss in order of complexity:\\

\noindent {\bf 1) Idealized hydrodynamical simulations:}\\
Early hydrodynamical simulations modeled isolated systems (e.g., single galaxies, galaxy mergers, and galaxy groups/ clusters; e.g., \citealt{Navarro1993, Hopkins2006, Moreno2015}). These provide key insights into gas dynamics and feedback but lack cosmological context, both spatially and temporally. \\

\noindent {\bf 2) Cosmological hydrodynamical simulations:}\\
Modern simulations (e.g., Illustris, EAGLE, IllustrisTNG and Simba; \citealt{Vogelsberger2014a, Vogelsberger2014b, Schaye2015, Nelson2018, Dave2019}) evolve large cosmological volumes across the vast majority of cosmic history, utilizing typical box sizes of $\sim$(100 Mpc)$^3$ from $z \sim 100$ to the present. These simulations form and evolve galaxies, groups, and clusters from initial conditions within the full cosmic environment.\\

\noindent {\bf 3) Zoom-in hydrodynamical simulations:}\\
Zoom-in simulations re-simulate selected regions of cosmological simulations at higher resolution (up to $\sim$pc scales; e.g., \citealt{Grand2017, Hopkins2018}). These enable detailed studies of feedback and ISM/CGM physics, and increasingly aim to model sub-grid processes from first principles.

\subsubsection{Summary of theoretical approaches}

Analytical theory provides the foundation for the background cosmology and for structure formation, but it fails in the non-linear regime. N-body simulations capture the dark matter backbone of structure formation, but exclude baryonic physics. SAMs provide an approximate bridge between N-body simulations and observations by forming and evolving galaxies within the N-body merger tree structure via relatively simple prescriptions tuned to key observations. 

Hydrodynamical simulations offer the most complete physical modeling, coupling gravitation of dark matter and baryons with the hydrodynamics of baryons, within an expanding background cosmology. There are three primary types of hydrodynamical simulation: (i) idealized, which are typically small volume and decoupled from the wider Universe; (ii) cosmological, which are typically large volume and account for cosmological evolution; and (iii) zoom-in, which re-simulate regions within cosmological simulations to higher resolution. Taken together, these theoretical approaches offer a powerful route for implementing the physics of galaxy formation and evolution, constructing predictions which can be tested with observations. In terms of quenching, each of these theoretical paradigms have led to important insights on the regulation and ultimate cessation of star formation within central and satellite galaxies (as will be discussed in detail in Sects.~\ref{s2} \& \ref{s3}).

\subsection{Scope of the review}\label{s15}

The goal of this review is to be expansive, but not exhaustive. As such, it is important to specify the scope clearly here. During this review we focus primarily on galaxies in the mass range: $10^9 \leq (M_*/M_\odot) \leq 10^{12}$. In practice, this means we are complete to high masses in our discussion, but not to low masses. In particular, this review does not discuss the quenching of low-mass dwarf galaxies. Since mass-quenching occurs at $M_* \gtrsim 10^{10.5}\,M_{\odot}$, this range is perfectly sufficient for a complete consideration. On the other hand, we miss the evolution of low-mass systems (like the satellites of the Milky Way), which may environmentally quench. 

Partly as a consequence of our stellar mass range, we also have an effective halo mass range of: $10^{11} \leq (M_{\rm Halo}/M_\odot) \leq 10^{15}$. Again, we are complete to high masses, but not low masses. Consequently, for environmental quenching we focus primarily on the quenching of satellites within relatively high mass groups, in addition to a full consideration of the cluster environment. This means we neglect satellite quenching in low mass groups.

This part of the review is also primarily focused on simulation results and direct observational tests at: $0 \leq z \leq 3$. In part I we also discuss observations at much higher redshifts ($z = 4 - 10$; see Sect. 7). However, in this part of the review we seek to make deep contact between theory and observations, where by far the most relevant data is at low-to-intermediate redshifts.

Ultimately, these restrictions on scope are mostly inherited from the limitations of wide-field galaxy surveys, in addition to the resolution limits within cosmological hydrodynamical simulations. On the other hand, in terms of quenching causes and mechanisms, we aim for completeness, regardless of their current favorability within the literature. That is, all quenching routes that have been widely discussed in the literature will be at least briefly assessed here. Finally, we will evaluate their relative success against observations in later sections.

\subsection{Outline of Part~II}\label{s16}

In Sect.~\ref{s2}, we review the theory of intrinsic galaxy quenching. In particular, we discuss star formation, supernova feedback, the formation of stable virial shocks in high mass haloes, AGN feedback in the high-Eddington ratio `quasar-mode', AGN feedback in the low-Eddington ratio `radio-mode', and dynamical stabilization processes. We also provide a detailed overview of how these processes are modeled within contemporary cosmological hydrodynamical simulations.

In Sect.~\ref{s3}, we review the theory of environmental galaxy quenching. In particular, we discuss galaxy mergers and the formation of groups and clusters, ram pressure stripping (at both high and low densities), dynamical stripping (from both satellite - central and satellite - satellite interactions), strangulation, and pre-processing. Additionally, we discuss the fascinating evidence for AGN-feedback modified environmental quenching within the cluster environment.

In Sect.~\ref{s4}, we review observational tests of the theoretically proposed causes of quenching in central galaxies. Specifically, we discuss observational evidence for ejective feedback from supernovae, ejective feedback from AGN, and preventative AGN feedback from relativistic jets. We also explore where AGN reside on the star forming main sequence, concluding that this is not a productive route to testing the AGN feedback paradigm. Alternatively, we argue for the value of using supermassive black hole mass as a useful estimate of the total energy injected via AGN feedback over the lifetime of a galaxy. At the end of this section, we review several machine learning tests of the AGN feedback paradigm.

In Sect.~\ref{s5}, we review observational tests of the theoretically proposed causes of quenching in satellite galaxies. Specifically, we review evidence for satellites quenching at a higher rate at fixed intrinsic properties than centrals. We go on to discuss direct observational evidence for both ram pressure and dynamical stripping within the group and cluster environments. Finally in this section, we review several machine learning tests of the environmental quenching paradigm.

In Sect.~\ref{s6}, we move on to tests of the specific mechanisms of galaxy quenching. In the first part of this section, we explore how the molecular and atomic gas content in the interstellar medium (ISM) changes during quenching. We also explore how the efficiency of star formation varies during the quenching sequence. In the second part of this section, we review tests utilizing stellar metallicity measurements on whether quenching operates via ISM ejection or via starvation (i.e., the prevention of gas replenishment into the system).

In Sect.~\ref{s7}, we explore the possible triggers of galaxy quenching. We start by reviewing various observational constraints on when galaxies quench and how long quenching takes. We go on to explore various strategies for `catching quenching in action' via observing green valley and post-starburst galaxies.

In Sect.~\ref{s8}, we summarize this part of the review series. Additionally, we attempt to answer the questions of how and why quenching occurs, separating this into intrinsic and environmental modes. Finally, we present a brief list of known unknowns to aid further research in this field.


\section{Quenching in theory and simulations I: Feedback}\label{s2}

In this section we review theoretically proposed intrinsic mechanisms for quenching central galaxies. We begin with a general discussion of the need for feedback processes within galaxy formation models, in order to accurately reproduce the galaxy stellar mass function. We also discuss the need for maintenance-mode feedback to achieve long term quiescence. We then explore stellar and supernova feedback, concluding that these processes are likely insufficient to explain massive central galaxy quenching. We go on to consider in detail three important theoretical paradigms: (i)~virial shocks and halo mass quenching; (ii)~quasar-mode AGN feedback; and (iii)~radio-mode AGN feedback. We then review the implementation of these physical processes within contemporary cosmological simulations. Finally, we discuss the potential role of dynamical stabilization for quenching galaxies with dense cores, independent of explicit feedback.

\subsection{The need for feedback processes in galaxy evolution}\label{s21}

\noindent In addition to cosmological expansion, self-gravitation, and fluid dynamics (which are typically modeled directly in contemporary simulations; see Sect.~\ref{s14} and, e.g., \citealt{Vogelsberger2020}), cosmological models must employ sub-grid recipes for baryonic heating and cooling (e.g., \citealt{Somerville2015}). Important cooling channels include: collisional excitation and ionization, recombination, bremsstrahlung (free–free) emission, inverse Compton scattering, and atomic line cooling (with metal-line cooling included where metallicity is tracked). Molecular cooling is generally approximated through density (and, potentially, metallicity) based criteria, rather than treated explicitly. 

Ultimately, it is the collisional nature of baryonic fluids which enables extraction of energy from the gravitational field as heat, subsequently lost through the various radiative processes outlined above. This allows baryons to progressively reduce their random motions and condense within galaxies, eventually reaching high enough densities for the formation of giant molecular clouds (GMCs). In turn, GMCs may finally collapse under their own self-gravity to form stars (e.g., \citealt{Jeans1902, Kennicutt1998, Larson1981}). See \cite{McKee2007} and \cite{Somerville2015} for particularly clear reviews of these processes within simulations.

In most simulations, star formation is modeled via a volumetric Kennicutt--Schmidt law \citep{Kennicutt1998}, parameterized as:

\begin{equation}
\dot{\rho}_\star = \mathcal{E}_\star \, \frac{\rho_\mathrm{g}}{t_\mathrm{ff}}, \qquad \text{where} \qquad
t_\mathrm{ff} = \sqrt{\frac{3\pi}{32\,G\,\rho_\mathrm{g}}},
\end{equation}

\noindent and where, $\dot{\rho}_*$ is the star formation rate per unit volume, $\rho_\mathrm{g}$ is the local gas density, and $\mathcal{E}_\star \,\, (\!\sim\! 0.01$) denotes the fraction of gas converted into stars per free-fall time ($t_\mathrm{ff}$, defined on the RHS above, see \citealt{Krumholz2012}). The corresponding stellar mass growth rate is $\dot{M}_\star = (1-R)\dot{M}_\mathrm{formed}$, where $R$ is the time-dependent return fraction (accounting for stellar mass loss via stellar evolution). Crucially, only gas satisfying a specific star formation criteria may form stars. Most frequently, a minimum density threshold is applied (i.e., $n_\mathrm{H} \approx 0.1 \, \mathrm{cm^{-3}}$), which is tuned to match observational scaling relations (e.g., \citealt{Kennicutt1998b}). Alternatively, a more sophisticated metallicity-dependent variant may be applied, following the Jeans stability condition for a pressurized fluid (e.g., \citealt{Schaye2015}). 

Note that although the microscopic efficiency of star formation is very low, owing to turbulence and magnetic support in molecular clouds, cosmological simulations absorb these effects into the sub-grid parameter ($\mathcal{E}_\star$). More stringent constraints on the simulations emerge from the global efficiency of baryon-to-stellar conversion within a halo, i.e., $\epsilon_* \equiv M_*/f_bM_H$ (see, e.g., \citealt{White1991, Kauffmann1993, Cole2000, Behroozi2013, Moster2013, Madau2014}).

Essentially all prescriptions which follow the above cooling and star formation procedures lead to serious tensions with observations. Specifically, they predict that the vast majority of all baryons will reside in stars by the present epoch, whereas in reality less than 10\% of baryons reside in stars by $z = 0$ (e.g., \citealt{White1991, Kauffmann1993, Cole2000, Baugh2006, Bower2006, Bower2008, Fukugita2004, Shull2012, Somerville2015}). Moreover, simple feedback-free models also predict that the stellar mass function of galaxies should trace the (baryon fraction scaled) halo mass function, which they do not (e.g., \citealt{Bell2003, Peng2010, Baldry2012}), and cannot explain the peak in the stellar mass \,--\,halo mass ratio at $M_H \sim 10^{12} M_{\odot}$ (e.g., \citealt{Moster2010, Behroozi2010, Moster2013, Behroozi2013}). See further discussion in Part~I of this review (especially Sects.~1.2 and 3.2).

\begin{figure}[ht]  
\centering
\includegraphics[width=0.64\textwidth]{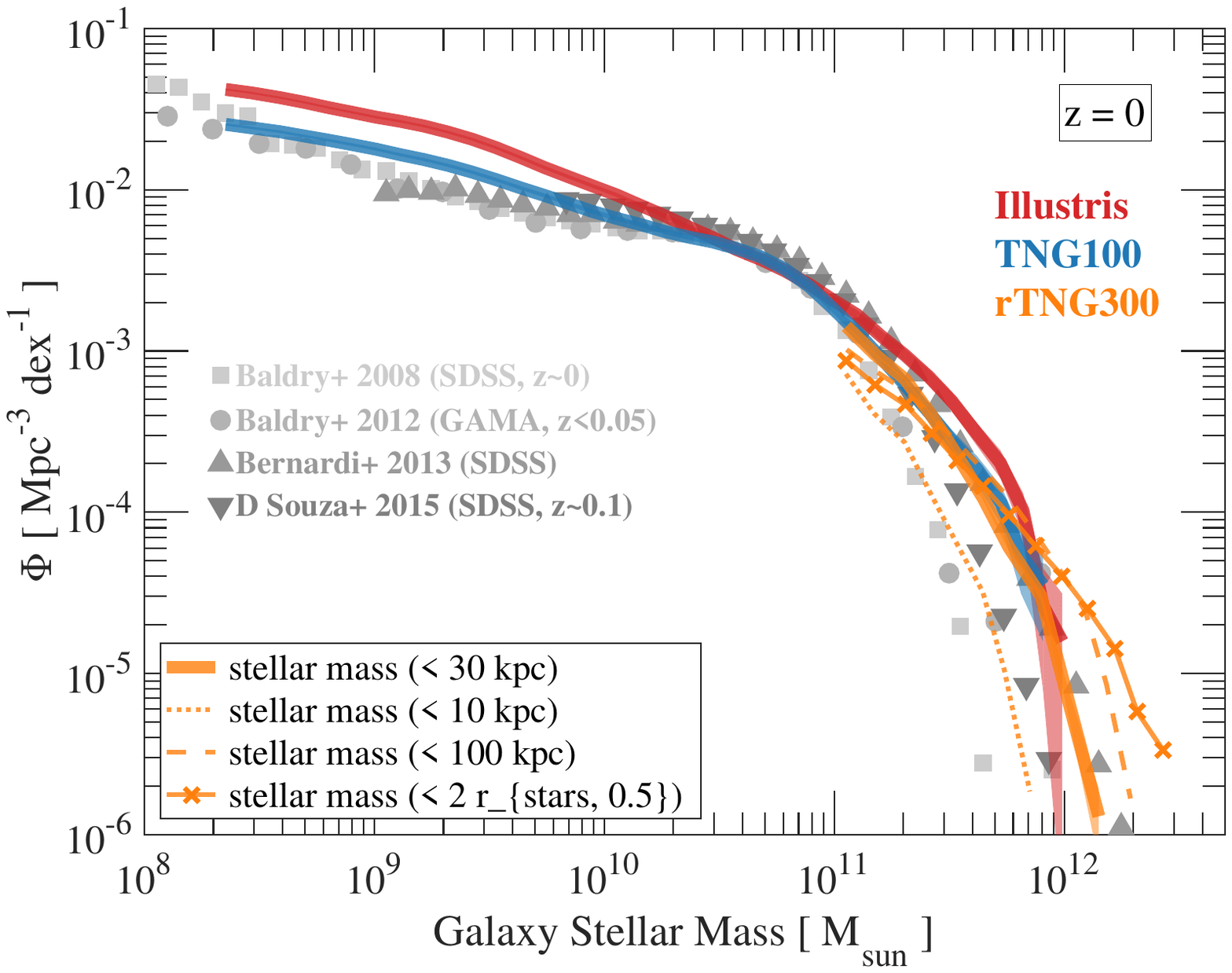}
\includegraphics[width=0.66\textwidth]{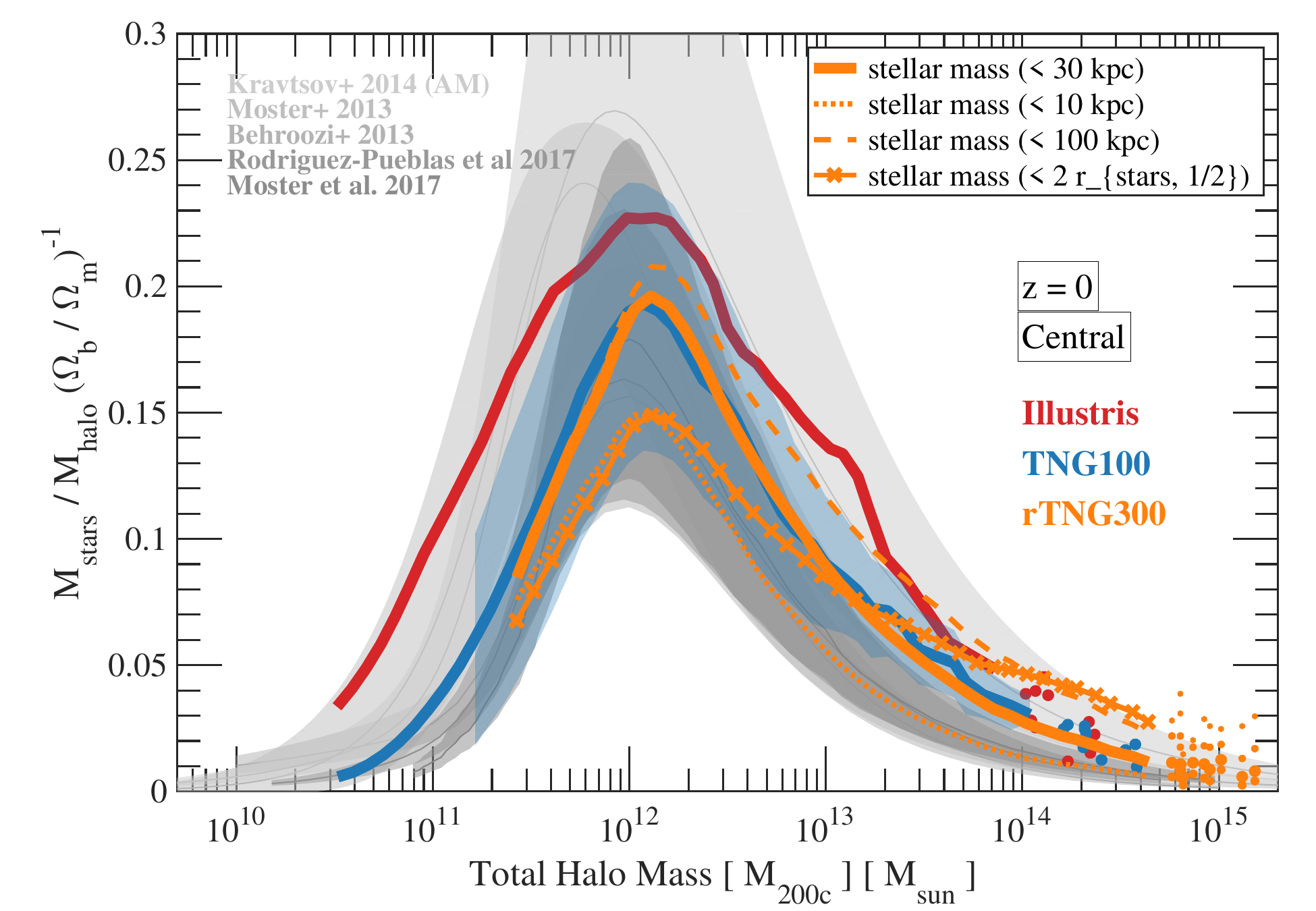}
\caption{{\it Top Panel:} Comparison of $z=0$ stellar mass functions from the Illustris and IllustrisTNG simulations to observational data from various sources (see legend). The simulations reproduce the observations reasonably well, achieving the characteristic power-law decline at low masses, followed by exponential fall off at high masses. Note that this high-mass cutoff is essentially mass quenching. {\it Bottom panel:} The halo star formation efficiency ($\epsilon_* \equiv M_* / f_b M_H$, where $f_b = \Omega_b/\Omega_M$) plot as a function of halo mass for the same simulations, compared to observational constraints (grey shaded regions). Again, the simulations achieve the observed shape of the relationship to a good approximation. In the simulations, the reduction in star formation efficiency at low masses is governed by supernova feedback; whereas, at high-masses, the reduction in star formation efficiency is governed by AGN feedback (in various modes). Both panels are reproduced from \cite{Pillepich2018}. }\label{f1}
\end{figure}

To combat these serious discrepancies with observations, theorists began to incorporate strong baryonic feedback processes which can: (i)~reduce the rate at which stars are formed within galaxies at low-to-intermediate masses, and (ii)~prevent star formation altogether in the highest mass systems. Note that only the latter is truly connected to quenching, although both aspects will be discussed here. Most generally, baryonic feedback may be defined as any astrophysical process arising from non-gravitational physics, which engenders a non-linear back-reaction on gas cooling and star formation. Feedback may in principle be either positive (enhancing star formation rates) or negative (reducing star formation rates). However, considering the problems outlined above, it is the latter which receives the most attention. Nonetheless, it is important to appreciate that feedback can lead to enhanced star formation in some cases (see, e.g., \citealt{Ciotti2007}).

Much emphasis has been placed on two fundamental types of feedback --- (i)~stellar and supernova feedback (i.e., originating from the unresolved nuclear physics of stars and stellar explosions), and (ii)~active galactic nuclei feedback (i.e., originating from the unresolved relativistic accretion of baryons onto supermassive black holes). 

The former is often broken down into feedback from supernovae Type II (core collapse), supernovae Type Ia (thermonuclear), and feedback from stellar winds, primarily in asymptotic giant branch stars (see, e.g., \citealt{Matteucci1986, Matteucci2006, Cole2000, Guo2011, Torrey2014, Henriques2015, Crain2015, Conroy2015, Dave2019}). 

The latter is often split into high-Eddington ratio `quasar-mode' feedback, which leads to galactic-scale winds and ISM ejection (e.g., \citealt{DiMatteo2005, Springel2005c, Hopkins2006, Hopkins2008, Crain2015, Maiolino2012, Cicone2012, Cicone2014}), and low-Eddington-ratio `radio-mode' feedback, which leads to CGM heating via relativistic jets (e.g., \citealt{Croton2006, Bower2006, Sijacki2007, Bower2008, Henriques2015, Fabian2012, Weinberger2017, Weinberger2018, Zinger2020}). 

We will consider all of these variants of feedback further in the subsequent parts of this section. Additionally, some theorists have incorporated feedback from cosmic rays, magnetic fields, the CMB, and even the possibility of dark matter annihilation into models of galaxy evolution (see \citealt{Vogelsberger2020} and references therein). However, these additional processes are not currently thought to be critical for star formation regulation and/or quenching.

In Fig.~\ref{f1} (top panel) we present the $z = 0$ stellar mass functions from the Illustris (\citealt{Vogelsberger2014a, Vogelsberger2014b, Torrey2014}) and IllustrisTNG  (\citealt{Pillepich2018, Nelson2018, Marinacci2018, Springel2018, Naiman2018}) simulations compared to observational constraints, reproduced from \cite{Pillepich2018}. In the bottom panel, we present the halo star formation efficiency ($\epsilon_*$) from the same simulation suites, compared to various observational constraints, also reproduced from \cite{Pillepich2018}. In both cases, the simulations recover the general observational trends to a good approximation, unlike in previous generations of feedback-free models.

In both simulations, this feat is achieved by supernovae (primarily Type~II) feedback at low masses (which strongly reduces the rate of star formation relative to feedback-free expectations), and via AGN feedback at high masses (which completely shuts down star formation within the highest mass haloes). The transition between these two feedback regimes leads to the characteristic peak in the halo star formation efficiency, at $M_{H} \sim 10^{12} M_{\odot}$ (as observed in, e.g., \citealt{Moster2010, Moster2013, Behroozi2010, Behroozi2013}). It is important to appreciate that low mass galaxies are not quenched by supernova feedback. The rate of star formation is reduced relative to that of a feedback-free model but, crucially, not halted altogether. On the other hand, the high-mass systems are quenched by AGN feedback, resulting in little-to-no star formation for many Gyr. This is necessary to be qualitatively in agreement with established observations (e.g., \citealt{Baldry2006, Peng2010, Peng2012}).

The details on the feedback used in Illustris and IllustrisTNG differ significantly with respect to AGN (see Sect.~~\ref{s25}). However, the ultimate cause of the deviation of the stellar mass functions from the underlying halo mass functions is identical between these two simulations. Moreover, in essentially all contemporary cosmological simulations, a similar use of supernova and AGN feedback is applied to achieve these fundamental properties of the galaxy population (see, e.g., \citealt{Bower2006, Bower2008, Vogelsberger2014a, Vogelsberger2014b, Somerville2015, Henriques2015, Schaye2015, Crain2015, Nelson2018, Pillepich2018, Dave2019}). Consequently, we will consider these feedback channels in some detail later in this section.

\subsection{Stellar mass loss and the maintenance of quenching}
\label{sec:stellar_quenching}

It is important to appreciate at the outset that it is not enough to remove gas from a galaxy to quench the system. Gas may be replenished by cold gas streams and CGM cooling flows. Furthermore, a critical, and often under-emphasized, aspect of galaxy quenching is that the cessation of cosmological gas accretion and the removal of the ISM does not, in itself, guarantee long-term quiescence. This is the case because, even in fully quenched/ passively evolving systems, stellar evolution provides a continuous and unavoidable source of gas injection into the ISM. This `stellar recycling' fundamentally alters the requirements for maintaining quenching over cosmological timescales. Hence, the physical processes responsible for initially triggering quenching need not be identical to those responsible for maintaining quiescence over cosmological timescales.

For an evolved stellar population, the mass return rate may be approximated as (e.g., \citealt{Ciotti1991, Mathews2003, Leitner2011}):

\begin{equation}
\dot{M}_* (t)  \simeq 1.5 \times 10^{-11} \bigg( \frac{t}{13\,\mathrm{Gyr}} \bigg)^{-1.3} \bigg( \frac{M_\star}{\mathrm{M_\odot}} \bigg)  \,\, \mathrm{M_\odot \,\, yr^{-1}} \, ,
\end{equation}

\noindent where, $M_*$ is the stellar mass and $t$ is the population age. For a typical massive early-type galaxy with $M_\star \sim 10^{11}~\mathrm{M_\odot}$, this yields:

\begin{equation}
\dot{M}_* \sim 1\text{--}2 \,\, \mathrm{M_\odot \,\, yr^{-1}} \,\,\,\, \text{at } z \sim 0,
\end{equation}

\noindent where the normalization is approximate and depends on the assumed IMF, stellar population model, and whether the scaling is expressed per unit stellar mass or optical luminosity. Furthermore, even higher rates are expected at earlier cosmic times. Integrated over a Gyr, this corresponds to $\sim 10^9~\mathrm{M_\odot}$ of recycled gas, comparable to the cold gas reservoirs of many star-forming galaxies.

This continuous injection implies that quenching is not a one-time event but, rather, it is a long-term process. Explicitly, quenching is ultimately a maintenance problem. Even if the ISM of a galaxy is initially removed (e.g., via ejective feedback or environmental stripping; see Sects.~\ref{s232} \& \ref{s3}) and gas inflows into the system are prevented (e.g., via radio-mode feedback or the removal of the CGM via stripping; see Sect.~\ref{s233} \& \ref{s3}), the evolution of the stellar population will rebuild a gaseous atmosphere. In the absence of efficient removal or heating, this gas accumulates, increasing the density of the hot ISM and correspondingly reducing its cooling time (e.g., \citealt{Binney2008}):

\begin{equation}
t_{\mathrm{cool}} = \frac{3 n k_B T}{2 n_e n_i \Lambda(T,Z)} \simeq \frac{3 k_B T}{n_e \Lambda(T,Z)} ,
\end{equation}

\noindent where, $T$ is the gas temperature, $n$ is the total particle number density, $n_e$ is the electron density, $n_i$ is the ion density, and $\Lambda(T,Z)$ is the cooling function. As gas density rises due to stellar mass return, $t_{\mathrm{cool}}$ decreases, inevitably driving the system toward a cooling instability. The above argument is a simplified overview of the classical cooling flow paradigm (see \citealt{Cowie1977} for full details). 

Ultimately, the hot atmospheres in elliptical galaxies and clusters undergo radiative cooling, with mass deposition rates inferred from their X-ray luminosities. However, the predicted accumulation of cold gas and star formation is not observed, nor are central supermassive black holes (SMBHs) sufficiently massive to account for the integrated mass return. This discrepancy constitutes the long-standing cooling flow problem (see the textbook by \citealt{Kim2012} for a detailed treatment of this critical issue).

The inclusion of stellar mass loss significantly augments the problem of galaxy quenching. Even in the absence of external accretion, galaxies are internally driven toward cooling flows and rejuvenation of star formation. Hence, a viable quenching mechanism must not only suppress initial star formation and prevent gas inflows into the ISM, but also continuously offset the radiative losses of an ever-replenished ISM through stellar evolution.

In lower-mass systems, heating from SN Ia may be sufficient to offset cooling and even drive galactic-scale outflows (see Sect.~\ref{s221}). To model this, the energy injection rate may be described by:

\begin{equation}
\dot{E}_{\mathrm{SN \, Ia}} = R_{\mathrm{SN \, Ia}} \, E_{\mathrm{SN}},
\end{equation}

\noindent where $R_{\mathrm{SNIa}}$ is the supernova rate and $E_{\mathrm{SN}} \sim 10^{51}\,\mathrm{erg}$ is the energy per event. Hydrodynamical models (e.g., \citealt{Ciotti1991}) demonstrate that in galaxies below a critical potential well depth, this heating can maintain a global wind, preventing gas accumulation.

However, in massive ellipticals, the deeper gravitational potential suppresses outflows, and SN Ia heating alone is likely insufficient to prevent the buildup of a dense, X-ray emitting atmosphere. In this regime, the system evolves toward a quasi-hydrostatic halo, which is prone to central cooling instabilities. Offsetting cooling from the CGM is discussed at length in Sect.~\ref{s23} \& \ref{s24}.

The continuous injection of stellar mass loss, combined with insufficient steady heating, leads naturally to a cyclic feedback scenario. As gas accumulates and the central cooling time drops below a critical threshold, a cooling flow ensues, triggering rapid inflow toward the SMBH and a subsequent episode of AGN feedback. This feedback reheats or ejects gas, temporarily restoring the hot atmosphere.

High-resolution hydrodynamical simulations (e.g., \citealt{Ciotti2007, Gaspari2013}) consistently find such limit-cycle behavior, characterized by alternating phases of quiescence and AGN outbursts. Importantly, the duty cycle evolves with time. As the stellar population ages and mass return rates decline, progressively longer intervals are required to rebuild the central gas density to the instability threshold.

A key consequence of this framework is the mismatch between the cumulative mass returned by stellar evolution and the observed masses of SMBHs. Over a Hubble time, stellar mass loss can supply a mass budget exceeding the present-day SMBH mass by one-to-two orders of magnitude. Yet, most of this material is neither converted into stars nor accreted onto the black hole.

Ultimately, this implies that efficient feedback must act not only to prevent star formation but also to regulate SMBH growth. In massive galaxies, AGN feedback is therefore required even in the absence of cosmological inflows or mergers, reinforcing its central role in maintaining quiescence. Moreover, as we will see in later subsections, AGN feedback is also critical for stabilizing halo-scale cooling flows around massive galaxies, groups, and (especially) clusters. Therefore, maintenance and prevention is at least as important as triggering quenching. AGN feedback has therefore become a near-ubiquitous component of contemporary theoretical models for maintaining long-term quiescence in massive galaxies (e.g., \citealt{Croton2006, Bower2006, Bower2008, Vogelsberger2014a, Vogelsberger2014b, Torrey2014, Crain2015, Schaye2015, Weinberger2017, Weinberger2018, Nelson2018, Pillepich2018, Dave2019}). In the next parts of this section we will carefully build up to this conclusion.

\subsection{Can star formation be self-quenching? The role of supernovae}\label{s22}

The processes of star formation, stellar evolution, and stellar death deposit vast quantities of energy into the ISM, especially for the most massive stars (particularly the O- and B-types). Consequently, it is natural to want to understand how this impacts future star formation and, indeed, to question whether star formation may be ultimately self-quenching. The observational fact that the highest stellar mass galaxies are most frequently quenched (see back to Fig.~\ref{f2a}) lends some initial credence to this idea, since these are the systems which must have experienced the greatest levels of stellar and supernova feedback (e.g., \citealt{Baldry2006, Peng2010, Peng2012}). 

In semi-analytic models (SAMs), the dominant form of stellar feedback arises from core-collapse (Type~II) supernovae. Following the classical formulation of \citet{White1991} and \citet{Kauffmann1993}, this is implemented as a galactic wind launched from the star forming disc, such that:

\begin{equation}
\frac{dM_\mathrm{ej}(t)}{dt} = \mathcal{E}_\mathrm{SN} \left( \frac{V_0}{V_c} \right)^{\alpha_\mathrm{SN}} \mathrm{SFR}(t),
\end{equation}

\noindent where $M_\mathrm{ej}$ is the mass of gas ejected from the disc, $V_c$ is the circular velocity of the galaxy, and $\mathcal{E}_\mathrm{SN}$ is the feedback efficiency. The exponent, $\alpha_\mathrm{SN}$, encodes the driving mechanism, where $\alpha_\mathrm{SN}=1$ corresponds to a momentum-driven wind, and $\alpha_\mathrm{SN}=2$ corresponds to an energy-driven wind. This prescription reflects the empirical and theoretical expectation that supernovae couple more efficiently to the interstellar medium in shallower gravitational potential wells. The efficiency, $\mathcal{E}_\mathrm{SN}$, is treated as a free parameter, tuned to reproduce the observed low-mass end of the galaxy stellar mass function (at $M_* \lesssim 10^{10.5} M_\odot$).

Ultimately, SN feedback functions by blowing out gas proportionally to the star formation rate. Hence, the faster a galaxy forms stars the stronger the feedback from supernovae becomes. This acts as an effective break on run-away star formation. In SAMs, the gas leaving the ISM may be either stored in the hot gas halo, or else deposited in an external reservoir, from which delayed re-accretion is possible (see, e.g., \citealt{Henriques2013, Henriques2015}). The net effect of strong feedback from supernovae in SAMs is to reduce the cosmological star formation efficiency ($\epsilon_*$). This has been shown to be highly effective at matching the low-mass end of the stellar mass function (e.g., \citealt{White1991, Kauffmann1993, Bower2006, Guo2011, Henriques2015, Somerville2015, Henriques2019}). It is important to emphasize that this is not quenching, since the low mass galaxies continue to form stars (just at much lower rates than in a feedback-free model).

However, as the mass of the galaxy (and, hence, the circular velocity) rises, winds become unable to escape the galaxy (and certainly cannot escape the dark matter halo); see \cite{Henriques2019, Dekel2019} for interesting discussions on this issue. Thereafter, feedback from supernovae actually hinders (rather than aids) reduction in star formation, due to gas building up in the disc, and the immediate vicinity of the disc in galactic fountains (e.g., \citealt{Marinacci2010, Armillotta2016}).

In hydrodynamical simulations of various types, supernova feedback is usually implemented as the key regulatory method for star formation in low mass galaxies, as in SAMs. Unlike in SAMs, a fixed energy per supernova is usually applied, where $E_\mathrm{SN} = \epsilon_\mathrm{SN} \, m_*$ (with $\epsilon_\mathrm{SN} \approx 10^{49}\,\mathrm{erg / M_\odot}$). Initially, energy injection into neighboring gas particles (or cells) was the most common route used to deploy this energy. However, this proved ineffective due to the rapidity of cooling of the ISM (see \citealt{Katz1996}). 

One method to combat this is to launch galactic winds via momentum (rather than thermal energy) injection, see \cite{Navarro1993, Springel2003}. Alternatively, in the EAGLE cosmological model, stochastic energy injection is employed (whereby most of the energy released from the supernova is channeled into a randomly chosen gas particle; see \citealt{Della2008, Schaye2015}). This is effective because the gas particles affected are heated far more than they otherwise would be, increasing their cooling time, and, hence, enabling sustained pressure to build on the ISM, ultimately driving a wind from the galaxy. 

Crucially for this review, all applications of core-collapse (Type II) supernova feedback in cosmological hydrodynamical simulations to date fail to quench the highest mass galaxies and haloes (see, e.g., \citealt{Vogelsberger2014a, Vogelsberger2014b, Schaye2015, Somerville2015, Nelson2018, Pillepich2018, Dave2019, Henriques2019}). These energetic processes reduce the rate of star formation by periodically blowing out gas from the ISM, ultimately resolving the cosmological star formation efficiency problem at low masses. However, at high masses, gas fails to escape the halo serving only to exacerbate the problem of why high-mass systems are quenched. Moreover, since low-mass galaxies are not quenched (but instead are simply forming stars at a lower rate than in a feedback-free Universe), SN II are unlikely to be relevant for quenching at all.

\subsubsection{The role of supernovae Type Ia in maintaining quiescence}\label{s221}

Since core collapse SN II occur contemporaneously to star formation, it is obvious that these events cannot account for long-term quiescence. At best they could be a trigger for quenching (and only in lower-mass systems where the ejecta can escape the halo; see the previous part of this sub-section). On the other hand, SN Ia (arising from thermonuclear explosions of C–O white dwarfs in binary systems) occur with a broad delay from the initial star forming episode, spanning from $\sim$100 Myr to many Gyr (e.g., \citealt{Matteucci1986, Maoz2012, Maoz2014}). Consequently, it is at least theoretically possible that they could be a relevant energy source for maintaining quiescence in some galaxies.  

Since $\alpha$-elements are predominantly produced by core collapse supernovae, while Fe is mostly produced by SN Ia (and, to a lesser extent, also by core collapse supernovae), the ratio of $\alpha$/Fe is a powerful diagnostic of the relative contribution of these two supernova types (e.g., \citealt{Matteucci1986, Matteucci1994, Pipino2004, Pipino2008}). As will be discussed in Sect.~\ref{s711}, high-mass elliptical galaxies show strong evidence of $\alpha$-enhanced stellar populations (e.g., \citealt{Cowie1996, Trager2000a, Trager2000b, Thomas2005, Thomas2010}). This implies that the primary star forming period of these systems must have been very short, such that stars formed out of $\alpha$-enriched gas from (near immediate) SN II, but without Fe-enhancement from the delayed SN Ia. Ultimately, $\alpha$-enhancement functions as a `clock' for the star formation episode, specifically implying that quenching must have happened within $\sim$100 Myr, or so.

Crucially, SN Ia must occur post-quenching in this scenario, enriching the Fe content of the ISM and CGM (as observed in, e.g., \citealt{Loewenstein1991, Renzini1993, Matthews2003, Humphrey2006, Pipino2011, Mernier2016a, Mernier2016b}). Therefore, something must prevent the gas produced from stellar evolution from forming stars within the ISM and, indeed, something must cause this enriched gas to enter the CGM (or ICM in the case of clusters).

It has been theorized that SN Ia may drive winds from the quenched ISM of elliptical galaxies, maintaining quiescence within the system (see, e.g., \citealt{Matteucci1994, Matteucci2006, Pipino2004, Pipino2008, Molero2023}). The essential idea is that the early starburst phase leads to rapid SN I production, which drives a powerful wind, depleting the ISM of gas and triggering quenching. Thereafter, SN Ia provide sufficient energy to unbind the residual gas (largely arising from stellar evolution). A key aspect of this model is that the SN Ia can couple more efficiently to the pre-heated ISM in a quenched elliptical (see full details in \citealt{Molero2023}).

Utilizing one-zone chemical evolution models, \cite{Molero2023} demonstrate that SN Ia provide sufficient energy to unbind the residual gas produced in the ISM of low-to-intermediate mass ellipticals during the post-quenching phase of their evolution. Moreover, they show that such a model can give rise to several fundamental scaling relations in close agreement to observations. In particular, the stellar mass - metallicity relation, the stellar mass - $\alpha$-enhancement relation, and the Fe-enrichment of the CGM may be simultaneously explained via SN Type Ia maintenance. Yet, at the very highest masses, \cite{Molero2023} find that other processes may be important, particularly ejective AGN feedback.

However, while SN Ia may be crucial for maintaining quiescence in the ISM, they are unlikely by themselves to be able to account for the stabilization of the hot gas halo surrounding massive galaxies, groups, and clusters. Ultimately, in cosmological simulations which incorporate SN Ia feedback, this is found to be insufficient to cause long-term quiescence (see, e.g., \citealt{Vogelsberger2014a, Vogelsberger2014b, Torrey2014, Crain2015, Schaye2015, Weinberger2017, Nelson2018, Pillepich2018, Dave2019}). The reason for this is that gas inflow into the ISM must be prevented to achieve full quenching. Since the vast majority of baryons within high-mass haloes do not reside within the ISM, the energy required to stabilize the halo is far greater than that needed to stabilize the ISM. Consequently, the required feedback to achieve halo stabilization has become the primary focus of most contemporary theoretical models of central galaxy quenching.

Nevertheless, some other theoretical works explore SN Ia as a quenching mechanism at the high-mass end (e.g., \citealt{Mohapatra2024}), with other works even considering the role of stellar feedback from massive stars as a long-term maintenance quenching mode (e.g., \citealt{Conroy2015}). It is important to appreciate that most large-volume cosmological simulations (e.g., Illustris and IllustrisTNG) include SN Ia only for chemical enrichment, not for energy or momentum feedback. Hence, there is more work to be done to fully rule out stellar and supernova feedback as a quenching mechanism, even at high masses.

In following parts of this section, we go on to consider several theoretical ideas for achieving long-term quiescence at the halo scale. However, before continuing, it should be emphasized that SN Ia are likely important for maintaining a quiescent ISM, provided that gas inflow into the system is halted by these more energetic processes. Indeed, chemical evolution models compared to observations suggest that, at the very least, SN Ia are crucial for depleting the ISM of Fe-enriched gas, and subsequently enriching the CGM (or ICM) surrounding quenched ellipticals.

\subsection{Three intrinsic quenching paradigms:}\label{s23}

It seems the year 2006 was very important for the theory of galaxy quenching. Three extremely influential papers were published (\citealt{Dekel2006, Hopkins2006, Croton2006}), which shape the way astronomers view the quenching of galaxies to this day. In this sub-section, we give an overview of each of these theoretical paradigms.

\subsubsection{I) Cosmic inflows, virial shocks and halo mass quenching}\label{s231}

Baryons are accreted into dark mater haloes along dark matter filaments within the cosmic web (e.g., \citealt{Peacock1999, Mo2010, Somerville2015}). The typical temperature of baryons within the IGM is $T_b \sim 10^{4} \, K$, which is set by the UV background and the Hydrogen line cooling floor (at 13.6\,eV; see \citealt{Somerville2002, Somerville2015}). Early galaxy formation theory posited that all baryons accreted into dark matter haloes are rapidly shock heated to the virial temperature of the halo (e.g., \citealt{Binney1977, Rees1977, Silk1977, White1991, Mo1998}). However, key insights from idealized hydrodynamical simulations of halo formation in \cite{Birnboim2003, Keres2005} revealed that this is likely not true. At low masses, shocks do not develop and galaxies are fed instead via cold gas streams. However, there is still some debate in the field as to whether cold gas streams make it all the way to the galaxy, with findings dependent upon numerical methods (see, \citealt{Nelson2013}). At higher masses, shocks do develop and a hot static atmosphere is formed, i.e., the CGM surrounding massive galaxies, groups, and clusters.

In \cite{Dekel2006}, building on these past results, a mass threshold for the formation of shocks is established, via both analytical arguments and direct demonstrations within idealized hydrodynamical simulations. The key idea is that below this mass threshold, galaxies acquire baryons through cold gas accretion from the IGM, but at higher halo masses a hot static atmosphere forms. Thereafter, further accretion of baryons results solely from cooling of the hot gaseous halo (primarily via bremsstrahlung free\,--\,free emission, with a cooling rate: $\Gamma_\mathrm{cool} \propto \rho^2 \, T^{1/2}$, where $T$ is the gas temperature and $\rho$ is the gas density; see, e.g., \citealt{Fabian2006, Fabian2012}). 

Of particular interest for the theory of quenching is the realization that this change in accretion mode occurs at approximately the `knee' of the stellar mass function ($M_* \sim 10^{10.5-11}\,M_{\odot}$), i.e., where galaxies begin to `mass-quench' in high numbers (e.g., \citealt{Baldry2004, Baldry2006, Peng2010}). Consequently, \cite{Dekel2006} argue that the fundamental change in galaxy properties observed to occur around the corresponding halo mass of $M_\mathrm{Halo} \sim 10^{12}M_{\odot}$ (i.e., the switch from blue, young, star forming discs to red, old, quiescent spheroids; e.g., \citealt{Strateva2001, Driver2006, Cameron2009a, Schawinski2014}) is intimately connected to the change in the thermodynamics of haloes at this mass threshold.

The fundamental criteria for stable shock formation is as follows (see \citealt{Dekel2006}):

\begin{equation}
t_\mathrm{cool} \geq t_\mathrm{comp} \sim \frac{r_\mathrm{shock}}{V_\mathrm{inf}} \sim \frac{R_{200c}}{V_{200c}} \sim t_\mathrm{dyn}
\end{equation}

\noindent i.e., shocks may stabilize when the cooling time becomes larger than the compression time, which is itself closely connected to the dynamical time of the halo. In the above expression, $t_\mathrm{cool}$ is the cooling time, $t_\mathrm{comp}$ is the compression time, and $t_\mathrm{dyn}$ is the dynamical time. Additionally, $R_{200c}$ and $V_{200c}$ represent the virial radius and velocity (respectively), $r_\mathrm{shock}$ is the radius of the shock front, and $V_\mathrm{inf}$ is the infall velocity. 

The logic behind this stability criterion is physically intuitive. As gas falls into the halo, it is accelerated due to gravity (e.g., for an idealized infall from $r = \infty$, the gas reaches a velocity of $\sqrt{2} \, V_{200c}$ at $R_{200c}$). As the infalling gas enters the halo, it interacts with a denser medium than the IGM. Consequently, shocks may form. What happens next depends on whether the shocks collapse or become stabilized. If the cooling time is short compared to the compression time, the pressure rapidly drops, resulting in the shock collapsing in on itself, and, hence, the infalling gas may continue as a cold stream into the galaxy. Conversely, if the cooling time is long relative to the compression timescale, a pressure gradient builds and the shock is stabilized, which results in the kinetic energy of the inflowing cold gas stream being converted into thermal energy of the CGM. See \cite{Birnboim2003} for the original derivation.

The cooling time of a hot gaseous halo is given by (e.g., \citealt{Dekel2006, Mo2010, Fabian2012}):

\begin{equation}\label{e4}
t_\mathrm{cool} \equiv \frac{\mathcal{E}}{|\partial \mathcal{E}/\partial t|} = \frac{3n k_B T}{2n_I n_e \Lambda(T,Z)} \sim \frac{T}{n\Lambda(T,Z)}
\end{equation}

\noindent where, $\mathcal{E}$ is the energy density of the hot gas halo, $n_I$ is the number density of ions, $n_e$ is the number density of electrons, and $n$ is the average number density of particles within the plasma. The cooling function is represented by $\Lambda$, which depends on both the temperature ($T$) and metallicity ($Z$) of the gas in the halo. This essentially results in two regimes of cooling for the ionized gas: (i) metal line emission; and (ii) bremsstrahlung free - free emission (in the fully ionized plasma).

The cooling rate drops precipitously at $T \sim 10^6 \, K$, where metal line cooling becomes ineffective due to complete ionization. This in turn causes the cooling time to rise dramatically, enabling the the hot gas halo to build up substantially for the first time. Future cold gas accretion into the system does not make it directly to the galaxy, but rather must pass through a hot gas phase (via shock heating to the virial temperature). Subsequent cooling is expected to be predominantly via free\,--\,free emission, and this yields the only remaining route for accreted baryons to make it from the virial radius to the galaxy, and, hence, provide fuel for future star formation. Consequently, a reduction in halo star formation efficiency is expected at around this critical virial temperature. This is known as `halo mass quenching', since the virial temperature is directly related to the total mass of the halo (see \citealt{Dekel2006} for the theory, and  \citealt{Woo2013} for observational support).

\begin{figure}[ht]  
\centering
\includegraphics[width=0.49\textwidth]{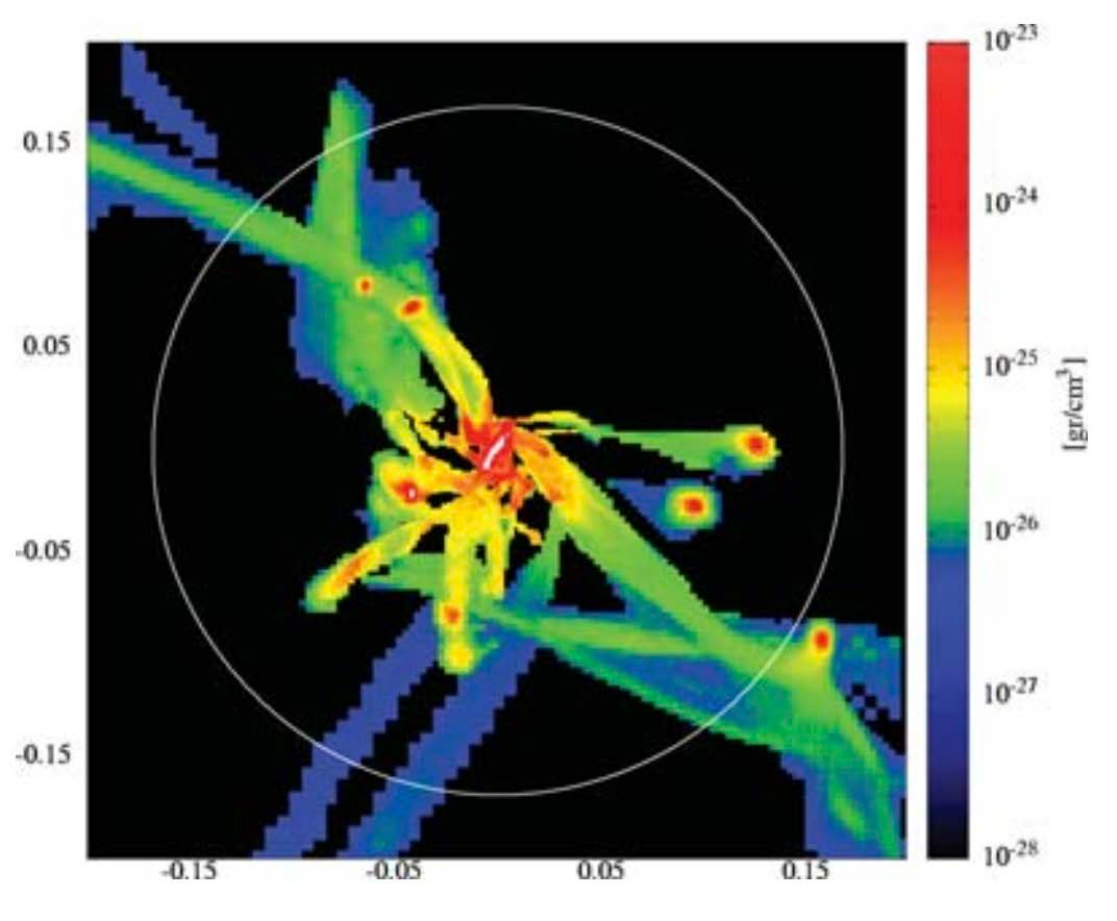}
\includegraphics[width=0.49\textwidth]{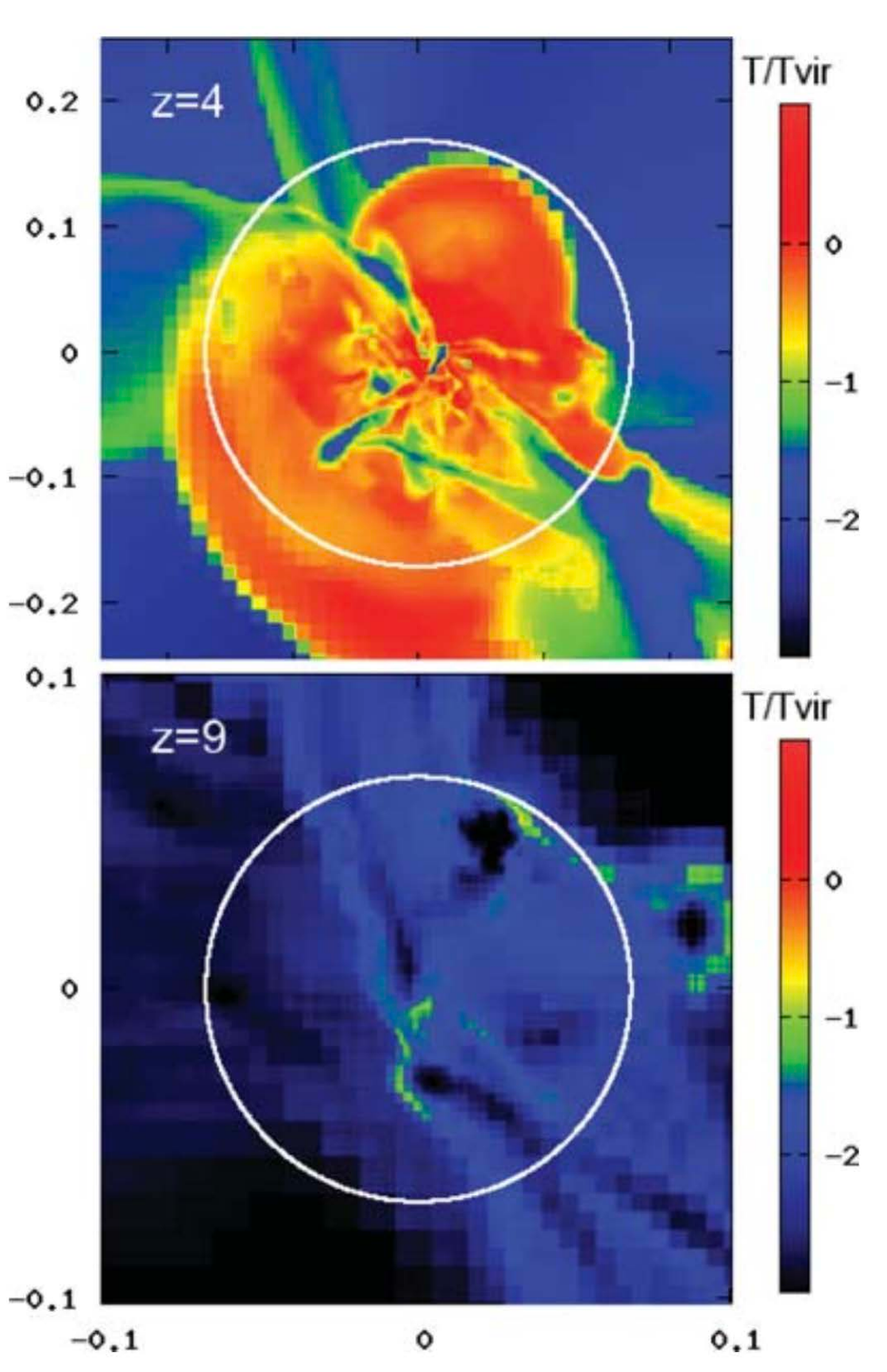}
\caption{\textit{Left-hand panel:} Projected snapshot from a hydrodynamical simulation showing the filamentary nature of cold gas inflows into a low mass halo. This panel is color coded by the density of the cold gas, and clearly shows that these cold streams make it to the center of the halo, and hence to the galaxy itself. \textit{Right-hand panels:} Projected snapshots from a hydrodynamical simulation of the evolution of the halo of a galaxy at $z=9$ when the halo mass is $M_\mathrm{Halo} = 2 \times 10^{10} M_{\odot}$ (lower panel) and at $z=4$ when the halo mass is $M_\mathrm{Halo} = 3 \times 10^{11} M_{\odot}$. Both panels are color coded by the gas temperature in units of the virial temperature. At high-redshift/ low mass, the temperature of the halo gas is much lower than the virial temperature, and the galaxy is fed by cold gas streams. Conversely, at lower-redshift/ high mass, the temperature of the halo gas is at (or around) the virial temperature, which prevents direct feeding of the ISM via cold gas streams. Nonetheless, cooling from the CGM may continue, unless stabilized through heating via additional feedback mechanisms. All panels in this figure are reproduced from \cite{Dekel2006}.}\label{f2}
\end{figure}

Accounting for all of the relevant cooling effects, and the evolution of the structure of dark matter haloes, \cite{Dekel2006} determine that the mass threshold between cold and hot mode accretion occurs at:

\begin{align}
M_\mathrm{Halo} \lesssim 10^{12} \, M_{\odot}  \,(1+z)^\alpha \, : \, \text{Cold Mode} \quad | \quad M_\mathrm{Halo} \gtrsim 10^{12} \, M_{\odot} \, (1+z)^\alpha \, : \, \text{Hot Mode}
\end{align}

\noindent where $\alpha \sim 1 - 1.5$, indicating that the mass threshold for transition to the hot-mode occurs at progressively higher values at earlier cosmic times. This is a direct result of the density of haloes increasing with increasing redshift, which results in enhanced cooling rates at earlier cosmic times (see \citealt{Dekel2006}). We show examples of cold gas streams into a low mass halo, and the bimodal temperature change in the gas halo either side of the the critical mass threshold, in Fig.~\ref{f2} (which is reproduced from \citealt{Dekel2006}).

Further theoretical support for this paradigm is provided in \cite{Dekel2009, Keres2009, Dekel2019}, primarily via idealized hydrodynamical simulations, which broadly confirm the transition in the thermodynamical state of haloes at approximately the predicted mass scale. Moreover, observational support of this picture was found in \cite{Woo2013}, who note that halo mass is a stronger correlator to quenching than stellar mass, clearly in line with this picture. However, a number of later studies confirm that halo mass is not as strongly constraining of quenching as central mass density, bulge mass, or central velocity dispersion (e.g., \citealt{Bluck2014, Woo2015, Bluck2016, Bluck2022}), potentially posing a challenge to this paradigm, at least in the absence of additional feedback processes.

A common misconception is that the criteria for quenching and hot gaseous halo stability are identical. This has led many astronomers to the naive assumption that \cite{Dekel2006} predicts quenching to occur at $M_\mathrm{Halo} \gtrsim 10^{12} \, M_{\odot} \, (1+z)^\alpha$. It is crucial to understand why this is not quite right. A more accurate way to link the thermodynamic stability of gaseous haloes to galaxy quenching would be to argue that intrinsic quenching cannot occur below this mass threshold (in isolated systems). This follows for two reasons: (i)~below this mass, galaxies are fed by cold gas streams which can replenish the ISM and, hence, enable future star formation; and (ii)~at higher masses, cold gas streams are prevented from reaching the galaxy, which engenders quenching via starvation if (\emph{and only if}) cooling from the hot static atmosphere is prevented (or sufficiently reduced). 

It is important to appreciate that the cooling time refers to the time for the \textit{entire} gaseous halo to condense into the galaxy. Furthermore, the hot static atmosphere rapidly grows in mass once the cooling time is longer than the dynamical time. Consequently, the baryonic mass of the CGM becomes much larger than the mass of the ISM in high mass haloes (as observed; see, e.g., \citealt{Fabian2006, McNamara2007, Fabian2012, HlavacekLarrondo2015}). Hence, star formation may still be fueled effectively, even within a halo with a hot gas cooling time greater than the Hubble time. This is further exacerbated at the highest halo masses (particularly within high-mass groups and clusters), where the cooling rates rise dramatically due to the increased density and temperature of the gas (which ultimately leads to the cooling catastrophe; see  \citealt{Fabian1994, Fabian1999, Fabian2006, Fabian2012}).

These limitations with halo mass quenching are explicitly acknowledged in \cite{Dekel2006}, where feedback from SN and/or AGN is suggested as a natural complement to this paradigm. That is, heating is required from an external source for the hot gas halo to maintain quiescent galaxies, even if this is not required to maintain a hot static atmosphere (at least close to the critical mass). At the highest masses, cooling rates increase progressively and this problem becomes even more severe, requiring ever more energetic feedback to prevent gas accumulation within galaxies, and inevitably future star formation (e.g., \citealt{Dekel2019, Henriques2019, Bluck2020a}). Indeed, essentially all large-volume cosmological simulations of galaxies precisely follow the shock heating of the massive gaseous haloes and yet do not achieve the observed-level of high-mass quenching without invoking some other mechanism (e.g., \citealt{Vogelsberger2014a, Vogelsberger2014b, Schaye2015, Nelson2018, Pillepich2018, Dave2019}).

As we have discussed previously, SN feedback is not effective in the most massive haloes, due to the deep potential wells being able to hold on to SN ejecta. Furthermore, the energy injected into the CGM via supernova and stellar feedback is insufficient to stabilize cooling in hot gaseous haloes (e.g., \citealt{Crain2015, Sijacki2007, Torrey2014, Vogelsberger2014a, Vogelsberger2014b, Weinberger2017, Weinberger2018, Zinger2020}). Therefore, theorists have turned en masse to considering AGN as the fundamental mechanism for CGM stabilization and, ultimately, central galaxy quenching. We consider the proposed mechanisms of AGN feedback in detail throughout the remainder of this section.

\subsubsection{Interlude: Modeling AGN feedback}

Quasars are observed to be the brightest objects in the Universe (e.g., \citealt{Schmidt1968, Greenstein1963, Schmidt1969, Baldwin1977}). It was quickly appreciated that these are extragalactic sources, with luminosities far exceeding that of stars. A theoretical picture to explain quasars was developed by \cite{Lynden-Bell1969, Shakura1973, Soltan1982, Rees1984, Silk1998}, whereby these sources are conjectured to be powered by accretion onto supermassive black holes. Later, a variety of similar objects were identified at lower luminosities (corresponding to lower accretion rates) and integrated into the broad, unified class of AGN (e.g., \citealt{Antonucci1993, Elvis1994, Urry1995}).

The basic idea behind AGN feedback is that baryons accelerating towards a supermassive black hole under gravity convert gravitational potential energy to kinetic energy, which may be thermalized in the dense media surrounding the black hole, and finally radiated away. This radiation may then impact the ISM (and potentially the CGM), either through momentum transfer (via radiation pressure), or else via heating the ISM gas (see, e.g., \citealt{Binney1995, Ciotti1997, Ciotti2001, Mathews2003}). Both of these types of feedback tend to drive ISM winds, and sometimes outflows from the galaxy (and even the halo). Additionally, ionized particles orbiting the supermassive black hole produce strong magnetic fields which can lead to powerful relativistic jets and synchrotron emission, which may impact both the ISM and the CGM (see, e.g., \citealt{Fabian2006, Fabian2012}). 

It should be noted that SMBH accretion involves more than just gas, including also the accretion of stars, dusts, and growth through black hole - black hole mergers. However, in our discussion we will focus primarily on gas accretion as this is the channel which is believed to release by far the greatest energy into the surrounding galaxy and halo, and, hence, impact galactic star formation and quenching.

In a quasi-Newtonian treatment, the kinetic energy of a particle reaching the event horizon of a black hole from infinity is, $E_{k} = (1/2)\,mc^2$.\footnote{To derive this simply set the modulus of the gravitational potential energy of a particle (of mass $M$) at the Schwarzschild radius equal to kinetic energy. Note that the kinetic energy released from infall at infinity at the Schwarzschild radius is independent of the black hole mass. This suggests that all black holes might be equally powerful. However, this turns out not be correct in practice, due to the Eddington limit on accretion rate (see Eq.~\eqref{e8} and associated text).} In a more complete general relativistic context, the total kinetic energy of accretion is $\sim0.06\,mc^2$ (for a Schwarzschild geometry) and $\sim0.42\,mc^2$ (for a maximally rotating Kerr geometry), see {\cite{Thorne1974} for the derivations. Note that even though the kinetic energy is directly proportional to the rest-mass energy in all of these cases, no mass-to-radiation conversion occurs in accretion, only potential-to-kinetic energy conversion. This is essential for baryon number conservation. The origin of the $mc^2$ factors arise from the definition of the Schwarzschild radius ($R_S = 2\,GM_\mathrm{BH}/c^2$).

It is worth pausing to note that accretion onto a compact object achieves the highest known efficiencies of energy generation, outside of matter - anti-matter annihilation (e.g., compare to the efficiency of nuclear fusion, $\epsilon_\mathrm{fusion} \sim 0.007$). Consequently, one expects that the high baryon energies engendered through accretion onto supermassive black holes may be significant to galaxies (at least in their cores, and potentially throughout the entire disc, and even the CGM). 

Most models of AGN feedback assume that some fraction of the kinetic energy of accreted particles is converted to radiation (the radiative efficiency, $\epsilon_r$), and some fraction of the radiated energy is available to do work on the galaxy or halo via feedback (the feedback efficiency, $\epsilon_f$). Hence (e.g., \citealt{Silk1998, King2003a, King2003b, Murray2005}),

\begin{equation}
\frac{d E_\mathrm{feedback}}{dt} = \epsilon_f L_\mathrm{AGN}, \quad \text{where} \quad L_\mathrm{AGN} = \epsilon_r \dot{M}_\mathrm{BH} \, c^2 \, ,
\label{e6}
\end{equation}

\noindent and where $E_\mathrm{feedback}$ is the feedback energy at a given time, $L_\mathrm{AGN}$ is the total bolometric luminosity of the AGN, and $\dot{M}_\mathrm{BH}$ is the mass accretion rate of the supermassive black hole. Most modern simulations take $\epsilon_r = 0.1$, and systematically vary $\epsilon_f$ to achieve agreement with observational constraints. 

By taking the temporal integral over all cosmic time, an important relationship between feedback energy and black hole mass is found (see \citealt{Soltan1982, Silk1998}):

\begin{equation}
E_\mathrm{feedback} \propto M_\mathrm{BH} \, .
\end{equation}

\noindent That is, the energy released from AGN feedback is expected to be directly proportional to the mass of the supermassive black hole. There are a number of caveats here. First, this does not account for black hole\,--\,black hole mergers, which are not expected to release usable feedback energy. Furthermore, at least in principle, both the efficiency terms may be time and mass dependent, which complicates this picture (see the appendix of \citealt{Bluck2020a} for a discussion).

In order to compute the feedback energy per time-step in a simulation, the accretion rate onto the supermassive black hole must be calculated. There exists a fundamental limit to the rate of accretion (the Eddington limit), which is found by setting the force from radiation pressure equal to the gravitational force for a single Hydrogen atom, assuming an isotropic matter distribution (i.e., spherically symmetric accretion). This leads to (see \citealt{Eddington1926, Rees1984}):

\begin{equation}\label{e8}
\dot{M}_\mathrm{Edd} = \frac{4\pi G \, M_\mathrm{BH} \, m_p}{\epsilon_r \, \sigma_T \, c},
\end{equation}

\noindent where, $m_p$ is the proton mass and $\sigma_T$ is the electron cross section. All other terms are as previously defined. Hence, the maximum allowably accretion rate is directly proportional to the mass of the black hole. Note that this limit may be violated in the case of highly non-isotropic feedback (e.g., in the case of jets). However, this is just a limit, not the actual accretion rate one would expect in any given situation. 

In most modern simulations the accretion rate of the supermassive black hole is set by Bondi-Hoyle-Lyttleton accretion (see \citealt{Hoyle1939, Bondi1944, Bondi1952}). This is defined by:

\begin{equation}
\dot{M}_\mathrm{Bondi} = \frac{4 \pi G^2 \, M_\mathrm{BH}^2 \, \rho_g}{(c_s^2 + v^2)^{3/2}} \, ,
\end{equation}

\noindent where, $\rho_g$ is the ambient gas density in the vicinity of the black hole, $c_s^2$ is the local sound speed of the baryon fluid, and $v$ is the relative velocity of the black hole to the ISM. The above expression is derived by first defining the Bondi radius, which is the radius at which gravitational acceleration is balanced by thermal motion (i.e., the sound speed of the gas). One then assumes that all material within this critical radius is accreted into the black hole following, $\dot{M}_\mathrm{BH} = 4\pi R_B^2 \, \rho_g c_s$, which is evidently true for a spherical accretion flow with $v = c_s$ and $r=R_{B}$. For completeness, we note that the Bondi radius is explicitly given by, $R_B \sim GM_\mathrm{BH}/c_s^2$.

This formulation does not account for a host of important physics, including angular momentum, magnetic fields, and general relativistic effects. As such, it is subject to some skepticism. More sophisticated models do exist, including the Shakura-Sunyaev accretion disc model, which is thought to be important when accretion rates are high (see \citealt{Shakura1973}), and the advection-dominated accretion flow (ADAF), which is thought to be important when accretion rates are vastly sub-Eddington (see \citealt{Narayan1994, Narayan1995}). Nonetheless, the Bondi model is by far the most common sub-grid model utilized for black hole accretion, and is used in essentially all contemporary idealized, and cosmological, simulations (e.g., \citealt{Sijacki2007, Torrey2014, Crain2015, Weinberger2017}).

With both the Eddington and Bondi accretion rates defined, the actual accretion rate of the supermassive black hole is set to be the lesser of the two, i.e.,

\begin{equation}
\dot{M}_\mathrm{BH} = \mathrm{min}\big( \dot{M}_\mathrm{Bondi}, \dot{M}_\mathrm{Edd}  \big).
\label{e10}
\end{equation}

Once the energy is extracted from accretion in a theoretically motivated way, one must decide what to do with it. There are two main variants for radiatively efficient accretion (i.e., the `quasar-mode'): (i)~momentum transfer, and (ii)~energy transfer. The former arises from photon pressure (itself arising from the radiation force from photons, $F_\mathrm{rad} = \dot{p} = L/c$, where $p$ indicates momentum), and the latter arises from thermal heating (e.g., \citealt{Ciotti2009, Shin2010, Ciotti2010}). Both methods of feedback have been shown to be effective at launching galactic-scale winds and driving outflows (e.g., \citealt{King2003a, King2003b, DiMatteo2005, Hopkins2006, DiMatteo2008a, Hopkins2008}). Hence, the main impact on galaxies from quasar-mode AGN feedback is the ejection of cold gas from the ISM.

\subsubsection{II) The quasar-mode AGN feedback paradigm}\label{s232}

The first hydrodynamical simulations to incorporate feedback from AGN were \cite{DiMatteo2005, Springel2005d}, in which black hole growth was governed by Eddington-limited, Bondi--Hoyle--Lyttleton accretion (as described above). Feedback energy was modeled exactly as in Eq.~\eqref{e6}, and coupled to the ISM via thermal injection (preserving energy conservation). These simulations demonstrate that quasar-mode AGN feedback can lead to galactic scale outflows, which are generated by expanding bubbles in the ISM, through rapid heating of nearby gas particles to the central black hole. Observational evidence supporting AGN-feedback driven ejective feedback is presented in Sect.~\ref{s4} (see, e.g., \citealt{Feruglio2010, Maiolino2012, Cicone2012, Cicone2014, Cicone2015, Fluetsch2019, Fluetsch2021}).

\begin{figure}[ht]  
\centering
\includegraphics[width=\textwidth,angle=0]{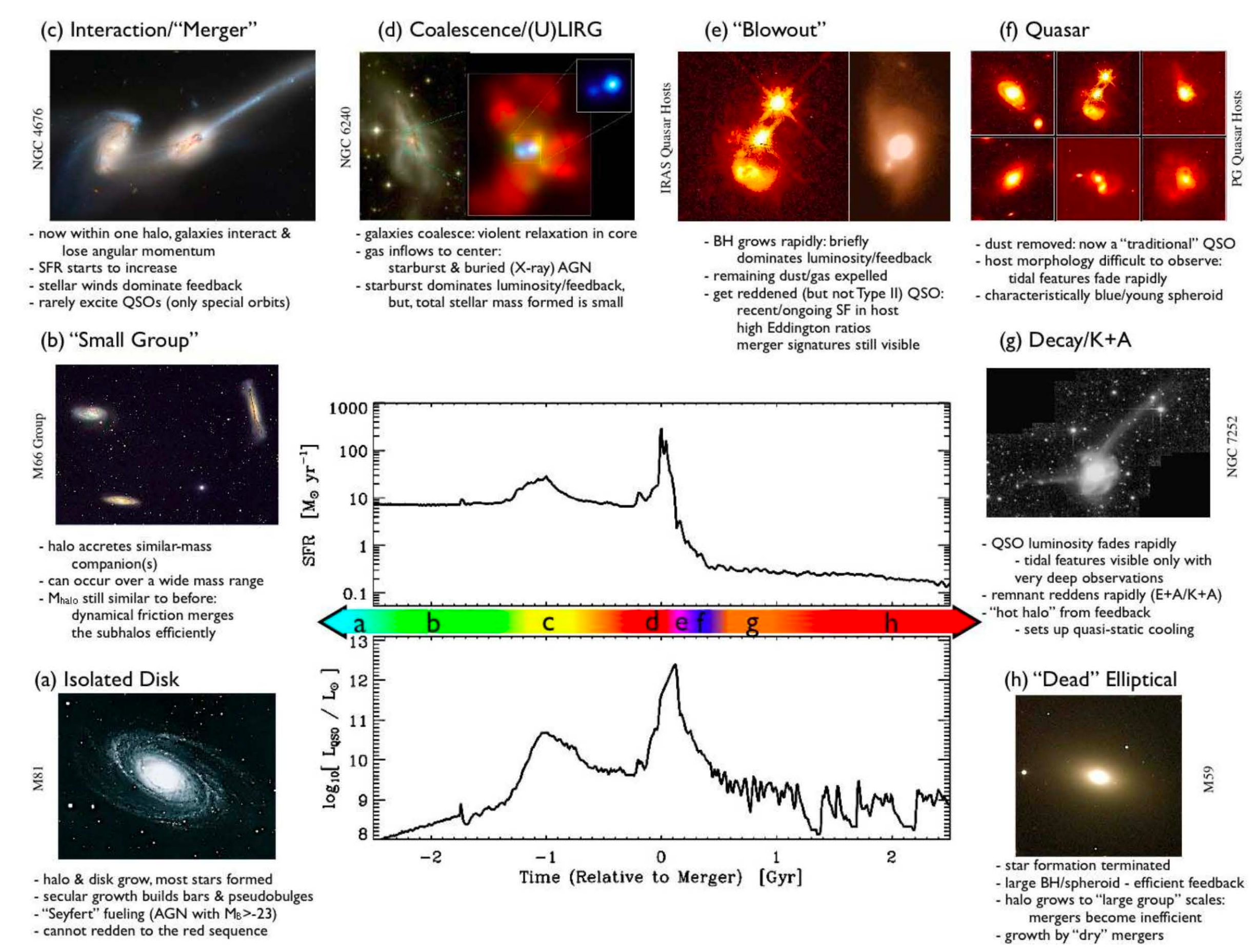}
\caption{The merger - inflow - starburst - quasar - outflow model of quenching, reproduced from \cite{Hopkins2008}. The center panels show the SFR (top panel) and AGN luminosity (bottom panel) as a function of time from an idealized hydrodynamical simulation of a galaxy - galaxy merger. Throughout the perimeter of the figure, the different stages in the evolution (labeled by letters a - h) are linked to astronomical observations. Galaxies begin as gas-rich isolated disks (a), which then accrete galaxies into their haloes forming small groups (b). Dynamical friction leads to host galaxy\,--\,satellite interactions, with a first pericentric passage (c), followed by eventual coalescence (d). Tidal torques induced by the merger drive cold gas flows to the center of the galaxy, leading to enhanced star formation rates and AGN luminosity. This ignites powerful quasar-mode feedback within the simulation, which drives galactic winds and halo-scale outflows (e), briefly revealing an unobscured quasar (f). Thereafter, the star formation and AGN luminosity decrease in a post-starburst, K+A, system (g), and eventually leave behind a quenched spheroid with a massive central black hole (h).}\label{f3}
\end{figure}

These simulations were developed further and expanded into a unified framework for the role of mergers and quasar feedback in galaxy evolution by \cite{Hopkins2006, Hopkins2008}. The key idea is to link the evolution of galaxy mass, structure, and morphology to the evolution of supermassive black holes via galaxy mergers. This leads to a possible solution for galaxy quenching, in addition to an explanation for a host of galactic scaling laws. See Fig.~\ref{f3} for a visual summary, which is reproduced from \cite{Hopkins2008}.

Briefly, hierarchical assembly of dark matter haloes leads to subhalo (and hence galaxy) mergers. Tidal torques induced within gas-rich galaxy mergers are expected to funnel gas towards the center of the system (e.g., \citealt{Toomre1972, Barnes1991, Mihos1996, Barnes2002}), increasing SFR, $\dot{M}_\mathrm{BH}$, and $L_\mathrm{AGN}$. This ignites powerful quasar-mode feedback (as outlined in Eqs.~\eqref{e6}\,--\,\eqref{e10} above), which in this model is injected thermally to gas particles neighboring the central black hole (see \citealt{Hopkins2006, Hopkins2008}). In turn, this drives an increase in gas pressure, which leads to an expanding bubble within the ISM. If the feedback is powerful enough, this can lead to halo-scale galactic outflows, which deplete (or destroy) the ISM, removing fuel for future star formation and, hence, quenching the galaxy. Since mergers are well known as a route to transform high-angular momentum discs into low-angular momentum spheroids (e.g., \citealt{Toomre1972, Barnes2002, DiMatteo2005, Hopkins2006, Hopkins2008, DiMatteo2008a}), one is left with a quiescent spheroid with a high-mass central black hole (as depicted in Fig.~\ref{f3}).

This is quite a remarkable model of galaxy evolution, which links together the formation of large scale structure (via hierarchical assembly), galaxy mergers, dynamical and structural evolution, supermassive black hole growth, the formation of quasars, and the quenching of galaxies. It is fair to say that this picture has left a big impression on the minds of many extragalactic astronomers. 

Among its many successes, this model explains the morphology\,--\,color relationship (e.g., \citealt{Driver2006, Cameron2009b, Wuyts2011}), the stellar mass\,--\,quenched fraction relationship (e.g., \citealt{Baldry2006, Peng2010}), and the apparent co-evolution of the cosmological star formation rate density and luminosity density of AGN (e.g., \citealt{Madau2014, Aird2010, Aird2015}). Additionally, numerous observational studies find enhanced AGN detection rates and/or luminosities in merging systems (e.g., \citealt{Ellison2008, Ellison2011, Ellison2013, Ellison2019, Ellison2024}), and identify galactic outflows associated with AGN (e.g., \citealt{Feruglio2010, Maiolino2012, Cicone2012, Cicone2014, Cicone2015, Fluetsch2019, Fluetsch2021}), both clearly in line with this picture.

Furthermore, the \cite{Hopkins2006, Hopkins2008} models also provides a deep solution for the existence of the black hole mass\,--\,galaxy scaling relationships (e.g., \citealt{Maggorian1998, Ferrarese2000, Haring2004, Saglia2016}). This works by assuming that once the feedback energy is greater than the binding energy of the gas in the center of the system, the gas is removed, which yields:

\begin{equation}
E_\mathrm{feedback} = \epsilon_f \, \epsilon_r M_\mathrm{BH} \, c^2  \sim  |E_\mathrm{bind}| = \frac{GM_\mathrm{gas} M_\mathrm{bulge}}{R} \ \rightarrow \ M_\mathrm{BH} \sim  M_\mathrm{bulge}
\end{equation}

\noindent where, $M_\mathrm{gas}$ is the gas mass within the bulge, $M_\mathrm{bulge}$ is the bulge mass, and $R$ is the characteristic radius of the system. This also leads to the $M_\mathrm{BH}$\,--\,$\sigma$ relationship in a more subtle manner. Ultimately, the scaling relationships arise from feedback-limited accretion in this paradigm.

So, are we done? Well, not quite! One of the weaker aspects of the \cite{Hopkins2006, Hopkins2008} paradigm is actually with respect to quenching (our principal concern). Even if every massive quiescent galaxy forms via major gas-rich mergers (which is far from certain since minor mergers, violent disc instabilities, and direct collapse are all likely important; see \citealt{Noguchi1999, Dekel2009b, Ceverino2010}), and these ignite quasars which in turn are successful at ejecting all of the ISM gas (which is also far from certain since quasar feedback is most effective at depleting the ISM of the central region within galaxies, often leaving the disc intact; see \citealt{Gabor2014, Costa2015, Angles2017}), a critical problem remains. Namely, \textit{what happens after the quasar activity subsides?}

In the real Universe (and in full cosmological simulations) one expects either cold gas accretion or hot halo cooling to occur (dependent upon the mass of the halo, see back to Sect.~\ref{s231}). This will lead to rejuvenation of star formation within the galaxy as the ISM is reformed. This process may be further enhanced by subsequent major or minor merging with gas-rich satellites, and through re-accretion of gas ejected in the original quasar event (especially in high mass haloes). Moreover, as discussed in Sect.~\ref{sec:stellar_quenching}, gas is continuously returned to the ISM through stellar evolution throughout the quiescent phase, which will also lead to rejuvenation without late-time maintenance feedback (likely via periodic AGN episodes).

Indeed, if one runs a closed-box hydrodynamical simulation (like in \citealt{Hopkins2006, Hopkins2008}) without any merger or quasar feedback, the star formation rate still reduces over time, ultimately `quenching' from starvation (once the gas in the simulation is depleted). The merger - quasar feedback only serves speed up this process. Yet, in a cosmological setting the abundant non-stellar baryon fraction of the Universe (at all cosmic times) can still cool and coalesce within galaxies, rendering the merger-driven quasar event `non-fatal'. The galaxy will reduce its SFR initially (relative to similar mass systems at the same epoch), but recover thereafter, typically within a few Gyr (dependent upon the cooling rate of the hot gaseous halo in high mass systems).

Consequently, one still needs to understand why quenched galaxies remain quenched for long cosmological timescales. The quasar quenching paradigm is not helpful here because quasars form in gas-rich systems, associated with star forming galaxies, and are not seen in gas-poor quiescent systems (e.g., \citealt{Hickox2009, Heckman2014, Piotrowska2022, Ward2022}). Moreover, one still needs to understand how the hot gas halo is stabilized in the highest mass haloes (i.e., the cooling problem) and, ultimately, why the cosmological star formation efficiency is so low. We turn to these deep questions in the next sub-section.

Finally, it is important to appreciate that there is a demographics problem in the merger - quasar - quench model. While merger-driven quasar feedback provides a compelling pathway for inducing rapid quenching in individual systems (e.g., \citealt{Hopkins2006, Hopkins2008}), there are tensions when considering the galaxy population as a whole. In particular, the evolving number density and duty cycle of luminous quasars, together with the observed rate of major, gas-rich mergers, may be insufficient to account for the high fraction of quenched $\gtrsim L^{\star}$ galaxies at low redshift (e.g., \citealt{Hopkins2007, Shankar2009, Conroy2013}). This problem is further exacerbated by the observation that a large fraction of massive galaxies ($M_{\star} \sim 10^{11}\,M_{\odot}$) are already quenched by $z \sim 1$--2 (e.g., \citealt{Bell2004, Faber2007, Ilbert2013, Muzzin2013}), implying that quenching mechanisms must operate early and with high efficiency. Taken together, these considerations suggest that the quasar-mode cannot represent the dominant quenching channel, further motivating the need for additional pathways.

\subsubsection{III) The radio-mode AGN feedback paradigm}\label{s233}

`Radio-mode' AGN feedback (which is also referred to in the literature as `mechanical' or `preventative' feedback) is thought to arise from radiatively inefficient accretion onto a supermassive black hole (when $\dot{M}_\mathrm{BH} \ll \dot{M}_\mathrm{Edd}$). The fundamental mechanism is believed to be the production of relativistic jets through magnetic field line twisting due to ionized particle accretion, which leads to direct mechanical energy injection into the CGM, as well as synchrotron radiation (e.g., \citealt{Fabian1994, Fabian1999, Fabian2006, McNamara2007, Fabian2012, HlavacekLarrondo2015}). The jets inflate cavities in the diffuse CGM, depositing energy as they expand via shocks, turbulence, mixing, and sound waves. 

Ultimately, the heat from radio-mode feedback has long been considered as a method to stabilize the hot gaseous haloes around massive galaxies, shutting down cooling flows, and ultimately quenching central galaxies via starvation of future gas supply (e.g., \citealt{Croton2006, Bower2006, Bower2008, Somerville2015}).

The first model to incorporate radio-mode AGN feedback was \cite{Croton2006}, in which a simple semi-analytical formulation was developed, which was shown to be effective at reproducing the high-mass end of the galaxy stellar mass function at low redshifts. This was a serious challenge for a generation of previous SAMs (stretching back to the early 1990s, e.g. \citealt{White1991, Kauffmann1993, Cole2000}). Unlike virial shocks and quasar-mode feedback (which were both first developed within the framework of idealized hydrodynamical simulations), radio-mode feedback was first developed within the semi-analytical approach. Many SAMs adopted some form of radio-mode feedback, and this has become a staple in essentially all contemporary cosmological models of this type (see, e.g., \citealt{Bower2006, DeLucia2007a, Bower2008, Guo2011, Henriques2013, Henriques2015, Somerville2015, Henriques2019}).

In \cite{Henriques2015} the original model from \cite{Croton2006} is developed into a particularly simple and intuitive form, which is successful at reproducing many high-mass features of the observed galaxy population. Explicitly, the accretion rate of the supermassive black hole in `hot-mode' is modeled as:

\begin{equation}\label{e12}
\dot{M}_\mathrm{BH} = k_\mathrm{AGN} \bigg(\frac{M_\mathrm{BH}}{10^{8} M_{\odot}}\bigg) \bigg(\frac{M_\mathrm{Hot}}{10^{11} M_{\odot}}\bigg)
\end{equation}

\noindent where, $M_\mathrm{Hot}$ is the mass of the hot gas halo, and $k_\mathrm{AGN}$ is a tuneable coupling constant. Hence, the rate of accretion is proportional to both the mass of the central black hole and the mass of the hot gas halo. This is clearly not a physical model of accretion (unlike, e.g., the Bondi accretion rate, discussed above), but it seeks to establish a plausible physical connection between hot-mode accretion and the availability of hot gas in the halo. Unlike the quasar mode, this mode is most effective when the baryonic content of the halo is dominated by an ionized plasma in the CGM. The essential idea is that, in this regime, supermassive black holes are not fed from cool ISM gas, but instead from cooling flows from the hot gaseous halo.

Much like in quasar-mode feedback, some fraction ($\epsilon_\mathrm{radio}$) of the rest-energy of an accreted particle is assumed to be available for feedback per unit time. Hence,

\begin{equation}
\dot{E}_\mathrm{radio} = \epsilon_\mathrm{radio} \, \dot{M}_\mathrm{BH} \, c^2
\end{equation}

\noindent where all terms are as defined previously. Unlike the quasar-mode, the energy accessed in the radio-mode is not used to either heat or push on the ISM directly. Alternatively, the energy is assumed to couple with the CGM, stabilizing gas cooling. Explicitly, the mass cooling rate into the ISM from the CGM is modified as (\citealt{Henriques2015}): 

\begin{equation}
\dot{M}_\mathrm{cool, \, eff} = \mathrm{max}\bigg( \dot{M}_\mathrm{cool} - \frac{2\dot{E}_\mathrm{radio}}{V^2_{200}} \, , \, 0 \bigg)
\end{equation}

\noindent where, $\dot{M}_\mathrm{cool}$ is the mass cooling rate in the absence of radio heating, and $V_{200c}$ is the circular velocity of the dark matter halo. Hence, the rate of cold gas accretion into the galaxy is offset by radio heating, up to a minimum of zero inflow rate (i.e., this mode does not drive outflows). The cooling time is computed as in Eq.~\eqref{e4}, under the assumption of bremsstrahlung, free\,--\,free emission (with $\Lambda \sim T^{1/2}$). The CGM is modeled as an isothermal sphere, with temperature set to the virial temperature. This enables a direct analytic calculation of the radius at which the cooling time equals the dynamical time, $t_\mathrm{dyn} \sim R_{200c}/V_{200c}$ (see \citealt{Henriques2015}). When, $t_\mathrm{cool} < t_\mathrm{dyn}$, the mass accretion rate into the galaxy is assumed to be:

\begin{equation}
\dot{M}_\mathrm{cool} = \frac{M_\mathrm{Hot}(r < r_\mathrm{crit})}{t_\mathrm{dyn}} \quad \text{where} \quad t_\mathrm{cool}(r_\mathrm{crit}) = t_\mathrm{dyn}(r_\mathrm{crit}),
\end{equation}

\noindent i.e., all gas with a cooling time less than the dynamical time is permitted to accrete into the galaxy as cold gas, with a rate set by the dynamical time.

\begin{figure}[ht]  
\centering
\includegraphics[width=0.65\textwidth]{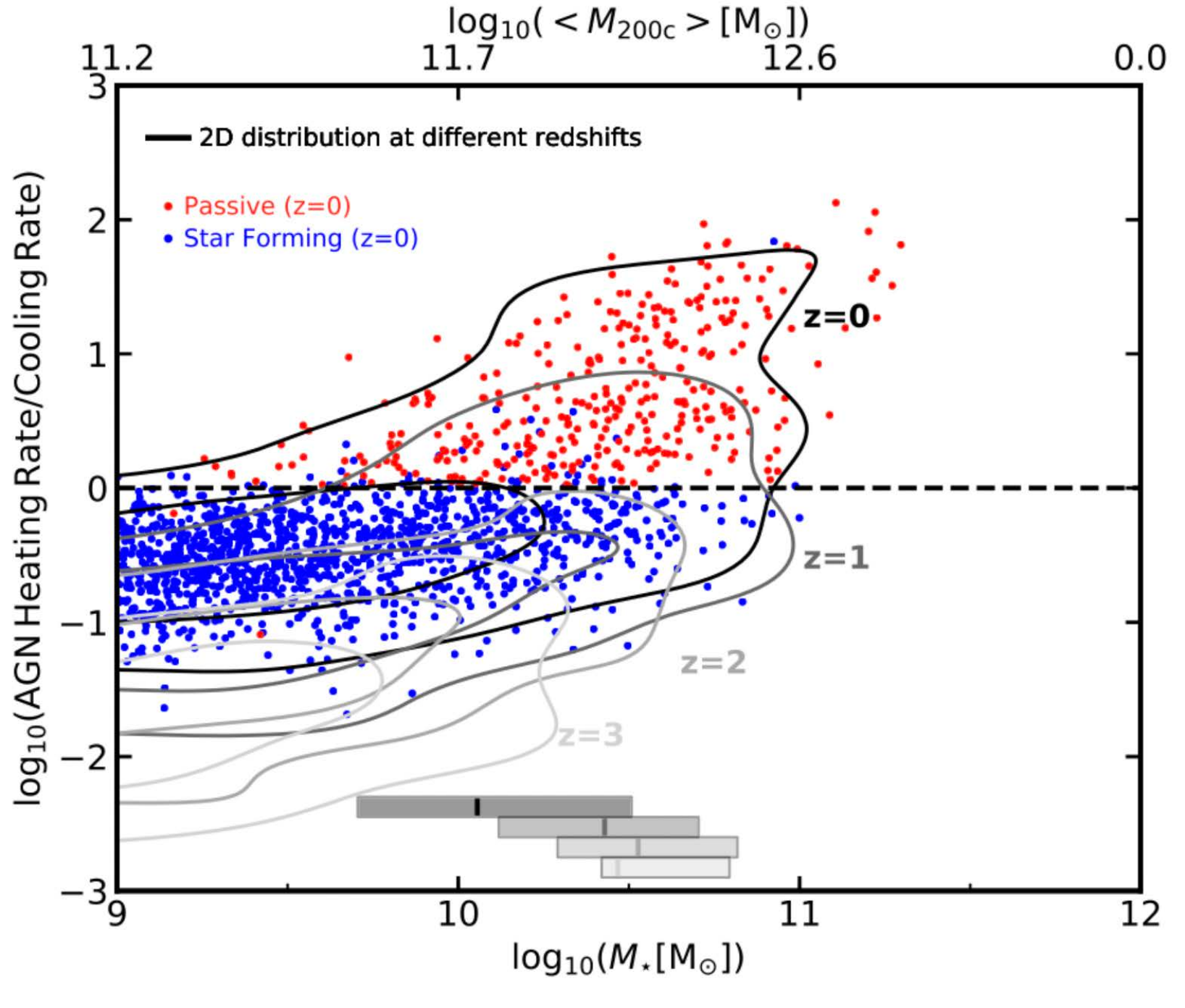}
\includegraphics[width=\textwidth]{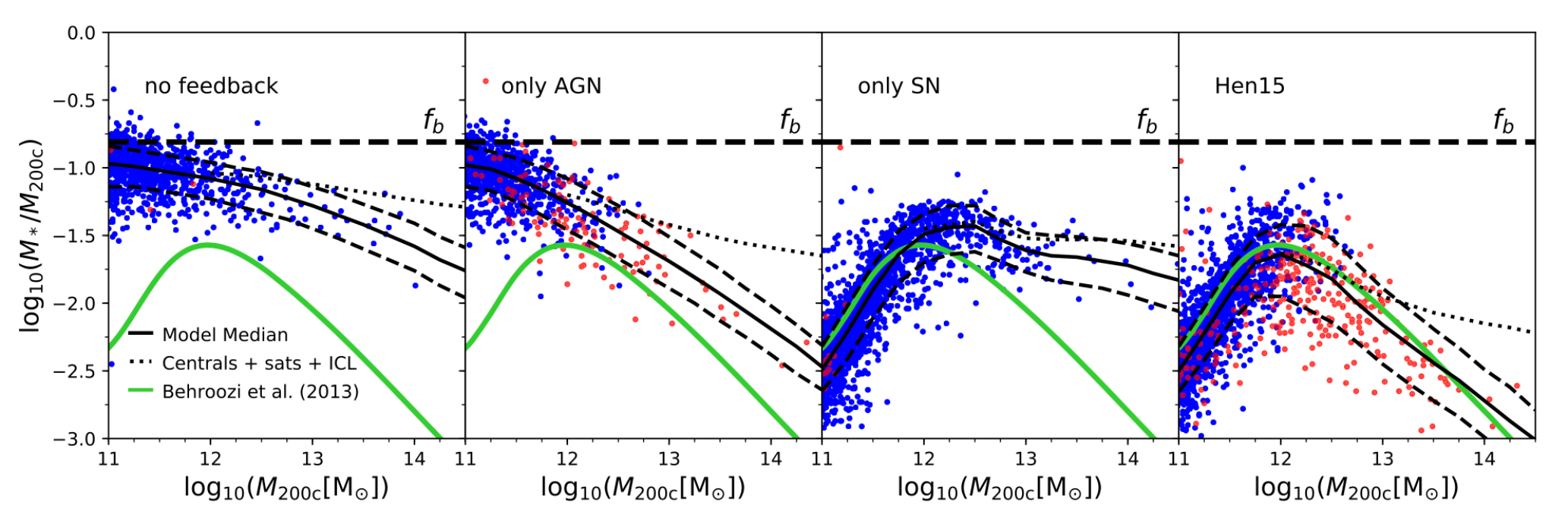}
\caption{\textit{Top Panel: } The ratio of maximum AGN heating rate to maximum cooling rate plot as a function of stellar mass. Contours show the distributions as a function of redshift, with points indicating $z = 0$ central galaxies (which are colored red for passive/ quenched and blue for actively star forming). It is clear in the model that quenching invariably involves the heating from radio-mode AGN feedback overcoming the cooling of the hot gaseous halo from free\,--\,free emission. \textit{Bottom panels:} The ratio of stellar-to-halo mass plot as a function of halo mass for a variety of model variants. From left to right: no feedback, only AGN feedback, only SN feedback, and the full model (incorporating both AGN and SN feedback). The observed relationship is recovered only in the presence of both types of feedback. Whilst AGN are essential in the model to quench central galaxies (shown as red points), obtaining the correct halo star formation efficiencies as a function of halo mass requires the two feedback channels to work in tandem. This figure is reproduced from \cite{Henriques2019}.}\label{f4}
\end{figure}

In Fig.~\ref{f4} (reproduced from \citealt{Henriques2019}) we present some consequences of this model. In the top panel, we show the ratio of radio-mode AGN heating to free\,--\,free cooling of the hot gas halo for central galaxies as a function of stellar mass. It is clear that quenched galaxies (shown in red, compared to star forming systems shown in blue) arise as a direct consequence of the heating rate becoming larger than the cooling rate. This model successfully resolves the cooling catastrophe in high mass groups and clusters, reproduces the form of the stellar mass function and its relationship to the halo mass function, and explains the existence of bimodality in the galaxy population (see also \citealt{Croton2006, Bower2006, Bower2008, Guo2011, Henriques2015, Henriques2019}). It is important to appreciate that this is the first quenching paradigm we have considered which simultaneously explains all of the big theoretical problems outlined in the Introduction to Part~I of this review series, albeit within a semi-analytical framework, lacking the detailed gas physics of hydrodynamics.

In the bottom panels of Fig.~\ref{f4}, we show the stellar-to-halo mass ratio plot as a function of halo mass in the model (from \citealt{Henriques2015}), compared to observations (from \citealt{Behroozi2013}), which is reproduced from \cite{Henriques2019}. From left-to-right, the SAM is run with: (i)~no feedback, (ii)~AGN feedback only, (iii)~SN feedback only, and (iv)~both feedback channels active. It is clear that both SN and AGN feedback are required to match observations in the model. It is worth exploring why this is the case, as it exposes some deep insights into the origin of the peak in star formation efficiency as a function of halo mass, which in turn sheds light on the peak in star formation rate density at cosmic noon (see back to Sect.~3 in Part~I).

In the absence of any feedback mechanisms, no central galaxies quench in the SAM. Moreover, central galaxies in low mass haloes transform the vast majority of their baryons into stars (i.e., they reside near the cosmic baryon fraction, which is shown as a dashed line in the lower panels of Fig.~\ref{f4}). AGN only impact galaxies with a hot gas halo that additionally have a substantial supermassive black hole (see back to Eq.~\eqref{e12}). As such, AGN feedback has no bearing on the low-mass end of the stellar-to-halo mass relationship. Some galaxies do quench at high masses, but they never reach the observed relationship in the absence of SN feedback. In the case of SN feedback only, the low-mass end of the relationship is recovered well. This is achieved via slowing the rate of star formation through SN-driven outflows. However, no central galaxies quench, and the high mass end of the relationship is poorly recovered. Finally, with both feedback mechanisms operating in the SAM, the fundamental $(M_*/M_H)$\,--\,$M_H$ relationship is recovered precisely, and quenched central galaxies are produced in approximately the right numbers (see \citealt{Henriques2013, Henriques2015, Henriques2019}).

So, how do these feedback processes work together to achieve the peak in star formation efficiency? In low mass haloes, SN are effective at ejecting gas from the system. However, once the kinetic energy of supernova ejecta equals (or is less than) the binding energy of the gas to the halo, SN feedback ceases to be effective at reducing star formation. In the model, this key transition occurs at $M_\mathrm{Halo} \sim 10^{12} \, M_{\odot}$, i.e., at around the same point that halo accretion from the ISM switches from cold to hot mode. Thereafter, galaxies are fed by cooling flows from the hot halo, and this is exacerbated by galactic fountains, caused by SN ejecta being bound to the halo.

In lieu of AGN feedback, central galaxies will continue to form stars until the hot halo gas is fully depleted, becoming over-massive in stars (relative to observational constraints). However, with radio-mode AGN feedback, quenching ensues via starvation, matching the observed relationships. Therefore, the peak in star forming efficiency occurs at the mass scale where both SN and AGN feedback are weak. The former reduces in effectiveness as mass increases, whereas the latter increases in effectiveness as mass increases. At the intermediate scale of $M_\mathrm{Halo} \sim 10^{12} \, M_{\odot}$, both forms of feedback are limited in their effectiveness, and, hence, star formation proceeds rapidly. Finally, the peak in the SFRD at cosmic noon is recovered because this is the time the number of haloes around the critical mass peaks (see back to Sect.~3 in Part~I of this review).

A very similar conclusion is reached in \cite{Dekel2019}, based on slightly different assumptions within a hydrodynamical context. Due to the enormous success of this paradigm, some version of radio-mode AGN feedback is incorporated within almost every current SAM (e.g., \citealt{Croton2006, Bower2006, Bower2008, Guo2011, Henriques2013, Henriques2015, Somerville2015}). Moreover, there is abundant observational evidence for relativistic jet heating of the CGM, which we discuss in detail in Sect.~\ref{s4} (see, e.g., \citealt{Fabian1994, Fabian1999, Fabian2006, McNamara2007, Fabian2012, HlavacekLarrondo2012, HlavacekLarrondo2015, HlavacekLarrondo2018, McNamara2007}). What remains is to incorporate this type of preventative feedback into hydrodynamical simulations (which we discuss in Sect.~\ref{s25}).

\subsection{Towards an integrated theoretical quenching paradigm}\label{s24}

\begin{figure}[ht]  
\centering
\includegraphics[width=\textwidth]{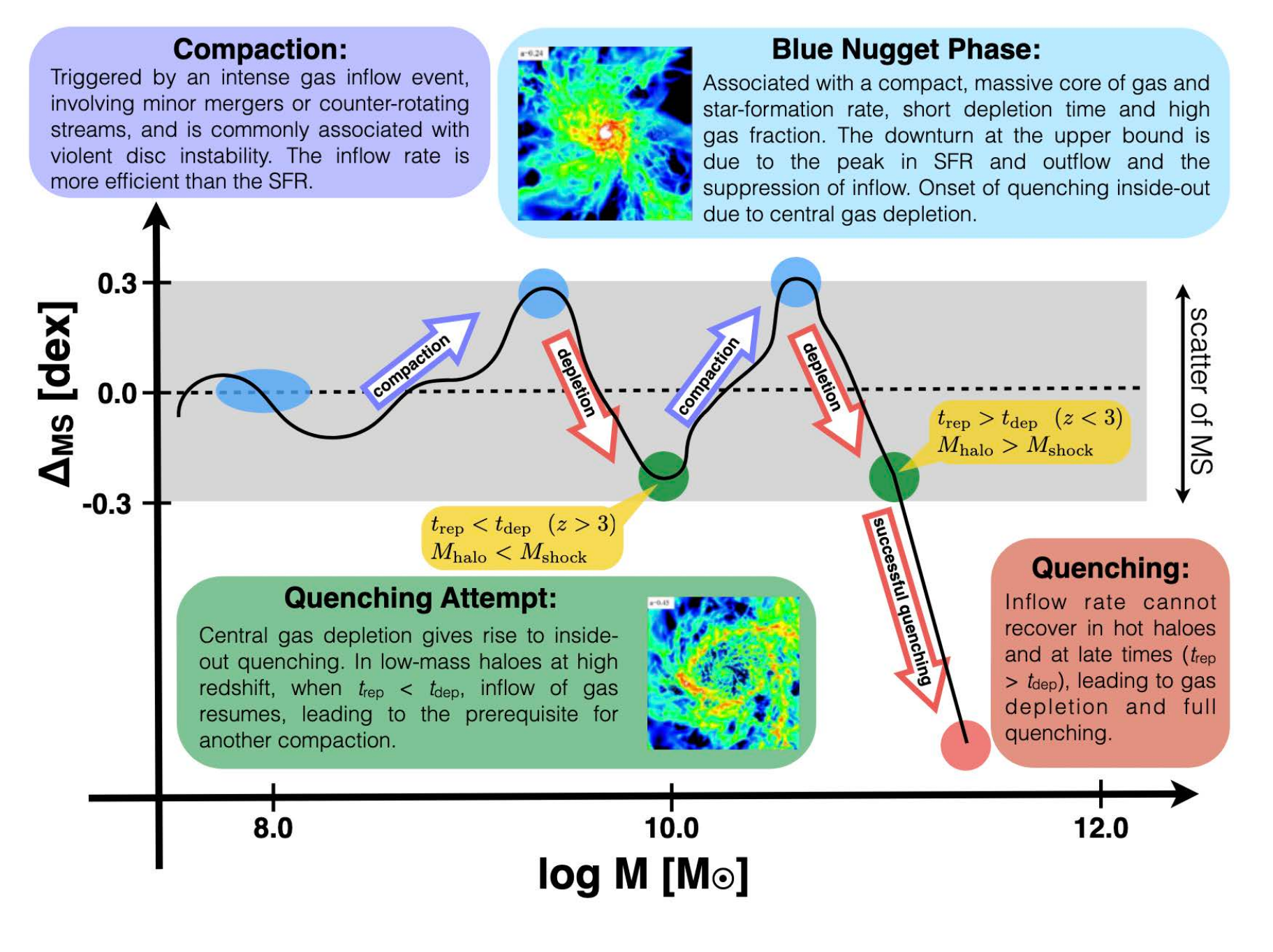}
\caption{Illustration of the lifetime of a galaxy, from active star formation in the early Universe to full quenching in the late Universe, reproduced from \cite{Tacchella2016}. The offset from the evolving star forming main sequence ($\Delta$MS) is plot as a function of stellar mass, which increases with cosmic time. Galaxies oscillate around the main sequence (shown by a dashed horizontal line) due to a number of factors, including major and minor mergers, violent disc instabilities, and clumpy cold gas accretion. Initially, these events cause the funneling of gas to the center of gas-rich systems, igniting star formation in their cores, and growing bulge structures (which is often referred to as `compaction'). Thereafter, due to gas depletion (and potentially also as a result of SN and quasar-mode AGN feedback), the star formation declines in a `quenching attempt'. What happens next depends on whether the ISM replenishment time ($t_\mathrm{rep}$) from the CGM is longer or shorter than its depletion time ($t_\mathrm{dep}$) from star formation. If $t_\mathrm{rep} < t_\mathrm{dep}$, the quenching event fails and the galaxy continues to form stars from newly accreted gas. However, if $t_\mathrm{rep} > t_\mathrm{dep}$, the quenching event succeeds and star formation is halted permanently via starvation of gas supply. The regulation of the replenishment time is governed by the effective cooling rate of the hot gas halo, which is reduced in the high mass ($M_\mathrm{Halo} \gtrsim 10^{12} M_{\odot}$) hot-mode (e.g., \citealt{Dekel2006}), and reduced further in the presence of radio-mode feedback (e.g., \citealt{Croton2006, Henriques2015}). Without preventative feedback acting to stabilize the hot gaseous halo, quenching attempts fail due to eventual gas replenishment.}\label{f5}
\end{figure}

The three quenching paradigms discussed in the previous sub-section (i.e., halo mass quenching, \citealt{Dekel2006}; quasar-mode AGN feedback, \citealt{Hopkins2006}; and radio-mode AGN feedback, \citealt{Croton2006}) clearly must interact with one another in the course of galaxy evolution. For instance, radio-mode AGN feedback posits a supermassive black hole accretion rate which scales with the hot gas halo mass, which in turn must be formed in some manner. Indeed, it is assumed in SAMs which utilize radio-mode feedback (e.g., \citealt{Bower2006, Bower2008, Henriques2015}) that a hot statistic atmosphere will develop in high mass haloes, essentially in accord with \cite{Dekel2006, Dekel2009, Dekel2019}. 

Looked at the other way around, the formation of a hot static atmosphere will reduce the star formation efficiency around the critical mass at which shocks start to form, but (crucially) at higher masses star formation will rejuvenate due to bremsstrahlung cooling (e.g., \citealt{Fabian1994, Fabian1999, Fabian2006, Fabian2012}). Hence, an additional heating source is required, even in high mass haloes, in order for the hot gas halo to maintain stability, and for quenching to hold across long cosmological timescales.

Considered in this manner, halo mass quenching and radio-mode AGN feedback are highly complementary to one another. Indeed, they are both required for an effective quenching mechanism. Hence, the odd one out in the previous subsection appears to be the merger - quasar - quenching paradigm of \cite{Hopkins2006, Hopkins2008}. Nevertheless, quasar-mode feedback may be important for \emph{triggering} quenching events, by causing galaxies to transition away from the star forming main sequence relationship. Initially, they will enhance their SFRs, before attempting quenching once the ISM is depleted (in part due to SN and/or quasar-mode AGN feedback). Nonetheless, IllustrisTNG explicitly demonstrates that low-Eddington ratio AGN-feedback alone is sufficient in a cosmological simulations to quench central galaxies with the appropriate relations as constrained by observations (see, \citealt{Nelson2018, Weinberger2018, Pillepich2018}).

In Fig.~\ref{f5}, we present an illustration of the lifetime of a galaxy, reproduced from \cite{Tacchella2016}. This represents an attempt at a holistic view of galaxy evolution, terminating in quenching. In a sense, this diagram unites the three paradigms discussed above. In the early Universe, the example galaxy forms stars due to cold gas accretion from the IGM. Due to mergers, violent disc instabilities, and clumpy gas accretion, gas gets funneled to the center of the system triggering starbursts, supernova feedback, and potentially quasar-mode AGN feedback (as in \citealt{Hopkins2006, Hopkins2008}). However, unless future gas replenishment is halted, the galaxy rejuvenates and returns to normal star formation thereafter. Only once a hot halo has developed (as in \citealt{Dekel2006}), and cooling is prevented by some form of heating (likely from radio-mode AGN feedback as in \citealt{Croton2006, Henriques2015}), does the galaxy fully quench. And even then, maintenance-mode feedback within the ISM is still required to prevent the gas produced from stellar evolution from forming stars.

In this unified picture, quenching occurs once stable virial shocks form a hot static atmosphere, which is prevented from cooling via heating from radio jets over long cosmological timescales. This is complemented by feedback within the ISM stabilizing cooling flows from stellar mass return. It is important to appreciate that these processes are sufficient to quench a galaxy conceptually and theoretically, although it remains to be seen whether in the right abundance to match the demographics of galaxy quenching from observations. Additionally, other processes like mergers and quasar-mode AGN feedback can modulate the rate of quenching. In this view, slow quenching occurs without a specific trigger, once the thermodynamics of the halo is stabilized and cyclic feedback within the ISM is initiated.

Alternatively, fast quenching may be triggered by ISM depletion or ejection, via mergers and radiatively efficient AGN feedback. But note that this is only successful in fully quenching the system if the thermodynamics of the hot gaseous halo is stabilized and maintenance mode feedback ensues in the ISM. Therefore, the formation of a hot static atmosphere, with subsequent radio heating to offset cooling flows, appears to be essential for intrinsic galaxy quenching in contemporary models. Alternatively, energetic feedback from the quasar-mode is an optional extra (potentially important for regulating quenching timescales).

In the next sub-section we turn to how these forms of feedback have been incorporated into contemporary cosmological hydrodynamical simulations.

\subsection{Modern implementations of AGN feedback in cosmological simulations}\label{s25}

Cosmological hydrodynamical simulations directly simulate the gravitation of dark matter and the gravitationally-coupled hydrodynamics (and sometimes magneto-hydrodynamics) of baryons simultaneously for a large cosmological volume, across the vast majority of cosmic history (see \citealt{Somerville2015, Vogelsberger2020} for reviews). Among the first cosmological hydrodynamical simulations to follow galaxy formation and evolution in a fully self-consistent manner, using advanced hydrodynamical solvers with sub-grid recipes calibrated to observations, were Illustris (\citealt{Vogelsberger2013, Vogelsberger2014a, Vogelsberger2014b, Torrey2014}) and EAGLE (\citealt{Schaye2015, Crain2015}). These build on a large history of earlier cosmological hydrodynamical simulations, which lacked some aspects of the later implementations (e.g., \citealt{Gnedin2000, Nagamine2004, Crain2009, Schaye2010, DiMatteo2012}). Additionally, other early cosmological hydrodynamical simulations were developed contemporaneously to EAGLE and Illustris (including, \citealt{Dubois2014, Hirschmann2014, LeBrun2014, Khandai2015}).

In this section we discuss how AGN feedback is implemented within these types of simulations. We focus on AGN feedback because this has proved critical in these types of simulations to achieve quiescence in central galaxies. The formation of virial shocks (as in \citealt{Birnboim2003, Dekel2006, Dekel2009}) is modeled directly in hydrodynamical simulations, but proves to be insufficient for quenching (although it is important for how feedback operates in practice). Additionally, the role of SN feedback is critical for regulating star formation in low-mass (generally star forming) systems, but proves incapable of producing quenching in high-mass systems. Only by employing powerful AGN feedback have contemporary cosmological simulations been able to resolve all of the key problems outlined in the Introduction to Part~I of this review.

Essentially all modern cosmological simulations seed supermassive black holes at $M_\mathrm{BH} \sim 10^{5-6} M_{\odot}$, within dark matter haloes with $M_H \sim 10^{10} M_{\odot}$. Hence, these simulations do not trace the formation of black holes and their initial growth to high masses. Thereafter, the most common (nearly ubiquitous) mode for black hole accretion is Eddington-limited, Bondi--Hoyle--Lyttleton accretion (as described in Sect.~\ref{s232}), sometimes with additional prescriptions to handle the angular momentum of accreted gas (e.g., \citealt{Dave2019}). Additionally, black holes may increase in mass through direct black hole\,--\,black hole mergers, which are determined by sub-grid recipes, initiated when the black hole particles become close enough (see, e.g., discussions in \citealt{Crain2015, Vogelsberger2014b}). The principal difference between simulations lies in how feedback from supermassive black holes is generated and coupled to the wider hydrodynamics of the simulation.

Perhaps the simplest approach is that adopted by EAGLE, which utilizes a single form of AGN feedback (\citealt{Schaye2015, Crain2015}). This is qualitatively similar to the quasar-mode feedback (discussed in Sect.~\ref{s232}). In EAGLE, a fraction of the rest-energy of accreted matter is made available for feedback within each time step (i.e., $\Delta E_\mathrm{BH} = \epsilon_f \epsilon_r \dot{M}_\mathrm{BH} c^2 \Delta t$). Ultimately, this energy will be transferred thermally to neighboring gas particles, as in \cite{Springel2005c, DiMatteo2005, Hopkins2006}. However, there is an important subtlety employed within the EAGLE prescription, which turns out to be critical for the effectiveness of this approach.

Unlike in previous hydrodynamical implementations of quasar-mode feedback, the thermal energy is not immediately released, but instead is stored (sub-grid) until a temperature change of $\Delta T = 10^{8.5} K$ is possible with a single neighboring gas particle. Thereafter, the energy is deposited stochastically, following a Monte Carlo method, with the probability of a randomly chosen gas cell receiving this energy boost given by \citep{Crain2015}:

\begin{equation}
P = \frac{E_\mathrm{BH}}{\Delta \epsilon_\mathrm{AGN} N_\mathrm{ngp} \langle m_g \rangle},
\end{equation}

\noindent where, $\Delta \epsilon_\mathrm{AGN}$ is the specific energy change due to the temperature increment, $N_\mathrm{ngp}$ is the number of neighboring gas particles, and $\langle m_g \rangle$ is the mean gas mass of the neighboring particles. This stochastic heating is shown to avoid excessive radiative losses (similar to the issue with the SN feedback prescription, discussed in Sect.~\ref{s22}). Ultimately, this feedback leads to increased pressure in the neighboring gas particles, which consequently expand, resulting in galactic-scale winds and outflows from the ISM. Furthermore, the ejected baryons may shock heat the CGM, providing a late-time heating source as well. Note that EAGLE does not incorporate any specific form of radio-mode heating. 

Whilst EAGLE does recover reasonable multi-epoch stellar mass functions (e.g., \citealt{Schaye2015, Crain2015}), there is evidence in the literature that high-mass systems within EAGLE rejuvenate their star formation in greater numbers than observed. This results in the highest mass galaxies in EAGLE not being fully quenched, unlike in observations (see, e.g., \citealt{Piotrowska2022, Goubert2024, Goubert2025}).

The first simulations to incorporate both quasar-mode and radio-mode AGN feedback were presented in \cite{Sijacki2007}. These prescriptions were adapted for use in the Illustris simulation (e.g., \citealt{Vogelsberger2013, Torrey2014, Genel2014, Vogelsberger2014a, Vogelsberger2014b}). The quasar-mode feedback operates essentially as discussed in Sect.~\ref{s232}, without the stochasticity introduced in \cite{Crain2015} for EAGLE. This is operative at $\dot{M}_\mathrm{BH}/\dot{M}_\mathrm{Edd} \geq 0.05$. In Illustris, this mode of feedback is not effective at stabilizing the hot gaseous halo, or quenching galaxies (see \citealt{Vogelsberger2014a, Vogelsberger2014b}). The lack of impact from the quasar-mode has often been argued to be due to excessive radiative losses from thermal coupling, which in EAGLE is ameliorated via stochastic bursts. 

Additionally, the thermal mode feedback in simulations like Illustris (and IllustrisTNG) is not effective because of its interaction with the ISM modeling, via the effective equation of state. In these models, star-forming gas is the component of the ISM which is most affected by quasar thermal mode feedback and yet its temperature is set by the ISM effective equation of state. Consequently, the extra internal energy accrued does not lead to star formation suppression (see discussion in \citealt{Pillepich2019}).

\begin{figure}[ht]  
\centering
\includegraphics[width=\textwidth]{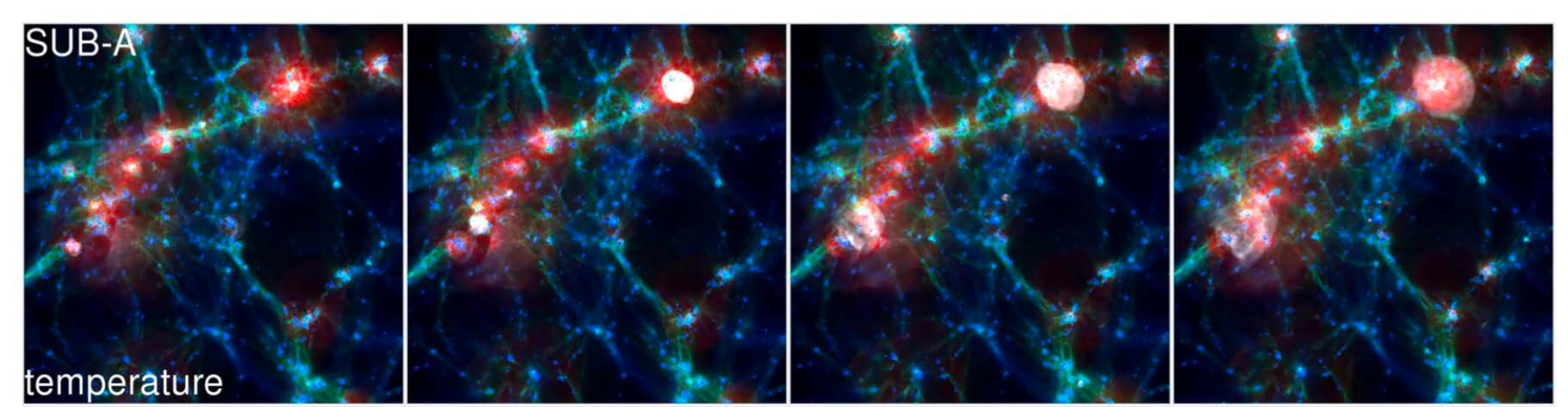}
\caption{Radio-mode AGN feedback in Illustris, reproduced from \cite{Vogelsberger2014b}. From left to right, different snapshots are shown from $z$ = 1.47 to 1.36. On each panel the gas is shown, with temperature indicated by color (blue~=~cold [$T < 10^{4} K$], green~=~warm [$T \sim 10^{4-6}K$], red~=~hot [$T > 10^{6} K$]). One can view the evolution of the radio bubble in the CGM, and see how thermal energy is dissipated. This is the primary mode of quenching operable within the Illustris simulation.}\label{f6}
\end{figure}

At lower values of the Eddington ratio ($\dot{M}_\mathrm{BH}/\dot{M}_\mathrm{Edd} \leq 0.05$), radio-mode feedback is initiated (which is qualitatively similar to introduced in Sect.~\ref{s233}, but very different in its practical implementation). Since cosmological hydrodynamical simulations do not have the resolution (or appropriate level of physics) to directly model the formation of relativistic jets, a simplified procedure is adopted. Mimicking the observed capacity of relativistic jets to create cavities in the hot CGM (e.g., \citealt{Fabian2006, McNamara2007, Fabian2012}), in Illustris, jet-feedback is modeled by stochastic bubble formation in the CGM (without a direct on-grid link to the supermassive black hole). 

Specifically, in the low-Eddington ratio regime, once the supermassive black hole increases its mass by at least 1\%, all of the energy is coupled thermally to a bubble formed in the CGM. The energy of the bubble is given by (\citealt{Sijacki2007, Vogelsberger2013}):

\begin{equation}
E_\mathrm{bubble} = \Delta E_\mathrm{radio} = \epsilon_{m} \epsilon_r \Delta M_\mathrm{BH} c^2, \quad \text{for} \quad \Delta M_\mathrm{BH} \geq 0.01 M_\mathrm{BH},
\end{equation}

\noindent where, the mechanical efficiency is set to $\epsilon_{m} = 0.35$ and the radiative efficiency is set to $\epsilon_r = 0.1$. The energy is all transferred to increased thermal energy of the bubble (i.e., $\Delta T$), and is uniformly distributed between gas cells therein. The size of the bubble is set by:

\begin{equation}
R_\mathrm{bubble} = R_0 \, \bigg( \frac{E_\mathrm{bubble}/E_0}{\rho_\mathrm{CGM}/\rho_0} \bigg)^{1/5},
\end{equation}

\noindent where, $R_0$, $E_0$ and $\rho_0$ are coupling constants, to be set via observational constraints. The exact form of this equation is motivated by the theory of radio cocoon expansion (e.g., \citealt{Scheuer1974}). Note that increased energy injections lead to larger bubbles, but increased densities in the CGM lead to smaller bubbles at a fixed total energy (as one might expect from simple thermodynamical considerations). This procedure approximates $PdV$-work done by the jet on the CGM, which forms a cavity (i.e., the bubble). In turn, this energy is dissipated (via expansion, bubble mergers, and shocks) into the wider CGM, providing a heating source to offset bremsstrahlung cooling of the hot gaseous halo. In Fig.~\ref{f6}, we show a series of snapshots of bubble formation, expansion, and dissipation from the Illustris simulation (reproduced from \citealt{Vogelsberger2014b}).

Radio bubbles are not formed at the location of the central black hole, but instead are set a faction of the virial radius away from the galaxy, in a random direction (see \citealt{Vogelsberger2014b} for full details). This is intended to mimic the effects of a large-scale relativistic jet, which typically does not strongly interact with the ISM, instead depositing vast quantities of energy into the CGM at large distances from the galaxy (see, e.g., \citealt{Fabian2012} for an observational review). 

Illustris is reasonably successful at reproducing the multi-epoch stellar mass functions from observations, as well as reproducing many other features of the galaxy population (e.g., \citealt{Vogelsberger2014a, Vogelsberger2014b}). Importantly for our present concern, Illustris successfully quenches high mass galaxies, although not in perfect accord with observations (e.g., \citealt{Bluck2016, Donnari2019, Donnari2021a, Donnari2021b, Piotrowska2022, Bluck2023}). However, the radio-mode feedback is overzealous in the sense that it tends to dramatically deplete the CGM of gas, resulting in very low X-ray emission from the high mass haloes of large groups and clusters (e.g., \citealt{Genel2014, Vogelsberger2014b, Suresh2015, Nelson2015}). This is in stark contradiction to observations (e.g., \citealt{Fabian2006, Fabian2012, HlavacekLarrondo2015, HlavacekLarrondo2018}).

Due in part to the issue with the radio-mode AGN feedback prescription in Illustris (and to discrepancies in the quenched fraction - stellar mass relation with observations), a series of follow-up simulations were produced by the same team, known as IllustrisTNG (see \citealt{Nelson2018, Pillepich2018, Springel2018, Naiman2018, Marinacci2018}). These simulations also differ by having three different volumes (with varying levels of resolution), and via the inclusion of magnetism into the hydrodynamical solver. However, for quenching, the main difference is that the radio-mode feedback of Illustris is replaced with a new `kinetic' feedback mode, which is also only operable at low Eddington ratios (as with radio-mode in Illustris). This is included again with the quasar-mode feedback, which remains essentially unchanged.

In IllustrisTNG a broader interpretation of mechanical feedback is taken to that of the classical radio-mode. In particular, observations of hot coronal winds, small-scale jets, and lower-power outflows were considered in building this model. Consequently, the low-accretion rate AGN-feedback mode of IllustrisTNG is not meant to replicate jet feedback (as observed in galaxy clusters and implemented in Illustris). Rather, it was phenomenologically inspired by red geysers (e.g., \citealt{Cheung2016}) and physically implemented as a subgrid model for high-velocity accretion-disk winds, or small-scale jets from low-luminosity SMBHs. 

Kinetic-mode feedback occurs in IllustrisTNG when \citep{Weinberger2017, Weinberger2018, Zinger2020}:

\begin{equation}
f_\mathrm{Edd} \equiv \frac{\dot{M}_\mathrm{Bondi}}{\dot{M}_\mathrm{Edd}} \leq \chi , \quad \text{where} \quad \chi = \mathrm{min} \bigg( 0.002\bigg(\frac{M_\mathrm{BH}}{10^{8} M_\odot}\bigg)^2, 0.1 \bigg).
\end{equation}

\noindent Hence, the kinetic-mode is suppressed at low SMBH masses, and becomes dominant at high SMBH masses. However, the radiatively efficient quasar-mode is still possible if the accretion rate is sufficiently high. The kinetic-mode in IllustrisTNG functions by a direct transfer of a fraction of the rest-energy accreted to pure kinetic energy of neighboring gas cells. Specifically, the total momentum injection from kinetic-mode AGN feedback is given by:

\begin{figure}[ht]  
\centering
\includegraphics[width=\textwidth]{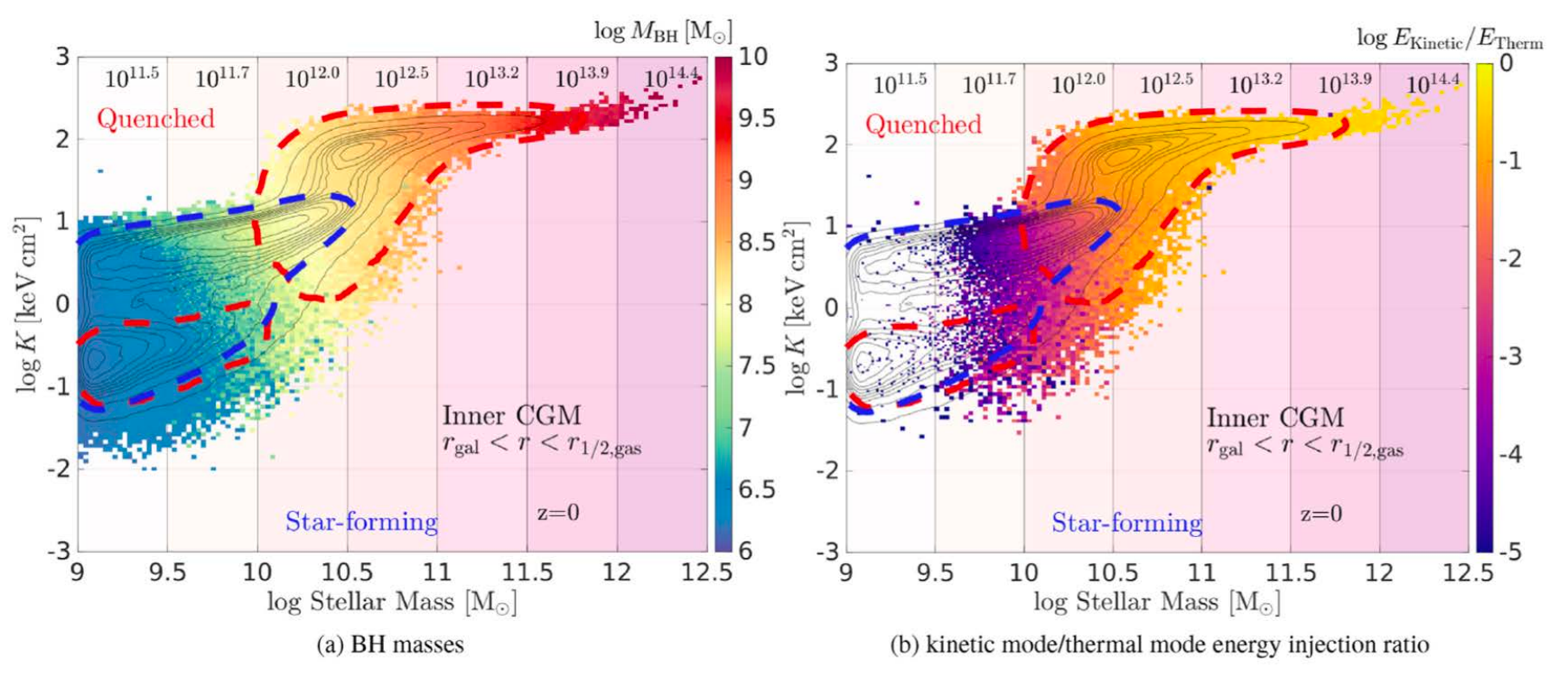}
\caption{Astrophysical entropy ($K = k_BT/n^{2/3} $) vs. stellar mass, color coded by supermassive black hole mass (left panel) and the ratio of cumulative energy from kinetic-to-thermal mode feedback (right panel), reproduced from \cite{Zinger2020}. Quenched and star forming contours are shown, as labeled on each plot. It is clear that quenched galaxies have higher CGM entropies, higher black hole masses, and a higher ratio of kinetic-to-thermal mode feedback. This demonstrates that it is the kinetic feedback which drives quenching in this simulation, and that this works (in part) by stabilizing the hot gaseous halo by increasing its entropy. Not shown, a similar increase in entropy is engendered in the ISM as well. Hence, the kinetic-mode feedback in IllustrisTNG quenches galaxies via both decreasing ISM gas fractions (due to reduced cooling flows from the CGM) and decreasing the star formation efficiency of the ISM (due to increased velocity dispersion and turbulence). }\label{f7}
\end{figure}

\begin{equation}
\Delta {\bf p}_\mathrm{inj} = \sum_i \left\{ m_i  \sqrt{\frac{2\Delta E_\mathrm{inj} \, w({\bf r})}{\rho_g}} \, \right\} {\bf n}, \quad \text{where} \quad \Delta E_\mathrm{inj} = \epsilon_k \dot{M}_\mathrm{BH} c^2 \Delta t,
\end{equation}

\noindent and where, the summation is performed over neighboring gas cells (labeled by $i$). The total energy available for injection into kinetic energy of gas cells ($\Delta E_\mathrm{inj}$) is defined above, and represents a fraction of the total rest-energy accreted (where $\epsilon_k = 0.2$). All of the momentum is given in the same (randomly assigned) direction, that of the unit vector (${\bf n}$). Whilst this method violates strict conservation of momentum, the constraint is recovered over long time periods due to the randomness of the direction of momentum injection. $w({\bf r})$ indicates a smoothing kernel, which is used to distribute the available energy between gas cells. Finally, $\rho_g$ is the density of the surrounding gas.

The above functional form may not look intuitive. To help with this, one can restructure the equation for a single gas cell as follows:

\begin{equation}
\Delta {\bf v}_\mathrm{cell} = \sqrt{\frac{2 E_\mathrm{cell}}{m_\mathrm{cell}}} \, {\bf n}, \quad \text{where} \quad E_\mathrm{cell} = \bigg( \frac{m_\mathrm{cell \, }w({\bf r}_\mathrm{cell})}{\rho_g} \bigg) \, \Delta E_\mathrm{inj},
\end{equation}

\noindent and where the momentum injection of a single cell is determined by the energy made available to the cell (conserving energy, i.e. $E_\mathrm{cell} = (1/2)M_\mathrm{cell}\, v_\mathrm{cell}^2$). The smoothing kernel is used together with the gas density and cell mass to set the fraction of the available energy each cell is given (see \citealt{Weinberger2017} for full details). Explicitly, the kernel is normalized such that: $\sum_i m_i w_i = \rho_g$. Similarly to radio-mode feedback in Illustris, the energy accumulated is not immediately dissipated. Instead, a minimum threshold is required to inject a momentum kick. Explicitly, this is set to $\Delta E_\mathrm{min} = f_\mathrm{burst}(1/2)m_\mathrm{enc}\,\sigma_\mathrm{DM}^2 $, where $\sigma_\mathrm{DM}$ is the 1D velocity dispersion of the dark matter halo and $m_\mathrm{enc}$ is the gas mass within the feedback region. This enables more powerful, though less frequent, kinetic feedback events to occur (regulated by the free `burstiness' parameter, $f_\mathrm{burst}$).

Unlike the radio-mode feedback in Illustris, the kinetic-mode feedback in IllustrisTNG originates around the supermassive black hole. Consequently, it impacts the ISM directly, in addition to the CGM (due to gas flows beyond the galaxy). This leads to a number of different characteristics compared to Illustris. This feedback mechanism has been shown to be highly successful at reproducing the multi-epoch stellar mass function, the properties of quenched galaxies, and it fully resolves the CGM gas depletion problem caused by radio-bubbles in Illustris (see \citealt{Nelson2018, Pillepich2018, Springel2018, Weinberger2018, Zinger2020, Piotrowska2022}).

In Fig.~\ref{f7}, we show the astrophysical entropy ($K = k_{B}T/n^{2/3}$, where $T$ is temperature and $n$ is particle number density) plot as a function of stellar mass for $z=0$ haloes from IllustrisTNG, reproduced from \cite{Zinger2020}. The left-hand panel color codes the plane by supermassive black hole mass, and the right-hand panel color codes the plane by the ratio of kinetic-to-quasar cumulative energies ($E_\mathrm{kin}/E_\mathrm{QSO}$). Additionally, the region where quenched and star forming galaxies reside is displayed by red and blue dashed contours, respectively. It is clear that in IllustrisTNG, quenching is achieved at high stellar masses, specifically when the central black hole mass and cumulative contribution from the kinetic-mode feedback are both high. Moreover, quiescent galaxies reside in haloes which have higher hot gas entropies than their star forming counterparts. Hence, the quenching of galaxies in IllustrisTNG is achieved (at least in part) by the stabilization of the hot gaseous haloes by entropy injection (see \citealt{Zinger2020} for full details). Actually, the entropy of the ISM is also increased by kinetic feedback in IllustrisTNG, reducing the star forming efficiency (SFE) in addition to the cold gas fraction (see \citealt{Zinger2020, Piotrowska2022}).

\begin{table}[ht]
\caption{Summary of AGN feedback mechanisms in cosmological simulations}\label{t1}
\begin{tabular}{l c cc ccc}
\toprule
\textbf{Simulation} & \textbf{Box [cMpc]} &
\multicolumn{2}{c}{\textbf{Quasar Mode}} &
\multicolumn{3}{c}{\textbf{Radio Mode}} \\
\cmidrule(lr){3-4} \cmidrule(lr){5-7}
& & \textbf{Thermal} & \textbf{Kinetic} & \textbf{Bubble} & \textbf{Kinetic} & \textbf{X-ray} \\
\midrule
Illustris         & 100      & \checkmark & --         & \checkmark & --         & -- \\
IllustrisTNG      & 50--300   & \checkmark & --        & --         & \checkmark & -- \\
EAGLE             & 100      & \checkmark & --         & --         & --         & -- \\
SIMBA             & 100      & --         & \checkmark & --         & \checkmark & \checkmark \\
Horizon-AGN       & 100      & \checkmark & --         & --         & \checkmark & -- \\
Magneticum        & 68--500   & \checkmark & --        & \checkmark & --         & -- \\
Cosmo-OWLS        & 100      & \checkmark & --         & --         & --         & -- \\
BAHAMAS           & 400      & \checkmark & --         & --         & --         & -- \\
FIREbox           & 40       & \checkmark & \checkmark & --         & \checkmark & -- \\
MassiveBlack-II   & 100      & \checkmark & --         & --         & --         & -- \\
Romulus           & 25       & \checkmark & --         & --         & --         & -- \\
FLAMINGO          & 1000--2800 & \checkmark & --       & --         & \checkmark         & -- \\
\bottomrule
\end{tabular}
\textbf{Notes:} Feedback is grouped by accretion mode: quasar-mode (high-Eddington ratio) and radio-mode (low-Eddington ratio). Check-marks indicate the inclusion of a given feedback mechanism. Below each general mode the precise implementation strategy is listed (e.g., radio-mode feedback through thermal bubble injection into the CGM).
\end{table}

In addition to the quasar and radio/ kinetic modes of Illustris and IllustrisTNG, both of these simulation suites also employ a `radiative-like' third feedback mode, whereby the cooling of the gas around SMBHs is modulated by the radiation of the SMBH (see, \citealt{Vogelsberger2013, Pillepich2018b}). However, this mode has little impact on the quenching of galaxies as a whole.

There are a host of other cosmological hydrodynamical simulations which incorporate AGN feedback, including: SIMBA \citep{Dave2019}, Horizon-AGN \citep{Dubois2014}, Magneticum \citep{Hirschmann2014}, BAHAMAS \citep{McCarthy2017}, MassiveBlack-II \citep{Khandai2015}, Cosmo-OWLS \citep{LeBrun2014}, Romulus \citep{Tremmel2017}, FREbox \citep{Feldmann2023}, and FLAMINGO \citep{Schaye2023}. These all use some combination of: (i)~radiatively efficient (quasar-mode) feedback and (ii)~radiatively inefficient (radio-mode/ mechanical) feedback, which are broadly similar to the prescriptions in the simulations we have already discussed (albeit with numerous subtle differences). 

Innovations beyond what has been discussed already in this section include the use of torque-limited and angular momentum dependent black hole accretion rates (see \citealt{Dave2019, Tremmel2017, Tremmel2019}) and the inclusion of X-ray heating ($L_X = \epsilon_X \dot{M}_\mathrm{BH} \, c^2$) from intermediate Eddington-ratio accretion (see \citealt{Dave2019}). Note also that some simulations model quasar mode feedback kinetically via radiation pressure, as opposed to the thermal injection used in the simulations discussed above. See Table~\ref{t1} for a summary of contemporary cosmological simulations, indicating which AGN feedback mechanisms they incorporate, and how these are implemented.

In summary, essentially all contemporary cosmological hydrodynamical simulations incorporate some form of AGN feedback (and several include multiple modes). Indeed, AGN feedback proves absolutely critical in simulations for stabilizing cooling flows in high mass haloes, and ultimately quenching central galaxies. In simulations which utilize separate AGN feedback modes as a function of Eddington ratio, it is most commonly the low accretion mode which dominates quenching in the simulation (e.g., \citealt{Vogelsberger2014b, Weinberger2018, Nelson2018, Dave2019, Zinger2020}). This may be counterintuitive, since low accretion rate implies low luminosity (or, equivalently, power). However, it is the total energy (i.e., the time integral of power: $E = \int\,P(t)\,dt$) which does work on the CGM to offset cooling flows. Whilst `radio-mode' style feedback is typically less powerful than the radiatively efficient `quasar-mode', it can operate for longer and, moreover, couple more effectively to the CGM. Crucially, the mode of energy injection (e.g., thermal vs. kinetic, or continuous vs. bursty) is as important as the total energy for achieving feedback-related outcomes. Identical energy budgets can lead to very different outcomes in terms of star formation and quenching.

\subsection{Dynamical stabilization and morphological quenching}\label{s26}

Before moving on to the additional avenues for quenching available to satellite galaxies (see next section), we pause here to consider an alternative quenching paradigm to that of AGN feedback, specifically dynamical (or morphological) quenching. Generally speaking, dynamical quenching refers to processes which may reduce star formation in galaxies as a result of stabilizing the cold ISM gas against gravitational collapse (see, e.g., \citealt{Martig2009, Gensior2020}). Before explaining some of the details of these processes, it is important first to understand what morphological quenching is \emph{not} capable of achieving. 

Morphological/ dynamical quenching does not expel gas from the ISM, and nor does it prevent gas accretion into the ISM from either cold gas streams or CGM cooling flows. As such, this type of quenching does not offer a solution to the group/ cluster cooling problem. Nonetheless, it may still be important for explaining why galaxies quench, and, hence, accounting for bimodality in galaxy properties. Furthermore, the reduction in star formation as a result of morphological and structural transformations within galaxies may help to explain some of the empirical results discussed in Part~I of this review series.

Following a radial plane wave perturbation of a thin disc, \cite{Toomre1964} derive the following criterion for gravitational collapse, and hence star formation (see also \citealt{Elmegreen2002, Kawata2007, Martig2009}):

\begin{equation}
\text{Stability  Criterion I:} \ Q > 1, \quad \text{where} \quad Q = \frac{\kappa(r) \sigma(r)}{\epsilon G \Sigma(r)},
\end{equation}

\noindent and where, $\Sigma$ is the surface mass density of the disc, $\sigma$ is the velocity dispersion in the disc, and $\kappa$ is the epicyclic (radial) frequency, which is related to the angular velocity of the disc ($\Omega(r) \equiv V_c(r)/r$, where $V_c(r)$ is the circular velocity of the disc at radius, $r$). Finally, $\epsilon$ is a constant, set by the structure of the disc (which is shown to be equal to precisely $\pi$ for a gas disc, and equal to 3.36 for a stellar disc, see \citealt{Toomre1964}). Although the derivation is complex, the result is physically intuitive --- collapse may only occur if gravity can overcome the non-rotational kinetic energy in the rotating frame (i.e., as a result of velocity dispersion of cool gas).

In \cite{Martig2009} another important stability criterion is introduced (based on prior work in \citealt{Hunter2001}). This is computed by considering the density of gas ($\rho_g$) required to overcome the local tidal force on a gas clump. Explicitly, this is given by (see \citealt{Martig2009}): 

\begin{equation}
\text{Stability Criterion II:} \ \rho_g < \rho_\mathrm{tidal}, \quad \text{where} \quad \rho_\mathrm{tidal} = \bigg( \frac{3 \Omega(r) \, r}{2\pi G} \bigg)  \bigg( \frac{\Omega(r)}{2r} + \frac{d\Omega(r)}{dr} \bigg),
\end{equation}

\noindent where all parameters are defined as above. The first term in the final bracket indicates the impact of the centrifugal force in the rotating frame, and the second term indicates the impact of the local shear force. This expression may be derived in a straightforward manner by recalling that the centripetal acceleration is given by $a_c = r\Omega(r)^2$, then set the differential tidal force equal to the self-gravity force of a clump within the disc, and finally re-phrase in terms of density (i.e., $a_\mathrm{self} = -(4/3)\pi \rho r$). Note that in \cite{Martig2009} only the shear term is retained (as it is expected to dominate in the most concentrated mass distributions).

Hence, for a giant molecular cloud (GMC) to collapse under its own weight, the non-rotational (dynamically hot) kinetic energy must be low enough to satisfy Criterion~I, and the density of the GMC must be higher than the critical tidal density to satisfy Criterion~II. In progressively more spheroidal galaxies, the stellar and gas dispersion rises (e.g., \citealt{Forster2006, Forster2009, Cappellari2013a, Brownson2022}), making bulge regions more stable to star formation at a fixed gas mass than discs. Furthermore, galaxies with substantial central mass concentrations produce stronger tidal forces on their discs, which can stabilize GMC collapse out to large galacto-centric radii (i.e., within the inner, and potentially even outer, disc).

In a series of zoom-in hydrodynamical simulations performed in \cite{Martig2009, Gensior2020, Gensior2021}, it is shown that systems with more massive bulge structures are progressively more stable to GMC collapse, and hence more likely to quench. This occurs despite no substantial change in the ISM cold gas content. Hence, it has been claimed that the dynamics of galaxies may explain quenching alone, and, in so doing, further explain the deep connections seen between the quenching of central galaxies and their morphologies, central mass densities, and central kinematics (e.g., \citealt{Driver2006, Cameron2009a, Wuyts2011, Bell2008, Bell2012, Wake2012, Omand2014, Bluck2014, Fang2013, Cheung2012, Lang2014, Bluck2022}).

So, was all the work we have done in trying to understand AGN feedback, SN feedback, and virial shocks in vain? Well, no! To help appreciate why, it is useful to recall that the dynamical quenching paradigm emerges in the context of idealized hydrodynamical simulations. This has both positive and negative impacts. On the one hand, the higher resolution in modern idealized simulations enable these dynamical effects to be traced with much greater accuracy than is possible in cosmological simulations. On the other hand, cosmological simulations are fully connected to the larger Universe and incorporate all of the relevant gas inflow physics (e.g., from CGM cooling or cold gas streams, in addition to major and minor mergers).

In the absence of any preventative feedback, CGM cooling (or cold gas streams at lower halo masses) will continue to replenish the ISM. In a scenario in which a galaxy is fully dynamically stabilized (and thus, quenched), the accreted gas will remain in a stagnant (non-star forming) form within the ISM. For a high mass halo, typically an order of magnitude more baryons reside in the CGM than in the ISM (e.g., \citealt{Fabian2006, Fabian2012}). Without preventing their cooling and collapse into a galaxy, one anticipates the ISM mass to increase dramatically, rising significantly compared to that of a star forming system at a similar stellar mass. However, contrary to this expectation, numerous observational studies demonstrate that the gas content of the ISM in quenched galaxies is substantially \textit{lower} than that of actively star forming galaxies (as will be discussed in Sect.~\ref{s61}; e.g., \citealt{Saintonge2016, Saintonge2017, Piotrowska2020, Brownson2020}). Hence, morphological quenching requires some other physical mechanism to reduce the gas fraction in the ISM, as explicitly seen in observations. 

Form a more theoretical perspective, even if the gas fractions of quenched galaxies were to rise, it is highly unlikely that dynamical stabilization would remain effective at these much higher gas masses. Furthermore, in the hypothetical scenario of a  gas-rich, non-star forming ISM, this equilibrium would be highly sensitive to major and minor mergers, which are well known to disturb gas from discs, generate galaxy bars, and drive central starbursts (e.g., \citealt{Springel2005c, DiMatteo2005, Hopkins2006, Hopkins2008}).

Therefore, morphological quenching alone cannot fully explain observations, and (as noted before) has no impact on the group/ cluster cooling problem. It is also highly limited in its ability to explain the low cosmological baryon-to-star conversion efficiency, since the vast majority of baryons do not reside in the ISM of galaxies (see \citealt{Shull2012, Fukugita2004}). Nonetheless, numerous observational studies do show that galaxies quench without fully depleting their cold gas ISM (e.g., \citealt{Saintonge2016, Saintonge2017, Lin2019, Ellison2020, Piotrowska2020, Brownson2020, Piotrowska2022}). Consequently, whilst morphological stabilization (from gas dispersion and/ or tidal torques) cannot be the whole story with respect to quenching, it may well play an important role in explaining the reduction in SFE along the quenching sequence. The potential role of dynamical stabilization is likely underestimated in contemporary cosmological simulations, due to a lack of resolution and the capacity to bury small-scale efficiency variations within the star formation prescription (see back to Sect.~\ref{s21}). We will revisit dynamical and morphological quenching again in Sect.~\ref{s6}, where we will compare theoretical expectations to a number of observational works.

\section{Quenching in theory and simulations II: Environment}\label{s3}

In this section we review many of the theoretically proposed mechanisms for quenching which operate explicitly as a function of galaxy environment. Most of the environmental quenching routes are only available to satellite galaxies. This is because satellites (by definition) are in orbit of central galaxies, moving with high relative velocities to the central galaxy's CGM (or for satellites in cluster, the ICM). As we will see, this enables various dynamical and hydrodynamical effects which may strip galaxies (and their surrounding haloes) of gas, which is required for future star formation. Additionally, simply by virtue of their location within their parent dark matter halo, satellites are expected to experience different evolutionary pathways, which may impact star formation and quenching in these systems. For dedicated reviews on galaxy environment see \citealt{Boselli2006, Blanton2009, Cortese2021}.

\subsection{Galaxy mergers and the formation of galaxy clusters}\label{s31}

The literature is divided on whether or not galaxy mergers should count as a potential intrinsic or environmental route to quenching. In the end, this is largely semantics and a strong case may be made either way. Indeed, we have already discussed mergers in the context of triggering quasar-mode ejective AGN feedback (see back to Sect.~\ref{s232}; e.g., \citealt{Springel2005c, DiMatteo2005, Hopkins2006, Hopkins2008}). However, while galaxy mergers can increase SMBH mass through coalescence and gas inflows, they may also lead to black hole displacement or ejection via gravitational-wave recoil, thereby introducing scatter into SMBH–galaxy scaling relations and complicating simple growth scenarios (see, \citealt{Ciotti2001a, Ciotti2007a}). Additionally, mergers may trigger starbursts, which invariably lead to gas depletion (through direct consumption of gas in star formation), as well as potentially quenching systems via enhanced SN feedback, which is expected to be more effective in lower mass haloes (see Sect.~\ref{s22}). 

Furthermore, mergers lead to increasing the mass of supermassive black holes (via black hole mergers and enhanced cold-mode accretion in gas-rich systems), and to a direct increase in the halo mass and stellar mass of galaxies. Increased halo mass has a direct bearing on the formation of the hot gaseous halo via virial shocks (e.g., \citealt{Dekel2006, Dekel2009}), and increased cental black hole mass has a direct bearing on the potential for triggering radio-mode feedback (e.g., \citealt{Croton2006, Bower2006, Bower2008, Henriques2015}). Furthermore, mergers of gas-rich systems tend to shock heat a significant fraction of the ISM gas to close to the virial temperature (e.g., \citealt{Cox2006, Moster2011, Torrey2012, Moreno2019}), helping to build the hot gaseous halo around massive galaxies.

Some of these processes are naturally interpreted as `intrinsic' (e.g., SN and AGN feedback), even though they may be triggered by a merger (which is itself, of course, environmental). On the other hand, some of the more direct consequences of mergers (like gas depletion in a starburst) are often considered to be environmental. It is important to appreciate that mergers are one route to galaxy evolution, the other being secular evolution. Ultimately, the principal difference between the two modes is simply how clumpy the accretion from the IGM into the galaxy is, i.e., whether it forms a continuous stream or arrives more discretely within bound sub-haloes (e.g., \citealt{Conselice2003a, Conselice2003b, Kormendy2004, Conselice2014}).

One might anticipate galaxy mergers to be more common in higher density environments, since these environments have (by definition) a higher density of galaxies and, hence, a smaller average galactic separation. However, this is only partially true. Of course, fully isolated galaxies in the field cannot merge. However, at the other extreme (i.e., in galaxy clusters), the typical relative velocity between satellites is extremely high ($v_\mathrm{rel} \sim 1000$\,km/s). This typically leads to the kinetic energy of satellite\,--\,satellite pairs being higher than their binding energy, and, hence, interactions no longer lead to merging (e.g., \citealt{Binney2008, Moreno2013}). Consequently, it is actually in intermediate density environments (i.e., galaxy groups) that mergers proceed most effectively (see, e.g., \citealt{Conselice2006, McIntosh2008, Bundy2009, Ellison2010, Lin2010, Moreno2013}).

\subsubsection{Hierarchical assembley \& the theory of cluster formation}

Within the $\Lambda$CDM paradigm, the growth of structure proceeds hierarchically, with dark matter haloes assembling through a combination of smooth accretion and mergers (see, e.g., \citealt{Frenk1988, Efstathiou1990, Cole2000, Springel2005a, Vogelsberger2020}). This framework makes quantitative predictions for galaxy merger rates as a consequence of the underlying merger rates of dark matter haloes. 

The overall accretion rate of matter into dark matter haloes is found in various simulations to scale as (e.g., \citealt{Dekel2009, Fakhouri2010, Lilly2013}):

\begin{equation}
\dot{M}_H \sim (M_H)^{1.1} \, \cdot \,  (1+z)^{2 - 2.5} \, ,
\end{equation}

\noindent whereby more massive haloes increase in mass faster than less massive haloes at a fixed redshift, and all haloes accrete much faster in the early Universe than at later cosmic times.

N-body simulations, SAMs, and cosmological hydrodynamical simulations find that major merger rates per unit cosmic time increase strongly with redshift, but only weakly with halo mass. For dark matter haloes, the mean major ($\mu \gtrsim 1/3$) merger rate per descendant halo is well described by (e.g., \citealt{Fakhouri2010, Genel2010}):

\begin{equation}
\frac{dN_{\rm merger}}{dt} \propto \left(\frac{M_{\rm h}}{10^{12}\,{\rm M_\odot}}\right)^{0.1-0.15} (1+z)^{2-2.5},
\end{equation}

\noindent although the precise normalization and exponent depend on the definition of
mass ratio, whether the rate is measured per unit redshift or per unit time,
and whether halo - halo or galaxy - galaxy mergers are considered.

Galaxy - galaxy merger rates in hydrodynamical simulations show broadly similar
redshift evolution. For example, Illustris finds a strong scaling with redshift,
approximately $(1+z)^{2.4-2.8}$, and a relatively weak dependence on descendant
stellar mass except at the highest stellar masses, where the mass dependence
steepens (e.g. \citealt{RodriguezGomez2015}).

These numerical results have important implications for quenching. At high redshift, frequent gas-rich major mergers can drive rapid black-hole growth, trigger compact starbursts, and potentially induce quasar-mode feedback. However, towards lower redshift the declining merger rate, combined with longer dynamical-friction and orbital decay timescales, makes major mergers unlikely to be the dominant quenching channel for the majority of galaxies. Instead, merger-driven quenching is more naturally associated with particular regimes, such as massive galaxies, gas-rich high-redshift systems, and group-scale environments where relative velocities remain low enough for mergers to proceed efficiently.

To understand why the satellites in high mass groups and clusters do not simply merge with their centrals, it is important to review why mergers occur at all. Ultimately, mergers arise from dynamical friction, which is caused by a mass moving through a medium (often referred to as a `sea') of small massive particles (i.e., the dark matter halo). 

The theory of dynamical friction was developed by S. Chandrasekhar (\citealt {Chandrasekhar1943}). Further developments were made to model satellite galaxy orbits around centrals (see \citealt{Binney2008} and references therein). The merger time is the orbital decay time over which dynamical friction removes orbital energy and angular momentum from the satellite, allowing it to sink toward the central galaxy. Explicitly, this is given by (e.g., \citealt{Binney2008, Boylan2008}):

\begin{align}
t_\mathrm{merger} = f(\ln(\Lambda))\, \bigg( \frac{M_\mathrm{cen}}{M_\mathrm{sat}} \bigg) \, \tau_\mathrm{dyn} \ \text{where} \ \tau_\mathrm{dyn} \approx \frac{R_{200c}}{V_{200c}} = \bigg( \frac{R_{200c}^3}{G M_\mathrm{cen}} \bigg)^{1/2} \sim \sqrt{\frac{1}{200G\bar{\rho}_c(z)}},
\end{align}

\noindent and where, $M_\mathrm{cen}$ is the central galaxy mass, $M_\mathrm{sat}$ is the satellite galaxy mass, $\tau_\mathrm{dyn}$ is the dynamical time of the parent halo, and $f(\ln{\Lambda})$ indicates a complex function of the Coulomb logarithm (see \citealt{Binney2008} for details). The dynamical time may be expressed in terms of the virial properties of the halo, which is ultimately a function of the background cosmology alone (specifically, the critical density, $\bar{\rho}_c(z)~\propto~H(z)^2$). 

The most interesting part about the merger timescale is its inverse dependence on the merger ratio ($\mu \equiv M_\mathrm{sat} / M_\mathrm{cen}$). More unequal mass mergers take longer to complete, such that interactions between very unequal mass galaxies may take a very long time indeed. This is further exacerbated by the dependence on the dynamical time, which increases as a function of cosmic time (due to the cosmological decrease in background density). This leads to mergers being less frequent at later cosmic times (as also seen in observations, e.g., \citealt{Bluck2009, Bluck2012, Conselice2009, Duncan2019}). Therefore, mergers are most effective between approximately equal mass galaxies at early cosmic times.

A necessary condition for a satellite to merge by a given epoch is thus given by:

\begin{equation}
t_\mathrm{merger} < t_\mathrm{univ.}(z) \approx \frac{1}{H(z)} \,\, \implies \,\, H(z)\, t_\mathrm{merger} \lesssim 1,
\end{equation}

\noindent where, $t_\mathrm{univ.}(z)$ is the age of the universe at a given redshift ($z$), and $H(z)$ is the evolving Hubble parameter.

As haloes grow more massive, the steepness of the halo and stellar mass functions implies that most subsequently accreted systems have relatively small mass ratios. These satellites experience long dynamical-friction timescales and, in massive clusters, high relative velocities, which both suppress efficient merging. Consequently, rich satellite populations can survive for many dynamical times in group- and cluster-scale haloes. It is within the dense CGM and ICM environments that environmental processes such as ram-pressure stripping, tidal stripping, harassment, and starvation become most effective. Thus, the emergence of dense environments is closely connected to hierarchical halo growth, while the survival of their satellite populations reflects the increasing inefficiency of dynamical friction for low mass-ratio systems.

\subsection{Quenching via ram pressure stripping}\label{s32}

One of the most important quenching avenues for satellites within clusters is that of ram pressure stripping (as first considered by \citealt{Gunn1972}). The essential idea is that as a satellite galaxy moves relative to the ICM hot gas, a pressure is exerted as follows (e.g., \citealt{Gunn1972, vandenBosch2008, Binney2008, Mo2010}):

\begin{equation*}
P_\mathrm{ram}(r) = {F}/A = \frac{d \, {p}(r(t))/dt}{A} \approx {v}(r) \cdot \bigg( \frac{d \, m(r(t))/dt}{A} \bigg) = {v}(r) \cdot \bigg( \frac{\rho_\mathrm{ICM}(r) A {v}(r)}{A} \bigg)
\end{equation*}

\begin{equation}\label{e26}
\implies P_\mathrm{ram}(r) \approx  \rho_\mathrm{ICM}(r) \, v^2_\mathrm{sat}(r),
\end{equation}

\noindent where, $P_\mathrm{ram}$ is the ram pressure, $F$ is force, $p$ is momentum, and $A$ is the surface area of the galaxy subject to ram pressure. It is often assumed that the velocity of the satellite ($v_\mathrm{sat}$) relative to the ICM (i.e., it's orbital velocity) is left unchanged, which leads to force being driven by mass transfer ($dm/dt$). This can then be written as a function of the density of the ICM ($\rho_\mathrm{ICM}$), which leads to the final expression (using, $dm = \rho A dr$). Therefore, ram pressure is most effective for systems moving with high velocities through dense media. 

Both the typical orbital velocity and the typical density of the ICM increase as a strong function of halo mass ($M_H$). Furthermore, for radial orbits, both the relative satellite velocity and the density of the ICM increase significantly as a function of decreasing distance from the center of the cluster ($r$). Hence, ram pressure becomes increasingly more effective closer to the center of higher mass haloes.\footnote{More quantitatively, $\langle v_\mathrm{sat} \rangle \sim V_c = \sqrt{GM_{200c}/R_{200c}} \propto M_{H}^{1/3}$. But, of course, for an elliptical orbit the actual velocity is a function of distance (explicitly given by the extended vis viva equation, $v_\mathrm{sat}^2(r) = GM_{H}(<r)((2/r) - (1/\langle r \rangle))$. Deriving the ICM density as a function of mass and radius is more complex in general. However, for an isothermal sphere (with constant, $T = T_{200}$), this is given simply by, $\rho_\mathrm{ICM}(r) \sim M_H^{2/3} / r^2$. Combining everything, we expect $P_\mathrm{ram}(r) \sim M_H^{5/3}/r^3$ (for a radial orbit through an isothermal sphere, with a centrally concentrated mass distribution).} 

There are two distinct forms of ram pressure stripping which are thought to be important in the quenching of satellite galaxies: (i)~high-density/ high-velocity ram pressure stripping of the ISM of satellite galaxies, which occurs primarily near the center of high mass groups and clusters; and (ii)~low-density / low-velocity ram pressure stripping of the CGM around satellite galaxies, which may occur throughout the full virial radius of groups and clusters.

\begin{figure}[ht]  
\centering
\includegraphics[width=\textwidth]{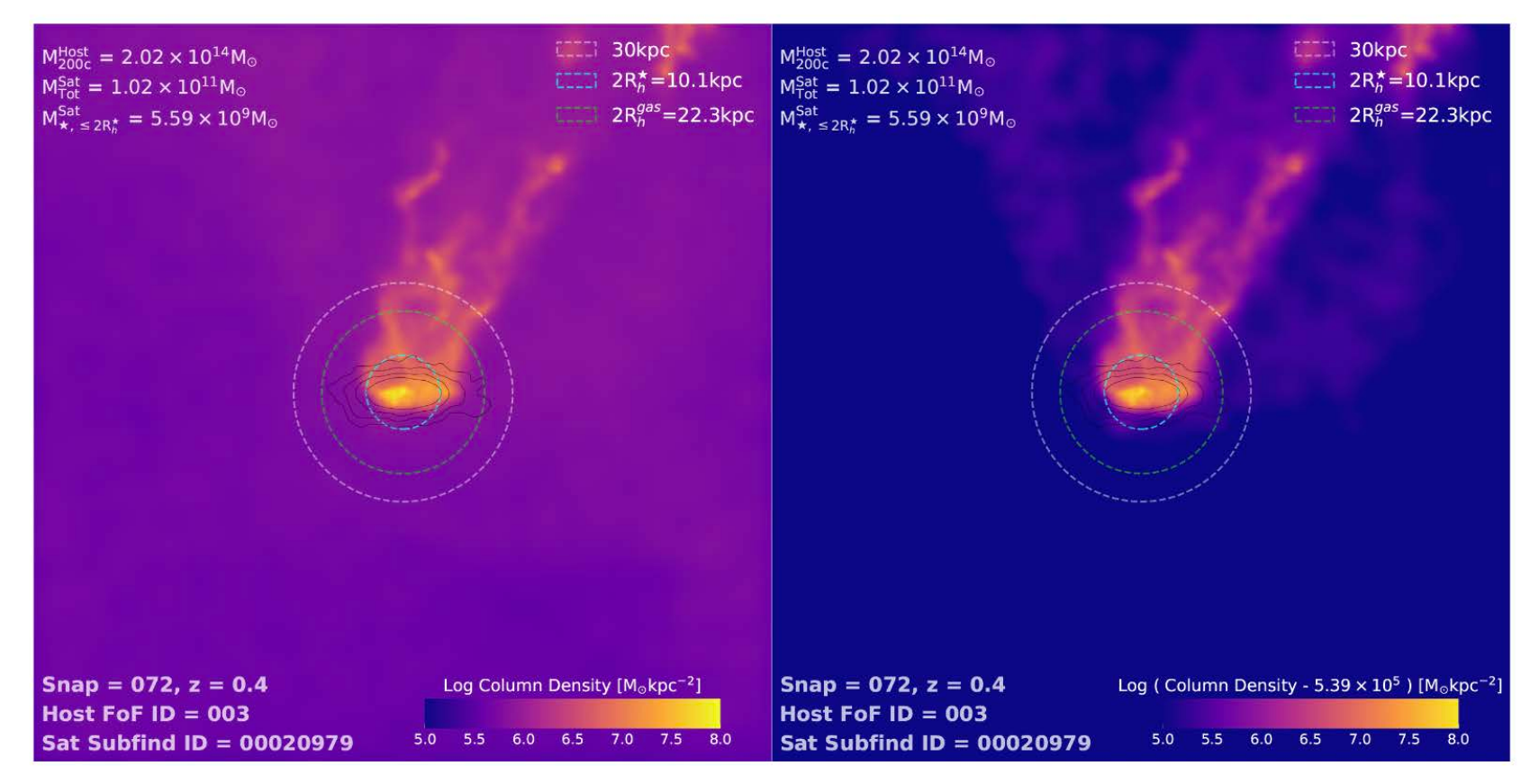}
\caption{An example `jellyfish' galaxy from the IllustrisTNG simulation, reproduced from \cite{Yun2019}. Both panels show the gas column density along the line of sight, with the right-hand panel being background subtracted. This is a satellite galaxy (with $M_* \approx 6 \times 10^{9} \, M_{\odot}$) residing within a cluster (with $M_H\,  = 2 \times 10^{14} M_{\odot}$). In the simulation, the gas tails are known to be induced by strong ram pressure stripping.}\label{f8}
\end{figure}

The instability criterion for high-density ram pressure to impact the ISM of satellite galaxy's disc may be given by (e.g., \citealt{Gunn1972}):

\begin{align}
P_\mathrm{ram}(r) = \rho_\mathrm{ICM}(r) \, v^2_\mathrm{sat}(r) >  P_\mathrm{grav} = \Sigma_g(R) \cdot g_z(R) \, \approx \, 2 \pi G \, \Sigma_*(R) \Sigma_g(R) \ \text{(Thin Disc)},
\end{align}

\noindent where, $P_\mathrm{grav}$ is the effective gravitational restoring pressure (given by the self-gravitational force of the disc divided by the effective ram pressure area), $\Sigma_g~\equiv~M_g/\pi R^2$, which represents the surface gas mass density in the disc, and $g_z(R)$ is the vertical gravitational acceleration just outside of a disk with radius, $R$. This may be written in terms of the stellar mass surface density ($\Sigma_* \equiv M_*/\pi R^2$), which is assumed to dominate over the gas term, to yield the final form (see \citealt{Gunn1972} for the full derivation, which assumes an infinitely thin massive circular sheet, with locally uniform $\Sigma_*$).

Alternatively, the instability criterion for low-density ram pressure impacting the CGM of a satellite galaxy is given by:

\begin{align}
P_\mathrm{ram}(r) = \rho_\mathrm{ICM}(r) \, v^2_\mathrm{sat}(r)  >  P_\mathrm{grav} = \frac{GM_\mathrm{dyn}(<R)M_g(R)}{\epsilon A_\mathrm{CGM}(R) R^2} \ \text{(Spherical CGM)},
\end{align}

\noindent where, $M_\mathrm{dyn}$ is the dynamical mass, $M_g$ is here taken to be the mass of the CGM, $A_\mathrm{CGM} = 4\pi R^2$, which represents the total  surface area of the CGM at a given radius ($R$), and $\epsilon$ is a geometric correction factor (enabling the ram pressure to act directionally). Note that because the CGM resides at much larger typical radii than the ISM (and hence has a much larger surface area as well), ram pressure stripping is much more effective at removing the CGM than the ISM.

Numerous studies have found evidence of ram pressure stripping in hydrodynamical simulations (in both the high and low density modes), including in idealized simulations (see, e.g., \citealt{Abadi1999, Roediger2005, Tonnesen2009, Jachym2007, Jachym2009}) and in large-volume cosmological simulations (see, e.g., \citealt{Bahe2017, Rhee2017, Tremmel2019, Yun2019}). The most beautiful example of which is the formation of `jellyfish' galaxies (e.g., \citealt{Yun2019, Lee2022, Goller2023, Rohr2023, Zinger2024}). These are satellite galaxies which exhibit gas tails streaming perpendicular to their discs (and, typically, anti-parallel to their velocities), as a result of high-density ram pressure stripping, which resemble jellyfish in their morphologies. In Fig.~\ref{f8}, we show an example of the production of a jellyfish galaxy within the IllustrisTNG cosmological hydrodynamical simulation (reproduced from \citealt{Yun2019}). 

Ram pressure stripping in the high-density mode leads to sudden satellite quenching via removal of the cold gas ISM, which is required as fuel for further star formation. On the other hand, ram pressure stripping in the low-density mode leads to slow-quenching via starvation, as a result of preventing further gas accretion into the ISM from the CGM (which is removed). Consequently, the timescales (and other observable signatures) between these two versions of ram pressure stripping are expected to be different. Additionally, it should be noted that either internal feedback (from SN or AGN) or else periodic stripping will still be required for long-term quiescence, in order to resolve the issue of gas return to the ISM from stellar evolution (see back to Sect.~\ref{sec:stellar_quenching}). We explore some of the observational tests to the ram pressure stripping scenario in Sect.~\ref{s5}.

\subsection{Quenching via dynamical stripping}\label{s33}

Another potential route to quenching satellite galaxies arises from the dynamics of interacting systems, which is known as `dynamical stripping' (see, e.g., \citealt{Hoemer1957, King1962, Johnston1995, Tormen1998, Hayashi2003, Read2006, Penarrubia2008}). Generally speaking, the dynamical stripping of satellites may be engendered either by satellite--central tidal interactions, or via satellite\,--\,satellite tidal interactions. The former tends to be a stronger effect (due to the central being the most massive galaxy in the group or cluster), but the latter can be a much more frequent event. As such, satellite\,--\,satellite dynamical stripping is often referred to as `harassment' in the literature, indicating the cumulative effect of many minor interactions. As with ram pressure stripping, both the ISM and the CGM may be impacted. Unlike with ram pressure stripping, the stellar and dark matter structures of satellite galaxies may also be disrupted dynamically. This is thought to be a leading cause of intra-cluster light, as well as the destruction of dark matter sub-haloes within high mass groups and clusters (see, e.g., \citealt{Richstone1983, Merritt1984, Murante2004, Murante2007, Pillepich2018}).

\begin{figure}[ht]  
\centering
\includegraphics[width=\textwidth]{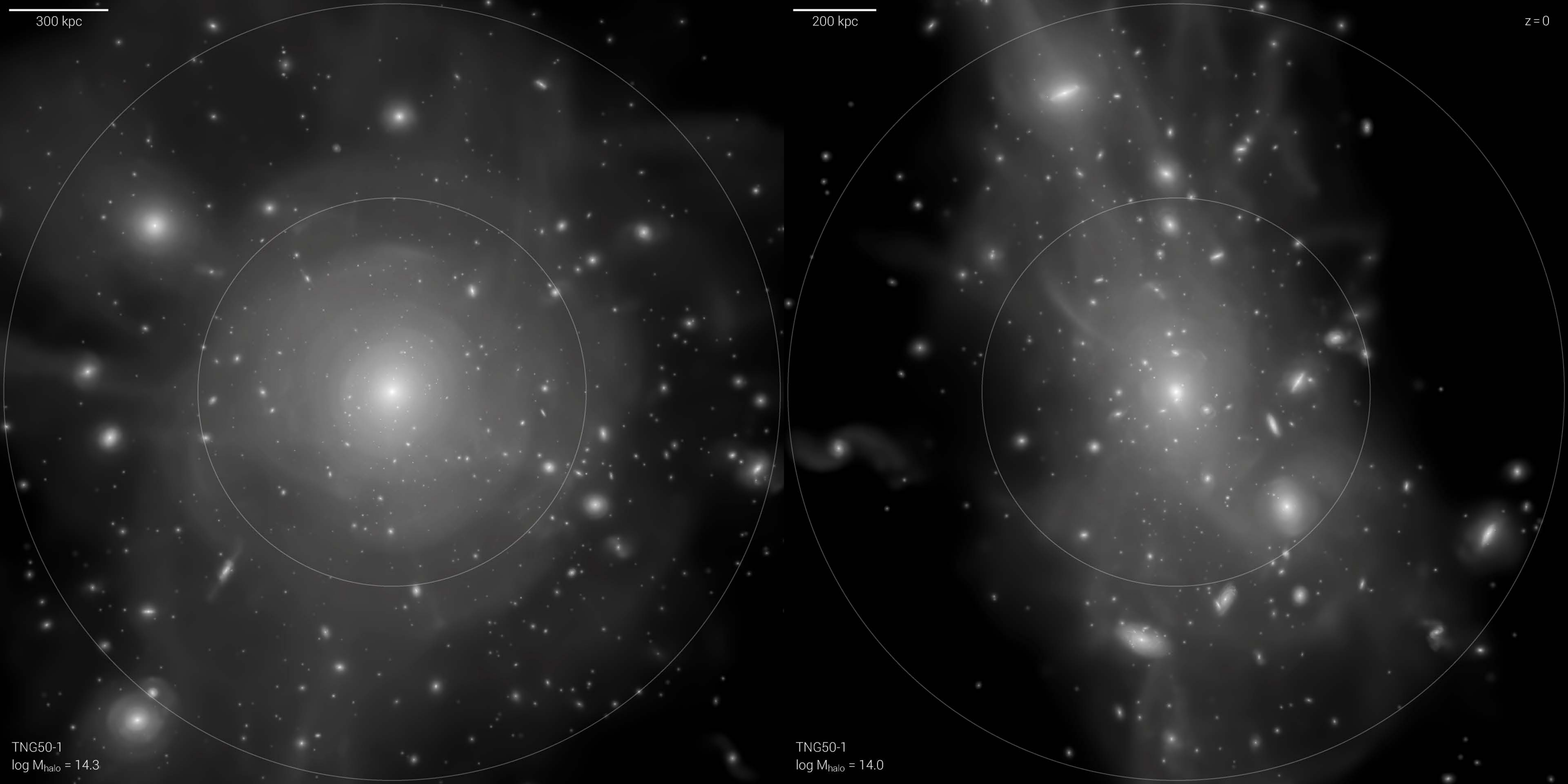}
\caption{The two most massive dark matter haloes from the high-resolution TNG50 simulation, from IllustrisTNG (\citealt{Nelson2018}), both of which are massive enough to count as medium sized galaxy clusters (with $M_H \gtrsim 10^{14} \, M_{\odot}$). In both panels the distribution in stellar mass is shown (brighter regions indicate higher mass densities). Faint white circles indicate 50\% and 100\% of the virial radius. Stellar mass not associated with galaxies is clearly visible in both panels (i.e., the intra-cluster `light'), which forms shells, streams, and filamentary structures. All of this non-galactic stellar mass structure is a direct result of dynamical stripping in this simulation. Although not explicitly shown here, a similar stripping of gas from the ISM and CGM of galaxies via tidal torques is also present in this simulation, which may have a strong impact on the quenching of satellite galaxies. This figure is reproduced from \cite{Joshi2020}.}\label{f9}
\end{figure}

To determine the distance between two galaxies (here taken to be a satellite and its central) at which tidal forces overcome gravitational binding, one may simply set the differential tidal force applied to the satellite (from the central) equal to the satellite's self-gravitational force. This sets the critical tidal radius (at which, $|F_\mathrm{tide}(r_\mathrm{tide})| = |F_{g}(r_\mathrm{tide})|$). Explicitly, this reduces to (e.g., \citealt{Binney2008, Mo2010}):

\begin{equation}
\bigg| \frac{\partial (GM_\mathrm{cen}/r^2)}{\partial r} \bigg|_{r=r_\mathrm{tide}} \cdot R_\mathrm{sat} = \frac{GM_\mathrm{sat}}{R_\mathrm{sat}^2}    \implies   r_\mathrm{tide} \approx R_\mathrm{sat} \, \bigg( \frac{2M_\mathrm{cen}}{M_\mathrm{sat}} \bigg)^{1/3},
\label{e29}
\end{equation}

\noindent where, $M_\mathrm{cen}$ is the mass of the central, $M_\mathrm{sat}$ is the mass of the satellite, and $R_\mathrm{sat}$ is the radius of the satellite galaxy. Note that more sophisticated derivations exist in the literature for various types of systems (see, e.g., \citealt{Binney2008}), but the above simple expression is sufficient for our purposes here.

This leads to a straightforward instability criterion for when dynamical stripping may ensue. Explicitly,

\begin{equation}
r_\mathrm{orbit}(t) < r_\mathrm{tide}(M_\mathrm{cen}, M_\mathrm{sat}, R_\mathrm{sat}).
\label{e30}
\end{equation}

\noindent That is, when the time variable distance between a satellite and its central becomes less than the tidal radius, stripping will occur. Note that stripping is more severe the more massive the central is relative to its satellite, and also increases with the size of the satellite galaxy. Accounting for both of these effects, dynamical stripping is most effective for low-mass satellites with extended (often low surface brightness) structures, which interact with high-mass centrals.

The basic criterion for the dynamical stripping of a satellite by another satellite follows the same structure as above (simply set `cen' $\rightarrow$ `sat-alt', in Eqs.~\eqref{e29} and \eqref{e30}). Typically, the effect of a single satellite\,--\,satellite interaction is marginal. However, there may be a great number of these interactions, especially towards the center of high mass clusters. The cumulative effect (i.e., `harassment'), therefore, is anticipated to be highly impactful. Ultimately, as with ram pressure stripping, dynamical stripping may remove gas (potentially in both the ISM and CGM), preventing star formation via either a direct lack of fuel, or a lack of replenishment of fuel over cosmic time.

While the dynamical stripping of stars and dark matter is well established, its role in quenching star formation is more indirect and depends sensitively on which gas components are affected. Tidal interactions can remove cold gas from the outskirts of galaxies, directly reducing the available fuel for star formation. However, in many cases the cold interstellar medium in the central regions remains largely intact, implying that dynamical stripping alone may not lead to immediate quenching. 

More importantly, tidal forces are highly effective at removing loosely bound gas in the CGM, thereby suppressing the future accretion of gas onto the galaxy. In this regime, dynamical stripping acts primarily as a form of environmental starvation, leading to a gradual decline in star formation over several Gyr (e.g., \citealt{Wetzel2015, Fillingham2015}). Additionally, tidal heating and morphological transformation may lower gas surface densities, further reducing star formation efficiency.

Consequently, dynamical stripping is unlikely to be the sole driver of rapid quenching in most systems, but instead operates in concert with other environmental processes, such as ram pressure stripping, to remove both the hot and cold gas reservoirs of satellite galaxies. The combined action of these mechanisms is thought to be responsible for the high quenched fractions observed in dense environments.

Many simulations show evidence of tidal stripping of satellite galaxies (e.g., \citealt{Cooper2010, Wetzel2015, Grand2017, Pillepich2018, Fattahi2020}). See Fig.~\ref{f9} for an example of the impact of dynamical stripping on the stellar component of satellite galaxies within the two most massive clusters from the TNG50 simulation within IllustrisTNG (reproduced from \citealt{Joshi2020}). We will review some of the observational evidence for tidal stripping leading to quenching of satellite galaxies in Sect.~\ref{s5}.

\subsection{Quenching via strangulation}\label{s34}

So far in our review of theoretically proposed environmental quenching processes we have focused on stripping (both as a result of ram pressure and gravitational dynamics), which results in the removal of gas from satellite galaxies (both from the ISM and CGM). An alternative route to the quenching of satellites is through `strangulation', which refers to the prevention of accretion of cold gas into the ISM of satellites (e.g., \citealt{Kawata2008, Wetzel2013, Bahe2015, Peng2015, Maier2019a, Maier2019b}).

There are two principal modes in which strangulation may operate within satellite galaxies: (i)~as a result of the removal of the CGM around massive satellites, preventing gas cooling and replenishment of the ISM (which is ultimately a result of low-density ram pressure stripping, discussed above); and (ii)~as a result of the location of satellites within their parent dark matter haloes (e.g., \citealt{Henriques2013, Henriques2015}). As central galaxies fall into clusters, becoming satellites, they transition from being the centrals of their own dark matter haloes to being in orbit of the group/ cluster central galaxy. Consequently, they are no longer fed directly by cold gas streams from the IGM (as in \citealt{Birnboim2003, Dekel2006, Dekel2009}). Consequently, the first mode is likely most important for massive satellites (which will likely have already formed a hot static halo prior to cluster infall) and the latter mode is likely most important for low-mass satellites (which are replenished primarily due to cold gas streams and typically lack hot gaseous haloes).

Compared to stripping, strangulation is most probably a slower quenching process (see, e.g., \citealt{Peng2015, Trussler2020, Bluck2020b}), which is expected to leave an observable imprint on the metalicities of galaxies (which we discuss within the context of observations in Sect.~\ref{s62}). Nevertheless, there is also evidence of ram pressure stripping occurring over many Gyr, complicating this simple picture (see, e.g., \citealt{Rohr2023}). Note also that strangulation is at base a form of starvation and, hence, shares similar observable properties with the central galaxy version (that of preventative, radio-mode AGN feedback; e.g., \citealt{Croton2006, Bower2006, Bower2008, Fabian2012}).

\subsection{Quenching via pre-processing}\label{s35}

The final environmental route to quenching we will discuss from a theoretical perspective is that of `pre-processing' (see, e.g., \citealt{Zabludoff1998, Fujita2004, Hou2014, Lopes2024}). This is not so much a distinct theoretical quenching mechanism as an acknowledgment that when we observe a quenched satellite galaxy within a cluster, we do not know (in general at least) whether it quenched in the cluster or prior to infall. As seems to often be the case with environmental quenching, there are two distinct cases to consider. 

First, the quenched satellite could have previously been a central and quenched via intrinsic means (e.g., via AGN feedback), prior to entering the cluster environment. This type of pre-processing is expected primarily in massive satellites, and one expects stronger correlations with intrinsic, rather than environmental, parameters in this case (which we discuss from an observational perspective in Sect.~\ref{s54}). Second, the quenched cluster satellite could have previously been a group satellite, quenching in the group environment (e.g., as a result of stripping, strangulation, or a combination of both effects). This case is more subtle than the first because one would expect environmental parameters to best predict quiescence, whether or not the satellite quenched in the cluster or the group environment. Moreover, this type of pre-processing can impact all masses of satellite galaxies (see, e.g., \citealt{Donnari2021b}). We will revisit this complex case periodically with respect to observations in the later sections of this review.

\subsection{AGN-feedback modified environmental quenching}\label{s36}

\begin{figure}[ht]  
\centering
\includegraphics[width=\textwidth]{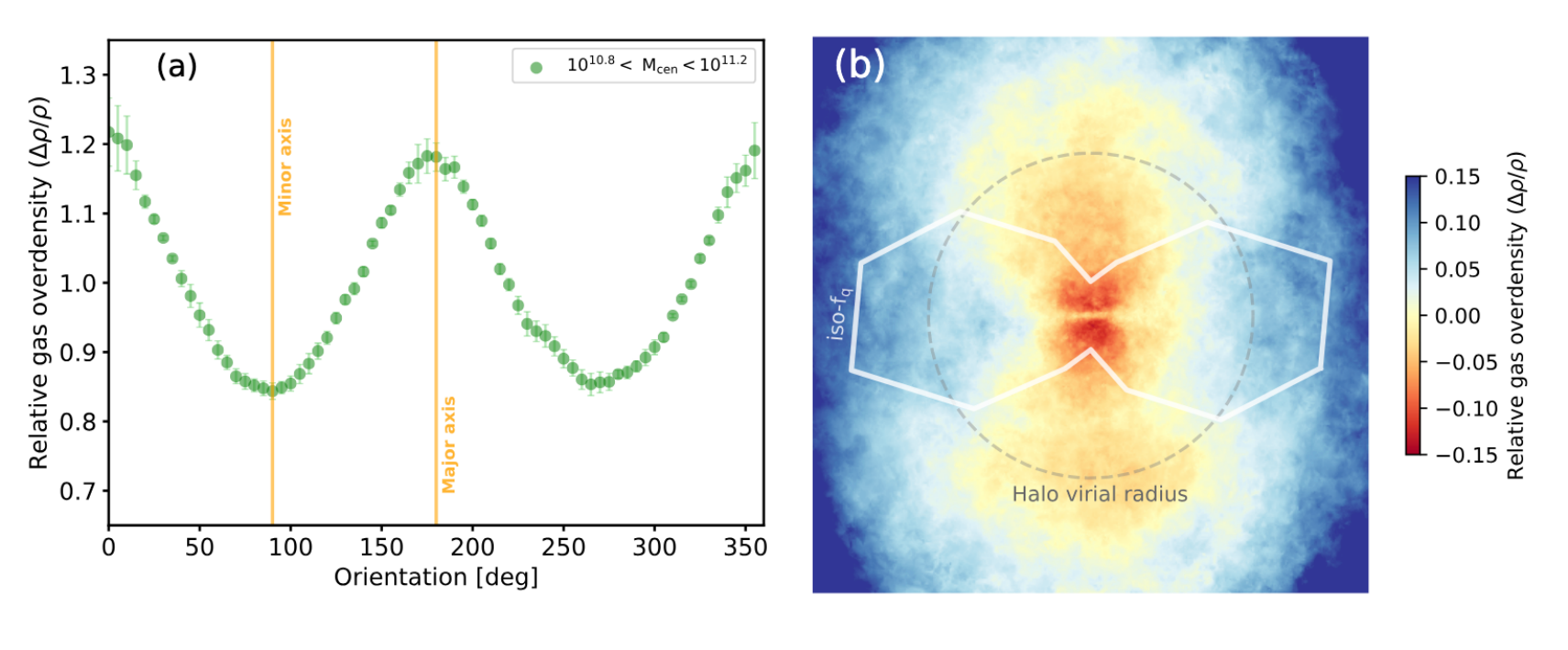}
\caption{Anisotropic quenching of satellite galaxies in clusters from the TNG100 simulation of IllustrisTNG, reproduced from \cite{Martin2021}. \textit{Left-hand panel: } The relative gas over-density plot as a function of orientation (measured relative to the semi-major axis of the cluster central galaxy). Note that the density reaches a minimum at the orientation of the semi-minor axis. \textit{Right-hand panel: } The relative gas over-density of an example cluster in TNG100. Note how the density is systematically decreased above and below the semi-minor axis of the central. This is a direct consequence of kinetic-mode AGN feedback in this simulation. Overlaid as a white contour is an iso-quenching line, for satellite galaxies, which dips in towards the center in the low density regions. In this simulation, this feature is a direct consequence of the lower gas density leading to less effective ram pressure stripping within the jet plume. Taken together, this figure provides evidence of AGN feedback and ram pressure stripping working together in a complex manner to quench galaxies within cluster environments.}\label{f10}
\end{figure}

\noindent Multiple quenching processes may interact with one another to create complex effects within and around galaxies. For example, in order for ram pressure stripping to be effective for satellite quenching in groups and clusters, a hot static atmosphere must develop and cooling must not significantly deplete it over cosmological timescales. This likely requires both virial shock heating (e.g., \citealt{Dekel2006, Dekel2009}) and late-time radio-mode heating (e.g., \citealt{Croton2006, Bower2006, Bower2008, Henriques2015}). Hence, whilst intrinsic and environmental quenching processes are often considered as separate channels in galaxy evolution, they do interact with one another (sometimes in surprising ways).

In the final part of this section we discuss a particularly remarkable example of this inter-connection, which combines intrinsic quenching (via radio-mode AGN feedback) and environmental quenching (via ram pressure stripping). In \cite{Martin2021}, the quenching of satellite galaxies is found to be anisotropic within both simulated (from TNG100 within IllustrisTNG) and observed (from the SDSS) galaxy clusters. Specifically, the authors find a significant (though weak in amplitude) sinusoidal dependence of satellite quenching fraction on orientation angle, measured relative to the position angle of the central galaxy (i.e., the semi-major axis). This operates such that the quenching of satellites is suppressed above and below the semi-minor axis of the central on halo scales. Similar observational results are found in \cite{Stott2022, Ando2023}, confirming this discovery and extending it out to $z \sim 1$.

In Fig.~\ref{f10} we show results from \cite{Martin2021} from the IllustrisTNG simulation, which is used as an explanatory tool to account for this anisotropy in satellite quenching within clusters. In the left-hand panel, the relative ICM gas over-density is plot as a function of orientation. Note that it reaches a minimum above and below the semi-minor axis (i.e., at $\theta$ = 90, 270 degrees), and a maximum either side of the semi-major axis (i.e., at $\theta$ = 0, 180 degrees). On the right-hand panel, the relative gas over-density in the ICM is shown for an example cluster. The distinct vertical structure in this plot is a result of kinetic-mode AGN feedback in the IllustrisTNG simulation, which operates by increasing the temperature and entropy of the gas (see \citealt{Weinberger2018, Zinger2020}). This results in a collimated reduction in gas density, which in turn causes a reduction in the effectiveness of ram pressure stripping within the jet plume (see back to Eq.~\eqref{e26}).

This result is remarkable because it clearly provides evidence for \textit{both} radio-mode quenching of central galaxies (via kinetic launched jets heating the CGM), and ram pressure stripping quenching of satellites, modulated by this AGN feedback. Moreover, there is excellent agreement with observations from the SDSS, which strongly suggests that radio-mode feedback modulates satellite quenching in nature as well (see \citealt{Martin2021} for full details).



\section{Testing the causes of central galaxy quenching}\label{s4}

In this section we turn our attention to observational tests of the theoretical quenching paradigms discussed in the previous two sections, specifically probing the fundamental causes behind central galaxy quenching. As discussed in Sect.~\ref{s2}, feedback is absolutely critical in simulations of galaxy formation and evolution to regulate star formation and ultimately quench high-mass galaxies. Fortunately, direct observational evidence for feedback exists and is plentiful (see Fig.~\ref{f11} for two striking visual examples). In this section we review some of the evidence for ejective feedback from SN and AGN, as well as preventative radio-mode AGN feedback. We also explore where AGN reside on the SFMS, how SMBH mass is related to quenching, and, finally, present some direct statistical tests of the paradigm of AGN-driven quenching for large populations of galaxies (at both low and high redshifts).

\subsection{SN-driven outflows from galaxies}\label{s41}

Galactic-scale outflows driven by SN feedback, particularly in starburst galaxies, are observed in a number of different types of observations (see Fig.~\ref{f11}(a) for a visual example). The  observational evidence for this feedback channel comes from a variety of complementary sources, including: (i)~direct morphological evidence of multi-phase gas outflows (see, e.g., \citealt{Shopbell1998, Strickland2009}); (ii)~X-ray emission from hot gas, produced in the SN itself and/or via shock heating of the ISM (see, e.g., \citealt{Strickland2000, Strickland2007, Grimes2005}; (iii)~blue-shifted absorption lines from metals within neutral gas flows, relative the the host galaxy (see, e.g., \citealt{Heckman2000, Martin2005, Weiner2009}); and (iv)~offset optical emission lines from ionized regions in the outflow, relative to the host galaxy (see, e.g., \citealt{Veilleux2005, Rich2010, Rich2011}). 

Typically, the rate of mass ejection from SN-driven outflows is found to scale with SFR (e.g., \citealt{Heckman2002, Veilleux2005, Heckman2017}), lending further credence to its origin. Furthermore, in some cases, the morphology of ejection may be traced back to individual star forming regions within starburst galaxies (see, e.g., \citealt{Strickland2009}). SN driven outflows are thought to be a critical route to reducing the efficiency of star formation in low mass haloes, as well as enabling the enrichment of the CGM with metals, and causing redistribution of metals throughout the ISM in galaxies (see, e.g., \citealt{Tumlinson2011, Werk2014}).

\begin{figure}[ht]  
\centering
\includegraphics[width=0.98\textwidth]{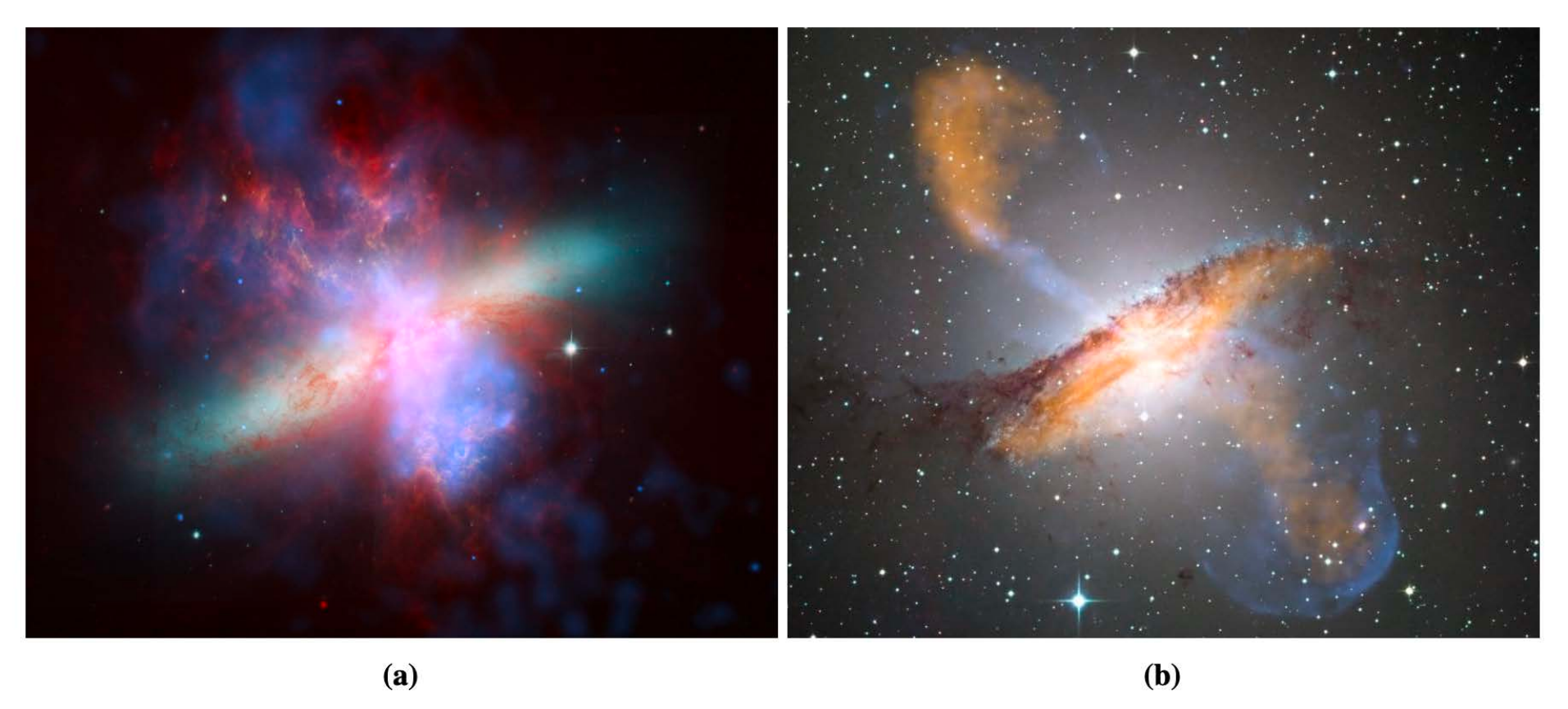}
\caption{\textit{Panel-(a)}: An example of ejective feedback from supernovae in M82, a local starburst galaxy, without evidence of an AGN. X-ray emission from high temperature gas is shown in blue, and infrared emission from dust is shown in red. Also visible is the stellar disc, shown in pale white-green, which is probed in optical. Note that the outflow is perpendicular to the disc, and extends well beyond the galaxy. Credit: X-ray: NASA/CXC/JHU/D. Strickland; Optical: NASA/ESA/STScI/AURA/The Hubble Heritage Team; IR: NASA/JPL- Caltech/Univ. of AZ/ C. Engelbracht. \textit{Panel-(b)}: An example of AGN jet feedback from the Centaurus A galaxy, which hosts a radio-loud AGN, as seen in X-ray (blue) and sub-mm (orange), with the stellar structure visible in optical (pale white). Again, the outflow is clearly perpendicular to the semi-major axis of the elliptical galaxy, and extends well beyond the galaxy. However, here, the outflow is much tighter, with a jet morphology. Credit: ESO/WFI (optical); MPIfR/ESO/APEX/A. Weiss et al. (sub-mm); NASA/CXC/CfA/R. Kraft et al. (X-ray).}\label{f11}
\end{figure}

Typically, winds produced via SN feedback are modeled in observations as either energy-driven or momentum-driven processes. These two cases are made explicit below:

\begin{equation}
\dot{E}_\mathrm{SN} = E_\mathrm{SN} \cdot \mathcal{R}_\mathrm{SN} \, ; \quad \dot{p}_\mathrm{wind} = \eta \, \text{SFR} \, v_\mathrm{wind} \,,
\end{equation}

\noindent where, the LHS expression indicates the energy-driven process, and the RHS expression indicates the momentum driven process. On the LHS, $\dot{E}_\mathrm{SN}$ is the rate of change in energy produced via SN, which is given as the product of the energy released per SN ($E_\mathrm{SN} \sim 10^{51}\, \mathrm{erg}$) multiplied by the rate of supernovae ($\mathcal{R}_\mathrm{SN}$, which is closely related to SFR). On the RHS, the rate of change of momentum of the wind ($\dot{p}_\mathrm{wind}$) is typically assumed to be the mass loading factor ($\eta$) multiplied by the SFR and the typical velocity of the wind ($v_\mathrm{wind} \sim 500\,\mathrm{km/s}$).

Using the canonical energy injection per SN ($\sim$$10^{51}\, \mathrm{erg}$; see \citealt{Chevalier1977, Thornton1998, Woolsey2001}) and the assumption that $\sim$$10\, M_{\odot}$ of material is ejected in the canonical case, one may infer the expected ejection velocity at the launch time. Moreover, one can then compare this to typical velocities observed for SN-driven outflows. Explicitly,

\begin{equation}
v_\mathrm{wind} \sim 500 \, \mathrm{[km/s]} \ll  v_\mathrm{launch} = \sqrt{\frac{2E_\mathrm{SN}}{M_\mathrm{ej}}} \sim 3000 \, \mathrm{[km/s]} \, ,
\end{equation}

\noindent where we have noted the `typical' wind velocity from SN-driven outflows above. But, in reality, this is highly component dependent, with hot gas reaching $v_\mathrm{wind} \sim$500 -- 1000\,km/s; warm gas reaching $v_\mathrm{wind} \sim$200 -- 500\,km/s; and cold has reaching $v_\mathrm{wind} \sim$50 -- 300\,km/s (see, e.g., \citealt{Chevalier1985, Heckman2000, Veilleux2005, Heckman2017, Cicone2014}). Since most of the mass is contained in the cold gas ejecta, our chosen wind velocity limit is a conservative estimate for the velocity per particle. In all cases, this is much lower than the launch velocity ($v_\mathrm{launch}$), computed by assuming full conversion of energy into bulk velocity (see RHS above). This is expected since the initial ejecta may sweep up material from the ISM, gaining mass but losing kinetic energy. Additionally, energy may dissipate, e.g., via heating the ISM through shocks, which may be ultimately radiated away. As such, it is the bulk wind velocity, rather than the launch velocity, which is most relevant for considering the global impact on the ISM, and ultimately on quenching.

To investigate the impact of SN-driven winds it is crucial to compare their observed velocities to the escape velocity at the radius from which they are launched. Since the majority of the mass in a given halo is in the form of an extended dark matter structure, one must assume a dark matter profile to make progress. Assuming an NFW profile (\citealt{Navarro1996, Navarro1997}), the escape velocity ($v_\mathrm{esc}$) at a given radius ($r$), is given by (e.g., \citealt{Lokas2001, Binney2008, Mo2010}):

\begin{equation}
v_\mathrm{esc}(r) \equiv \sqrt{2 \big| \Phi(r) \big|} = \sqrt{ \bigg( \frac{2 G M_\mathrm{Halo}}{r} \bigg) \cdot 
 \frac{\ln \left( 1 + c \, \bigg[\dfrac{r}{R_\mathrm{vir}}\bigg] \right)}{\ln\big(1 + c\big) - \bigg[\dfrac{c}{1 + c}\bigg]} } \, , \text{where} \ c \equiv R_\mathrm{vir} / R_s
 \label{e33}
\end{equation}

\noindent and where $\Phi(r)$ is the gravitational potential at $r$, $M_\mathrm{Halo}$ is the halo mass, $R_\mathrm{vir}$ is the virial radius of the halo, and $c$ is the concentration parameter, defined above in terms of the virial velocity and the NFW scale length ($R_s$). Note that the function under the square-root sign factorizes into the expected function for a point mass, and a complex ratio of logarithmic terms, which accounts for the structure of the dark matter halo (see \citealt{Navarro1996, Navarro1997}). By assuming a concentration ($c$) and a redshift ($z$), we can compute the escape velocity for various halo masses and launch radii. See Fig.~\ref{f12} for some example values ranging from $M_\mathrm{Halo} = 10^{10} - 10^{15}\,M_{\odot}$ and $R_\mathrm{launch} = 1 - 10\,$kpc. Note that this varies as a strong function of halo mass, but as a much weaker function of launch radius. This is because any distance within the galaxy is typically only a very small fraction of the total virial radius of the halo.

\begin{figure}[ht]  
\centering
\includegraphics[width=\textwidth]{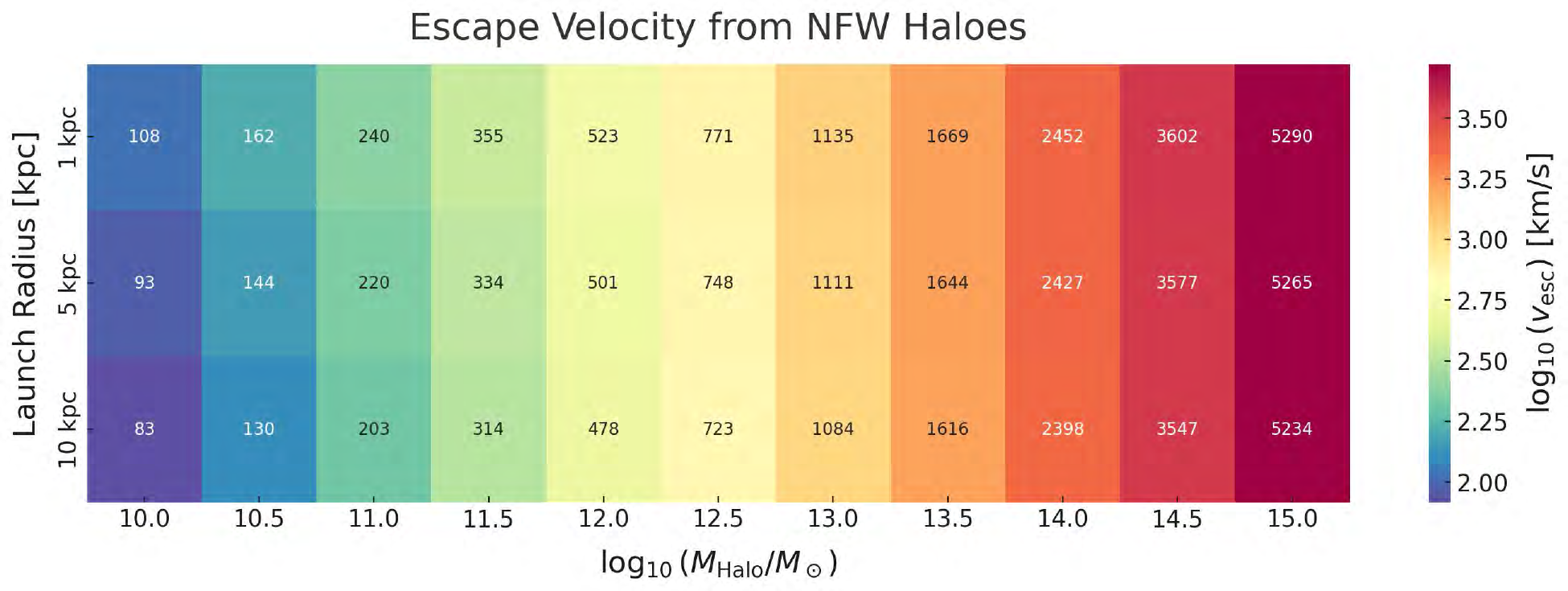}
\caption{Escape velocity as a function of halo mass and launch radius, assuming an NFW profile (\citealt{Navarro1996, Navarro1997}) with $c = 10$ at $z = 0$ (see Eq.~\eqref{e33}). Note that escape velocities rise with redshift and, hence, these values represent minimum escape velocities as a function of cosmic time (for a given halo mass and launch radius). Escape velocities are shown in linear units ([km/s]) within each cell, color coded by logarithmic units (see color bar).  }\label{f12}
\end{figure}

Combining our tabulated data (in Fig.~\ref{f12}) with the observed wind velocities (discussed above) we may identify a critical halo mass at which SN-driven winds become incapable of escaping the halo (see, e.g., \citealt{Henriques2019, Dekel2019}):

\begin{equation}
M_\mathrm{Halo} \, \bigg|_\mathrm{SN \, crit.} = M_\mathrm{Halo} \, \big[{v_\mathrm{esc}(M_\mathrm{Halo}\,,\,R_\mathrm{launch}) \, = \, v_\mathrm{wind}}\big] \, \sim \,  10^{12} \, M_{\odot} \,,
\end{equation}

\noindent where, the critical mass is identified by matching the escape velocity (from a given halo at a given launch radius) to the typical SN-driven galactic wind velocity (discussed above). This implies that for halo masses $M_\mathrm{Halo} \lesssim 10^{12}\,M_{\odot}$, SN-driven winds may lead to effective ISM depletion, as a result of wind velocities being sufficiently fast to escape the halo. However, at $M_\mathrm{Halo} \gtrsim 10^{12}\,M_{\odot}$, SN-driven winds become ineffective at removing gas from the halo, and the ejecta are ultimately expected to be re-accreted into the system via galactic fountains (e.g., \citealt{Shapiro1976, Bregman1980, Fraternali2006, Boosma2008, Marasco2012, Marasco2013}).

In terms of quenching, low-mass isolated galaxies are typically star forming (e.g., \citealt{Baldry2006, Peng2010, Peng2012, Woo2013, Bluck2014, Bluck2016}), albeit at much lower rates than predicted by theory in the absence of SN feedback (see, e.g., \citealt{Cole2000, Bower2006, Bower2008, Somerville2015}). Hence, at least using the definitions for quenching adopted in this review, SN feedback does not quench low-mass systems. Nonetheless, one should appreciate that it clearly slows the rate of star formation, dramatically lowering the efficiencies from the high values expected in feedback-free models. Alternatively, high-mass isolated galaxies are typically quenched (e.g., \citealt{Baldry2006, Peng2010, Peng2012, Woo2013, Bluck2014, Bluck2016}), even though SN-driven winds cannot remove gas from the halo effectively. Consequently, to explain quenching one must look beyond SN-feedback (as also concluded via several theoretical arguments presented in Sect.~\ref{s2}; see also \citealt{Cole2000, Croton2006, Bower2006, Bower2008, Henriques2015}).

\begin{figure}[ht]  
\centering
\includegraphics[width=0.32\textwidth]{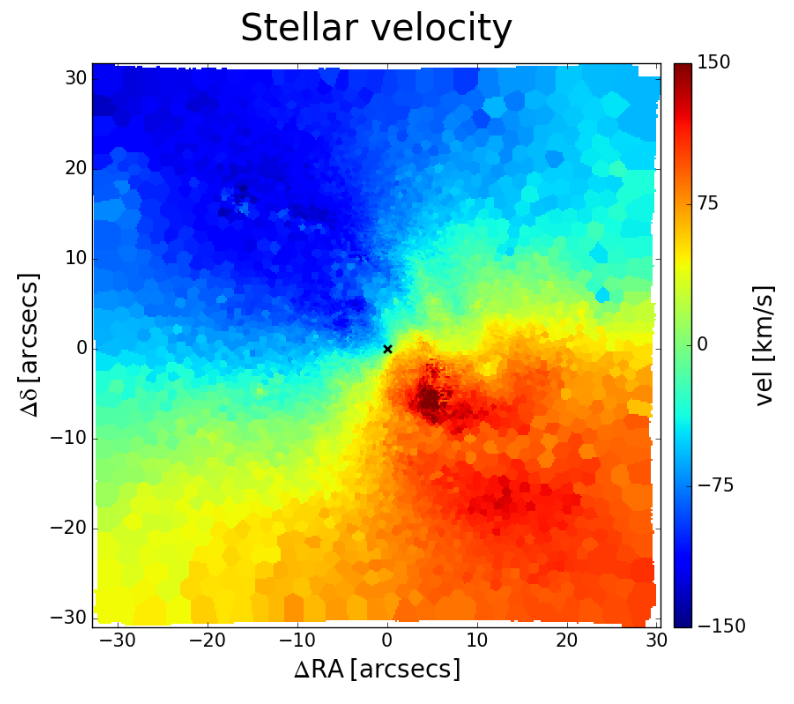}
\includegraphics[width=0.32\textwidth]{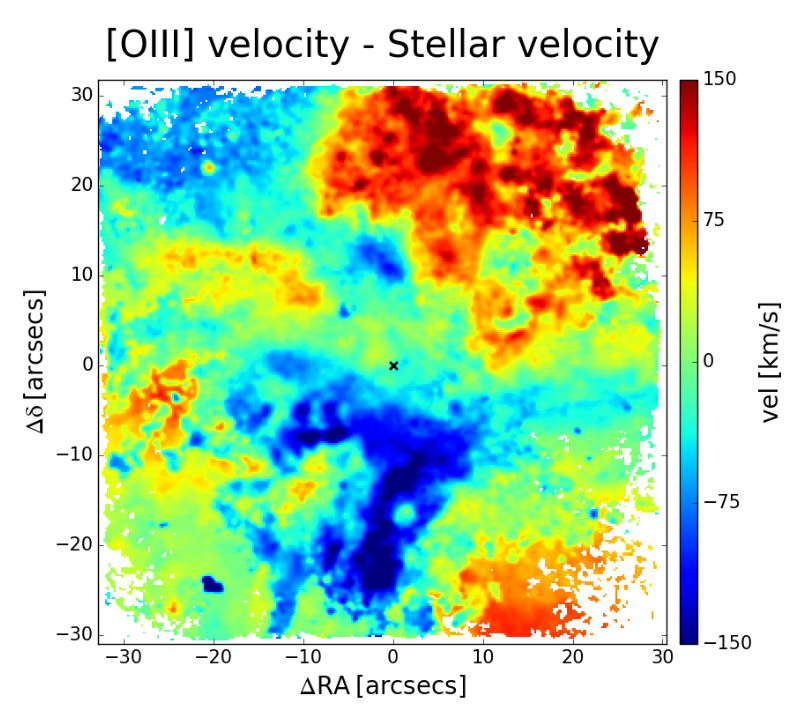}
\includegraphics[width=0.33\textwidth]{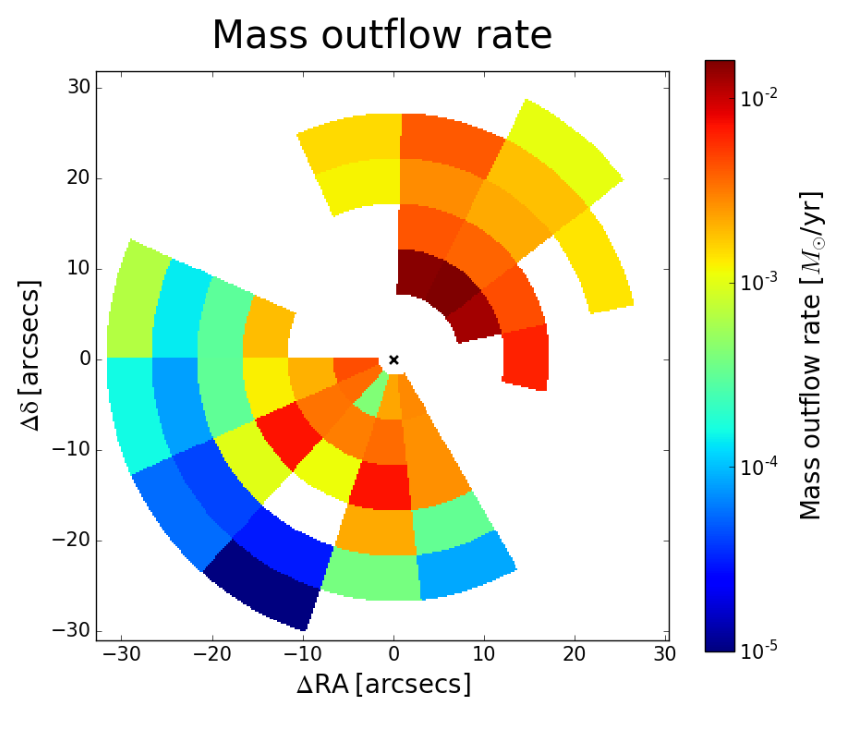}
\caption{Stellar velocity map (left-hand panel), subtracted [OIII] - stellar velocity map (center panel), and mass outflow rate map (right-hand panel) from VLT-MUSE observations of NGC1365 (a nearby Seyfert galaxy). Clearly, the [OIII] velocity map is offset perpendicular to the rotational velocity of the galaxy, and exhibits a conical-like structure. This is a very strong indicator of a substantial ionized outflow from this system (with wind velocity, $v_\mathrm{wind} \sim$150\,km/s). From modeling of the line intensity and velocity maps, the mass outflow rate is determined, as shown on the right-hand panel. The total mass of the outflow is estimated to be $M_\mathrm{out} = 9.2 \times 10^5 \, M_{\odot}$. This figure is reproduced from \cite{Venturi2018}.}\label{f13}
\end{figure}

\subsection{AGN-driven outflows from galaxies}\label{s42}

AGN outflows driven by radiatively efficient accretion around SMBHs are observed in a variety of forms. Specifically, evidence for AGN-driven outflows is observed in: (i)~ionized gas (e.g., \citealt{Harrison2014, Bae2014, Kakkad2016, Venturi2018}); (ii)~molecular gas (e.g., \citealt{Feruglio2010, Feruglio2017, Maiolino2012, Cicone2012, Cicone2014, Cicone2015, Cicone2018, Fluetsch2019, Fluetsch2021}); (iii)~neutral gas (\citealt{Rupke2013a, Rupke2013b, Coil2011, Perna2017}), and (iv)~X-ray in ultra-fast outflows (UFOs; e.g., \citealt{Reeves2009, Tombesi2010, Nardini2015}). Typically, AGN-driven outflows have been observed in quasars or high-luminosity Seyferts, where AGN luminosities far exceed the power from SN feedback. Outflows are typically identified via kinematic offsets from the galaxy in integral-field unit spectroscopy (see Fig.~\ref{f13} for an example), or else via broad wings out to high velocities in emission lines (see Fig.~\ref{f14} for an example).

\begin{figure}[ht]  
\centering
\includegraphics[width=\textwidth]{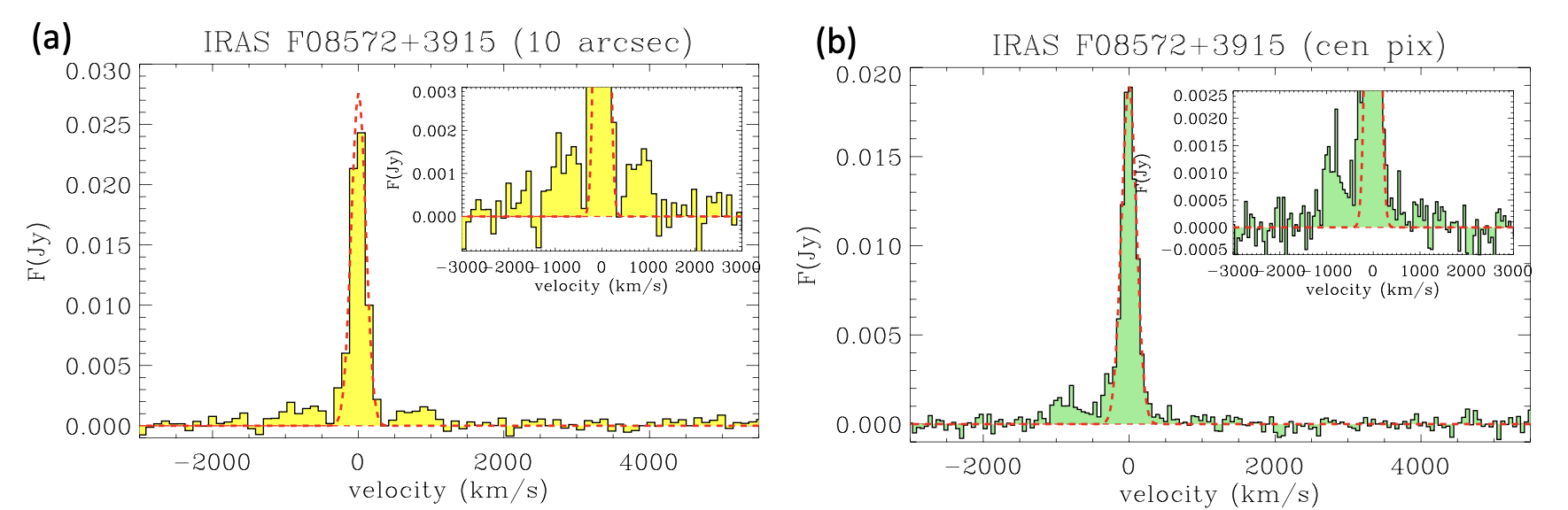}
\caption{Continuum subtracted IRAM spectroscopy of the CO(1-0) line from IRAS F08572+3915, a known obscured quasar. In panel-(a) the flux is integrated across a 10 arcsec field of view, whereas in panel-(b) only the central pixel is utilized. In both panels, there is a clear peak at $v \sim 0$\,km/s, which represents the gas content of the galaxy. Additionally, there is clear broad-wing structure reaching out to $v \sim 1000$\,km/s (or more), which is indicative of a high velocity AGN-driven molecular outflow from this system. This is particularly evident in the zoom-in regions, shown as insets within each panel. This figure is reproduced from \cite{Cicone2014}.}\label{f14}
\end{figure}

As with SN-driven winds, AGN-driven winds may be momentum or energy driven. In the case of a momentum conserving AGN wind, the rate of change of momentum is given by:

\begin{equation}
\dot{p}_\mathrm{momentum-driven} = \dot{M}v_\mathrm{wind} \approx L_\mathrm{AGN}/c
\end{equation} 

\noindent where, $\dot{M}$ is the mass outflow rate, $v_\mathrm{wind}$ is the wind velocity, and $L_\mathrm{AGN}$ is the bolometric AGN luminosity. The above expression assumes all photons recoil off surrounding matter, transferring momentum with `maximal coupling' (which may be reasonable only in very optically thick media).

Alternatively, the photons may heat the surrounding gas, and subsequently launch winds via gas expansion. Energy launched winds are typically modeled heuristically as:

\begin{equation}
\frac{1}{2} \dot{M} \, (v_\mathrm{wind})^2 = \dot{E}_\mathrm{AGN} \approx \eta_w \,  L_\mathrm{AGN}  \quad \implies \quad \dot{p}_\mathrm{energy-driven}  = \frac{2 \eta_w L_\mathrm{AGN}}{v_\mathrm{wind}}
\end{equation}

\noindent where, the rate of change in kinetic energy for a fixed velocity wind is set equal to a fraction of the rate of change of energy from the AGN, which is set by $\eta_w$ (with a typical values of $\sim$0.05 -- 0.1). In the second term above, we define the rate of change of energy-driven momentum directly. Comparing the two effective forces we have:

\begin{equation}
\frac{\dot{p}_\mathrm{energy-driven}}{\dot{p}_\mathrm{momentum-driven}} = \frac{2\eta_w \, c}{v_\mathrm{wind}} \sim 10 - 50,
\end{equation}

\noindent which, for typical wind velocities demonstrates that energy injection is more effective than momentum injection for launching galactic-scale winds (see, e.g., \citealt{Hopkins2006, Hopkins2008, Cicone2014}). Moreover, observations frequently show evidence of AGN winds exceeding the momentum limit and, hence, conclude that AGN outflows are often energy-driven (e.g., \citealt{Cicone2014, Fluetsch2019}).

Measured wind velocities show scaling relations with AGN luminosity and Eddington ratio (which further establishes their AGN origin). However, the actual observed outflow rates, and the specifics of these relationships, vary strongly as a function of the outflow type. Consequently, in Table~\ref{t2} we present a summary of observational results separated into measurement type. In this table, we summarize the observed scaling between wind velocity and AGN luminosity (if any), the typical extent of the outflow, and the typical velocity of the outflow.

\begin{table}[t!]
\centering
\caption{Summary of AGN-driven outflow properties by gas phase}\label{t2}
\begin{tabular}{lllll}
\toprule
\textbf{Gas Phase} & \textbf{Tracer Lines} & \textbf{Scaling} & \textbf{Speeds} & \textbf{Scales} \\
\midrule

Neutral        & Na\,\textsc{i} D, Mg\,\textsc{ii} (abs.)  & weak   & 100--1000 km\,s$^{-1}$ & $\sim$1--10 kpc    \\\\

Molecular      & CO(1--0), [CII]    & $v_\mathrm{w} \sim (L_{\mathrm{AGN}})^{0.3}$ & 100--1000 km\,s$^{-1}$ & $\sim$0.1--5 kpc   \\\\

Ionized        & [O\,\textsc{iii}], C\,\textsc{iv}, H$\alpha$  & $v_\mathrm{w} \sim (L_{\mathrm{AGN}})^{0.5}$ & 100--2000 km\,s$^{-1}$ & $\sim$0.1--10 kpc  \\\\

X-ray UFOs     & Fe\,\textsc{xxv}/\textsc{xxvi} (abs.)  & $v_\mathrm{w} \sim (L_{\mathrm{AGN}})^{0.6}$ & 0.1--0.3$c$            & $\sim$0.01--0.1 pc \\

\bottomrule 
\end{tabular}
\label{tab:agn_outflows}
\end{table}

Except in the case of UFOs, which incorporate only a very small amount of gas relative to the ISM, the majority of quasar-driven winds have peak velocities of $\lesssim$1000\,km/s. This is considerably higher than for SN-driven winds ($\sim$500\,km/s; see previous sub-section). Nonetheless, by comparing to escape velocities from Fig.~\ref{f12}, we see that at halo masses $M_\mathrm{Halo} \gtrsim 10^{13}\,M_{\odot}$, even the fastest observed AGN-driven winds are not capable of escaping the halo. Hence, quasar-mode feedback is likely most effective in a relatively narrow halo mass range (i.e., $M_\mathrm{Halo} \sim 10^{12} - 10^{13}\,M_{\odot}$). At lower masses, powerful AGN are rarely observed, which is likely due to these galaxies hosting low-mass SMBHs, with accretion limited via Eddington (e.g., \citealt{Maggorian1998, Ferrarese2000, Haring2004, Saglia2016}). Alternatively, at higher masses, the deep gravitational potential wells in which galaxies reside will effectively hold on to all quasar-driven outflows, enabling efficient re-accretion into the ISM (see, e.g., \citealt{Fluetsch2019, Fluetsch2021}). 

Even more importantly, quasars are rare events, which typically happen once in the lifetime of a galaxy (e.g., \citealt{Shen2007, Shankar2008, Chen2018, Pizzati2024}). They also appear to be primarily associated with actively star forming galaxies at high redshifts (e.g., \citealt{Hickox2009, Hickox2014, Heckman2014, Ward2022}). Hence, even for a system which is initially quenched via a dramatic AGN-driven outflow, there is plenty of opportunity and time for it to re-accrete material (from either direct accretion from the IGM or cooling of the CGM) and rejuvenate thereafter. It is important to appreciate the similarity here with SN-driven outflows at low halo masses (see previous sub-section). Ultimately, these events are capable of reducing the rate of star formation in galaxies, but not in fully quenching these systems in the long term, because they may recover after an outflow via re-accretion of gas into the ISM, and also via gas produced via stellar evolution.

Therefore, we must still explore further options to understand both the long-term quiescence of essentially all high-mass galaxies, and the quenching of any kind for the central galaxies within high-mass groups and clusters. Nonetheless, observations of ejective AGN feedback do confirm that this process may be effective at \textit{triggering} quenching within galaxies with intermediate halo masses, as envisaged in some theoretical models (see Sect.~\ref{s232}; and, e.g., \citealt{DiMatteo2005, Springel2005c, Hopkins2006, Hopkins2008, Crain2015}).

\subsection{Radio jet heating of the ICM and CGM}\label{s43}

Radio jets from AGN are observed across a phenomenal range in scales (from pc to Mpc), jet power ($P_\mathrm{jet} =  10^{42} - 10^{47}\,\mathrm{erg/s}$), and radio luminosity ($L_\mathrm{radio} = 10^{38} - 10^{45}\,\mathrm{erg/s}$), see Fig.~\ref{f11}(b) and, e.g., \cite{Fanaroff1974, Carilli1991, Owen2000, McNamara2005, McNamara2007, Fabian2000, Fabian2003, Fabian2006, Croston2007, Fabian2012, Heckman2014}. Powerful radio jets are observed to inflate X-ray cavities in the ICM of clusters, transferring significant thermal energy via bubble expansion and shocks (see, e.g., \citealt{Birzan2004, McNamara2007, Fabian2012, HlavacekLarrondo2012, HlavacekLarrondo2015, HlavacekLarrondo2018, Heckman2014, Croston2011, Hardcastle2020}). 

In Fig.~\ref{f15}, we show a striking example of radio-mode feedback in action within a galaxy cluster, reproduced from \cite{McNamara2007}. The radio lobes, produced via synchrotron emission in the relativistic jet, are seen on cluster-scales (shown in red). Moreover, these jets form cavities in the ICM, as seen in X-ray emission (shown in blue). Clearly, radio-mode AGN feedback has a big impact on the thermodynamics of galaxy clusters. In principle, if the heating from radio jets is sufficient to offset cooling from bremsstrahlung in the ICM, this may lead to quenching of central galaxies via starvation. As such, we will explore this process in some detail here.

\begin{figure}[t!]  
\centering
\includegraphics[width=0.55\textwidth]{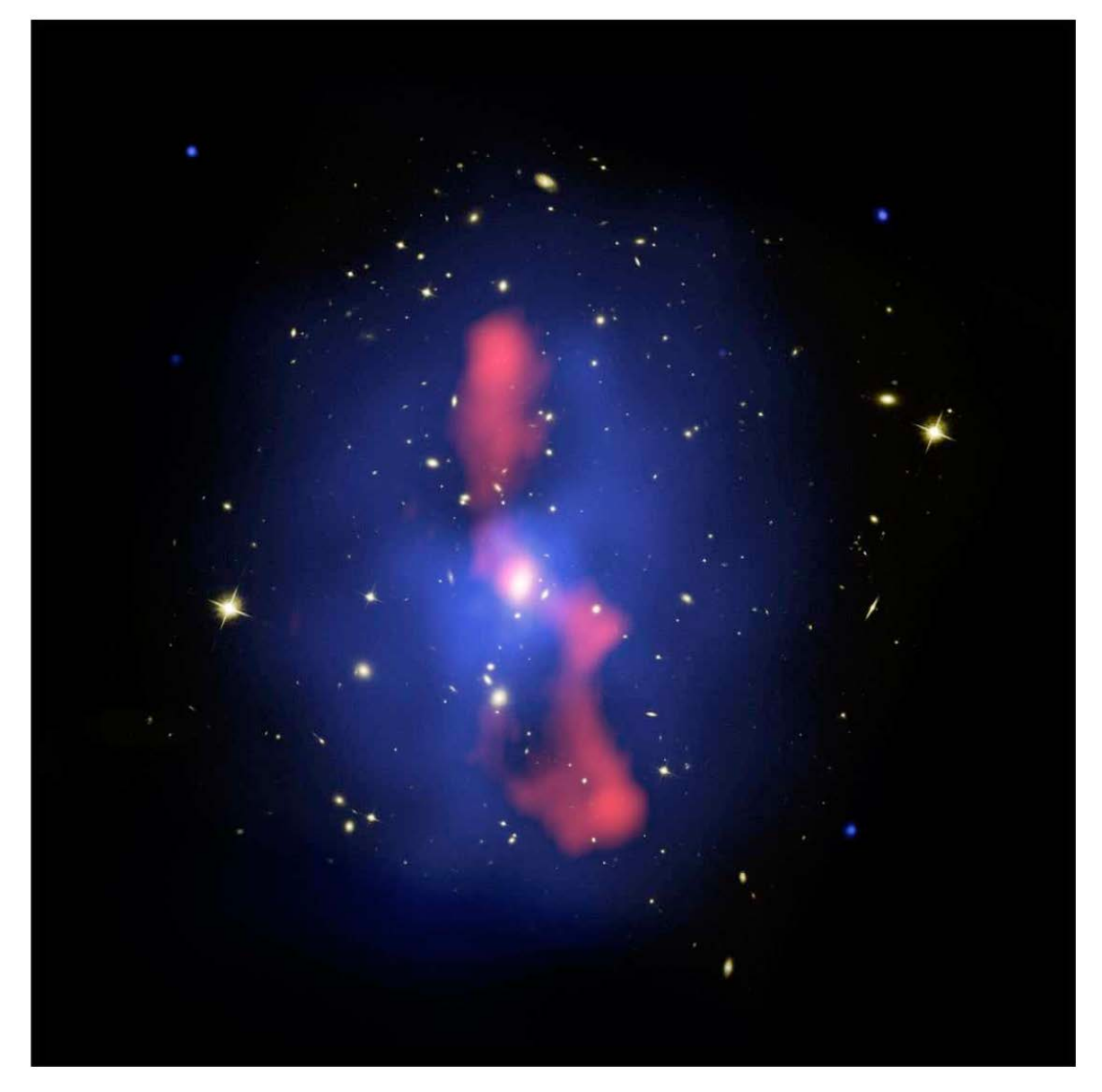}
\includegraphics[width=0.44\textwidth]{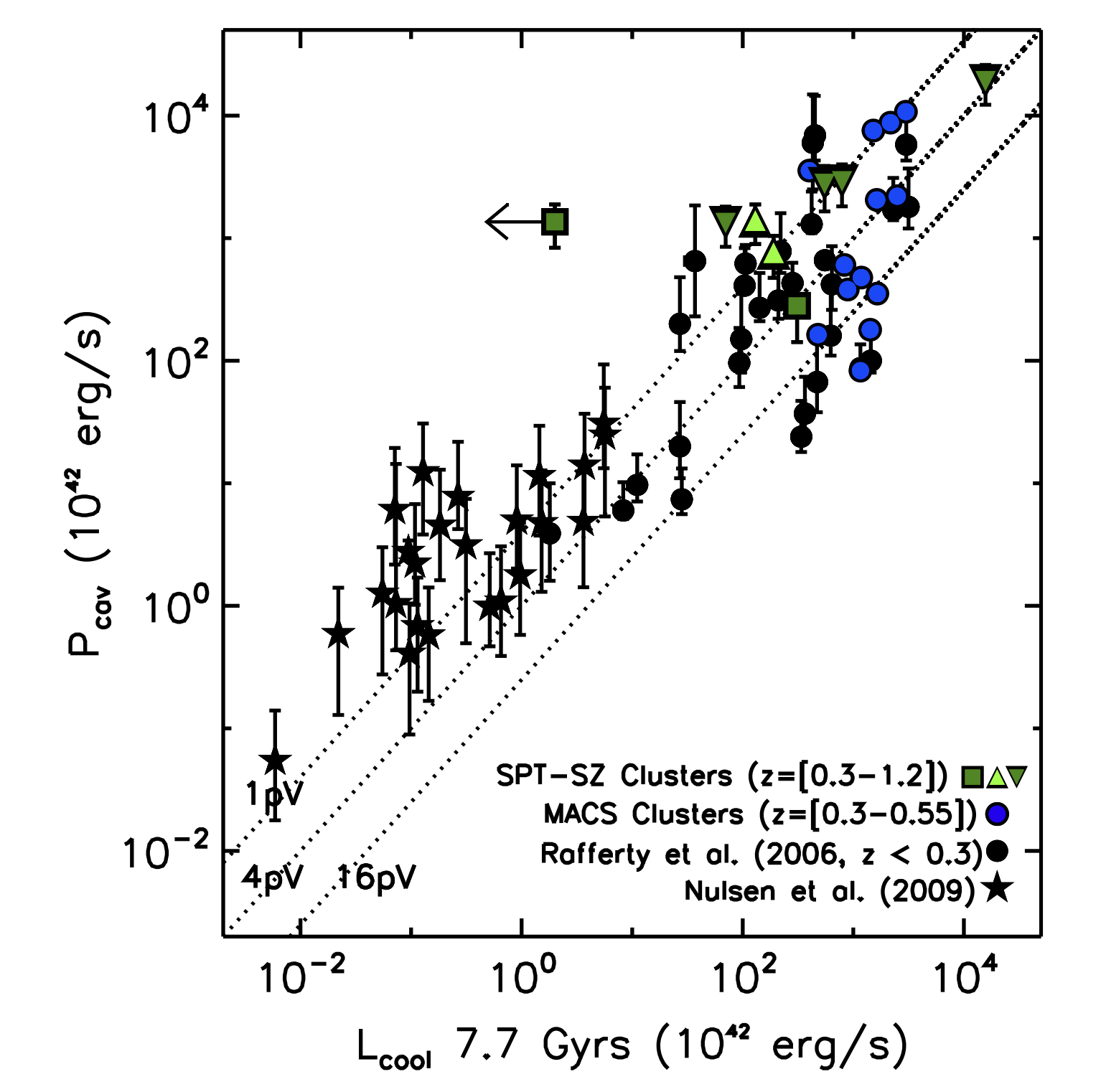}
\caption{\textit{Left-hand panel:} Observed radio-jet feedback in the MS0735.6+7421 galaxy cluster. Optical HST imaging of the cluster galaxies is superimposed with Chandra X-ray imaging (shown in blue) and VLA radio interferometry at 330 MHz (shown in red). The radio lobes, formed around the relativist jets via synchrotron emission, are spatially co-located with vast ($\sim$200\,kpc scale) X-ray cavities in the ICM. This is interpreted as the jets creating the cavities, which transfers thermal energy to the hot gas ICM via shocks during expansion. Modeling reveals that these jets have been inflating the cavities for $\sim$10$^8$\,yr, displacing $\sim$$10^{12}\,M_{\odot}$ of ICM gas, with an estimated jet power of $\sim$$10^{46}\,$erg/s. This image is reproduced from \cite{McNamara2007}. \textit{Right-hand panel:} Cavity power vs. cooling luminosity for a wide range of clusters. The cooling luminosity is computed within a sphere with cooling time $t_\mathrm{cool } < 7.7\,\mathrm{Gyr}$. Clearly, power from cavities increases with increasing cooling luminosity. Dotted lines indicate equality between heating and cooling, assuming: $1pV$, $4pV$ and $16pV$ total enthalpies. From this, it is apparent that the total cavity power is sufficient to offset cooling within clusters at all redshifts observed, with the canonical $4pV$ enthalpy injection. This plot is reproduced from \cite{HlavacekLarrondo2015}.}\label{f15}
\end{figure}

The total energy associated with an X-ray cavity may be computed as (e.g., \citealt{McNamara2005, McNamara2007, Fabian2012, HlavacekLarrondo2015}):

\begin{equation}\label{e38}
E_\mathrm{cav} = H  =  E + pV = \bigg( \frac{\gamma}{\gamma-1} \bigg) \, pV \approx 4pV,
\end{equation}

\noindent where, $\gamma \equiv c_p/c_V$ (i.e., the ratio of specific heats). The total energy of the cavity is taken to be its enthalpy ($H$; i.e., the internal energy of the cavity plus the work done to inflate the cavity to a volume, $V$, against an external ICM pressure, $p$). Assuming an ideal gas (for which $E = (\gamma-1)^{-1}\,pV$), this yields the central term in Eq.~\eqref{e38}, which is true in general in this case. Evaluating this term in the relativistic limit (where $\gamma = 4/3$) yields, $E_\mathrm{cav} = 4pV$. 

The cavity energy is often converted to a power (to enable comparison to the cluster cooling luminosity, see below). This is accomplished by dividing by a timescale, usually taken to be the buoyancy time. Explicitly, this is given by (e.g., \citealt{Churazov2001, Fabian2012, HlavacekLarrondo2015}):

\begin{equation}
P_\mathrm{cav} = \frac{E_\mathrm{cav}}{t_\mathrm{cav}} \quad \text{where} \quad t_\mathrm{cav} \approx t_\mathrm{buoyancy} = R_\mathrm{cav} \sqrt{\frac{A_\mathrm{cav}C_D}{2g(R_\mathrm{cav})V_\mathrm{cav}}}.
\end{equation}

\noindent On the right-hand side above, $R_\mathrm{cav}$ is the cluster-centric radius of the center of the cavity, $A_\mathrm{cav}$ is the cross-sectional area of the cavity, $V_\mathrm{cav}$ is the volume of the cavity, $g(R_\mathrm{cav})$ is the gravitational field strength at the center of the cavity, and $C_D$ is the drag coefficient. This is approximately given by the ratio of cavity size to expansion terminal velocity, $R_\mathrm{cav}/V_\mathrm{term}$ (which may be derived by setting the buoyancy force equal to the drag force, for a given drag coefficient).

In order to compare to the cooling luminosity in clusters it is conventional to first compute the cooling time, and then consider only the luminosity of X-ray emission within a region which could feasibly cool (from the time of observation until the present age of the Universe). Specifically, the cooling time may be defined as (e.g., \citealt{Rafferty2006, Nulsen2009, HlavacekLarrondo2015}):

\begin{equation}\label{e40}
t_\mathrm{cool}(r) \equiv \frac{\mathcal{E}(r)}{|\partial \mathcal{E}(r)/\partial t|} = \frac{5 n(r) K_B \, T(r)}{2L_X(r)},
\end{equation}

\noindent where, the cooling time is defined as normal in this review, and the total energy per unit volume is evaluated as the enthalpy for an ideal gas. Explicitly, $\mathcal{E}(r)$ is the energy density in the ICM, $n(r)$ is the particle number density, $T(r)$ is the temperature, and $L_X$ is the bolometric X-ray luminosity, all evaluated at a given radius, $r$. All cooling is assumed to be performed by X-ray emission from bremsstrahlung, free\,--\,free emission.

The cooling luminosity is typically defined out to the radius at which the cooling time is equal to 7.7 Gyr (see \citealt{Rafferty2006, Nulsen2009, HlavacekLarrondo2015}). This is really just a normalization, which ensures that only the part of the hot ICM which may cool in the time from $z \sim 1$ to the present is incorporated into the calculation. Explicitly, this is computed as:

\begin{equation}
L_\mathrm{X, \, cool, \, 7.7\,\mathrm{Gyr}} = \int_0^{R(t_\mathrm{cool} = 7.7\,\mathrm{Gyr})} \mathcal{E}_X(r) \, dV \quad \text{where} \quad \mathcal{E}_X(r) \equiv \frac{dL_X(r)}{dV},
\end{equation}

\noindent and where, $\mathcal{E}_X$ is the X-ray emissivity of the ICM at a given cluster-centric radius (i.e., the bolometric luminosity per unit infinitesimal volume, see definition on the right above). This is integrated from the center of the cluster out to the radius at which the cooling time (defined in Eq.~\eqref{e40}) is equal to 7.7\,Gyrs. It is assumed that beyond this radius gas cooling would not be relevant for clusters out to $z \sim 1$. Ultimately, this leads to an important stability criterion:

\begin{equation}
\text{ICM Stability Criterion:} \quad P_\mathrm{cav} \gtrsim L_\mathrm{X, \, cool,\, 7.7\,\mathrm{Gyr}}
\end{equation}

\noindent where, in order for radio-jets to offset cooling in a cluster, the power in X-ray cavities must exceed (or at least approximately equal) the cooling luminosity on relevant cooling time scales.

In Fig.~\ref{f15}(b), we show cavity power vs. cooling luminosity for a variety of galaxy clusters with observed X-ray cavities, reproduced from \cite{HlavacekLarrondo2015}. The heating power from radio-mode feedback induced X-ray cavities rises steeply with the hot gas ICM cooling luminosity. Lines of heating - cooling equality are plot for a range of possible cavity energies as a multiplier of $pV$. For the canonical relativistic enthalpy (i.e., $4pV$, see above), it is clear that radio-mode feedback can offset cluster cooling in essentially all cases analyzed here. This is a really important result because radio-mode heating is the only known mechanism to prevent CGM cooling and condensation into galaxies within the highest mass haloes. Ultimately, this may quench central galaxies via starvation of gas supply (see, e.g., \citealt{Fabian2006, McNamara2007, Fabian2012, HlavacekLarrondo2012, HlavacekLarrondo2015, HlavacekLarrondo2018}). 

Interestingly, radio-mode feedback is also observed in the lower mass haloes of groups and isolated elliptical galaxies (e.g., \citealt{McNamara2007, Croston2008, David2009, Baldi2009a, Baldi2009b, Nulsen2009, Gitti2010, OSullivan2011, Fabian2012, Heckman2014}). These results strongly suggest that radio-mode AGN feedback may be effective at stabilizing the CGM around massive galaxies at all halo masses above $M_\mathrm{Halo} \sim 10^{12}\,M_{\odot}$, where hot static atmospheres start to form (e.g., \citealt{Dekel2006, Dekel2009, Dekel2019}).

However, it is fair to acknowledge that the evidence for radio-mode AGN feedback stabilizing the thermodynamics of the hot gas halo is most strong in clusters, and high-mass groups. Lower-mass galaxies than the centrals of clusters (e.g., $M_* \sim 10^{11} M_\odot$) also need to be maintained quenched for long cosmological time periods. While relativistic jets are clearly observationally compelling, their relevance for the demographics of high-mass galaxies across $\sim$1.5 dex of galaxy stellar mass still needs stronger observational support. Other forms of AGN feedback (and, indeed, other processes altogether) may still be required to fully explain the phenomenon of mass-quenching as a whole. That said, in the highest mass haloes, stabilization of cooling flows from radio-mode feedback appears to be both theoretically essential and well-supported by contemporary observations.

\subsection{Where do AGN reside on the SFMS?}\label{s44}

So far in this section, we have discussed some of the compelling evidence for SN and AGN-driven galactic outflows, as well as direct evidence for radio jets heating the ICM and CGM around massive central galaxies. There can be little doubt that feedback from stellar evolution and SMBH accretion occurs within galaxies, broadly as implemented in simulations and models. However, so far we have not directly linked observations of feedback to quenching in galaxies, and, crucially, we have not yet considered the impact of feedback on galactic populations as a whole. In the remaining parts of this section, we will address these outstanding issues.

Given that AGN feedback is predicted by theory and simulations to be the leading cause of high-mass galaxy quenching (see Sect.~\ref{s25}; and, e.g., \citealt{Hopkins2006, Hopkins2008, Croton2006, Vogelsberger2014a, Schaye2015, Dave2019}), it is natural to consider where AGN reside within the SFMS. Somewhat naively, one may guess that, if AGN quench galaxies, they ought to be abundant in quenched systems (or at least in quenching systems). This question has garnered intense research interest over the past couple of decades, and has produced much confusion in the field.

Some works claim that AGN are commonly found in quenched, or quenching (i.e., green valley) galaxies, presumably in support of a potential connection between AGN and quenching (e.g., \citealt{Nandra2007, Salim2007, Bundy2008, Georgakakis2008, Trump2015}). Alternatively, many other works identify no link between AGN and quenching, and even find that luminous AGN reside predominantly within star forming galaxies (e.g., \citealt{Hickox2009, Silverman2009, Aird2012, Rosario2013, Mullaney2012, Hickox2014, Stanley2015, Ellison2016, Hickox2018, Florez2020}). However, it is important to note that this does not rule out quenching via AGN feedback for several reasons, which we will discuss in the next two subsections. Indeed, many simulations predict that the most powerful AGN should reside predominantly within the star forming main sequence, with no obvious link between AGN and quenched systems, despite AGN feedback manifestly causing quenching within the simulation (see \citealt{Piotrowska2022, Ward2022, Bluck2023, Bluck2024, Lim2025}). 

There are at least two issues which are important in understanding why this confusion exists, and how to move beyond it. First, there are many ways in which to identify AGN, which depend upon the intrinsic bolometric luminosity of the SMBH accretion disc, its surrounding environment (e.g., the presence of a molecular torus), and the orientation at which it is viewed from Earth (e.g., \citealt{Urry1995, Hickox2009, Heckman2014}). Hence, it is at least possible that certain types of AGN are more prevalent in certain types of galaxies than others. Second, AGN are notoriously highly stochastic, varying substantially in luminosity on short timescales. Furthermore, the duty cycle (i.e., how long an active phase lasts) is very short compared to the age of a galaxy (e.g., \citealt{Kelly2009, Hickox2014, Aird2017, Sartori2018}). Consequently, the misaligned timescales of star formation and SMBH accretion pose serious challenges to `catching AGN-driven quenching in the act'.

\begin{figure}[ht]  
\centering
\includegraphics[width=0.85\textwidth]{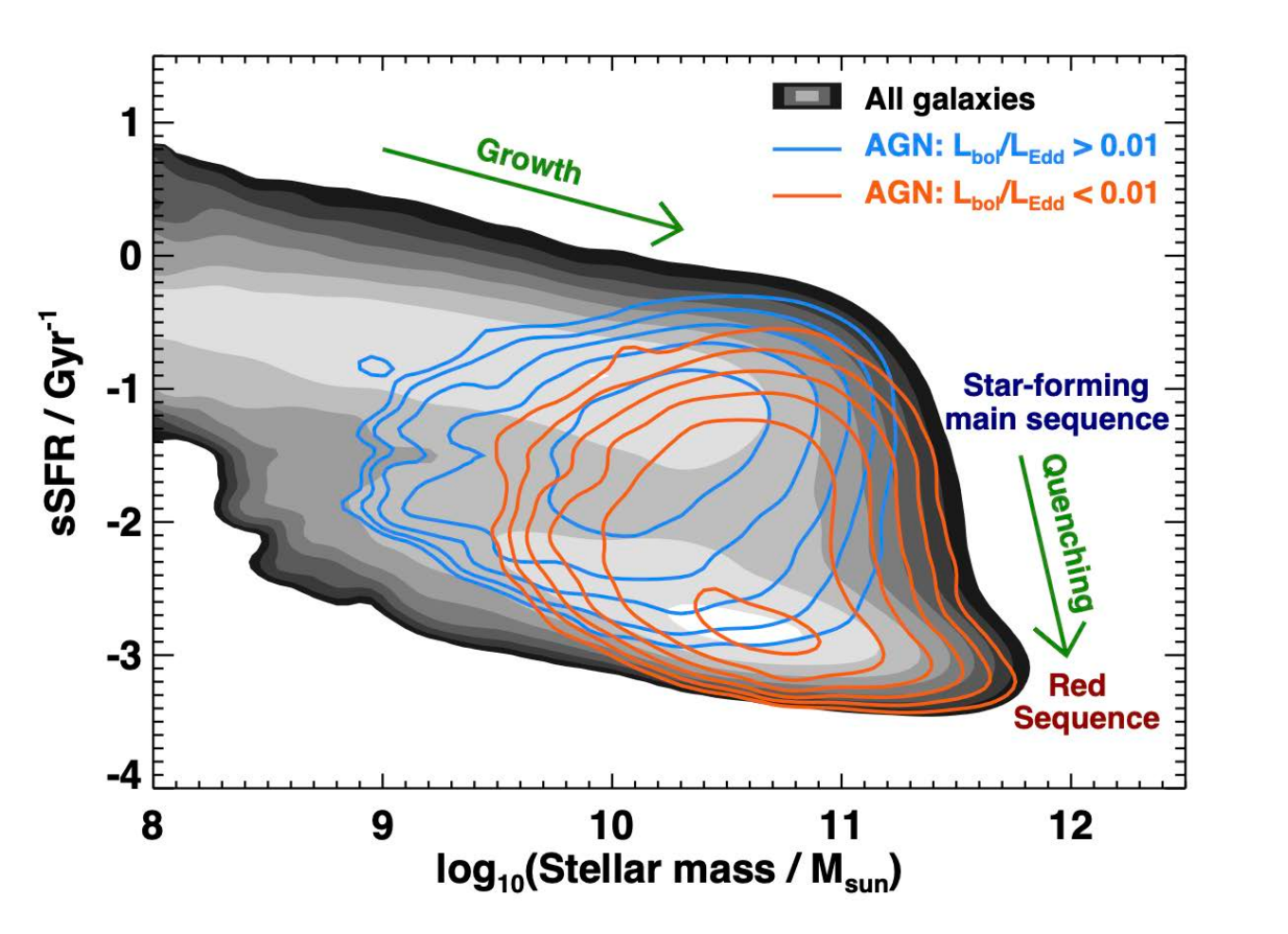}
\caption{The sSFR\,--\,$M_*$ relationship for local galaxies, reproduced from \cite{Heckman2014}. As with the SFMS, local galaxies separate out into two peaks on this diagram (as seen by grey-scale contours). Additionally, the location of radiatively efficient ($L_\mathrm{AGN}/L_\mathrm{Edd} > 0.01$; shown in blue) and radiatively inefficient ($L_\mathrm{AGN}/L_\mathrm{Edd} < 0.01$; shown in red) AGN are overlaid as open contours. In the local Universe, the most radiatively efficient AGN (which typically present with X-ray, optical, and MIR emission) tend to reside in actively star forming systems, although the population extends also into the green valley. Conversely, radiatively inefficient AGN (which tend to present primarily with radio emission) reside predominantly on the quenched `red sequence', although these also extend throughout the green valley region. }\label{f16}
\end{figure}

Both of the above issues severely impact what one might expect to find in the AGN population with respect to star formation, even if AGN are ultimately the cause of quenching. For example, it remains unclear how long after an ejective phase an AGN is expected to be visible (e.g., \citealt{DiMatteo2005, Hopkins2006, Hopkins2008}), and hence if any observable connection between AGN identification and star formation reduction is even predicted in contemporary simulations (e.g., \citealt{Piotrowska2022, Ward2022}). Finally, it is important to appreciate that the term `AGN' is not truly a physical class in any objective sense. Any SMBH may accrete matter and radiate energy, whether or not this constitutes an AGN is really just a matter of observer convention --- it is highly dependent upon both the waveband and depth of observation, the level of obscuration, and the background emission expected from `normal' galaxies at the relevant frequencies. Moreover, throughout the lifetime of an `average' galaxy, it is likely to host an AGN (by one definition or another) at various stages. This may well leave an imprint on its evolution, including its capacity to form stars. But this is highly likely to be decoupled from whether it currently presents as an AGN or not (in much the same way that the stellar mass of a galaxy is decoupled from its current SFR, especially in quenched systems). 

In Fig.~\ref{f16}, we show the sSFR\,--\,$M_*$ relationship for low-$z$ galaxies (as grey-scale contours), reproduced from \cite{Heckman2014}. This is essentially just a rotation of the SFMS, and contains identical information content to the canonical relation. The plane is clearly divided into two peaks --- an actively star forming peak (at high sSFR) and a quenched peak (at low sSFR). Additionally, in Fig.~\ref{f16}, the location of radiatively efficient AGN (shown in blue; $L_\mathrm{AGN}/L_\mathrm{Edd} > 0.01$) and radiative inefficient AGN (shown in red; $L_\mathrm{AGN}/L_\mathrm{Edd} < 0.01$) are displayed by open contours. Radiatively efficient AGN (which are typically identified in X-ray, optical, and mid-IR observations) are predominantly located within actively star forming, high-mass galaxies, with a long tail to lower star formation rates. Conversely, radiatively inefficient AGN (which are primarily identified in radio observations) are predominantly located within high-mass, quenched galaxies, but with a long tail to higher star formation rates. This general picture is broadly supported by, e.g., \citealt{Hickox2009, Best2012, Mullaney2012, Rosario2013, Hickox2014, Ellison2016, Hickox2018}. Hence, where AGN reside on the SFMS depends critically upon which type of AGN one is referring to.

In Fig.~\ref{f17}(a), we show bolometric AGN luminosity vs. FIR luminosity (a good tracer of SFR), reproduced from \cite{Hickox2014}. A very strong correlation is identified, which clearly implies that the most luminous AGN reside in the most actively star forming galaxies. However, in Fig.~\ref{f17}(b), we show this relationship the other way around, i.e., FIR luminosity vs. AGN luminosity. Perhaps surprisingly, the results are very different. Since the points shown are not measurements of individual galaxies, but rather binned measurements containing many galaxies, there is no logical requirement that the two versions of the plot be identical. However, the underlying reason for this change (proposed in \citealt{Hickox2014}) is a result of short AGN duty cycles and high stochasticity.

If one bins by FIR luminosity, which is a consequence of re-radiated starlight, one has a stable baseline. With this approach, it is clearly visible that AGN luminosity rises systematically with SFR. However, if one bins by AGN luminosity (which is highly stochastic) one does not have a stable baseline. For example, a measurement of low luminosity AGN will include systems which are temporarily `off', in addition to systems which are fundamentally low accreting (regardless of the exact time of measurement). This results in the true trends being diluted with noise. The lines on the two panels of Fig.~\ref{f17} show model predictions for the \textit{same} underlying relationship, just with AGN stochasticity incorporated. Hence, this narrative is clearly effective at explaining these interesting differences between binning strategies.

\begin{figure}[ht]  
\centering
\includegraphics[width=\textwidth]{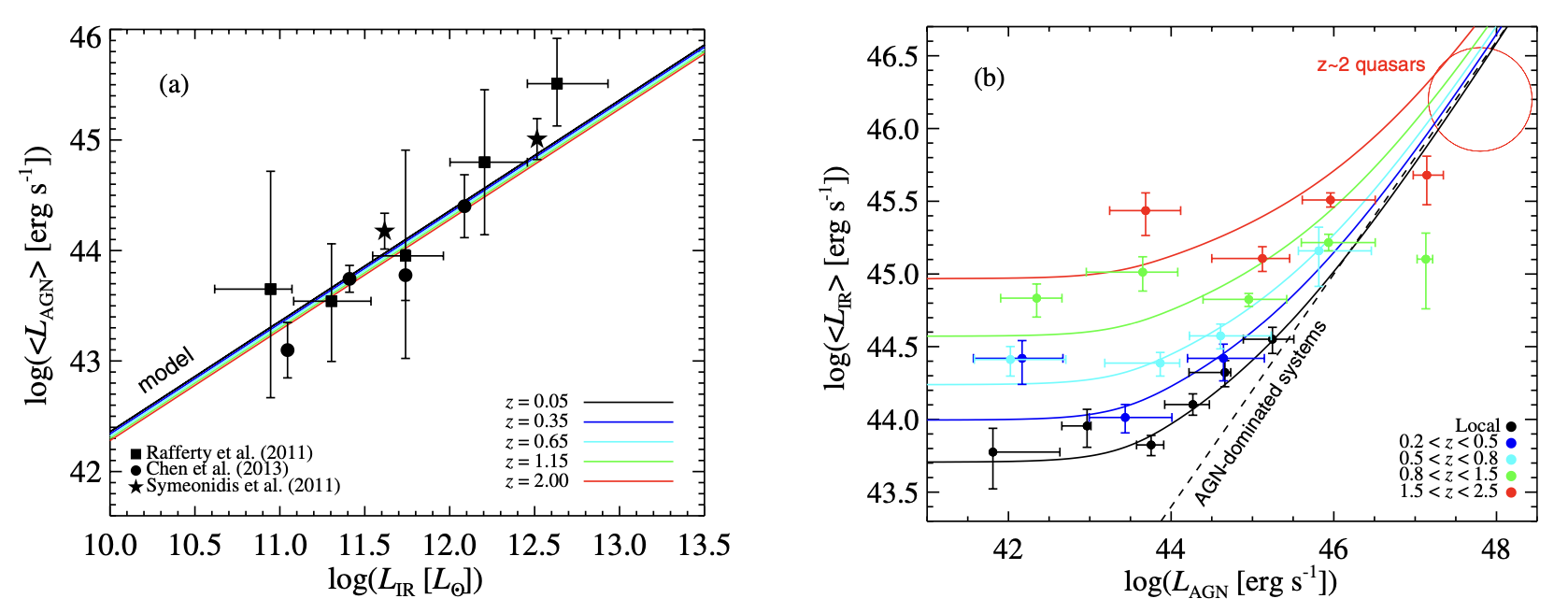}
\caption{\textit{Panel-(a):} AGN luminosity plot as a function of FIR luminosity (a good tracer of SFR). A clear positive trend is identified, indicating co-evolution between galaxies and their SMBHs, likely as a result of co-fueling. \textit{Panel-(b):} FIR luminosity plot as a function of AGN luminosity, i.e., the other way around from panel-(a). No strong trends are seen in this parameterization. This is possible because both panels show binned data within each point. In the case of binning by AGN luminosity, high stochasticity results in an unrepresentative sampling. Put simply, low luminosity AGN may be just going through an `off' period, which is normal for all types of galaxies. Lines on both panels show the expected results for strongly correlated SFR - $L_\mathrm{AGN}$, when applying AGN stochasticity, which clearly show that the observed results are explicable, given this scenario. This figure is reproduced from \cite{Hickox2014}.}\label{f17}
\end{figure}

Taken together, Figs.~\ref{f16} and \ref{f17} provide strong evidence for co-evolution of galaxies and AGN, whereby high SFRs and high $L_\mathrm{AGN}$ go hand in hand. Ultimately, this is likely a result of co-fueling, whereby both SMBHs and galaxies require cool gas in the ISM for efficient star formation and SMBH accretion. So, what does this have to do with quenching? Well, so far nothing at all! There appears to be an AGN main sequence which runs parallel to the SFMS, and for most systems this does not seem to cause quenching (at least contemporaneously to the detection of the AGN). It remains possible that at the very high-mass end of the SFMS quasar-mode feedback may become energetic enough to engender the triggering of quenching. Yet, if this is the case, the AGN luminosity must rapidly decline thereafter to be consistent with contemporary observations. This follows because the majority of green valley systems (and the vast majority of quenched galaxies) do not host radiatively efficient AGN. Therefore, the observations of radiatively efficient AGN neither support nor reject the AGN ejective feedback quenching paradigm.

On the other hand, radiatively inefficient AGN are prevalent in quiescent systems (see \citealt{Hickox2009, Heckman2014, Hickox2014, Best2014, Gurkan2014, Ellison2016, Barisic2017, Mulcahey2022}). This implies that radio-mode AGN feedback is at least a plausible mechanism to keep high-mass galaxies quiescent, on a population basis, according to contemporary observations. Moreover, it is crucial to appreciate that preventative feedback is all that is required to quench a galaxy, from a conceptual point of view. Once gas replenishment from cold gas inflows is shut off due to the formation of a hot static atmosphere, and cooling from the CGM is stabilized via radio-mode feedback, galaxies will quench via starvation, once they deplete their ISM via star formation. 

Nonetheless, it still remains to be seen whether this mode of feedback is sufficient to quench central galaxies at the right rates and fractions (i.e., with the correct demographics) as observed at varying cosmic epochs. Additionally, maintenance-mode feedback within the ISM will still be required to offset cooling flows via gas return from stellar evolution, as discussed in Sect.~\ref{s2}. Of course, additional processes may speed this up (e.g., mergers, quasar-driven outflows, or internal dynamical stabilization), but these processes are not \textit{essential} for quenching to occur, only for modifying the rate at which it transpires. 

Nonetheless, the observations are only \textit{consistent} with radio-mode AGN feedback being connected to quenching on a populations basis. A perfectly valid alternate narrative would be that, just as high luminosity AGN and high SFR go together via co-fueling, so do low luminosity AGN and low SFR via a lack of fuel (caused by whatever mechanism). Hence, these observations are far from decisive. Ultimately, the issue with this approach arises from a lack of control samples, which enable ruling out alternative scenarios.

At the end of this sub-section, despite many points of confusion, hopefully at least one thing is clear --- that the question of where AGN reside on the SFMS is not nearly as decisive a test of AGN-driven quenching as one might have assumed. Indeed, even which galaxies deserve the moniker `AGN' is rather subtle and ambiguous. We turn to more promising avenues for testing the AGN feedback paradigm on a population basis in the final parts of this section.

\subsection{Quenching dependence on supermassive black hole mass}\label{s45}

Ultimately, the problem with interpreting the location of AGN on the SFMS arises because of three underlying issues: (i)~AGN are highly stochastic and relatively short lived; (ii)~AGN present in profoundly different forms, which are observed in different types of observations; and (iii)~whether a galaxy is observed to be an AGN at the time of observation is not constraining of whether it was an AGN in the past. To combat all of these issues simultaneously, ideally one would like to extract the total energy released from AGN feedback over the lifetime of a galaxy. Fortunately, there exists an excellent observable proxy for just this parameter, SMBH mass (see \citealt{Soltan1982, Silk1998, Bluck2014, Bluck2016, Terrazas2016, Terrazas2017, Bluck2020a, Piotrowska2022, Bluck2022, Bluck2023, Bluck2024}). Explicitly,

\begin{equation}
E_\mathrm{AGN} \, \sim \, \int \epsilon_f \,  L_\mathrm{AGN}(t) \, dt \, \sim \, \epsilon_f \epsilon_r c^2 \int \dot{M}_\mathrm{BH}(t) \, dt \, \sim \, \epsilon_f \epsilon_r c^2 M_\mathrm{BH} 
\label{e43}
\end{equation}

\begin{equation}
\implies E_\mathrm{AGN} \propto M_\mathrm{BH} \quad \text{where} \quad L_\mathrm{AGN} = \epsilon_r c^2 \dot{M}_\mathrm{BH}
\label{e44}
\end{equation}

\noindent and where, the total energy from AGN feedback ($E_\mathrm{AGN}$) is given by the temporal integral of bolometric AGN luminosity, multiplied by the feedback efficiency ($\epsilon_f$). In turn, the bolometric AGN luminosity is given by accretion rate multiplied by $\epsilon_r\,c^2$ (see Sect.~\ref{s232}). This leads to a direct proportionality between SMBH mass and total feedback energy, assuming constant efficiencies (as first pointed out in \citealt{Soltan1982, Silk1998}). In reality this is an over-simplification. Both efficiencies may vary as a function of SMBH mass (and hence implicitly time; see \citealt{Bluck2020a} for further discussion). Nonetheless, this is an excellent baseline to help interpret how one might anticipate SMBH mass to relate to quenching. Explicitly, if galaxies quench via AGN feedback, quenched galaxies must host more massive SMBHs than their star forming counterparts. This follows because more massive SMBHs indicate greater integrated injection of feedback energy, over the lifetime of a galaxy.

\begin{figure}[ht]  
\centering
\includegraphics[width=0.49\textwidth]{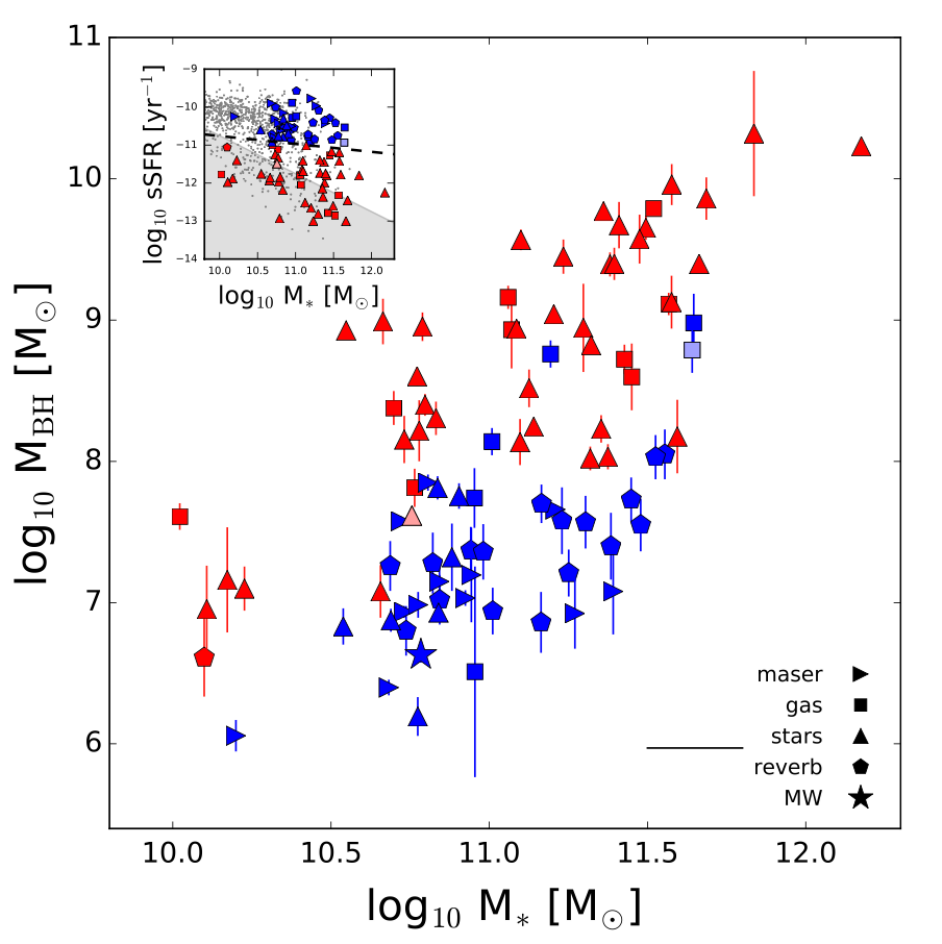}
\includegraphics[width=0.49\textwidth]{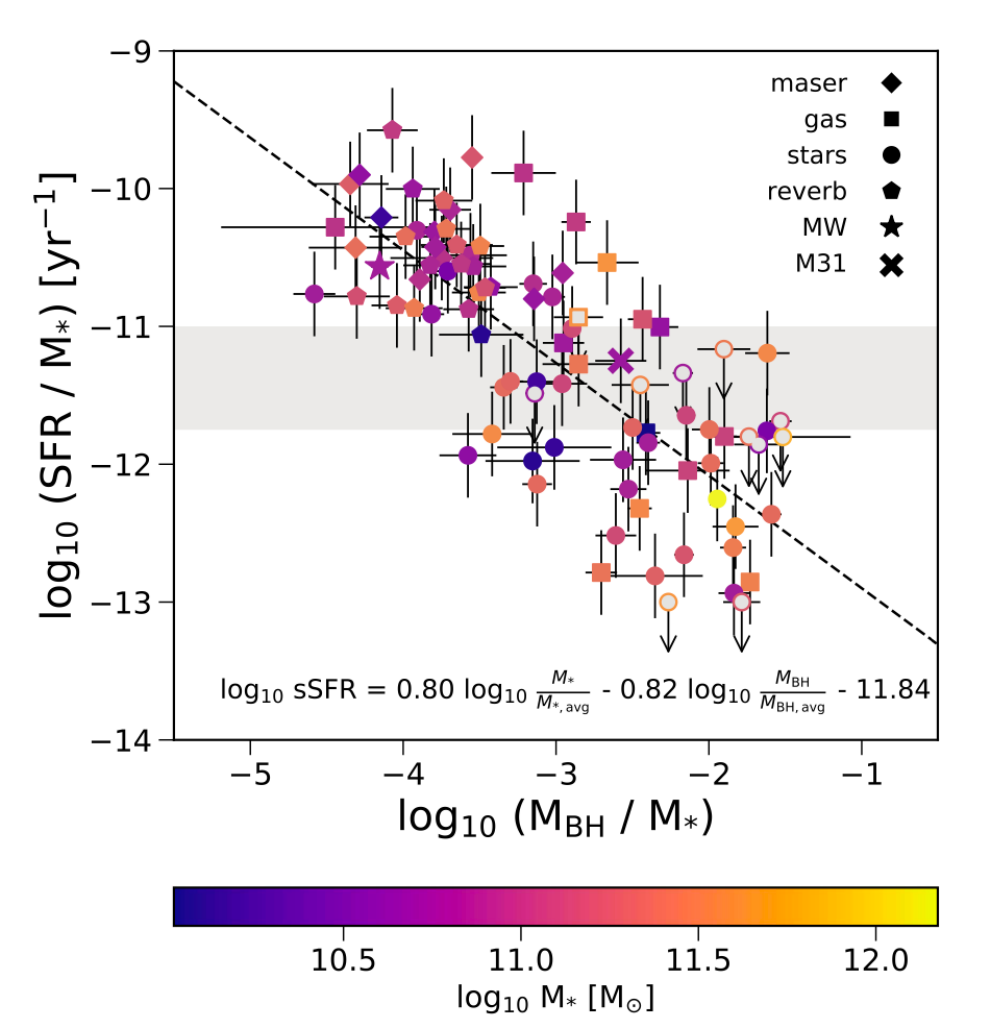}
\caption{\textit{Right-hand panel: }Dynamically measured SMBH mass plot as a function of host galaxy stellar mass, reproduced from \cite{Terrazas2016}. Each data point in the plot is color coded by star forming state, with blue indicating star forming systems and red indicating quenched systems. It is immediately apparent that quenched galaxies tend to host higher mass SMBHs than their star forming counterparts, at a fixed stellar mass. This result may be interpreted as implying that historic AGN feedback (traced by $M_\mathrm{BH}$) is important for quenching central galaxies. \textit{Left-hand panel: }sSFR plot as a function of the ratio of SMBH mass to stellar mass, for galaxies with dynamically measured central black hole masses, reproduced from \cite{Terrazas2017}. A clear sequence is apparent, whereby galaxies with higher $M_\mathrm{BH}/M_*$ ratios form stars at a lower rate for their stellar mass. This indicates a plausible causal connection between historic AGN feedback and central galaxy quenching. }\label{f18}
\end{figure}

Using a sample of 91 central galaxies with dynamically constrained SMBH masses, \cite{Terrazas2016} investigate how quenching relates to SMBH mass. In Fig.~\ref{f18} (left-hand panel), we show SMBH mass vs. stellar mass for this sample, color coded by whether galaxies are star forming (blue) or quenched (red), reproduced from \cite{Terrazas2016}. Clearly, quenched galaxies host more massive SMBHs at a fixed stellar mass than star forming systems. This is certainly consistent with the AGN quenching paradigm for centrals. Furthermore, similar results are found for much larger data sets ($\sim$500\,k systems), utilizing estimated SMBH mass (e.g., via the $M_\mathrm{BH}$\,--\,$\sigma$ relation) in \cite{Bluck2014, Bluck2016, Piotrowska2022, Bluck2022}. We will discuss these large statistical samples in more detail in the next sub-section, in the context of direct tests of cosmological simulations.

On the right-hand panel of Fig.~\ref{f18}, we show the sSFR\,--\,$M_\mathrm{BH}/M_*$ relation from \cite{Terrazas2017}. A clear anti-correlation between sSFR and the ratio of black hole to stellar mass is evident. This relationship is interpreted in \cite{Terrazas2016, Terrazas2020} as being a direct consequence of AGN feedback causing quenching, which requires more energy to offset cooling in more massive haloes (which in turn is approximated by stellar mass in these data). Ultimately, the advantage of testing the AGN quenching paradigm with SMBH mass, rather than AGN detection or luminosity, is that the former gives a stable (integrated) estimate of the entire accretion history, not just a single snapshot thereof. Moreover, the results are really quite striking. Clearly, observations support a picture in which central galaxies quench once they host a sufficiently massive SMBH for their halo, in line with heuristic theoretical expectations.

\subsection{Direct tests of the AGN feedback quenching paradigm}\label{s46} 

We now turn to direct tests of cosmological simulations with observational data, for large statistical samples of galaxies. We start with a relatively simple visual comparison of a semi-analytic model to SDSS data, and then move on to more sophisticated machine learning tests applied to cosmological hydrodynamical simulations.

In Fig.~\ref{f19}, we present the SMBH mass--halo mass relationship from the L-Galaxies semi-analytic model (left-hand panel; see \citealt{Henriques2015}) and observed galaxies from the SDSS DR7 (right-hand panel; see \citealt{Abazajian2009}), reproduced from \cite{Bluck2022}. For the SDSS, halo masses are estimated via an abundance matching approach (performed in \citealt{Yang2007, Yang2009}), and SMBH masses are estimated via the $M_\mathrm{BH}$\,--\,$\sigma$ relation (see \citealt{Ferrarese2000, Saglia2016, Piotrowska2022}). This enables over 400\,k central galaxies to be included in the observational analysis, compared to over a million systems in the model. Both panels are color coded by the fraction of quenched galaxies within each region of the parameter space (as indicated by the color bar). Additionally, density contours are overlaid in white, to enable one to see where the majority of the sample resides.

\begin{figure}[ht]  
\centering
\includegraphics[width=\textwidth]{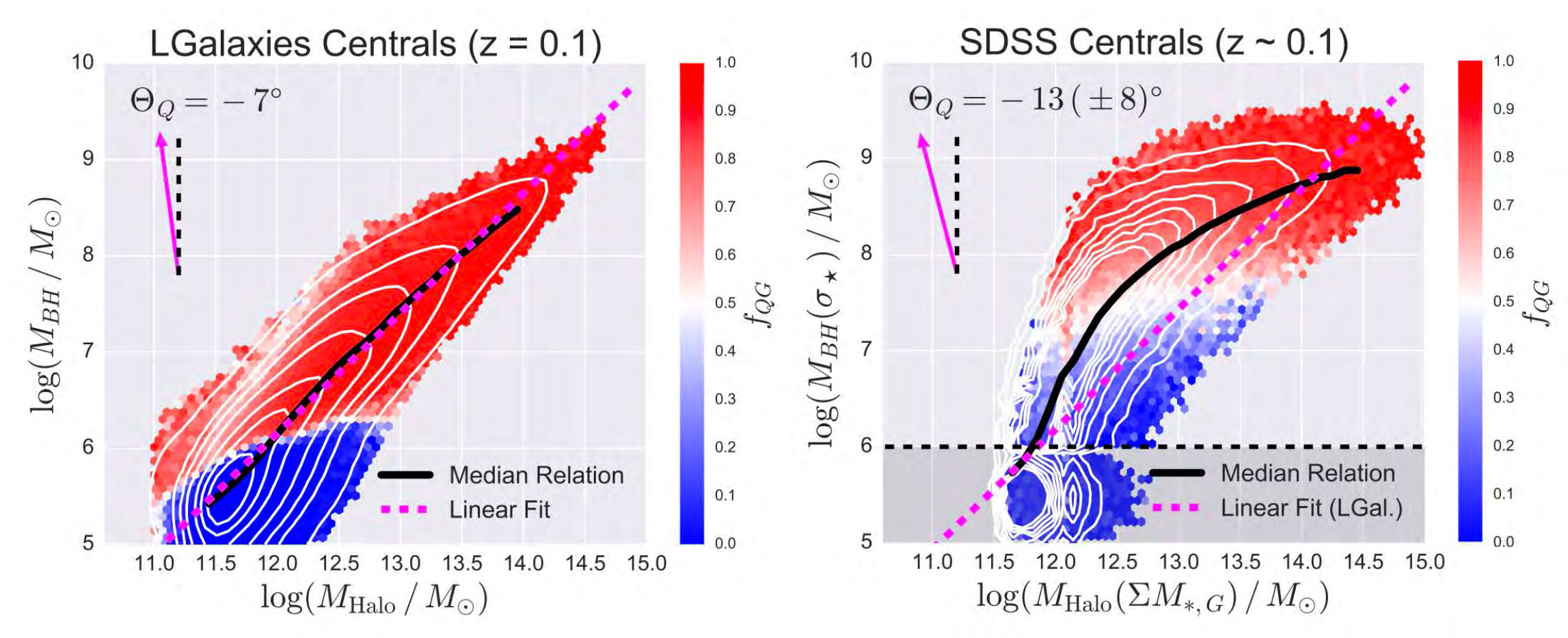}
\caption{SMBH mass vs. halo mass, color coded by quenched fraction for central galaxies in the L-Galaxies semi-analytic model (left-hand panel) and the SDSS (right-hand panel). In the SDSS, halo masses are inferred via abundance matching and SMBH masses are inferred via the $M_\mathrm{BH} - \sigma$ relation. Clearly, the quenched fraction rises as a steep function of SMBH mass in both panels, at a fixed halo mass. We quantify the optimal route to quench galaxies in the plane via the quenching angle. In both the model and the data, this shows a primary quenching dependence on SMBH mass at a fixed halo mass, with a weak anti-correlation with halo mass at a fixed SMBH mass (see arrows on both panels). Ultimately, SMBH mass may be taken as a proxy for radio-heating and halo mass as a proxy for CGM cooling (which is explicitly true in the model). This figure is reproduced from \cite{Bluck2022}.}\label{f19}
\end{figure}

It is clear in both the L-Galaxies semi-analytic model and the SDSS observations that quenching proceeds as a strong function of SMBH mass (in agreement with observations of small dynamically measured samples, see \citealt{Terrazas2016, Terrazas2017}). Conversely, little dependence of quenching on halo mass is evident, which strongly implies that halo mass quenching by itself cannot cause massive galaxy quenching (see also \citealt{Bluck2014, Bluck2016, Bluck2022, Piotrowska2022} for similar conclusions).

Furthermore, we quantify the optimal route to quench galaxies through the plane via the quenching angle statistic (for details on the method see \citealt{Bluck2020a}). In both the model and the observed data, this reveals that quenching proceeds primarily via a strong correlation with SMBH mass (at a fixed halo mass) and via a much weaker anti-correlation with halo mass (at a fixed SMBH mass). Qualitatively, the agreement between L-Galaxies and the SDSS is excellent in this respect (although there are numerous features which are not well reproduced, see \citealt{Bluck2022} for further discussion). 

In the L-Galaxies model, quenching is caused explicitly by radio-mode AGN heating, which scales fundamentally with SMBH mass (see Sect.~\ref{s233}), which gives rise to the strong positive correlation with quiescence (at a fixed halo mass). Alternatively, increasing halo mass increases the cooling rate from free\,--\,free emission (see Sect.~\ref{s231}), giving rise to the weak anti-correlation seen at a fixed SMBH mass. Ultimately, more radio-mode heating is required to stabilize a higher mass halo. The close agreement with observations in the direction of quenching arrows through this plane implies that this is a highly plausible conceptual model for central galaxy quenching.

In Fig.~\ref{f20}, we show a host of random forest (RF) classification analyses to predict when galaxies will be quenched or star forming, reproduced from \cite{Piotrowska2022}. Full details on this method can be found in \cite{Bluck2022}. In the top panels, SMBH mass, stellar mass, halo mass, and a random variable are passed to the classifier as features. The left-hand panel shows results for three cosmological hydrodynamical simulations (EAGLE, \citealt{Schaye2015, Crain2015}; Illustris, \citealt{Vogelsberger2014a, Vogelsberger2014b}; and IllustrisTNG, \citealt{Nelson2018, Pillepich2018, Springel2018, Marinacci2018, Naiman2018}). 

In all simulations, SMBH mass is found to be overwhelmingly the most predictive feature for predicting when galaxies quench. This is in close alignment with the simple argument presented in Eqs.~\eqref{e43}\,--\,\eqref{e44}, above. Given that these simulations all utilize AGN feedback to quench galaxies, but have very different mechanisms in detail, this result points towards a clear method to test the entire paradigm of AGN feedback quenching, largely independent of the exact mode of operation.

On the top right-hand panel in Fig.~\ref{f20}, the same test is performed for observed central galaxies from the SDSS (see \citealt{York2000, Abazajian2009}). Halo masses are estimated via abundance matching (performed in \citealt{Yang2007, Yang2009}) and stellar masses are inferred via SED fitting (performed in \citealt{Mendel2014}). Five different calibrations are used to infer SMBH masses on a population basis (as outlined in the legends). These include calibrations with central velocity dispersion, bulge mass, and combinations of the two. Additionally, separate scaling relations for early- and late-type galaxies, as well as pseudo- and classical-bulges, are explored.

\begin{figure}[ht]  
\centering
\includegraphics[width=\textwidth]{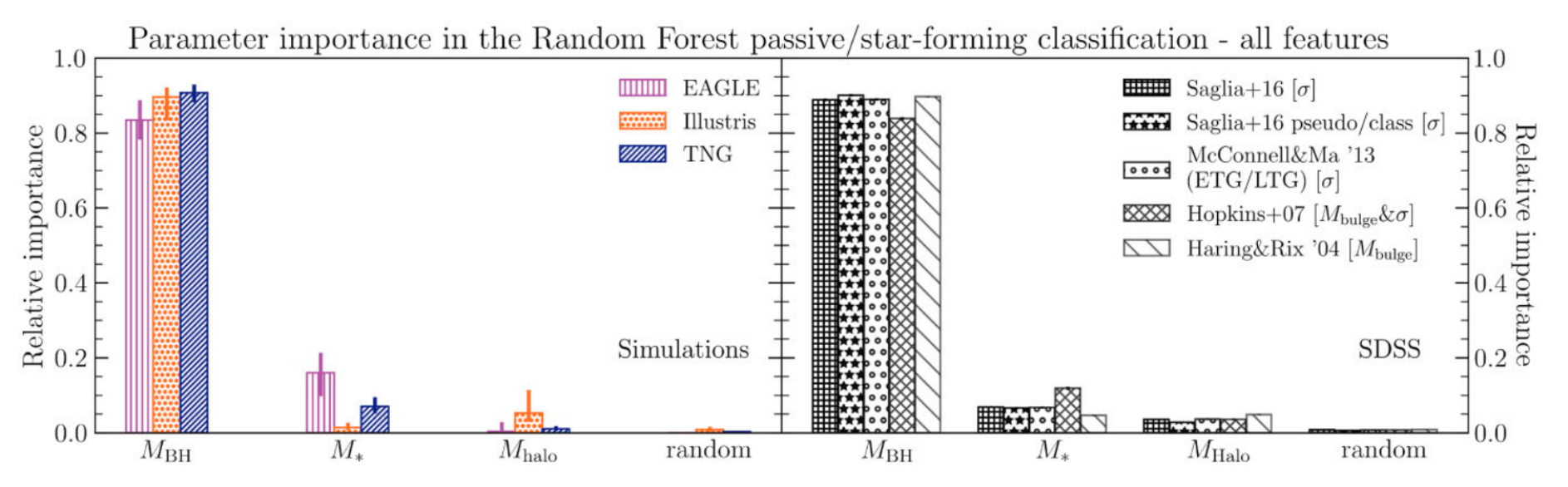}
\includegraphics[width=0.65\textwidth]{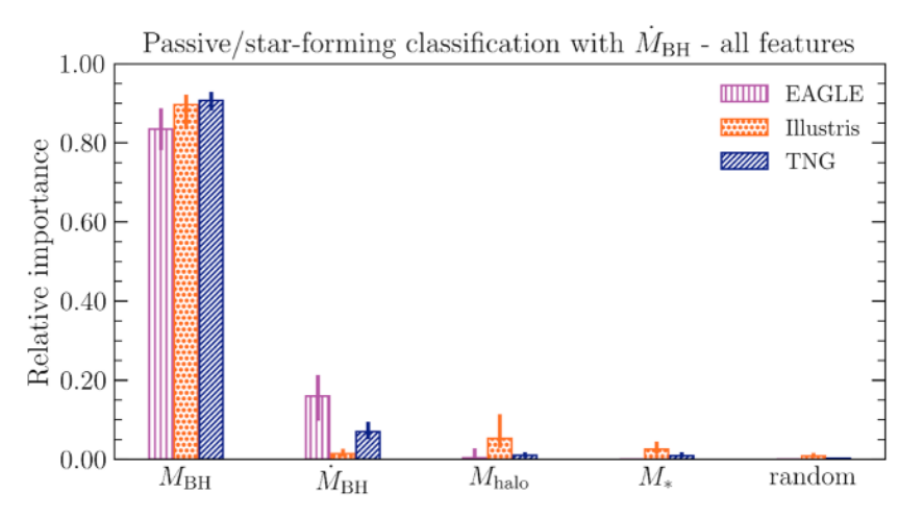}
\caption{A series of random forest (RF) quenching classification analyses to predict whether galaxies are star forming or quenched on the basis of various input parameters. On all panels, the variables used in the classification are listed along the X-axis, and the relative importance is presented as the bar height on the Y-axis. In the top row, results are shown for SMBH mass, stellar mass, halo mass, and a random number, for three cosmological simulations (left panel) and for observations from the SDSS DR7 (right panel). In the SDSS, halo masses are inferred via abundance matching, and SMBH masses are inferred via five different calibrations (utilizing central velocity dispersion, bulge mass, and combinations thereof). In all simulations studied, and in the observational data (regardless of calibration choice), SMBH mass is clearly found to be by far the most predictive parameter of central galaxy quenching. In the bottom panel, SMBH accretion rate is added as a parameter for the simulations. Interestingly, SMBH accretion rate is of negligible predicted power over quenching, compared to SMBH mass. This figure is reproduced from \cite{Piotrowska2022}.}\label{f20}
\end{figure}

Regardless of the calibration method, SMBH mass is found to be by far the most effective predictor of quenching in observations, exactly as predicted by cosmological simulations. Furthermore, \cite{Piotrowska2022} also establish that a qualitatively similar result is found for the \cite{Terrazas2016} sample of dynamically measured SMBH masses (discussed above). This test clearly demonstrates that it is black hole mass (not stellar or halo mass) which is truly connected to central galaxy quenching (see back to Fig.~\ref{f2a}).

In the bottom panel of Fig.~\ref{f20}, we show an expanded RF analysis for simulations, here including the accretion rate of SMBHs (which is directly proportional to the bolometric AGN luminosity). In all simulations considered, accretion rate is of negligible predictive power when compared to SMBH mass. Moreover, if SMBH mass is removed, accretion rate does not even come in second place. This actually goes to stellar mass (which is more strongly correlated with SMBH mass than with the current SMBH accretion rate). This result is of paramount importance for the observational testing of the contemporary AGN feedback paradigm: \textit{cosmological simulations do not predict a strong connection between quenching and current accretion rate and, hence, current AGN activity.}

To see this more directly, in Fig.~\ref{f21} (reproduced from \citealt{Piotrowska2022}) we show distributions in AGN luminosity and SMBH mass for quenched and star forming systems, for the three simulations analyzed above. In terms of SMBH mass, there are clear offsets seen in all simulation suites, whereby quenched galaxies host much more massive SMBHs than their star forming counterparts. This provides a very effective way to classify galaxies (as in the RF analysis) and, moreover, offers a remarkably clear test of the AGN feedback paradigm in observations. Crucially, this test is passed in the local Universe via observations from the SDSS (see, e.g., \citealt{Piotrowska2022, Bluck2022}).

\begin{figure}[ht]  
\centering
\includegraphics[width=\textwidth]{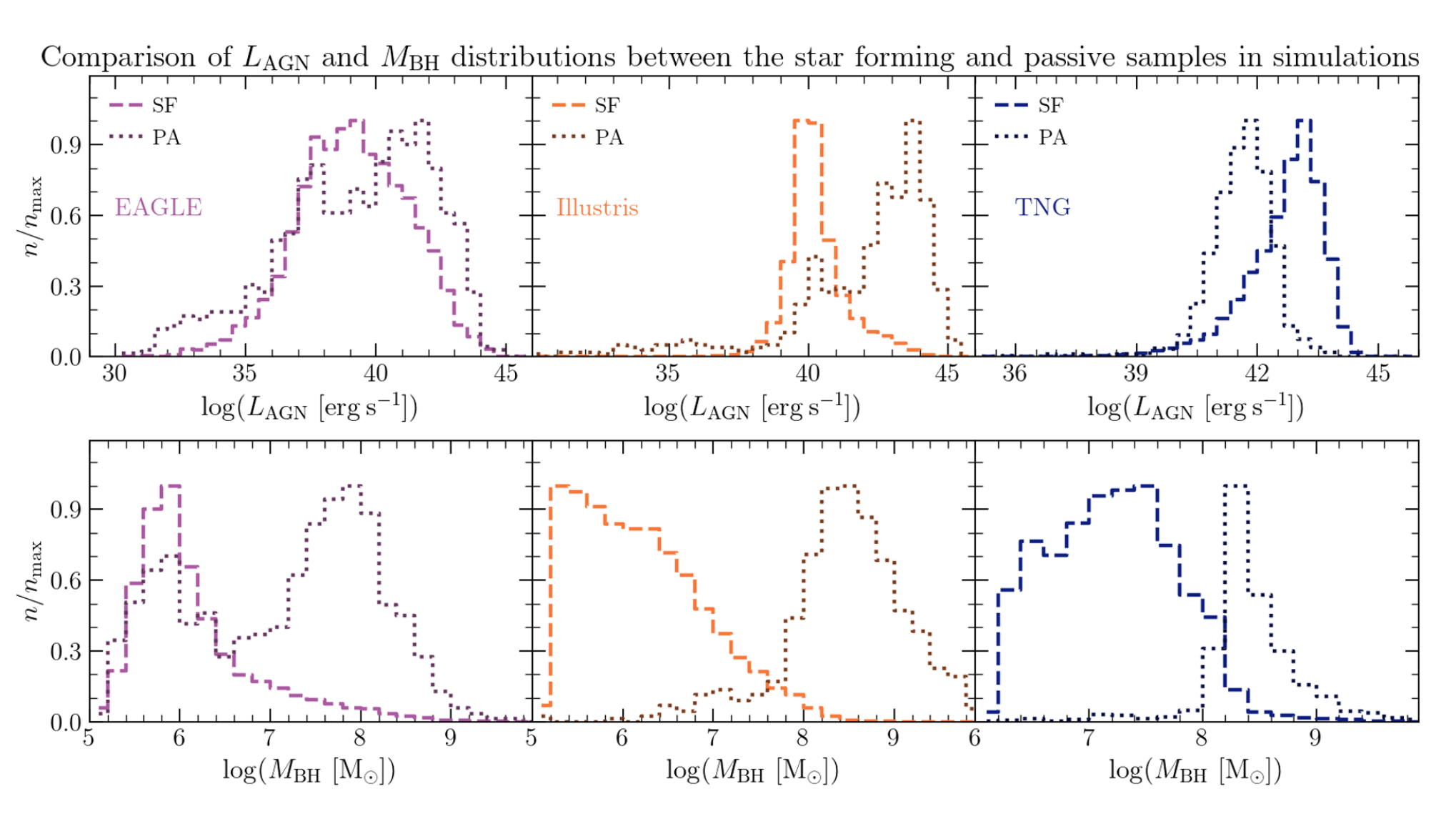}
\caption{\textit{Top panels: }Distributions of bolometric AGN luminosity separated for star forming (dashed lines) and quenched (dotted lines) galaxies from the EAGLE, Illustris, and IllustrisTNG cosmological hydrodynamical simulations (from left to right). \textit{Bottom panels: }Distributions of SMBH mass for the same simulations, also split by star forming and quenched types. Quenched galaxies typically host more massive SMBHs than their star forming counterparts within all simulations analyzed here. However, the distributions in AGN luminosities show much less difference between star forming and quiescent systems. Moreover, in EAGLE little difference is noted at all, in Illustris quenched galaxies have on average higher $L_\mathrm{AGN}$, and in IllustrisTNG quenched galaxies have on average lower $L_\mathrm{AGN}$. Hence, the only consistent and universal way to test the AGN feedback paradigm is via the historic record (i.e., $M_\mathrm{BH}$), not the current level of activity (i.e., $L_\mathrm{AGN}$). This figure is reproduced from \cite{Piotrowska2022}.}\label{f21}
\end{figure}

On the other hand, the distributions of bolometric AGN luminosities are much less different between star forming and quenched systems. Moreover, how they differ is highly model dependent. In EAGLE, essentially no meaningful statistical difference is seen in AGN luminosity between star forming and quenched systems. In Illustris, quenched galaxies tend to have higher AGN luminosities than star forming systems (as many observers have assumed; e.g., \citealt{Nandra2007, Trump2015}). However, in IllustrisTNG the opposite is true, whereby quenched systems have lower AGN luminosities than star forming systems (as observed in, e.g., \citealt{Hickox2009, Heckman2014}). This analysis demonstrates that the connection between quenching and SMBH mass is essentially universal in contemporary AGN feedback models, but the connection between quenching and current AGN luminosity is much more uncertain, varying significantly with the details of the AGN feedback prescription. Consequently, the only way to test the paradigm of AGN feedback quenching (rather than merely one instantiation thereof), is to utilize the fossil record of past accretion, i.e., SMBH mass (see also \citealt{Bluck2016, Bluck2020a, Terrazas2020, Piotrowska2022, Bluck2022, Ward2022, Lim2025} for similar conclusions).

It is important to appreciate that many other quenching mechanisms are effectively ruled out via these observational results. For instance, the total energy injected via supernovae must scale with the total stellar mass of the galaxy (see \citealt{Bluck2020a} for a simple derivation), and the total energy input from virial shock heating must scale with halo mass (see \citealt{Dekel2006, Woo2013}). The fact that both of these parameters have essentially zero predictive power over quenching in observations, once SMBH mass is controlled for, reveals that their well-known correlations with quenching are spurious (i.e., of no underlying causal origin). 

These machine learning test have been extended up to cosmic noon in \cite{Bluck2022, Bluck2023}, and all the way to cosmic dawn in \cite{Bluck2024}. At all epochs, contemporary cosmological simulations posit that central galaxies quench due to AGN feedback, which results in SMBH mass being the best predictor of quenching. Crucially, accretion rate (and, hence, bolometric AGN luminosity), stellar mass, halo mass, and various other parameters (including environment and morphology) are all found to be of no importance for predicting central galaxy quenching in simulations (once SMBH mass is controlled for). This holds across an astonishing $\sim$13\,Gyr of cosmic history. Direct observational tests, utilizing a variety of well-motivated proxies for SMBH mass (including central velocity dispersion, bulge mass, and the stellar potential: $\phi_*~=~M_*/R_h$), agree extremely well with the precise predictions from these cosmological simulations, when restricted to the available observational parameters.

In conclusion, contemporary cosmological simulations quench massive galaxies via AGN feedback, which leaves a clear and unambiguous imprint in the galaxy population: \emph{quenched galaxies host more massive SMBHs than star forming galaxies}. This statement is true for central galaxies in an absolute sense, as well as holding at a fixed stellar or halo mass as well. Perhaps surprisingly, the current accretion rate, bolometric AGN luminosity, and (hence) the detection of AGN are all not predicted to be strongly connected to quenching. Therefore, looking for correlations with these properties in observational data is not a productive method to test the AGN feedback paradigm. Most importantly, observations at $z \sim 0 - 8$ are remarkably consistent with the prediction that quenched galaxies host more massive SMBHs than star forming systems (at fixed internal properties), precisely in line with the key testable prediction from AGN feedback prescriptions incorporated within large-volume cosmological hydrodynamical simulations.

\section{Testing the causes of satellite galaxy quenching}\label{s5}

In this section we turn our focus to observational evidence for environmental quenching of satellite galaxies. As discussed in Sect.~4 of Part~I, central galaxies have at most a weak dependence of quenching on environment (e.g., \citealt{Peng2012, Woo2013, Bluck2014, Bluck2016, Bluck2022}). However, for satellites, environment becomes much more significant, especially for low-mass systems (e.g., \citealt{Peng2012, Woo2013, Goubert2024, Goubert2025}). We start this section by reviewing the key distinction between intrinsic and environmental quenching, i.e., enhanced quenching of satellites relative to centrals at fixed intrinsic parameters. We go on to explore observational evidence for both ram pressure and dynamical stripping in high density environments. Finally, we present tests of the environmental quenching paradigm via a machine learning approach, in analogy to the tests of AGN feedback presented in the previous section.

\subsection{Enhanced satellite quenching compared to centrals}\label{s51}

In the previous section we established that SMBH mass is the most predictive parameter of quenching for central galaxies, in precise agreement with predictions from AGN feedback driven quenching in cosmological simulations (see \citealt{Piotrowska2022, Bluck2022, Bluck2023, Bluck2024, Lim2025}). As such, SMBH mass is the key parameter for establishing the level of intrinsic quenching a galaxy is expected to experience. Therefore, if environment impacts galaxies beyond spurious inter-correlations with intrinsic parameters, it ought to be evident at a fixed SMBH mass.

\begin{figure}[ht]  
\centering
\includegraphics[width=\textwidth]{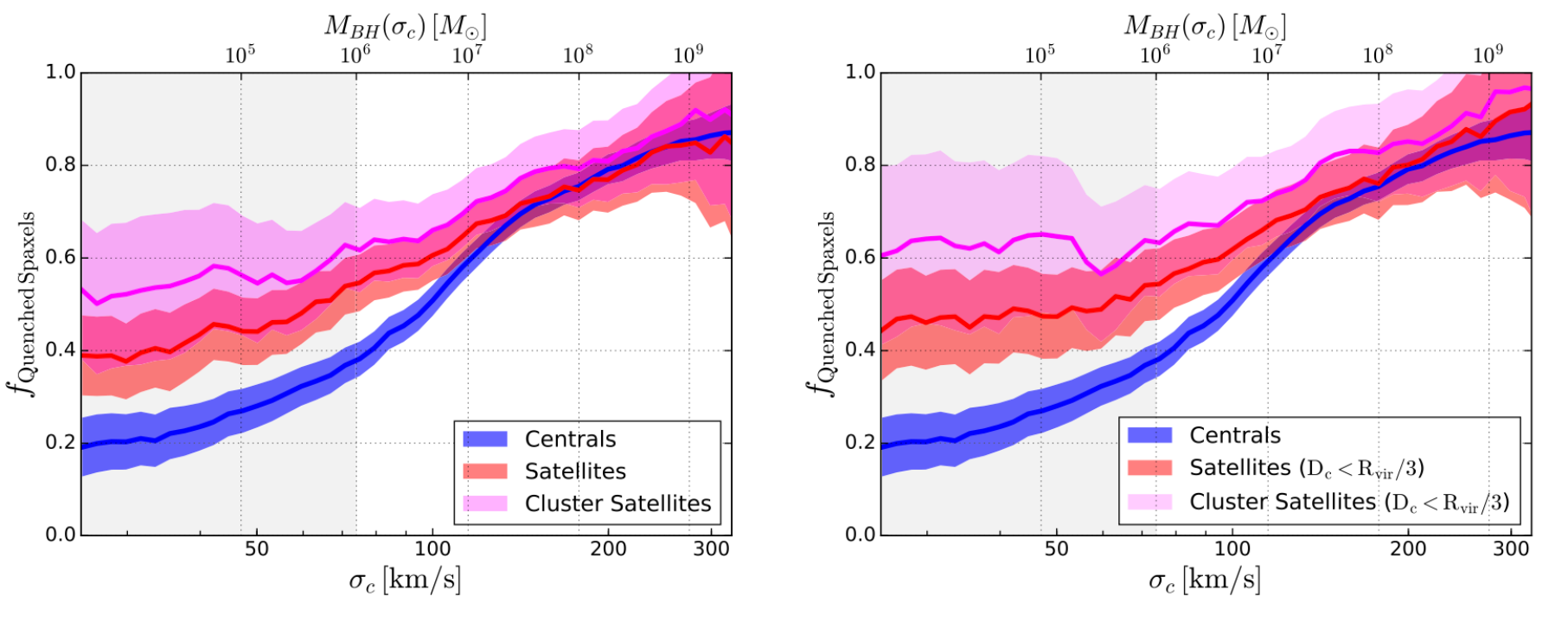}
\caption{The fraction of quenched regions (spaxels) within galaxies as a function of central velocity dispersion (and estimated SMBH mass), from the MaNGA survey (see \citealt{Bundy2015}). Results are presented separately for central galaxies (shown in blue), all satellite galaxies (shown in red), and cluster satellite galaxies (shown in magenta; with $M_\mathrm{Halo} \geq 10^{14} \, M_\odot$). The left-hand panel shows satellites within one virial radius of their central galaxy, and the right-hand panel shows satellites within one third of a virial radius of their central galaxy. All galaxies show evidence of increasing quenched fraction with central velocity dispersion (and, hence, SMBH mass). At low SMBH masses, satellites are more frequently quenched than centrals, and cluster satellites are even more frequently quenched than the general satellite population (see both panels). Furthermore, for satellites observed close to their centrals (right-hand panel), these general trends are amplified. This figure is reproduced from \cite{Bluck2020b}. }\label{f24}
\end{figure}

To test this hypothesis, in Fig.~\ref{f24} (left-hand panel) we show the fraction of quenched regions within galaxies from the MaNGA survey (see \citealt{Bundy2015}) as a function of central velocity dispersion (bottom horizontal-axis) and as a function of estimated SMBH mass (top horizontal-axis; estimated via the $M_\mathrm{BH} - \sigma$ relation). This figure is reproduced from \cite{Bluck2020b}. Central galaxies show a strong increase in quenched fraction as a function of SMBH mass, as also seen in \cite{Terrazas2016, Bluck2016, Terrazas2017, Terrazas2020, Piotrowska2022, Bluck2022, Bluck2023, Bluck2024}. Most importantly for our present concern, satellites are offset to higher quenched fractions at a fixed SMBH mass, particularly at low SMBH masses. Conversely, at high SMBH masses, satellites and centrals have essentially identical quenched fractions. 

Additionally, in the left-hand panel of Fig.~\ref{f24} we show the quenched fractions of regions within clusters satellites. This population is offset to higher quenched fractions at lower masses than the general satellite population. However, at high masses, the quenched fraction of cluster satellites is essentially identical to that of centrals (and to the complete satellite population). On the right-hand panel of Fig.~\ref{f24}, the same analysis for satellites is reproduced for systems residing within one third of a virial radius from their centrals (i.e., focusing on core satellite galaxies). Here, the general trends are qualitatively similar to the full virial radius analysis (shown on the left-hand panel), but are amplified to even higher quenched fractions in satellites at lower masses, compared to centrals. This indicates that satellite quenching evolves as a strong function of both the mass of the halo and the location within the halo, at fixed estimated SMBH mass.

Therefore, we have established three important observational results: (i)~satellites require additional quenching avenues to AGN feedback at low masses; (ii)~these additional quenching avenues must be correlated with both the mass of the halo and the location within the halo at which the satellite resides; and (iii)~at high masses, there is no requirement for additional quenching processes of satellites evident. Similar results are found in \cite{Bluck2016, Goubert2024, Goubert2025}. Note also that these results build upon important prior work at a fixed stellar mass (e.g., \citealt{Baldry2006, Peng2010, Peng2012, Woo2013, Bluck2014}), but expand upon this approach by quantifying the need for additional quenching avenues via environment at fixed values of the most important intrinsic quenching parameter.

\subsection{Ram pressure stripping observed: Jellyfish galaxies}\label{s52}

One of the leading theoretical routes to environmental quenching is ram pressure stripping (which we have discussed in Sect.~\ref{s32}). There is bountiful observational evidence for ram pressure impacting galaxies by removing gas from the ISM and CGM. For a particularly striking example, see Fig.~\ref{f25}, reproduced from \cite{Friedman2017}. This image shows a vast X-ray emitting hot gas tail flowing away from the disc of a cluster satellite (ESO 137-001). This is interpreted as gas ejected from the system via strong ram pressure stripping, as a result of the galaxy's high relative motion through the dense ICM. The X-ray emission is further explained via shock heating of the ejected gas tail.

\begin{figure}[ht]  
\centering
\includegraphics[width=0.75\textwidth]{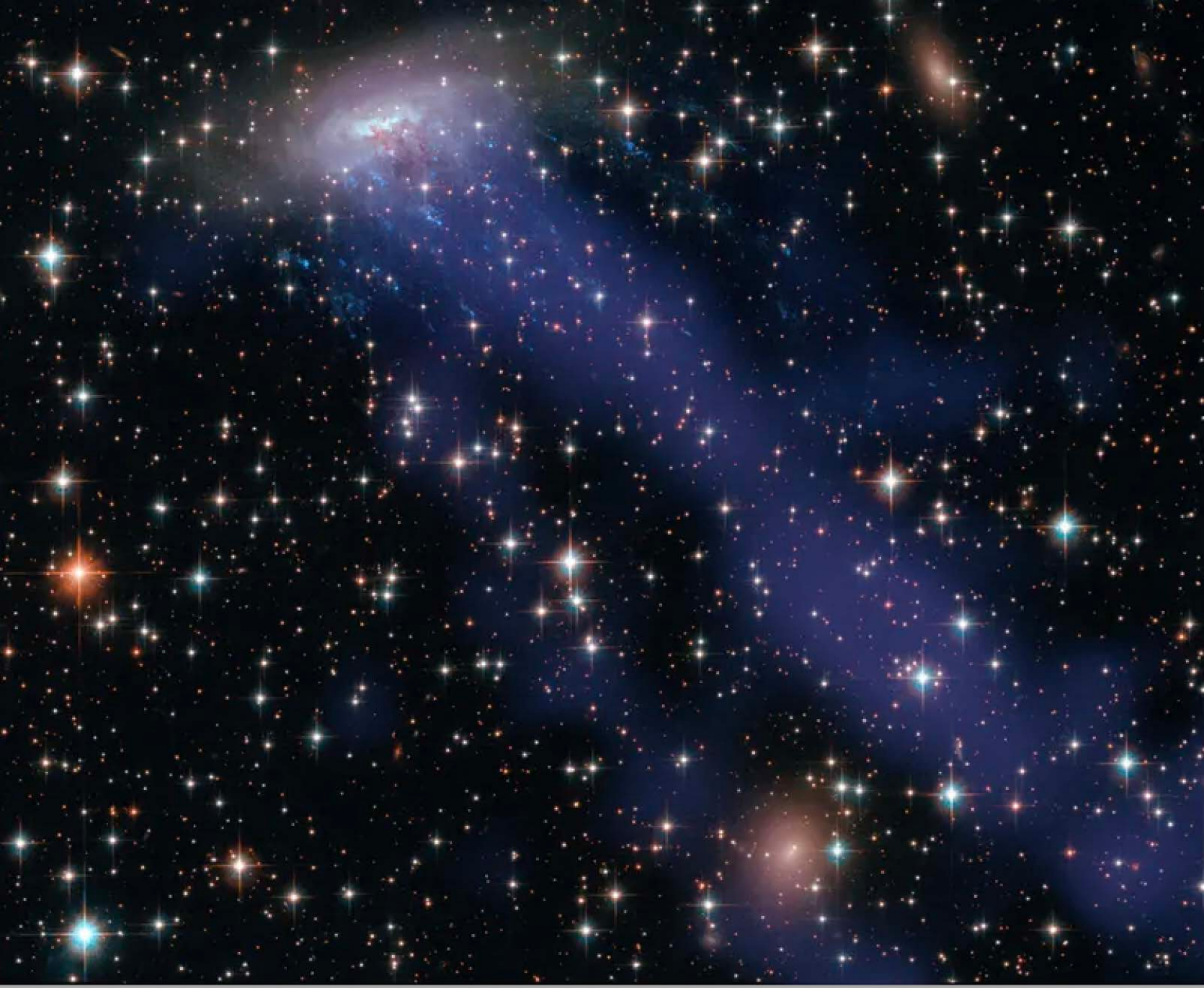}
\caption{An extraordinary image of a `jellyfish' galaxy (ESO 137-001), showing ram pressure stripping in action, reproduced from \cite{Friedman2017}. The image combines NASA/ESA Hubble Space Telescope observations of the cluster and background galaxies (shown in white - orange) with data from the Chandra X-ray Observatory tracing the hot gas tail from the galaxy (shown in electric blue). The hot gas tail is believed to originate from strong ram pressure stripping in conjunction with shock heating as the galaxy moves at high relative velocity through the ICM. Image Credit: NASA, ESA, CXC.}\label{f25}
\end{figure}

The evidence for ram pressure stripping comes in a variety of forms, which we will briefly summarize here. Following the first theoretical work on ram pressure stripping (see \citealt{Gunn1972}), the first observational hint of this process was found in \cite{Davies1973, Giovanelli1985}, who find that spiral galaxies in clusters have lower HI gas masses than field systems of similar types. These works were extended by \cite{Warmels1988, Cayatte1990}, who find asymmetric HI gas distributions around cluster satellites, and demonstrate the existence of a morphology - density relation in gas (in addition to stellar structure; see \citealt{Dressler1980}). The first observational evidence for ram pressure stripping in ionized gas came in \cite{Vollmer2001}, who identify extra-planar H$\alpha$ regions in satellites of the Virgo cluster.

The modern evidence for ram pressure stripping spans a large array of observations, from radio to X-ray (see, e.g., \citealt{Abramson2011, Ebeling2014, Jaffe2015, Fossati2016, Jaffe2016, McPartland2016, Poggianti2017, Jaffe2018, Waldron2023, Luo2023}). Generally, this evidence centers around identifying displaced gas from satellite galaxy discs within galaxy clusters (which is observed in molecular, atomic, and ionized forms). Additionally, researchers have identified significant kinematic offsets between stellar and gaseous dynamics in cluster satellites. Unlike in intrinsic feedback driven outflows (which typically exhibits bi-conical structure; e.g., \citealt{Veilleux2005, Cicone2014, Venturi2018}), the displaced gas from satellites via ram pressure stripping are typically observed to be asymmetric, flowing away from the direction of motion of the satellite galaxy through the cluster medium (e.g., \citealt{Ebeling2014, Jaffe2015, Poggianti2016, Poggianti2017}). These properties are exactly as expected for ram pressure stripping in hydrodynamical simulations (see back to Sect.~\ref{s32} for a discussion). 

In Table~\ref{t3}, we present a comprehensive summary of the observational evidence for ram pressure stripping of satellite galaxies within clusters. We include notes on the physical evidence, the observed wavelengths of measurements, and provide key references to important publications which present each piece of evidence.

Taken as a whole, the observational evidence summarized in Table~\ref{t3} establishes beyond reasonable doubt that ram pressure stripping impacts satellite galaxies within dense environments, by causing removal of gas from both the ISM and CGM (exactly as expected in simulations, see Sect.~\ref{s32}). Consequently, we know that ram pressure stripping is a viable route to quenching at least some satellite galaxies (specifically, low-mass systems within high mass clusters). Higher mass galaxies are more effective at holding onto their ISM due to it being more tightly bound to the system (e.g., \citealt{Fillingham2016, Steinhauser2016, Cortese2021}). Additionally, both the density of the ICM and the typical relative velocity of satellites increase systematically as a function of halo mass (see \citealt{Pasquali2012, Fossati2013, Woo2013, Roberts2019, Goubert2024}), resulting in ram pressure stripping of the ISM being mostly observed in the cluster satellite population (e.g., \citealt{Poggianti2016, Poggianti2017, Gullieuszik2020}).

\begin{table}[t]
\caption{Summary of observational evidence for ram pressure stripping.}\label{t3}
\begin{tabular}{l l l}
\toprule
\textbf{Observation Type} & \textbf{Waveband(s)} & \textbf{Refs.} \\
\midrule
Displaced Neutral Gas (HI) & Radio (HI 21cm)  & (1) \\ 
Ionized Gas Tails (H$\alpha$) & Optical (H$\alpha$)  & (2) \\ 
X-ray Stripped Gas & X-ray  & (3) \\ 
Star Forming Tails & UV, Optical, IR  & (4) \\ 
Truncated Gas Disks & CO, HI 21cm, H$\alpha$ & (5) \\ 
Kinematic Distortions & Optical IFU  & (6) \\ 
Molecular Gas Displacement & Sub-mm & (7) \\ 
UV Continuum Asymmetry & UV & (8) \\ 
Jellyfish Galaxy Catalogs & Multi-Wavelength & (9) \\ 
Polarized Radio Continuum & Radio  & (10) \\ 
\bottomrule
\end{tabular}
Refs.: (1)~\cite{Chung2007, Jaffe2015} (2)~\cite{Poggianti2017, Gavazzi2001} (3)~\cite{Sun2010, Machacek2006} (4)~\cite{Sun2007, Gullieuszik2020} (5)~\cite{Koopmann2004, Vollmer2001} (6)~\cite{Fossati2016, Bellhouse2019} (7)~\cite{Jachym2014, Souchereau2025} (8)~\cite{Boselli2006, Yoshida2008} (9)~\cite{Ebeling2014, Poggianti2016} (10)~\cite{Vollmer2004a, Vollmer2004b}
\end{table}

Moreover, unlike for centrals, the removal of the ISM is much more effective as a quenching mechanism in satellites than in high-mass centrals. This is the case because satellites are (by definition) in orbit of central galaxies and, hence, are no longer fed fuel for star formation via cold gas streams, unlike low-mass centrals (e.g., \citealt{Aragon-Calvo2016, Darvish2017, Malavasi2017, Kraljic2020}). Additionally, for high-mass satellites, which will have formed hot static atmospheres within their subhaloes prior to cluster infall, ram pressure stripping is especially efficient at removing these extended structures (e.g., \citealt{Yoon2013, Zhang2013, Burchett2018, Burchett2019}). Therefore, ram pressure stripping (acting on both the ISM and CGM), in conjunction with cosmic location, provides a highly plausible route to satellite galaxy quenching in clusters, with much supporting observational evidence. 

Nevertheless, the issue of mass return from stellar evolution persists (see back to Sect.~\ref{sec:stellar_quenching}). An important empirical constraint on quenching models is provided by the ubiquitous presence of X-ray emitting atmospheres in early-type galaxies, including large populations of satellite systems. The observed X-ray luminosities of these haloes are broadly consistent with expectations from the continuous injection of gas via stellar mass loss, indicating that recycled material from evolved stellar populations constitutes a non-negligible component of the hot interstellar medium (see, \citealt{Kim2012}). 

While this internal source of gas is typically sub-dominant compared to cosmological accretion from the CGM, it nonetheless ensures that even quiescent systems are persistently replenished. In the absence of efficient heating, the atmospheres surrounding these satellites are expected to undergo catastrophic radiative cooling, leading to renewed star formation and enhanced SMBH fueling (e.g., \citealt{Pellegrini2018, Pellegrini2025}). However, steady, low-level (maintenance-mode) AGN feedback is likely sufficient to offset this process in high mass satellites. In low-mass satellites, heating from SN Ia may be sufficient without addition AGN contributions. Ultimately, this reinforces the view that quenching in (even satellite) galaxies is fundamentally a time-dependent maintenance problem, requiring recurrent feedback to achieve long-term quenching.

\subsection{Dynamical stripping observed: Streams, shells and asymmetry}\label{s53}

An alternative quenching route to ram pressure stripping is dynamical stripping (e.g., \citealt{Mihos2005, Lisker2009, MartinezDelgado2010, Toloba2014, Fillingham2015, Fillingham2016}), which we have discussed from a theoretical perspective in Sect.~\ref{s33}. Here we will review the observational evidence for dynamical stripping of satellite galaxies within both groups and clusters, including its potential for quenching these systems. There are two primary modes of dynamical stripping: (i)~satellite - central interactions (e.g., \citealt{Tal2009, MartinezDelgado2010, Wetzel2015}); and (ii)~satellite - satellite interactions (e.g., \citealt{Moore1996, Stierwalt2015, Pearson2016, Paudel2018}). The latter is often referred to as `harassment' in the literature, highlighting the potential for numerous interactions over time leading to a cumulative impact on the satellite. Both of these modes are observed, and may potentially be important for satellite galaxy quenching. 

Much as with ram pressure stripping, tidal stripping is primarily identified morphologically or kinematically in observations. Tidal stripping often presents as tidal streams, tails, bridges, shells, and other asymmetric structures around central, or nearby satellite, galaxies (see Fig.~\ref{f26} for examples). Unlike ram pressure stripping, the morphological signature of dynamical stripping tends to be a much more complex distribution of matter, which is characteristically found close to a central (or another satellite) galaxy, rather than being evident in relatively isolated satellites within clusters (see, e.g., \citealt{Johnston2008, Tal2009, MartinezDelgado2010, Bilek2016}). 

Moreover, tidal stripping is most frequently detected in the stellar component (unlike ram pressure, which does not impact stars directly, and is seen near exclusively in gas). Nonetheless, tidal stripping is still expected to impact the gas in the ISM as well (see, \citealt{Barnes1992, Barnes1996, Mayer2006, Bahe2017, Jackson2021}). Evidence for dynamical stripping of massive satellites is abundant within group environments, unlike ram pressure stripping (see \citealt{Tal2009, Wetzel2013, Pearson2016, Montes2018, Janssens2019}). That said, one should note that ram pressure stripping can impact the ISM, and ultimately quench, low-mass group satellites as well as tidal stripping (see, e.g., \citealt{Engler2023}). Yet, in this review we focus primarily on galaxies with $M_* > 10^9 \, M_\odot$, where this is much rarer.

\begin{figure}[ht]  
\centering
\includegraphics[width=0.9\textwidth]{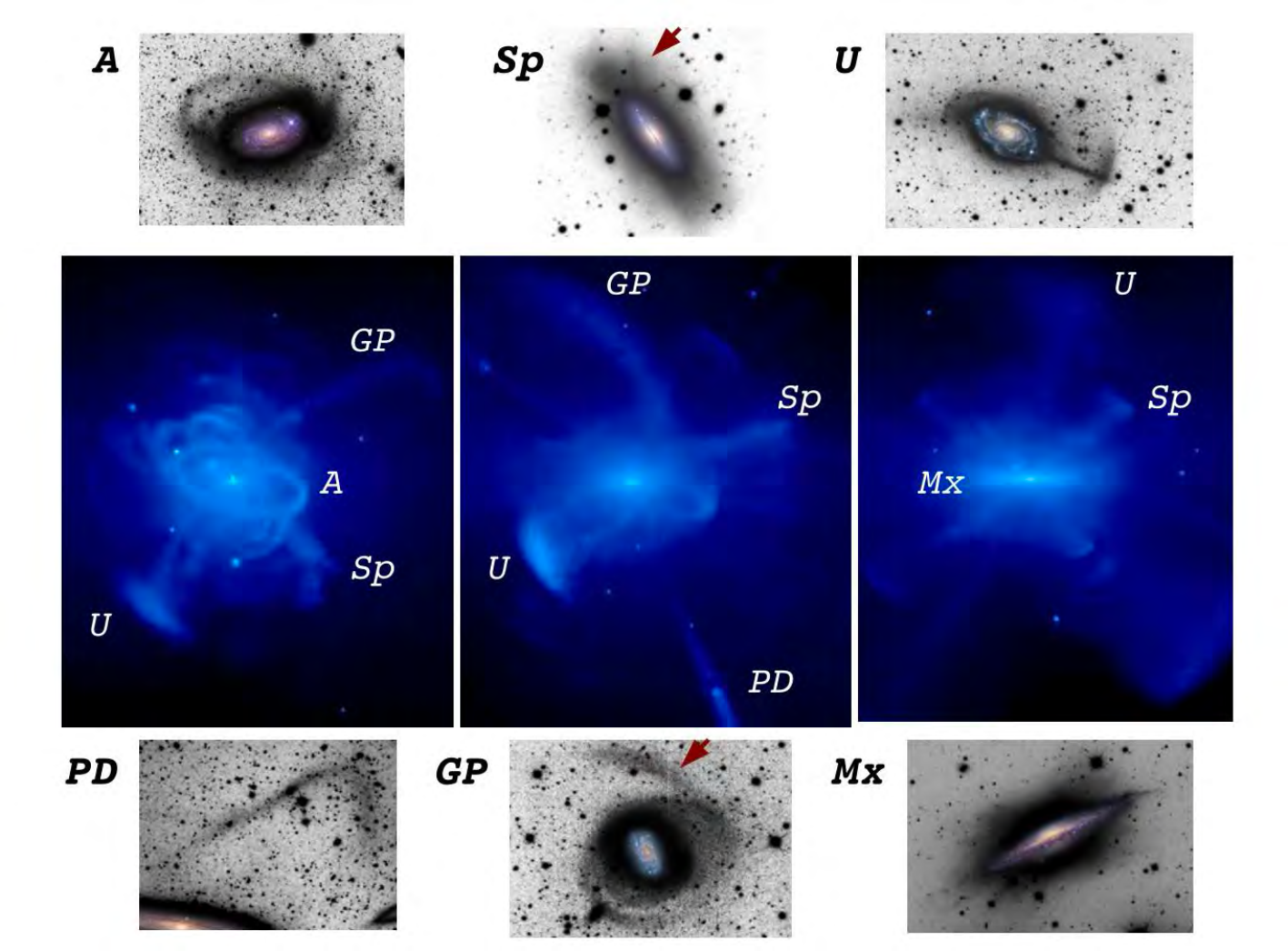}
\caption{Evidence of dynamical stripping of satellite galaxies in observations (grey-scale images around the edge) compared to simulations (from \citealt{Johnston2008}; blue images in center of figure), reproduced from \cite{MartinezDelgado2010}. Tidal features are labeled to identify similar observed structures to those found in the simulations. The labels are as follows;  A: great circles features (Messier 63); Sp: spikes (NGC 5866); U: umbrella-shaped structures (NGC 4651); PD: partially disrupted satellites (NGC 4216); GP: giant plumes (NGC 1084); and Mx: mixed-type streams (NGC 1055). Note that the morphological signature of dynamical stripping is much more complex than that of ram pressure stripping (compare to Fig.~\ref{f25}). Although the observations shown in this figure directly probe stellar stripping, this is also expected to impact the gas within satellite galaxies as well, potentially causing quenching.}\label{f26}
\end{figure}

In terms of quenching, numerous studies have linked dynamical stripping to cessation of star formation via various different approaches (e.g., \citealt{Paudel2018, Stierwalt2015, Wetzel2013, Pearson2016, Fillingham2015, Fillingham2016}). For example, \cite{Wetzel2013} find that satellite quenching in groups is a two stage process. This starts with slow quenching as a result of starvation (for satellites in the outer region of groups), and transitions to fast quenching when satellites reach the peri-center of their orbit, consistent with dynamical stripping driven quenching via satellite - central interactions. Additionally, \cite{Pearson2016} find that there is a stronger impact on the gas content of satellites via satellite - satellite interactions when this occurs close to the central. This introduces the idea of gas `parking', whereby gas ejected from satellite - satellite interactions may be re-accreted in the outskirts of groups, but will be tidally sheared away by the central at closer distances. Furthermore, \cite{Rhee2017} find through numerical simulations that quenched satellites preferentially reside on radial orbits, broadly in line with these observational results.

In essentially all cases, satellite quenching via dynamical stripping is found to vary as a strong function of satellite stellar mass, whereby low-mass satellites are much more prone to quenching than higher mass systems (e.g., \citealt{Geha2012, Wetzel2013, Wheeler2014, Fossati2017}). This is particularly interesting because at low masses this yields an anti-correlation between quenching and stellar mass, unlike the strong (though ultimately spurious) positive correlation between stellar mass and quenching at higher masses. The transition appears to occur at, $M_* \sim 10^{9.5}\,M_{\odot}$ (see, e.g., \citealt{Slater2014, Phillips2014, Goubert2024, Goubert2025}). Ultimately, the change in stellar mass dependence provides very clear evidence for a fundamental transition in how galaxies quench, revealing (at least) two important channels. As discussed in Part~I of this review series, the two regimes are often referred to as `mass-quenching' (for high-mass systems, which ultimately depends upon SMBH mass) and `environmental-quenching' (for low-mass systems, where environment dominates).

In group environments, the majority of studies agree that dynamical stripping (particularly in the satellite - central mode) dominates over ram pressure stripping, see, e.g., \cite{Rasmussen2006, Cortese2006, Kawata2008, Fillingham2015, Fillingham2016, Yun2019}. This is thought to be the case due to the CGM around group centrals not reaching the needed density for effective strong ram pressure stripping of the ISM of relatively massive satellites (although it is still expected to impact the CGM, at least within high-mass groups). Conversely, tidal interactions are observed to occur within groups of all masses, leading to effective gas, stellar, and dark matter stripping, with the former leading to quenching. 

Alternatively, in clusters, both strong ram pressure stripping and tidal stripping is observed (e.g., \citealt{Ebeling2014, Poggianti2016, Poggianti2017, Mihos2005, Murante2007, Montes2018}). However, although the debate is not completely settled, most contemporary publications emphasize the critical importance of ram pressure over dynamical stripping in clusters (see \citealt{Boselli2006, Chung2009, Poggianti2017, Bellhouse2017, Bellhouse2022}). These processes are often distinguished by gas-only stripping (in the case of ram pressure) vs. both gas and stellar stripping (in the case of tidal forces).

In summary, the field appears to be reaching a tentative consensus whereby ram pressure stripping of the ISM is the dominant trigger of massive satellite quenching in clusters, whereas dynamical stripping of the ISM is the dominant trigger of massive satellite quenching in groups. In both of these cases, long-term quiescence is achieved via starvation, which is ultimately a consequence of the location of satellites within their dark matter haloes (leading to no direct cold gas streams) and the efficient removal of the satellite's CGM by a mix of low-density ram pressure and tidal stripping. Of course, some maintenance feedback will also be required to offset cooling flows from stellar evolution as well. Furthermore, we have also reviewed the strong evidence for environmental quenching of satellites being anti-correlated with stellar mass (within the low-mass regime), in stark contrast to the strong, positive correlation seen in central galaxy quenching. Perhaps more than any other observational result we have seen so far, this provides unambiguous evidence for the need for multiple quenching channels as a function of galaxy class.

\subsection{Direct tests of the environmental quenching paradigm}\label{s54}

In the final part of this section we turn to direct observational tests of the environmental quenching paradigm from simulations, via a machine learning approach. The analyses presented in this sub-section are analogous to the results discussed in Sect.~\ref{s46} for centrals, but here extended to both high- and low-mass satellites, as well as incorporating environmental metrics.

\begin{figure}[htbp]  
\centering
\includegraphics[width=\textwidth]{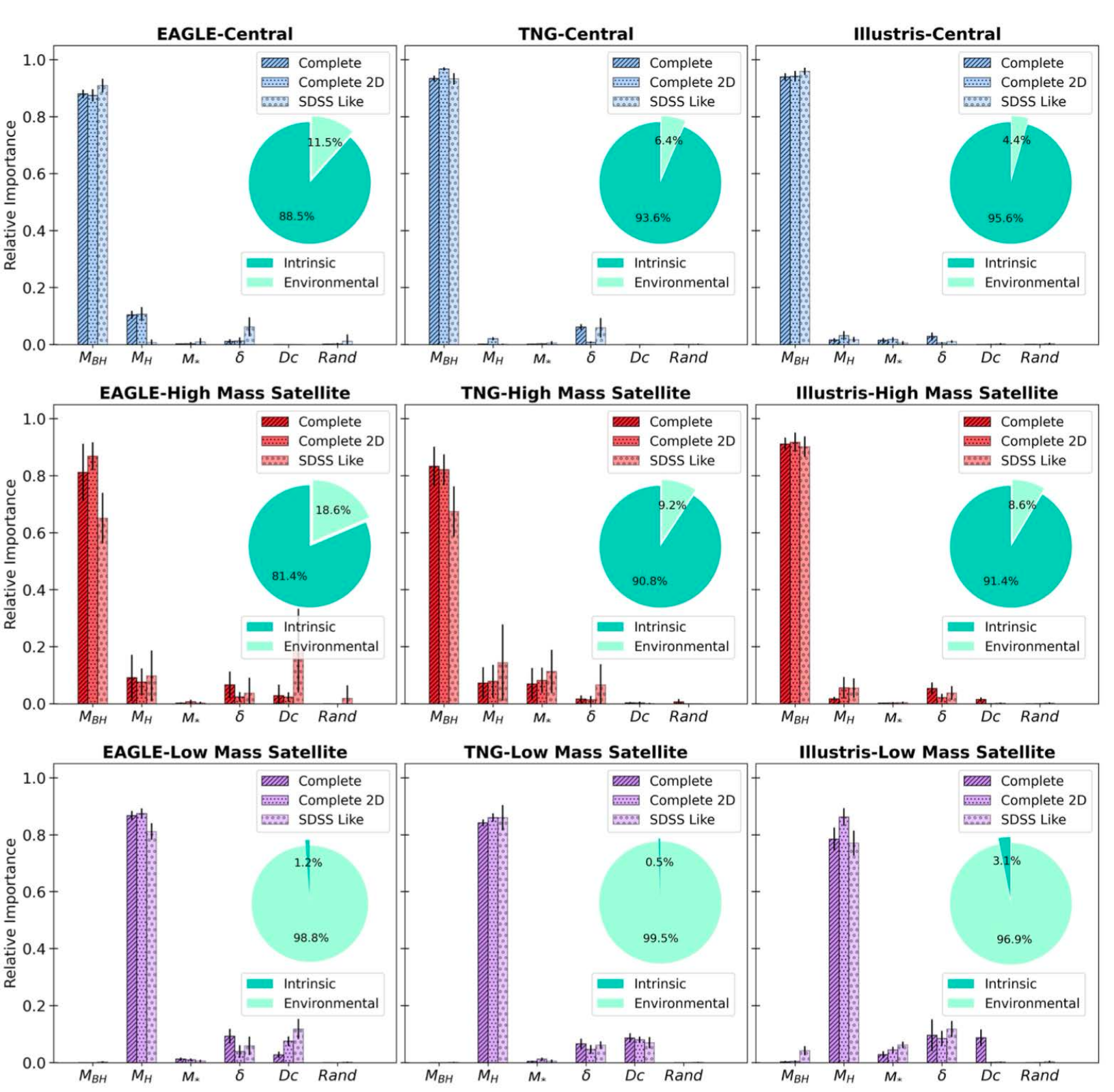}
\caption{A series of random forest (RF) classification analyses to predict whether simulated galaxies are star forming or quenched, on the basis of various input parameters (listed along the horizontal axis in each panel). The bar heights display the relative importance of each feature. Columns display results for (from left to right): EAGLE, IllustrisTNG, and Illustris. The top row displays results for central galaxies, the middle row displays results for high-mass satellites, and the bottom row displays results for low-mass satellites. Various bar types display the raw results from the simulations, and two versions of observational realism (2D and 2D `SDSS-like'). It is clear that both centrals and high mass satellites have quenching primarily dependent upon SMBH mass and, hence, upon intrinsic parameters (see pie plot insets). Conversely, low-mass satellites have quenching primarily dependent upon halo mass and, hence, environmental parameters (see pie plot insets). This is remarkably consistent between all simulations studied. Therefore, high-mass satellites are predicted to behave more like centrals than low-mass satellites in terms of quenching. This figure is reproduced from \cite{Goubert2024}.}\label{f27}
\end{figure}

In Fig.~\ref{f27}, we present a series of random forest classification analyses to predict whether galaxies of various types are star forming or quenched from cosmological simulations, reproduced from \citealt{Goubert2024}. In particular, we show results for EAGLE (left-hand column; \citealt{Schaye2015, Crain2015}), IllustrisTNG (center column; \citealt{Nelson2018, Pillepich2018, Springel2018, Marinacci2018, Naiman2018}), and Illustris (right-hand column; \citealt{Vogelsberger2014a, Vogelsberger2014b, Torrey2014}). 

Results are shown separately for centrals (top row), high-mass satellites ($M_*~>~10^{10.5}\, M_\odot$; middle row), and low-mass satellites ($M_* < 10^{10}\, M_\odot$; bottom row). In all panels, SMBH mass ($M_\mathrm{BH}$), halo mass ($M_H$), stellar mass ($M_*$), local galaxy over-density ($\delta$)\footnote{Note that the relative importance for local galaxy over-density ($\delta$) is computed as the sum of importances for over-densities evaluated at 3rd, 5th, and 10th nearest neighbor distances.}, distance to the nearest central in units of the virial radius ($D_c$), and a random variable (Rand) are utilized as features within the classifier. Within each panel, three versions of the results are shown, explicitly for: (i)~the complete sample (directly extracted from the simulations); (ii)~a~2D complete sample, with environmental metrics computed within random 2D projections; and (iii)~`SDSS-like', which reproduces the observed stellar mass function of the SDSS in simulations, in addition to random 2D projection.

For central galaxies, SMBH mass is found to be the strongest predictor of quenching in all simulations (and data samples). This is in agreement with Fig.~\ref{f20} (and, e.g., \citealt{Piotrowska2022, Bluck2022, Bluck2023}), but here also controlling for various environmental parameters. This clearly establishes that there is little-to-no impact on central galaxy quenching from environment. Interestingly, high-mass satellites have a very similar quenching dependence on SMBH to centrals. This indicates that high-mass satellites are predominantly quenched via intrinsic, AGN-feedback driven quenching in all of these simulations. Conversely, low-mass satellites have quenching most accurately predicted via halo mass (unlike centrals and high mass satellites), which clearly links to environmental quenching mechanisms. All simulations agree on the change from intrinsic quenching (governed by historic AGN feedback, as traced by $M_\mathrm{BH}$) for centrals and high-mass satellites to environmental quenching (governed by halo mass) for low-mass satellites. This provides a very clear set of predictions to test in observations.

In Fig.~\ref{f28}, we show the same random forest analysis as performed in simulations in Fig.~\ref{f27}, but here applied to over 500\,k galaxies observed as part of the SDSS (see \citealt{York2000, Abazajian2009}). This figure is also reproduced from \cite{Goubert2024}. Exactly as predicted in cosmological simulations, both centrals and high-mass satellites have their quenching most accurately predicted by SMBH mass (with negligible contributions from all other investigated parameters, once SMBH mass is controlled for). In observations, SMBH mass is estimated via the $M_\mathrm{BH} - \sigma_\star$ relationship for all galaxy types in \cite{Saglia2016} (but these results are not strongly sensitive to calibration choice, see \citealt{Bluck2016, Piotrowska2022}). Ultimately, the results shown in Fig.~\ref{f28} strongly imply that high-mass satellites behave similarly to centrals in terms of quenching, and, furthermore, that AGN-feedback is the most probable cause of their quenching (as predicted by simulations).

For low-mass satellite galaxies, the most important observational parameter is found to be local galaxy over-density. This clearly implicates environment as the causal origin of low-mass satellite quenching (as in simulations). However, the dominant parameter is different to that predicted in the simulations analyzed in Fig.~\ref{f27} (i.e., halo mass). Hence, at a qualitative level, the shift from intrinsic to environmental quenching occurs in observations of the local Universe much as predicted in cosmological simulations (as illustrated by the pie plot insets in Figs.~\ref{f27} and \ref{f28}, which show the breakdown of intrinsic vs. environmental importance to quenching). However, the details are different, with observations favoring local galaxy over-density and simulations favoring halo mass.

\begin{figure}[ht]  
\centering
\includegraphics[width=0.9\textwidth]{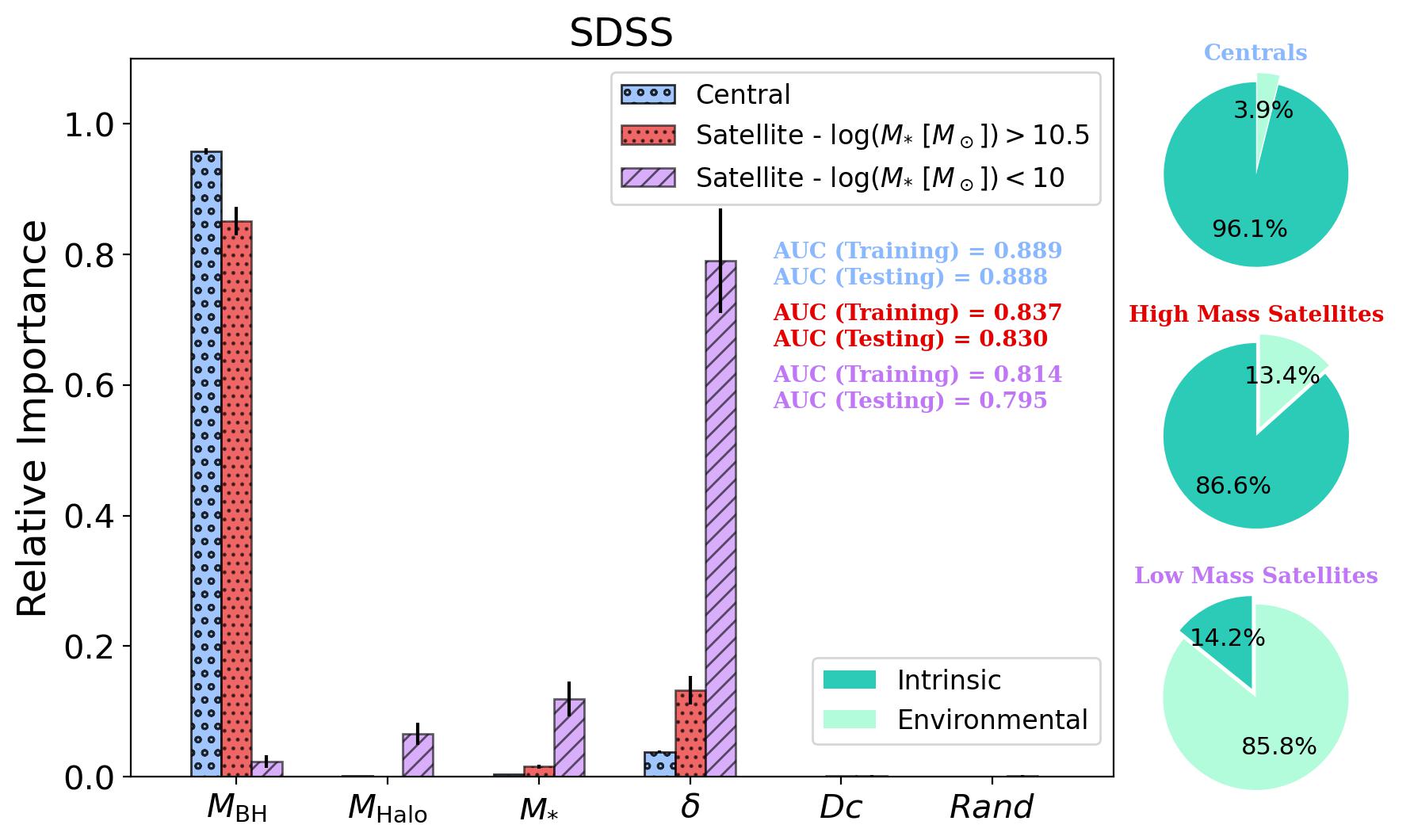}
\caption{A series of random forest (RF) classification analyses to predict whether observed galaxies from the SDSS are star forming or quenched, on the basis of various input features (listed along the horizontal axis). The bar heights display the relative importance of each feature for solving the classification problem. As in simulations, both centrals and high-mass satellites quench primarily as a function of SMBH mass, and, hence, are intrinsically quenched (see pie plots on right of figure). Conversely, low-mass satellites quench environmentally (as in simulations; see bottom right pie plot). However, the optimal parameter for environmental quenching is found in observations to be local galaxy over-density, not halo mass as predicted by cosmological simulations. This figure is reproduced from \cite{Goubert2024}.}\label{f28}
\end{figure}

There are two possible explanations for this discrepancy. First, the simulations may fail to account for some important satellite quenching processes which occur in nature. For instance, taken at face value, a greater importance of quenching on local galaxy density over halo mass may plausibly indicate a greater importance of satellite\,--\,satellite interactions over halo-wide (or satellite\,--\,central) effects. This in turn may be interpreted as a greater importance of dynamical stripping over ram pressure stripping, since the former is most closely linked to galaxy interactions and the latter is primarily connected to the halo as a whole. However, it is not immediately clear why the simulations would be missing the importance of galaxy\,--\,galaxy interactions, especially given that the analyses presented here are limited to $M_* > 10^{9}\,M_{\odot}$, where systems are well resolved in simulations. Furthermore, since no sub-grid models are explicitly relevant here, it would be hard to tweak simulations to better reproduce the observations in a straightforward manner. 

Second, the discrepancy could arise from an issue with the observations. Halo masses are inferred indirectly in Fig.~\ref{f28} via abundance matching (see \citealt{Yang2007, Yang2009}). Hence, there are reasonably high uncertainties on these measurements, of $\sim$0.5\,dex (e.g., \citealt{Yang2007, Yang2009, Woo2013}). In \cite{Goubert2024}, it is found that these uncertainties can lead to erroneously favoring local galaxy density over halo mass, explaining the observed discrepancy. Crucially, however, the uncertainties on halo masses are not sufficiently large to explain the superiority of SMBH mass over halo mass for both centrals and high-mass satellites, which is much more stable to uncertainty (as also established via various uncertainty tests in \citealt{Piotrowska2022, Bluck2022, Brownson2022}). 

Therefore, the results for centrals and high-mass satellites are robust and clearly in line with theoretical expectations. Alternatively, for low-mass satellites, some uncertainty remains as to the optimal parameter driving quenching. Nonetheless, low-mass satellite quenching is found to depend almost exclusively on environment, in both observations and simulations. At the very least, this establishes a broad agreement between theory and observations with respect to the underlying causes behind quenching of various types of galaxies. To move beyond this, greater accuracy in halo mass estimation for wide-field galaxy surveys will be required (see, e.g., \citealt{Hahn2024, Bluck2025} for examples of novel approaches in this direction, which show improvement over abundance matching).

Finally, essentially identical results are found in cosmological hydrodynamical simulations at all epochs up to cosmic noon (see \citealt{Goubert2025}). That is, at all redshifts, high-mass satellites are expected to behave like centrals in terms of quenching, with SMBH mass expected to be the dominant parameter. Conversely, low-mass satellites are expected to quench environmentally, resulting in a very strong dependence on halo mass. Testing environmental quenching at cosmic noon has been fraught with difficulty because there is a relative dearth of dense spectroscopic sampling over large volumes at early cosmic times (which is crucial for accurate tracing of environment). Fortunately, in the coming few years these issues will be resolved with an unprecedented wide-field spectroscopic survey targeting cosmic noon: VLT-MOONRISE (see \citealt{Maiolino2020, Cirasuolo2020}). Moving forwards, it will be highly instructive to compare the results on satellite quenching from this survey to the extant predictions from various cosmological simulations.



\section{Testing the mechanisms of galaxy quenching}\label{s6}

In the previous two sections we explored various observational tests of the underlying causes of quenching, i.e., which physical processes are most likely responsible for the transition of galaxies from the SFMS to the quiescent `red sequence'. In this section we turn our focus to the question of how these processes actually achieve the needed quenching of galaxies in practice, i.e., the mechanisms at work in quenching. Recalling Fig.~4 from Part~I of this review series, one can separate quenching processes along two axes --- the cause (as discussed in the previous sections) and the mechanism (which is discussed here). Multiple mechanisms may be available to the same cause, and multiple causes can achieve quenching through the same mechanism. Hence, in order to fully understand quenching we must isolate both the case and the mechanism.

\subsection{Fuel vs. efficiency: How is sSFR reduced during quenching?}\label{s61}

In the broadest sense, there are only two viable ways to quench a galaxy, either (i)~reduce the gas content of the ISM (which is required as fuel for further star formation), or else (ii)~reduce the efficiency with which extant fuel is converted into stars (e.g., \citealt{Saintonge2016, Saintonge2017, Piotrowska2020, Brownson2020, Ellison2020, Ellison2021}). This insight may be summarized by the use of a simple equation:

\begin{equation}\label{e45}
\mathrm{sSFR} \equiv \frac{\mathrm{SFR}}{M_*} = \bigg( \frac{\mathrm{SFR}}{M_g} \bigg)   \bigg( \frac{M_g}{M_*} \bigg) \equiv \mathrm{SFE} \cdot f_{g} \qquad \text{where} \qquad \mathrm{SFE} \equiv \frac{1}{\tau_\mathrm{dep}} \, ,
\end{equation}

\noindent and where, $M_g$ refers to gas mass (typically taken to be the molecular gas mass), and SFE is the efficiency of star formation (explicitly, the inverse of the depletion time, $\tau_\mathrm{dep} = M_g/\mathrm{SFR}$). By use of this mathematical identity, sSFR is decomposed into a gas content term and an efficiency term. However, in the case of a super-linear KS relation, this decomposition strategy will fail to fully remove the impact of gas mass on SFE (see, e.g., \citealt{Kennicutt1998b}).

\begin{figure}[ht]  
\centering
\includegraphics[width=\textwidth]{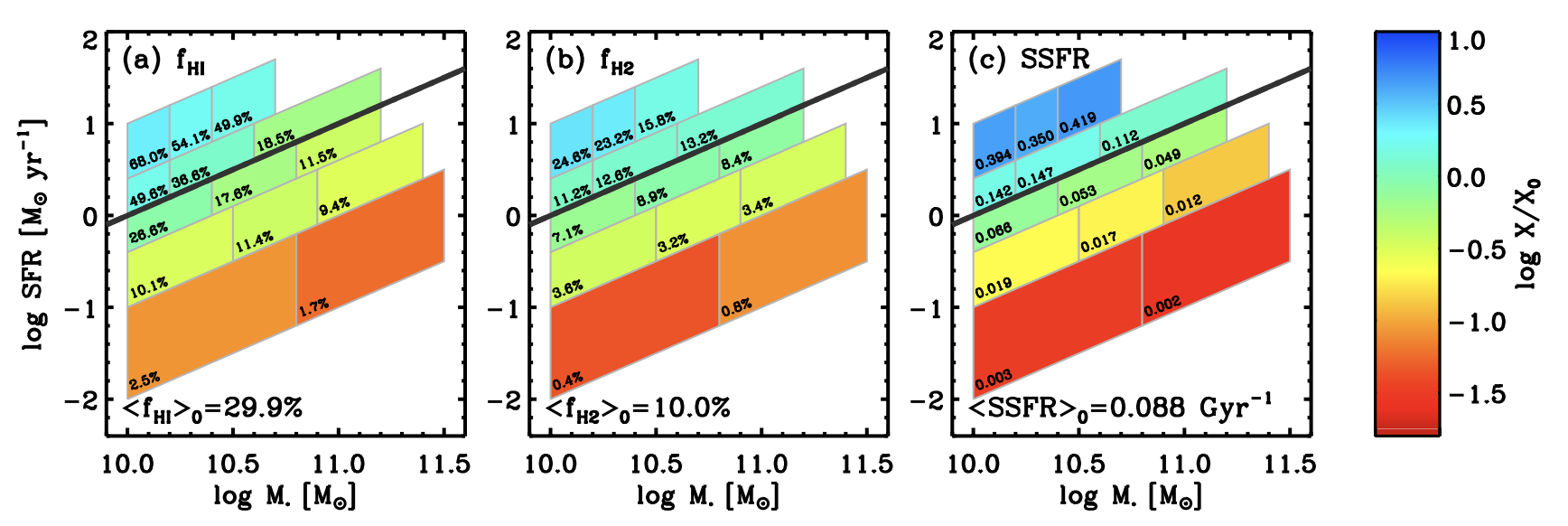}
\caption{Variation in HI gas fraction (left panel), H$_{2}$ gas fraction (center panel), and sSFR (right panel) across the SFR\,--\,$M_*$ plane. HI gas masses are inferred from Arecibo 21\,cm observations for 760 galaxies, and H$_2$ gas masses are inferred from IRAM CO(1-0) observations (assuming an $X_\mathrm{CO}$ conversion factor; see \citealt{Accurso2017}) for 350 galaxies. These data are released as part of the ALFALFA, GASS and COLD GASS surveys. Galaxy spectra are stacked within boxed regions of the parameter space to increase the S/N of the emission lines. Each region is color coded by the offset from their fiducial values, set at sSFR = 0.1\,Gyr$^{-1}$ (as labeled by the color bar). Both HI and H$_2$ gas fractions systematically decline as sSFR reduces below the main sequence, decreasing by an order of magnitude (or more). However, this is not a sufficient decrease to fully explain the reduction in sSFR (compare color shades across panels). This figure is reproduced from \cite{Saintonge2016}. }\label{f29}
\end{figure}

In Fig.~\ref{f29}, we show how HI and H$_2$ gas fractions (defined as $f_\mathrm{H1} \equiv M_\mathrm{HI}/M_*$ and $f_\mathrm{H_2} \equiv M_\mathrm{H_2}/M_*$, respectively) vary on the SFR\,--\,$M_*$ plane, reproduced from \cite{Saintonge2016}. HI gas masses are inferred from Arecibo 21\,cm observations, and H$_2$ gas masses are inferred from IRAM CO(1-0) observations, assuming an $X_\mathrm{CO}$ conversion (see \citealt{Bolatto2013, Accurso2017}). Due to frequent lack of detections in both HI and CO in the quiescent population (see, \citealt{Saintonge2016, Saintonge2017}), regions within the SFR\,--\,$M_*$ plane are grouped together and stacked to achieve higher S/N in emission lines (see boxed regions in Fig.~\ref{f29}).

From the SFMS to the quenched population, there is over an order of magnitude reduction in both HI and H$_2$ gas content revealed in Fig.~\ref{f29} (see left and center panels). This clearly demonstrates that quenching involves the reduction in gas mass of the ISM, in both the atomic and molecular phase, at a fixed stellar mass. To explain these observations, a mechanism to prevent gas accretion into quiescent systems is required\footnote{Some readers may wonder whether ejective feedback could also explain this result. The answer is most probably, no. This is the case because ejective feedback in central galaxies is observed primarily in actively star forming systems, and does not continue into the quenched population. Consequently, without preventative feedback, there is no way to account for lower gas fractions which persist for long cosmological times. Whether via CGM cooling (at high halo masses) or cold gas inflows (at low halo masses), the ISM will be replenished if nothing prevents this from happening.}. Without such a mechanism, the ISM can and will be replenished, violating direct observational constraints. The needed prevention of gas accretion may be effectively explained via radio-mode AGN feedback, stabilizing the hot gas haloes around centrals (as discussed in Sects.~\ref{s2} and \ref{s4}). Crucially, in the absence of preventative feedback, one actually anticipates the gas content of the ISM to \textit{increase} during quenching, due to ISM gas loss into stars being halted, but CGM cooling into the ISM continuing. This is further exacerbated by gas return to the ISM via stellar evolution.

However, the reduction in gas content is not by itself fully able to explain the reduction in sSFR in Fig.~\ref{f29}. This is evident because the boxed regions are color coded by fractional offsets from a fiducial value of each parameter, calibrated at sSFR~=~0.1\,Gyr$^{-1}$. Comparing the color gradients across all panels of Fig.~\ref{f29}, it is evident that neither HI nor H$_2$ gas content reduction fully accounts for the reduction in sSFR. Hence, leveraging our insight from Eq.~\eqref{e45}, we immediately identify the need for reduction in SFE as well. 

In \cite{Saintonge2017}, the sample size of CO measurements is almost doubled, and a more robust galaxy selection methodology is applied. In this work, a clear decrease in SFE is observed, in addition to a clear decrease in gas content, throughout the quenching sequence. Preventative feedback alone cannot explain this reduction in efficiency. Note also that pure ejective feedback is also (by itself) incapable of explaining this trend, since this only leads to gas reduction.

To investigate these interesting observational results further, it is helpful to expand to large sample sizes and make direct comparisons with cosmological simulations. One method to achieve this is to leverage an observed correlation between the surface density of gas and the level of dust reddening along sight lines within the Milky Way (see, e.g., \citealt{Bohlin1978, Gudennavar2012}). These early works were extended for use on external galaxies in \cite{Piotrowska2020}, by adding a metallicity dependence (accounting for more metal rich systems harboring more dust for their gas mass, and hence exhibiting greater dust attenuation). Explicitly, \cite{Piotrowska2020} present the following calibration:

\begin{equation}
\Sigma_\mathrm{H_2} = \bigg( (3.16\pm1.0)A_V + 1.0 \bigg) \, \bigg[ \frac{Z_{\odot}}{Z_g} \bigg] \,\, [M_\odot \, \mathrm{pc}^2] \,\, ; \,\,\,\, \sigma_\mathrm{rms} = 0.30\,\mathrm{dex} \, ,
\label{e46}
\end{equation}

\noindent where, $\Sigma_\mathrm{H_2}$ is the molecular gas mass surface density, $A_V$ is the V-band extinction (discussed below), and $Z_g$ is the gas-phase metallicity (estimated in this work via the O3N2 calibration; see \citealt{Marino2013, Kumari2019}). The above calibration yields a very good reproduction of molecular gas masses as inferred from the $X_\mathrm{CO}$ method, with an rms dispersion of $\sim 0.3$\,dex (see \citealt{Piotrowska2020}). The extinction at a given wavelength ($A_\lambda$) may be estimated via measuring the Balmer decrement, i.e., the ratio of $F_{\mathrm{H}\alpha}/F_{\mathrm{H}\beta}$ to its theoretically expected value (here taken to be 2.83 for Case B recombination; see \citealt{Hummer1987}). Explicitly,

\begin{equation}
A_{\lambda} = \bigg( \frac{K_\lambda}{K_\mathrm{H\beta} - K_\mathrm{H\alpha}} \bigg) \, \cdot \, 2.5 \, \log_{10}\bigg\{ \frac{F_\mathrm{H\alpha}/F_\mathrm{H\beta}}{2.86} \bigg\} \, ,
\label{e47}
\end{equation}

\noindent where, $K_\lambda$ ($\equiv A_\lambda/A_{V}$) may be extracted from an extinction law of choice (taken in \citealt{Piotrowska2020} from \citealt{Cardelli1989}). For a straightforward derivation of the above expression, see \cite{Binney1998}. Ultimately, the value of these calibrations is that they may be applied to much larger data sets than the more direct measurements, which utilize CO transition lines.

\begin{figure}[ht]  
\centering
\includegraphics[width=0.55\textwidth]{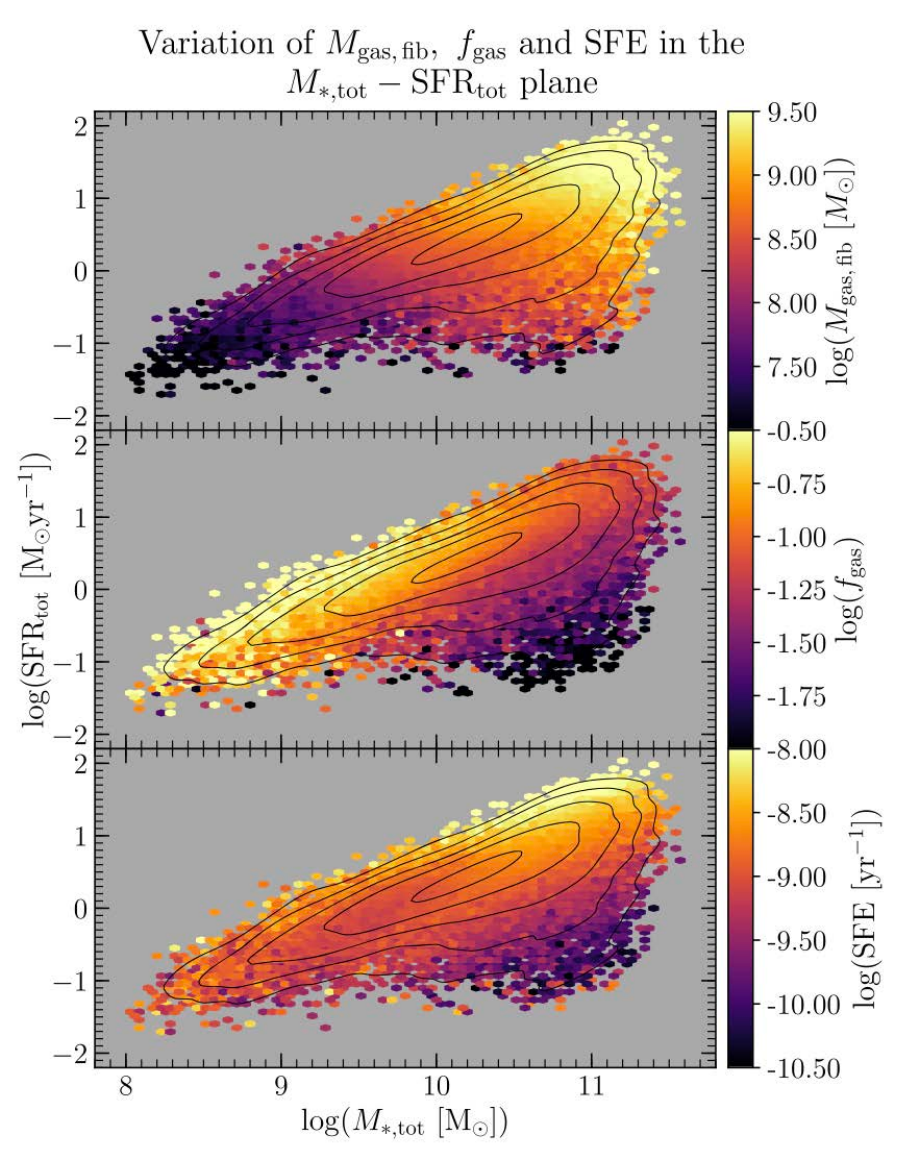}
\includegraphics[width=0.43\textwidth]{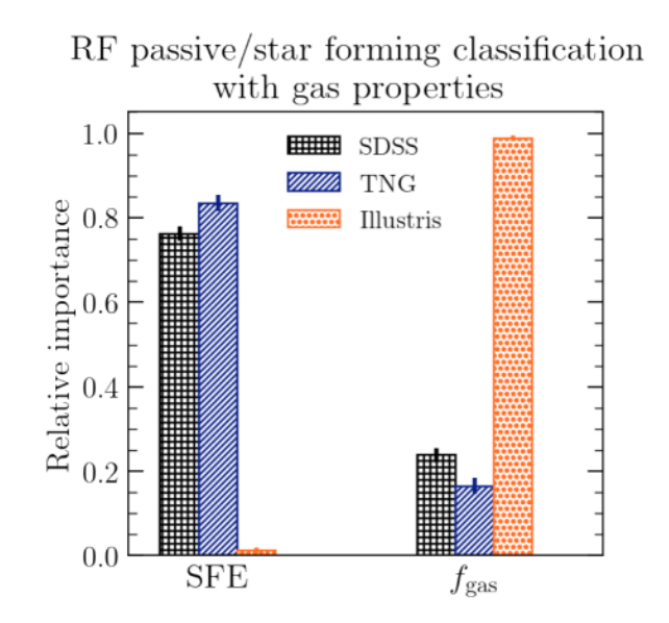}
\caption{\textit{Left-hand panel:} The SFR\,--\,$M_*$ plane for emission line galaxies from the SDSS, color coded by molecular gas mass (top panel), molecular gas fraction (center panel), and SFE ($\equiv 1/\tau_\mathrm{dep}$; bottom panel). Gas masses are estimated via Eq.~\eqref{e46}, enabling application to $\sim$60,000 galaxies. Along the SFMS, gas mass increases with increasing stellar mass and SFR. Away from the SFMS, both the gas fraction and SFE decline with decreasing sSFR. This panel is reproduced from \cite{Piotrowska2020}. \textit{Right-hand panel:} A random forest classification analysis to predict star forming and quenched galaxy classes in the SDSS emission line sample (black bars), compared to similar samples extracted from the Illustris (orange bars) and IllustrisTNG (blue bars) cosmological hydrodynamical simulations. Variation in SFE is found to dominate the quenching of emission-line galaxies in both the SDSS and IllustrisTNG. However, in Illustris the opposite trend is found, whereby gas fraction reduction dominates quenching. This panel is reproduced from \cite{Piotrowska2022}.}\label{f30}
\end{figure}

In Fig.~\ref{f30} (left-hand panels), we show the SFR\,--\,$M_*$ plane for SDSS emission-line galaxies, color coded by molecular gas mass (top panel; inferred via Eqs.~\eqref{e46} and \eqref{e47}), molecular gas fraction (center panel), and SFE (bottom panel). This figure is reproduced from \cite{Piotrowska2020}. Results are shown for over 60,000 systems, increasing the number of galaxies analyzed in gas content by over two orders of magnitude, compared to prior studies. As in \cite{Saintonge2016, Saintonge2017}, a clear increase in gas mass is found moving along the SFMS from low mass/ low SFR to high mass/ high SFR systems, which suggests that the main sequence may be primarily a result of gas trends (which we discuss in more detail in Sect.~6 of Part~I of this review). Moreover, both the molecular gas fraction and SFE decrease as one moves away from the SFMS into the quiescent population. Again, this suggests that both prevention of gas accretion into the ISM and reduction in star forming efficiency are required to quench local galaxies.

However, there are two critical caveats here, which must be appreciated to correctly interpret these results. First, this analysis can only be performed on emission-line systems (in order to measure the Balmer decrement and gas-phase metallicity). This results in the sample missing the majority of truly quenched galaxies, which do not present with strong emission lines (e.g., \citealt{Brinchmann2004, Bluck2020a}). Indeed, in the sub-mm analysis of \cite{Saintonge2017}, almost all fully quenched systems are found to be non-detections in CO, and the majority are non-detections in HI as well. This implies that the importance of SFE-driven quenching is over-estimated in the current analysis, and indeed in any analysis which does not account for the non-detections. Nonetheless, the results for the emission-line SDSS sample are interesting and clearly imply that at least some of quenching arises from reduction in SFE. 

Second, SFE is expected to decline with decreasing gas mass as a result of self-gravitation in a super-linear KS relation (see \citealt{Kennicutt1998b}). However, many works find that the KS relation for molecular gas is close to linear (e.g., \citealt{Bigiel2008, Leroy2008, Lin2019}), mitigating this issue. Additionally, some works have accounted for this potential issue directly by quantifying offsets in the empirical KS relation, i.e., probing $\epsilon_\mathrm{SF}$ rather than SFE (see \citealt{Ellison2020, Ellison2021}). Encouragingly, these works also find a strong dependence of quenching on efficiency, establishing the reliability of this result to the functional form of the KS relation. 

In the right-hand panel of Fig.~\ref{f30}, we show a set of three random forest classification analyses to predict whether galaxies will be star forming or quenched in the SDSS emission-line sample, compared to the Illustris and IllustrisTNG cosmological simulations. This figure is reproduced from \cite{Piotrowska2022}. Crucially, the simulations are matched in SFR completeness to the SDSS emission line sample, enabling a robust comparison (see full details in \citealt{Piotrowska2022}). Clearly, SFE variation dominates the quenching in observed emission-line systems. In IllustrisTNG, a similar key dependence on SFE is found to observations. However, in Illustris the opposite trend is evident, i.e., a dominant dependence of quenching on gas fraction (unlike in observations).

In the simulations, the origin of this difference is well understood. Illustris quenches galaxies via radio-mode feedback, which is modeled via remote thermal bubble injection into the CGM (see Sect.~\ref{s25}; \citealt{Sijacki2007, Vogelsberger2013, Vogelsberger2014a, Vogelsberger2014b}). Consequently, the Illustris quenching mechanism has no direct impact on the ISM. Alternatively, quenching in IllustrisTNG occurs via a different form of radio-mode feedback, modeled as kinetic gas cell `kicks', which originate at the center of the galaxy, strongly impacting both the ISM and CGM (see Sect.~\ref{s25}; \citealt{Weinberger2017, Weinberger2018, Nelson2018, Zinger2020}). Given that the latter is in much better agreement with observations than the former, we conclude that, if AGN feedback is the sole cause of central galaxy quenching, it must impact the ISM in addition to offsetting cooling in the CGM. 

This is an important result because it establishes that radiatively inefficient (radio-mode) AGN feedback most probably operates in nature via multiple mechanisms. Explicitly: (i)~prevention of gas accretion into the ISM is required to account for lowering gas fractions (which is achieved in simulations via heating the CGM); \textit{and} (ii)~reduction in the capacity of the ISM to form stars is required to account for lowering efficiencies (which is achieved in simulations via turbulence injection into the ISM). See \cite{Weinberger2017, Weinberger2018, Zinger2020, Terrazas2020, Piotrowska2022} for further discussion.

To summarize this sub-section, many works have investigated the gas content of galaxies across the SFR\,--\,$M_*$ plane (including, \citealt{Saintonge2016, Saintonge2017, Piotrowska2020, Piotrowska2022, Brownson2020, Ellison2020, Ellison2021}). There is a consensus among these works that \textit{both} the gas content (in atomic and molecular form) and the efficiency (computed via several methods) decline within galaxies as sSFR decreases. To explain the reduction in gas content, some form of CGM stabilization is required in high mass centrals. This may be effectively explained via radiatively inefficient, radio-mode feedback (e.g., \citealt{Croton2006, Bower2006, Bower2008, Sijacki2007, Weinberger2017, Weinberger2018, Zinger2020}). Yet, to explain the reduction in efficiency, one must evoke a mechanism whereby the ISM is impacted, along with the CGM. We have shown that IllustrisTNG achieves this through its kinetic implementation of radio-mode feedback (see \citealt{Weinberger2017, Weinberger2018, Zinger2020}). An alternative explanation for declining efficiencies in quenching galaxies may be provided via dynamical stabilization (see Sect~\ref{s26}; e.g., \citealt{Martig2009, Gensior2020, Gensior2021}). However, to explain the observed reduction in gas fractions, one must still couple this with radio-mode AGN feedback, to prevent re-accretion of gas into the ISM from the CGM.

\subsection{Ejection vs. starvation: A test with stellar metallicities}\label{s62}

In this sub-section we explore observational evidence for the critical role of starvation in quenching both high- and low-mass galaxies. In particular, we review a fundamental test of quenching mechanisms via stellar metallicities (see \citealt{Peng2015, Spitoni2017, Trussler2020, Trussler2021, Bluck2020b}).

\begin{figure}[ht]  
\centering
\includegraphics[width=\textwidth]{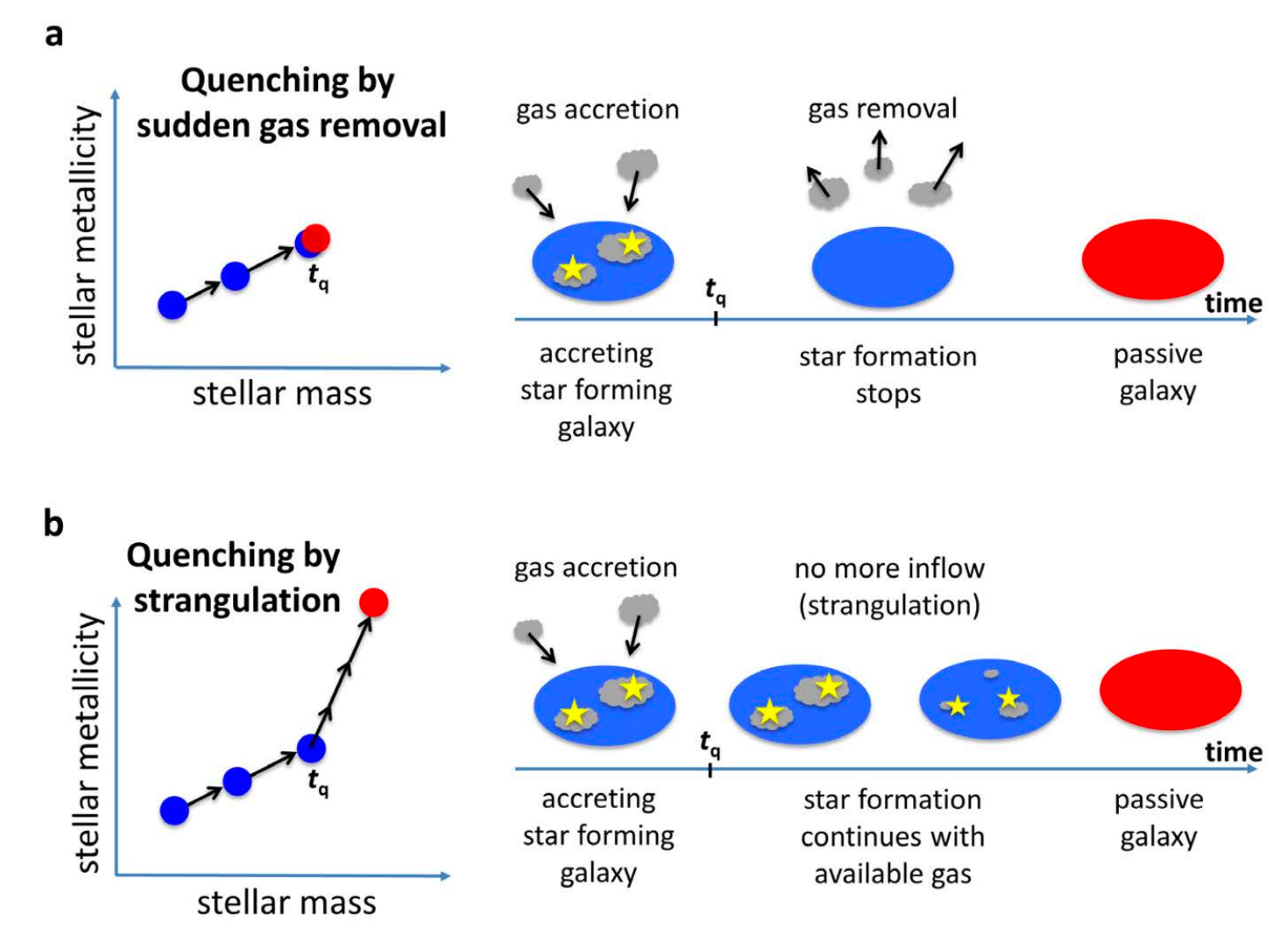}
\caption{Cartoon illustrations of the impact of quenching on stellar metallicity via sudden gas removal (top panels (a)) and via strangulation/ starvation (bottom panels (b)). In the case of quenching via sudden ejection of the ISM, the metallicity in stars in quenched systems is expected to be essentially identical to the metallicity in stars in star forming systems, at the same stellar mass. However, in the case of quenching via prolonged starvation/ strangulation, stars may form from the enriched ISM during the quenching process (which is no longer diluted via gas inflows from the IGM or CGM). This leads to increased metallicities of quenched galaxies compared to star forming systems, at a fixed stellar mass. This figure provides a simple framework for interpreting Fig.~\ref{f32}. Explicitly, the significant offsets in stellar metallicity highlight the importance of starvation as a quenching mechanism. This figure is reproduced from \cite{Peng2015}.}\label{f31}
\end{figure}

In Fig.~\ref{f31}, we show a cartoon illustration of how sudden quenching via gas removal (top panel) and slow quenching via starvation/ strangulation (bottom panel) are expected to impact the stellar metallicities of galaxies. This figure is reproduced from \cite{Peng2015}. In the case of rapid quenching via sudden depletion of the ISM (e.g., via quasar-driven outflows in centrals or strong ram pressure stripping in satellites), one expects the stellar metallicity of the quenched galaxy to be essentially identical to the star forming progenitor. Therefore, on a population basis, one expects the metallicities of quenched and star forming galaxies, evaluated at a fixed stellar mass, to be essentially identical. This follows because quenching occurs at a characteristic stellar mass (see, e.g., \citealt{Peng2010, Peng2012}).

Alternatively, in the case of starvation, gas inflows into the system are shut off entirely, but the ISM may continue to form stars until it is fully depleted. Typically, this will lead to slower quenching of galaxies. Crucially, the metallicities of the stellar population of quenching galaxies will continue to rise as a consequence of stars forming out of enriched gas in the ISM, undiluted by cosmic inflows (or CGM cooling). Therefore, in the case of starvation, one anticipates the metallicities of quenched galaxies to be significantly higher than that of star forming systems, at a fixed stellar mass. Furthermore, this process is expected to leave a strong mass-dependent signature, whereby the offset in stellar metallicity between quenched and star forming systems is higher at lower stellar masses. This follows because the higher the mass of the progenitor the lower the relative contribution of new stellar mass is to the whole stellar population.

\begin{figure}[ht]  
\centering
\includegraphics[width=\textwidth]{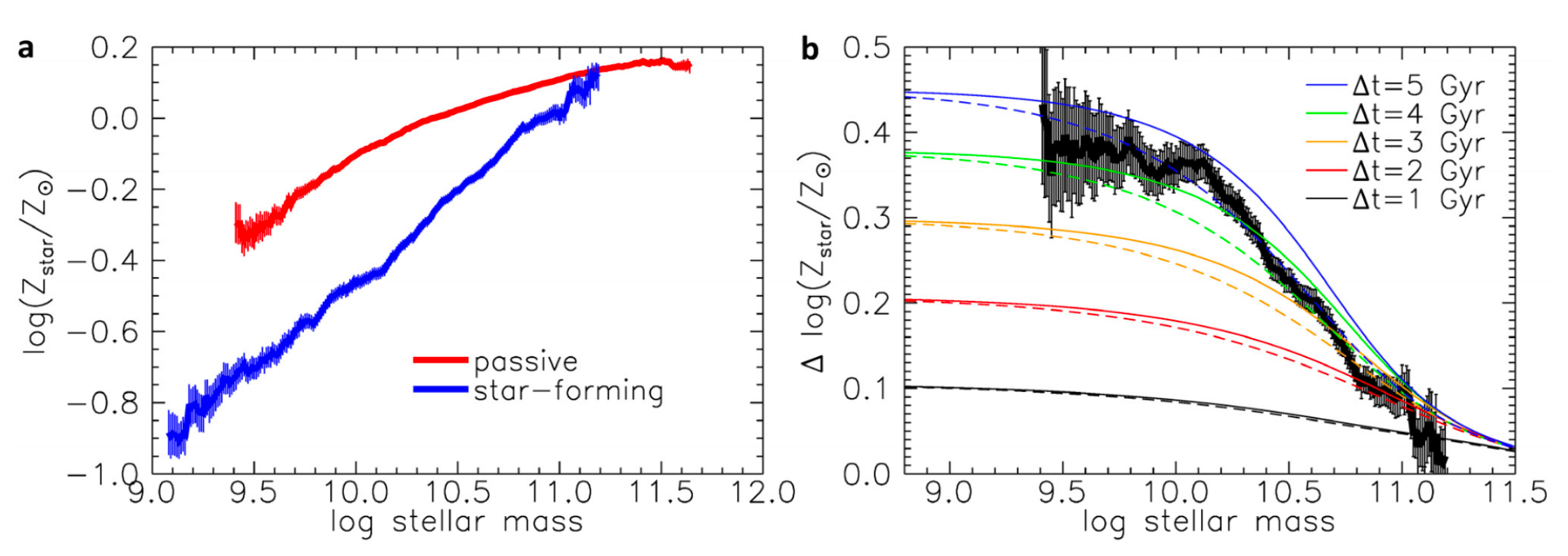}
\caption{\textit{Left-hand panel (a):} Stellar metallicity plot as a function of stellar mass for star forming (blue) and quenched (red) galaxies. A clear offset to higher metallicities in quenched systems is evident at a fixed stellar mass across a wide range in masses, but the offset becomes systematically smaller with increasing stellar mass. \textit{Right-hand panel (b):} The offset in stellar metallicity between quenched and star forming systems plot as a function of stellar mass. Colored lines indicate the expected trends from closed-box enrichment models with various quenching timescales (see legends). Solid lines indicate the final stellar mass, with dashed lines indicating the mass at onset of starvation. A constant quenching time of $\sim$4\,Gyr provides a reasonable fit to the data at all stellar masses. This figure is reproduced from \cite{Peng2015}.}\label{f32}
\end{figure}

In Fig.~\ref{f32} (left-hand panel), we show stellar metallicity as a function of stellar mass for star forming (blue) and quenched (red) systems, reproduced from \cite{Peng2015}. It is clear that quenched systems are significantly offset to higher stellar metallicities compared to star forming systems, at the same stellar mass. Additionally, the offset decreases systematically as a function of stellar mass. These results are qualitatively in line with quenching involving significant starvation, for all masses of galaxies.

These qualitative insights may be made more quantitative by constructing a simple model for the evolution of galaxies, including their change in ISM gas content, star formation rates, gas-phase and stellar metallicities. To begin, the rate of change in ISM gas mass ($M_g$; taken as total atomic plus molecular gas mass) is given by (e.g., \citealt{Peng2015, Trussler2020}):

\begin{equation}\label{e48}
\frac{dM_g(t)}{dt} = -(1-R)\, \Psi(t) - \Lambda(t) + \Phi(t)
\end{equation}

\noindent where, $R$ is the return fraction, $\Psi$ is the SFR, $\Lambda$ is the gas outflow rate (taken to be set by: $\Lambda(t) = \lambda_\mathrm{eff} \Psi(t)$, where $\lambda_\mathrm{eff}$ is the effective mass loading factor), and $\Phi$ is the gas accretion rate into the ISM, with all parameters evaluated at a time ($t$). This is a simple mass continuity equation, which enables the gas mass of the ISM to increase from accretion and decrease via both star formation and outflows from the system. The former is modeled via an instantaneous return of high mass stars (set by $R$), with the latter governed by a simple proportionality with star formation rate (set by $\lambda_\mathrm{eff}$; see, e.g., \citealt{Lilly2013, Peng2015} for further discussion). 

In the case of pure starvation, $\Phi(t) = 0$, and in the case of no outflows which fully escape the system, $\lambda_\mathrm{eff} = 0 \, \implies \, \Lambda(t) = 0$. Note that in this model, quasi-static equilibrium is not assumed (unlike in many `bathtub' models), and hence in general, $dM_g(t)/dt \neq 0$.

Next, the rate of change of gas-phase metallicity ($Z_g$) may be determined via (e.g., \citealt{Peng2015, Trussler2020}):

\begin{equation}
M_g(t) \, \frac{dZ_g(t)}{dt} = (1-R)\, y \, \Psi(t) - \big(Z_\Lambda(t) - Z_g(t)\big) \, \Lambda(t) + \big(Z_\Phi(t) - Z_g(t)\big) \, \Phi(t)
\end{equation}

\noindent where, $y$ is the net yield of star formation, i.e., the mass-weighted fraction of newly formed metals released into the ISM per unit new mass in stars formed (and locked up in low mass stars). $Z_{\Lambda}$ is the metallicity of outflowing gas, which in this model is assumed to equal the metallicity of the ISM. This results in the second term on the RHS evaluating to zero. Finally, $Z_\Phi$ is the metallicity of inflowing gas to the ISM, which is assumed to be (much) lower than the metallicity within the ISM. Hence, the final term evaluates to a negative number, causing dilution whilst gas inflows persist. 

Therefore, during the starvation phase, the rate of change in gas-phase metallicity is given simply by:

\begin{equation}
\frac{dZ_g(t)}{dt} = \frac{(1-R)\, y \, \Psi(t)}{M_g(t)} \, ,
\label{e50}
\end{equation}

\noindent where all terms are defined above. Hence, the gas-phase metallicity rises monotonically during starvation, as a function of metals added to the ISM via ongoing star formation and stellar evolution. Finally, the rate of change of stellar metallicity ($Z_*$) is given by:

\begin{equation}
\frac{dZ_*(t)}{dt} = \bigg( \frac{(1-R)\Psi(t)}{M_*(t)} \bigg) \, \big(Z_g(t) - Z_*(t) \big)   \, ,
\label{e51}
\end{equation}

\noindent where, $M_*(t)$ is the stellar mass of a galaxy at time, $t$, which is governed by the following differential equation: $dM_*(t)/dt = (1-R)\Psi(t)$. In the above expression, the first term on the RHS is the ratio of new stellar mass added per unit time to the total stellar mass, and the second term on the RHS is the difference in metallicity between the ISM (from which new stars form) and the established stellar population. Because the stellar population was formed some arbitrary time in the past, this lags behind the gas-phase metallicity. Hence, one anticipates the stellar metallicity to rise as a delayed function of the gas-phase metallicity during starvation.

To solve Eqs.~\eqref{e50} and \eqref{e51} numerically, one must first evaluate the functional form of $\Psi(t)$. This is achieved by analytically solving Eq.~\eqref{e48} in the case of pure starvation, where $\Phi(t)$ = 0. In order to do this, a link between SFR and gas mass must be assumed. Leveraging the assumption of a linear KS relation (e.g., \citealt{Kennicutt1998, Lin2019}), this is given simply by: $\Psi(t) = \epsilon_\mathrm{SF} \, M_g(t)$. The final result is then: 

\begin{equation}
\Psi(t) = \Psi_0 \, e^{-t/\tau_q}\, , \quad \textrm{where} \quad \tau_q = \frac{1}{\epsilon_\mathrm{SF} (1-R + \lambda_\mathrm{eff})} \, ,
\end{equation}

\noindent and where, $\tau_Q$ indicates the e-folding time of the quenching process. In the case of pure starvation, one sets $\lambda_\mathrm{eff} = 0$ (i.e., no outflows; see \citealt{Peng2015}). Additionally, one may model different strengths of outflows as desired (see \citealt{Trussler2020}).

In Fig.~\ref{f32} (right-hand panel), the observed offset in stellar metallicity between star forming and quenched galaxies is plot as a function of stellar mass (reproduced from \citealt{Peng2015}). Additionally, model lines derived via the above analytical approach are shown for various quenching timescales, assuming no outflows (see legends). The dashed lines show results for stellar mass at the time starvation begins, and solid lines show the final stellar mass at which galaxies are fully quenched. Remarkably, a fixed quenching timescale of 4\,Gyr is found to reproduce the results across a wide range in stellar masses. This indicates that quenching can plausibly be accounted for via pure starvation at masses above $M_* = 10^{9.5}\, M_{\odot}$. Moreover, these results clearly demonstrate that starvation must be involved in quenching (even if some outflows do occur) in order to account for stellar metallicity rising significantly throughout the quenching phase.

This work is expanded upon in \cite{Trussler2020}, who add a factor of three increase in sample size from the SDSS, perform more sophisticated modeling, and include comparison to progenitors. In this later work, the offset in stellar metallicity between star forming and quenched galaxies is found to be even more pronounced, once one accounts for the evolution in the metallicities of star forming systems at a fixed stellar mass. This generally strengthens the prior result from \cite{Peng2015}. Moreover, even at the highest masses analyzed, a clear offset in stellar metallicity is witnessed, significantly strengthening the claim of starvation in the highest mass systems.

Most importantly, \cite{Trussler2020} simultaneously match the increase in stellar metallicity and the reduction in SFR during quenching, allowing the mass loading factor ($\lambda_\mathrm{eff}$) to vary as a free parameter. They find that $\lambda_\mathrm{eff}$ declines as a strong function of stellar mass, such that low-mass galaxies are required to have both starvation and outflows during quenching, but high-mass systems are found to quench predominantly via starvation. This may be a little counterintuitive, since the offsets in metallicity are higher at low masses. Yet, this is driven solely by high mass systems requiring greater star formation during quenching to cause the same offset in metallicity as lower mass systems (see normalization in RHS of Eq.~\eqref{e51}). 

Additionally, \cite{Trussler2021} investigate the impact of environment on these results, finding that this effect is generally quite weak. However, they do note enhanced offsets in stellar metallicities in quenched and green valley satellites over that of centrals, indicating a plausible enhancement of starvation in dense environments. Finally, a significant offset in stellar metallicities is observed on kpc-scales within galaxies in \cite{Bluck2020b}, whereby quenched regions have significantly higher stellar metallicities than star forming regions, evaluated at a fixed stellar mass surface density (and total stellar mass of galaxy). This clearly implies that regions within galaxies must continue to form stars during the quenching phase, consistent with starvation being a critical aspect of quenching galaxies in general.

In summary of this sub-section, there exists strong observational evidence for the stellar metallicities of galaxies rising significantly during quenching, at essentially all stellar masses. This is incompatible with sudden ejection of the ISM, followed by rapid quenching. Conversely, this is exactly as expected from simple models tracing metal enrichment in galaxies which quench via starvation (see \citealt{Peng2015, Spitoni2017, Trussler2020}). More detailed analyses suggest that high-mass galaxies quench predominantly via starvation, with little need for significant outflows; whereas, low-mass galaxies are found to require both significant outflows and starvation to fully explain their quenching (see \citealt{Trussler2020, Trussler2021}). Furthermore, evidence for starvation leading to enhanced stellar metallicities during quenching is also found on spatially resolved scales, further establishing the reliability of these conclusions (see \citealt{Bluck2020b}). Therefore, the quenching of high-mass galaxies operates primarily via starvation, which most likely occurs via radio-mode AGN feedback preventing CGM cooling (see Sects.~\ref{s2} and \ref{s4} for further discussion and evidence). On the other hand, low-mass quenching involves both starvation and ejection/ stripping. The former may be accounted for in low-mass satellites by their location within the dark matter halo, and the latter via a mix of ram pressure and dynamical gas stripping (see Sects.~\ref{s3} and \ref{s5} for further discussion and evidence).

\section{Testing the triggers of galaxy quenching}\label{s7}

\subsection{Timing the quenching process}\label{s71}

In this sub-section we explore two important questions: First, {\it when do galaxies quench?} and, second, {\it how long does quenching take?} Additionally, we explore how these two timescales depend on the properties of galaxies and their environments. This is a big topic in the literature, with a long history (see, e.g., \citealt{Matteucci1986, Trager2000a, Trager2000b, Thomas2005, Thomas2010, Carnall2019, Belli2019, Leja2019, Tacchella2022}). The most important tool for ascertaining the timescales of both star formation and quenching is stellar population synthesis modeling (see \citealt{Conroy2013, Carnall2019, Leja2019} for excellent dedicated reviews). In particular, there are two levels of sophistication which may be applied: (i)~the use of SSP models, to identify the best-fit effective single stellar population consistent with a given galaxy spectra (e.g., \citealt{Matteucci1986, Trager2000a, Thomas2005, Thomas2010}); and (ii)~the use of CSP models, to identify the full star formation history (SFH) of a galaxy, under various assumptions (e.g., \citealt{Belli2019, Leja2019, Carnall2019, Tacchella2022}). We will review key results relevant for galaxy quenching from both of these approaches in the following parts of this sub-section.

\subsubsection{Results utilizing SSP models: Downsizing}\label{s711}

Although not explicitly related to quenching, \cite{Thomas2005, Thomas2010} investigate when local early-type galaxies (which are predominantly quenched) form their stars and, additionally, how long the star forming period lasts. Their approach is to fit spectral indices from observed galaxy spectra to a grid of SSP model spectra (from \citealt{Thomas2003, Thomas2004}). The SSP models are constructed in a three dimensional grid, varying: (i)~stellar population age ($t_\mathrm{SSP}$), (ii)~stellar metallicity ([Z/H], for Z = Fe and other heavy elements), and (iii)~$\alpha$-enhancement ([$\alpha$/Fe], for $\alpha =$ O, Ne, Mg, Si, etc.). Unlike many more recent models, dust extinction is not included. However, this is much less of an issue for early type galaxies, which are observed to be both gas and dust poor in the local Universe. 

The spectral indices used come in three categories: age-sensitive indices (e.g., H$\beta$, H$\gamma$, H$\delta$), metallicity-sensitive lines (e.g., Fe5270, Fe5335), and $\alpha$-sensitive lines (e.g., Mg\,b). Additionally, composite indices are formed from the above to help break degeneracies in the age - metallicity space (see \citealt{Thomas2005} for details).

The redshift at which stars form may be estimated by identifying the age of the best-fit SSP model, as follows: 

\begin{equation}\label{e53}
z_\mathrm{form} = z \, \big(t_\mathrm{Universe}(z_\mathrm{obs.}) - t_\mathrm{SSP}\big)
\end{equation}

\noindent where, $t_\mathrm{SSP}$ is the age of the best-fit stellar population model, and $t_\mathrm{Universe}(z_\mathrm{obs.})$ is the age of the Universe at the time the galaxy is observed. The formation redshift is then determined by assuming a set of cosmological parameters. Individual SSP models are essentially a delta-function in time. However, in \cite{Thomas2005, Thomas2010} the formation of stars is not taken to occur all at once at the formation redshift. Instead, the formation of stars is spread out from the formation time across a star formation period, $\tau_\mathrm{SF}$, to earlier cosmic times. This is inferred by considering the level of $\alpha$-enhancement of a galaxy (i.e., the $\alpha$-to-Fe abundance ratio, relative to solar). We will discuss more modern techniques utilizing SFHs within CSP models in the next part of this sub-section.

The essential insight for why $\alpha$-enhancement can help to estimate the duration of the star forming episode is that $\alpha$-elements are produced more or less instantaneously via high-mass stars undergoing core-collapse supernovae (i.e., SN Type II). On the other hand, lower mass stars may undergo supernova much later in binary systems through Roche accretion, once one member evolves into a White Dwarf star (producing SN Type Ia). Whilst both types of supernovae produce iron (though in significantly different amounts), only Type II supernovae produce $\alpha$-elements. This is because $\alpha$-elements are synthesized in the advanced burning stages of massive stars, and Type~Ia progenitors lack the high-mass stellar cores required for these processes. Hence, the $\alpha$-to-Fe abundance ratio serves as a clock, reflecting the duration of the star formation period. In the case of a short star forming burst, one would expect very high $\alpha$-to-Fe abundance ratios in stars, which form out of $\alpha$-enriched ISM gas. On the other hand, a very extended star forming period would yield very low $\alpha$-to-Fe abundance ratios, as many stars will form from ISM gas which is heavily Fe-enriched.

From SSP models, a simple relationship between star formation duration ($\tau_\mathrm{SF}$) and and $\alpha$-enhancement  is found (see, \citealt{Thomas1999, Thomas2005}). Explicitly,

\begin{equation}\label{e54}
\tau_\mathrm{SF} \approx 1\, \mathrm{Gyr} \times 10^{-1.5 \, [\alpha/\mathrm{Fe}]} \, .
\end{equation}

\noindent Therefore, the expected duration of a star formation event may be estimated directly from the measured $\alpha$-enhancement in observations.

\begin{figure}[ht]
\centering
\includegraphics[width=0.85\textwidth]{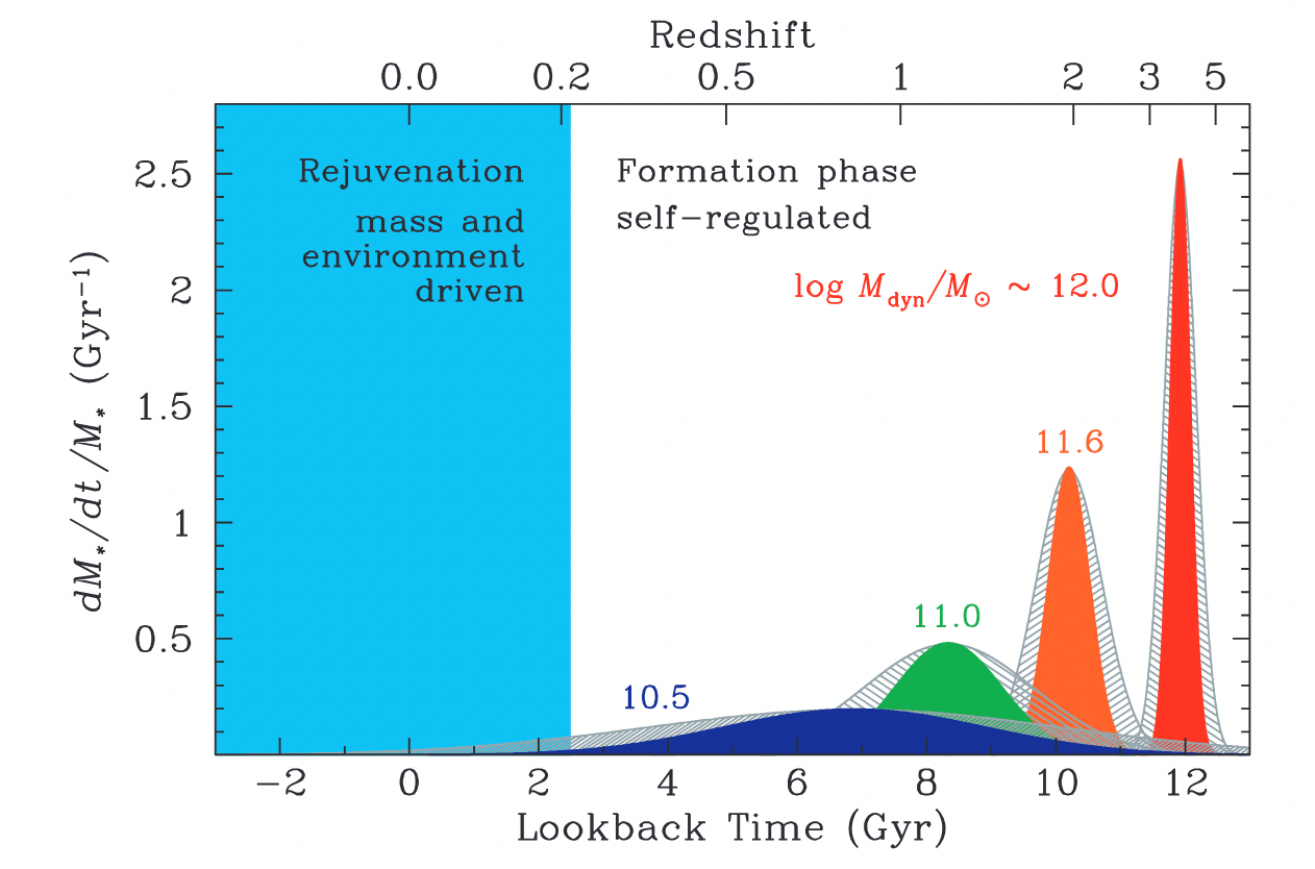}
\caption{Specific star formation rate plot as a function of lookback time (and redshift) for early type (essentially `quenched') galaxies computed for various stellar mass bins (see labels on figure). This figure is reproduced from \cite{Thomas2010}. High mass galaxies form the vast majority of their stellar populations early in the history of the Universe, with lower mass galaxies forming their stars systematically later in cosmic time. Additionally, high mass galaxies tend to form their stars in shorter bursts, followed by rapid reduction in star formation. Conversely, lower-mass systems have systematically longer star forming and quenching timescales. In the late Universe ($z < 0.2$) further star formation is assumed to arise from environmental effects (e.g., via gas-rich minor mergers). }\label{f33}
\end{figure}

In \cite{Thomas2005, Thomas2010}, effective star formation histories for local early-type galaxies are inferred via determining the star formation time (via Eq.~\eqref{e53}) and the star formation duration (via Eq.~\eqref{e54}). In Fig.~\ref{f33}, we show sSFR as a function of lookback time (and redshift) for these galaxies in bins of stellar mass (see labels), reproduced from \cite{Thomas2010}. It is clear that the most massive early-type galaxies form early in the history of the Universe, with lower mass galaxies forming at progressively later cosmic times. Furthermore, higher-mass systems tend to have very short star formation periods, with lower-mass systems having progressively longer star formation periods. These characteristic trends are supported by numerous other key works (e.g., \citealt{Cowie1996, Trager2000a, Trager2000b, Gallazzi2005, McDermid2015}).

The above results have been interpreted as evidence for cosmic `downsizing' (see, \citealt{Cowie1996, Trager2000a, Trager2000b, Thomas2005, Thomas2010, McDermid2015}). That is, the most massive galaxies form early in the history of the Universe, with lower mass systems forming much later. In terms of quenching, this further implies that high mass systems quench early, with lower mass systems quenching at progressively later cosmic epochs. At first pass, this scenario appears to contradict the theoretical paradigm of hierarchical assembly (whereby high mass haloes form out of mergers between lower mass haloes), which is very well established in $\Lambda$CDM theory (see, e.g., \citealt{Press1974, Efstathiou1985, Efstathiou1990, Frenk1988, Lacey1993, Springel2005a, Fakhouri2010}). However, these two concepts are not mutually exclusive because star formation can break the underlying structural hierarchy.

In $\Lambda$CDM, dark matter haloes form hierarchically via mergers, but the response of baryons to this underlying structural evolution is more complex. The key to resolving the apparent discrepancy is two-fold. First, in cosmological simulations, the baryon accretion rate into haloes rises with redshift, specifically as: $\dot{M}_b \sim f_b \, M_\mathrm{Halo} \, (1+z)^{2.5}$ (e.g., \citealt{Dekel2013,  VanDeVoort2011, Tacchella2022}). Hence, for a given halo mass, the inflow of baryons into the system is much higher at earlier cosmic times. Therefore, even though more massive haloes continue to form in greater abundance at later cosmic times, they are no longer efficiently fed by baryons from the IGM. Furthermore, the cooling time of gas within dark matter haloes increases with increasing age of the Universe (as a result of the virial radius for a fixed halo mass increasing as, $R_\mathrm{vir}(z) \propto H(z)^{-2/3}$). These effects combine to make early star formation much more efficient than late star formation, even in a hierarchical Universe. Ultimately, massive haloes at early cosmic times produce a lot of stars, whereas massive haloes at later times form significantly less stars. This explains qualitatively why the star formation peaks are higher and earlier for more massive systems in Fig.~\ref{f33}.

However, the above reasoning does not explain why star formation does not persist in high mass galaxies to late cosmic times, and ultimately why so few accreted baryons ever make it into stars. To account for this, there must be a cutoff in star formation at an approximate stellar mass threshold, i.e. mass-dependent quenching (e.g., \citealt{Baldry2006, Peng2010, Peng2012}). Combining `mass-quenching' with efficient early gas accretion and cooling leads to observed downsizing within a hierarchical assembly of dark matter haloes (see, \citealt{Keres2005, Birnboim2003, Dekel2006, Croton2006, Fontanot2009, Oser2010, Somerville2015}). Indeed, many recent cosmological simulations explicitly predict that more massive galaxies form their stars (and quench) at earlier cosmic times than their less massive counterparts (see, \citealt{Vogelsberger2014b, Genel2014, Furlong2015, Trayford2016, Nelson2018, Donnari2019}). Hence, despite the apparent contradiction, galaxy downsizing is actually an expected consequence of dark matter hierarchical assembly, once one accounts for a mass-dependent quenching threshold (as has been well established observationally in Part~I of this review). On the other hand, simulations and observations find much more diversity in terms of star formation (and quenching) durations, which we discuss in the context of CSP modeling in the next part of this sub-section.

\subsubsection{Results utilizing CSP models: Complexity}\label{s712}

In reality, star formation histories (SFHs) may be much more complex than considered in the previous part of this sub-section. For instance, there may be many star forming episodes in the lifetime of a galaxy and their timescales, total number of stars formed, and the functional forms of the star forming episodes may vary widely (e.g., \citealt{Conroy2013, Carnall2018, Carnall2019, Belli2019, Leja2019, Tacchella2022}). Hence, while a single star formation event may be a reasonable approximation for some galaxies, it is unlikely to be sufficient to model all types of galaxies. Furthermore, the prior SSP-based analyses do not directly constrain quenching. This follows because star formation events are assumed to be a simple step-function for a duration set by the $\alpha$-enhancement. Hence, essentially all quenching is instantaneous at the end of the single star forming period, which is unphysical.

All of the above issues may be resolved with the use of composite stellar population modeling (see \citealt{Conroy2013} for an excellent introduction). As a reminder, the essential idea is that the spectrum of a galaxy may be modeled as a sum over individual SSP models, with a weighting given by the star formation history (SFH $\equiv$ SFR(t)$/M_*$). In essence, this is accomplished via:

\begin{equation}\label{e55}
L_\mathrm{CSP} \, (\lambda) \, \bigg|_{t_\mathrm{obs}} \sim \int_0^{t_\mathrm{obs}} \Psi (t_\mathrm{obs} - t') \, L_\mathrm{SSP}(\lambda, t') \, dt'
\end{equation}

\noindent where, $L_\mathrm{CSP}$ is the luminosity of the CSP model, $L_\mathrm{SSP}$ is the luminosity of the SSP model (with a fixed age, $t'$), and $\Psi$ is the time evolving star formation rate (i.e., the SFH). Note that the SFH is a function of $t \equiv t_\mathrm{obs} - t'$, where $t'$ is the age of the stellar population (which is integrated over) and $t_\mathrm{obs}$ is the age of the Universe at the time of observation. This results in the SFH depending on time (i.e., age of the Universe at the star formation event), as expected. Alternatively, the SSP luminosity depends on the age of the stellar population, not the age of the Universe. Consequently, Eq.~\eqref{e55} is a convolution of the SFH over SSP model space. The above expression is not complete as written because one can also include a metallicity function, a dust extinction correction, and a dust re-emission term (e.g., \citealt{Conroy2013, Leja2019, Carnall2019}). However, this form is sufficient to understand the essential idea. 

In Fig.~\ref{f34}, we show an example CSP model spectrum produced with {\small BAGPIPES}, reproduced from \cite{Carnall2018}. The top panel shows a high resolution region of the model spectrum, with many spectral features visible. The inset within this panel shows the full wavelength range of the spectrum, showing contributions from both stars and dust. In the bottom panel, the values for the UltraVista wavebands are shown for the model CSP, with the inset on this panel showing the underlying SFH for this CSP model. Determining the SFH (in either a parametric or non-parametric form) is a crucial aspect of all CSP modeling. Consequently, an important output from this approach is the full history of star formation in any given model. By finding the best-fit CSP model to an observed spectrum or set of waveband measurements (or else via marginalizing over the posteriors in the Bayesian case), one can infer the history of star formation in any given galaxy, without the simplifying assumptions of the previous sub-section.

\begin{figure}  
\centering
\includegraphics[width=\textwidth]{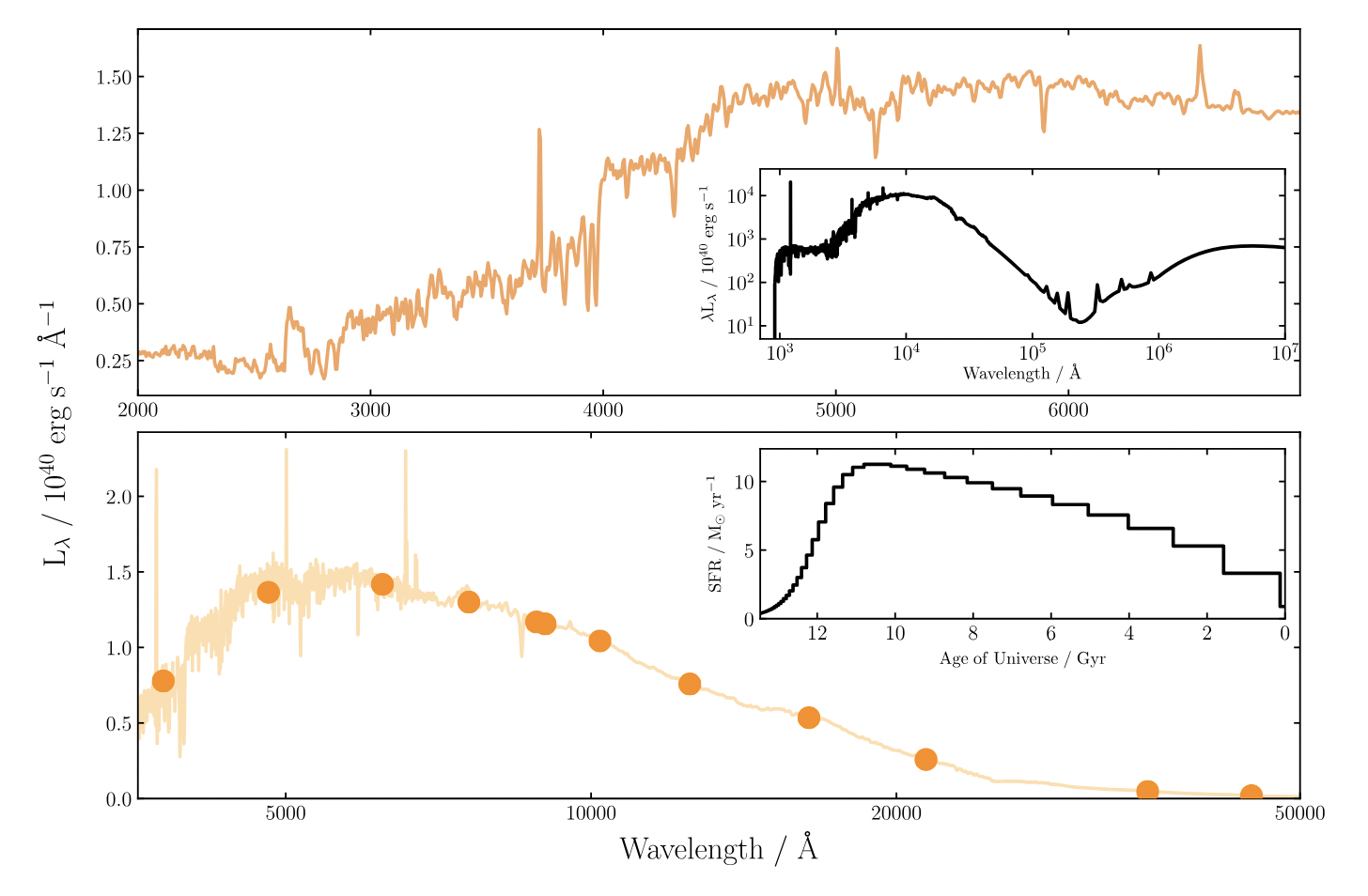}
\caption{An example CSP model spectrum from {\small BAGPIPES}, reproduced from \cite{Carnall2018}. The top panel shows a high-resolution region of the model spectrum, with the inset in this panel showing the full spectral range (including contributions from both stars and dust). The bottom panel shows the UltraVista wavebands extracted from this CSP model. In the inset of this panel, the underlying star formation history (SFH) is shown for the model, which shows a slow rise in star formation, followed by relatively fast quenching at late cosmic times. Via the CSP modeling approach a highly diverse range in star formation and quenching timescales may be effectively modeled, enabling sophisticated insights to be extracted from galaxy spectra and multi-band photometry.}\label{f34}
\end{figure}

Many publications have explored when galaxies quench, and how long quenching takes, via the CSP modeling approach (see Fig.~\ref{f35} for an example, reproduced from \citealt{Tacchella2022}; and, e.g., \citealt{Smethurst2015, Pacifici2016, Leja2017, Rowlands2018b, Carnall2018, Iyer2019, Belli2019}). The majority of works confirm cosmic downsizing, whereby the most massive galaxies form their stars, and then quench, early within the history of the Universe, with lower-mass systems forming stars significantly later (see, e.g., \citealt{Pacifici2016, Carnall2018, Tacchella2022}). This supports the general picture established via the more simplistic SSP modeling in the previous part of this sub-section (see, e.g., \citealt{Cowie1996, Thomas2005, Thomas2010}). However, contemporary publications frequently emphasize a large diversity of formation times for galaxies across a wide range in stellar masses, and the overall trends with mass are typically weaker than seen in effective SSP analyses.

\begin{figure}  
\centering
\includegraphics[width=0.8\textwidth]{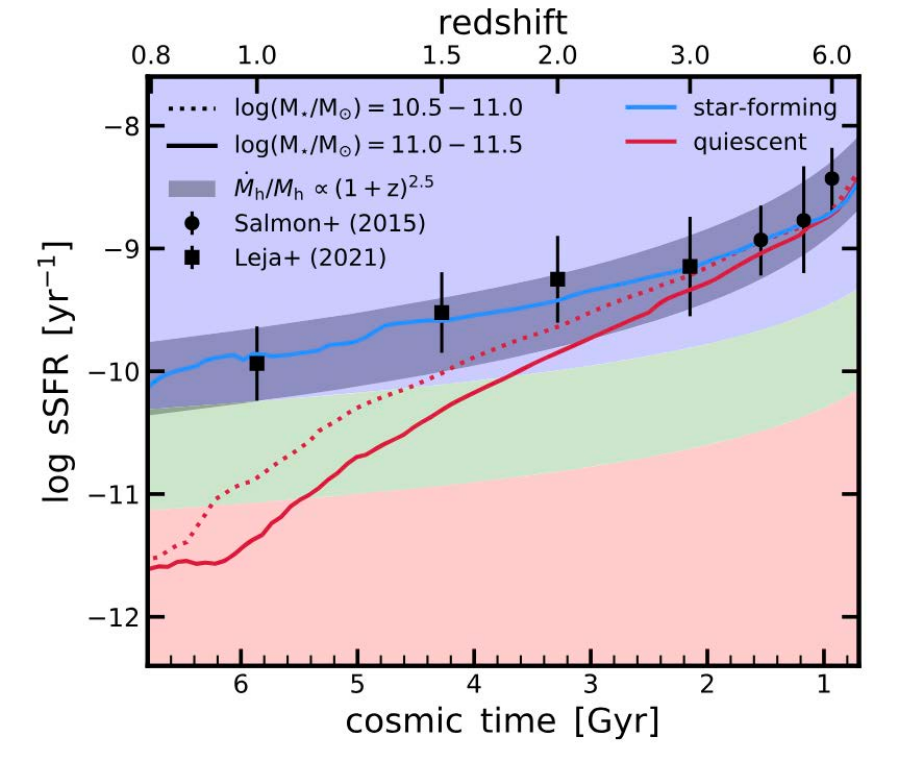}
\caption{The star formation histories of 161 massive galaxies at $z \sim 0.8$ from  Keck/DEIMOS spectra observed as part of the HALO7D survey. The cosmic history of star formation for individual galaxies is inferred via SED fitting to CSP models. Data points show prior observations for star forming galaxies from \cite{Salmon2015, Leja2022}, which trace the expected relation for an accretion-limited sSFR evolution (see shaded region). The averaged SFHs of quenched galaxies are shown within two stellar mass bins (see legend). The green region indicates the cosmological green valley ($1/3 < \mathrm{sSFR(z)} \cdot t_H(z) < 1/20$), with the star forming region residing above this (shaded blue) and the quenched regions relying below this (shaded red). Galaxies in this survey tend to quench at $z \sim 1 -1.5$, with more massive systems quenching slightly earlier than less massive systems. This figure is reproduced from \cite{Tacchella2022}.}\label{f35}
\end{figure}

In terms of quenching timescales, the literature is much more varied. Several works note a bimodality in quenching timescales (e.g., \citealt{Smethurst2015, Carnall2018, Belli2019, Tacchella2022}), where quenching separates out into fast vs. slow variants. Some publications find generally faster quenching in higher mass systems (e.g., \citealt{Carnall2018, Tacchella2022}), some publications find a clearer trend with morphology than stellar mass (e.g., \citealt{Schawinski2014, Smethurst2015, Iyer2019}), and some other works find the opposite trend, whereby higher mass systems quench slower, particularly at late cosmic times (see, \citealt{Pacifici2016, Belli2019}). Furthermore, the quenching timescales are generally found to be significantly longer than naively estimated via SSP modeling, with fast quenching taking $\sim$0.5 -- 1\,Gyr, and slow quenching taking $\sim$1 -- 5\, Gyr (e.g., \citealt{Rowlands2018b, Smethurst2017, Carnall2018, Tacchella2022}).

Taken in concert, the current observational constraints on when galaxies quench are broadly consistent with the older idea of downsizing, with the most massive galaxies (typically spheroids) forming their stars early in the history of the Universe, and lower mass systems (typically discs) forming their stars at later epochs. From a theoretical perspective, mergers and quasar-mode feedback are often invoked to explain fast-quenching of high mass spheroids (e.g., \citealt{Smethurst2015, Carnall2019, Belli2019, Tacchella2022}), with slow-quenching of low-mass discs frequently attributed to preventative AGN feedback in centrals and environmental effects in satellites (e.g., \citealt{Wetzel2013, Schawinski2014, Hirschmann2014, Balogh2016, Fossati2017, Tacchella2022}).

In comparison to cosmological simulations, most contemporary models reproduce the general downsizing trend (see, e.g., \citealt{Teklu2017, Wright2019, Dave2019, Correa2019, Donnari2021a, Donnari2021a}). However, in terms of quenching durations, the results are more diverse (as in current observational constraints). For example, EAGLE and SIMBA both reproduce fast-quenching via ejective AGN feedback, which leads to decreased quenching times in more massive systems (see, \citealt{Wright2019, Dave2019}). On the other hand, IllustrisTNG exhibits the opposite trend, whereby more massive galaxies quench slower than less massive systems (see, \citealt{Donnari2021a, Donnari2021b}). This is due to quenching occurring almost exclusively via preventative radio-mode kinetic feedback in this simulation, without significant outflows via quasar-mode feedback (see, \citealt{Weinberger2017, Weinberger2018, Zinger2020}).

In summary, leveraging the more sophisticated CSP modeling approach, cosmic downsizing is now essentially confirmed. However, a much greater diversity in quenching timescales is found than anticipated in simple SSP modeling. This is interpreted in the literature as evidencing multiple quenching pathways, which separate out as a function of stellar mass, morphology, and galaxy type (i.e., cental vs. satellite). While there is not full consensus, several works find that higher mass/ spheroidal systems quench fast, whereas lower mass/ disc-dominated systems quench slow. Furthermore, satellites are typically found to quench slower than centrals, at a fixed stellar mass. It is important to appreciate that these works specifically constrain the triggers of quenching (i.e., what causes galaxies to first depart the SFMS). In order for galaxies to maintain quiescence, further processes are required (see discussions in Sects.~\ref{s2}, \ref{s4} and \ref{s6}).

Finally, in terms of the triggers of quenching, it is often conjectured that some form of ISM removal is needed in addition to starvation in order to explain the fastest quenching timescales (e.g., via AGN-driven outflows in centrals or via gas stripping in satellites). However, given that the depletion time of the ISM deceases towards earlier cosmic times (as $t_\mathrm{dep} \sim (1+z)^{-(0.5 - 1.0)}$; see \citealt{Lilly2013, Tacconi2018, Baker2023}), one may also anticipate faster quenching to occur in systems which quench at earlier cosmic times, even within a purely preventative quenching scenario (e.g., \citealt{Feldmann2015, Peng2015, Tacconi2018, Baker2023}). This is potentially very important because it could explain the strong observational evidence for short quenching times in high-mass galaxies (which quench early in the history of the Universe) in conjunction with the equally strong evidence for starvation-driven quenching in these systems (see Sect.~\ref{s62}, and \citealt{Peng2015, Trussler2020, Bluck2020b}). However, it is fair to say that there is no accepted explanation for the diversity in quenching timescales at present, with more work needed to constrain precisely what this diversity depends on at a fundamental level.

\subsection{Catching quenching in action}\label{s72}

A popular idea in the contemporary quenching literature is to `catch quenching in action'. That is, to identify a population of galaxies which are currently undergoing quenching, determine what separates these systems from their star forming counterparts, and, hence, constrain the mechanisms driving the transition. This is, of course, highly logical. However, there are a number of challenges and caveats which must be addressed. First, this method is implicitly sensitive primarily to the {\it triggers} of galaxy quenching, and (by design) is much less sensitive to the process(es) which maintain long-term quiescence in galaxies. As discussed previously (see Sects.~\ref{s2} and \ref{s4}), provided replenishment of the ISM is halted (e.g., via preventative AGN feedback or location within the cosmic web), quenching will inevitably follow, modulo the ubiquitous requirement for maintenance feedback to offset gas return from stellar evolution. Alternatively, without preventing cooling flows into galaxies, quenching will never occur, although oscillations around the main sequence are still expected. Therefore, there is a risk with focusing on quenching triggers that the fundamental cause of quenching is lost, or under-appreciated.  

Second, since all astronomical observations are a snapshot in time, one cannot know with certainty that a given galaxy is quenching, only that it is currently forming stars lower than it was in the past. It remains possible that some (or even most) of these systems will rejuvenate their star formation significantly post observation. If this is the case, what is interpreted as quenching may in reality be simply main sequence oscillations, which are a part of the normal life-cycle of star forming galaxies (see Sect.~\ref{s24} for a discussion). Therefore, one must be careful extracting insights about quenching from populations of galaxies with uncertain futures.

Nonetheless, galaxies which are genuinely quenching must transition in star formation between the active and passive phases. Hence, at least some of the galaxies presenting with intermediate levels of star formation must be quenching and, hence, insights from this population can be highly valuable. Moreover, simulations and models may be used to mitigate the effect of contamination within this population. 

The two main observational routes to identify galaxies which are undergoing quenching are: (i)~to select systems in the `green valley' (e.g., \citealt{Strateva2001, Salim2007, Schawinski2014, Smethurst2015, Bluck2016, Trussler2020}), and (ii)~to select `post starburst' systems (e.g., \citealt{Wild2009, Whitaker2012, Wild2016, Almaini2017, Rowlands2018a, Wild2020, Almaini2025}). The former are galaxies with intermediate star formation rates and/or colors between the star forming (blue) and quenched (red) populations. The latter are galaxies which exhibit evidence for A-type stars (with no evidence of O- and B-type stars), which implies a recent and abrupt decline in their star formation rates. There is significant overlap between these two transition populations, but they are not identical. 

In the subsequent parts of this sub-section, we will discuss both of these techniques in turn, outlining the numerous insights these populations have revealed on the triggers of galaxy quenching throughout cosmic time.

\subsubsection{The green valley and compaction}\label{s721}

As discussed in Part~I of this review series, galaxies separate out in color (primarily UV-to-NIR) and sSFR into bimodal distributions, with the region between the two peaks often being referred to as the `green valley' (see, e.g., \citealt{Strateva2001, Brinchmann2004, Salim2007, Wyder2007}). Given that star forming galaxies are more abundant at earlier cosmic times and quiescent galaxies are more abundant at later cosmic times, the dominant route through the green valley must be towards the quenched population (see, e.g., \citealt{Faber2007, Kriek2008, Whitaker2012, Muzzin2013, Whitaker2014, Madau2014}). However, this does not imply that all green valley galaxies are quenching. Nonetheless, many studies make this assumption as a starting point and attempt to elucidate evidence for how quenching occurs utilizing this population.

In Fig.~\ref{f36} we show the location of early- and late-type systems in ($\text{NUV} - u$) - ($u - r$) color space, reproduced from \cite{Schawinski2014}. Both morphological types occupy essentially the same position in ($u - r$) color, but separate out significantly in ($\text{NUV} - u$) color, with green valley early-type galaxies having much reader UV-to-optical colors than green valley late-types. This is interpreted as early-type galaxies having fewer (or no) O- and B-type stars, and, hence, being closer to fully quenching than late-type systems with intermediate optical colors.

\begin{figure}[ht]
\centering
\includegraphics[width=0.9\textwidth]{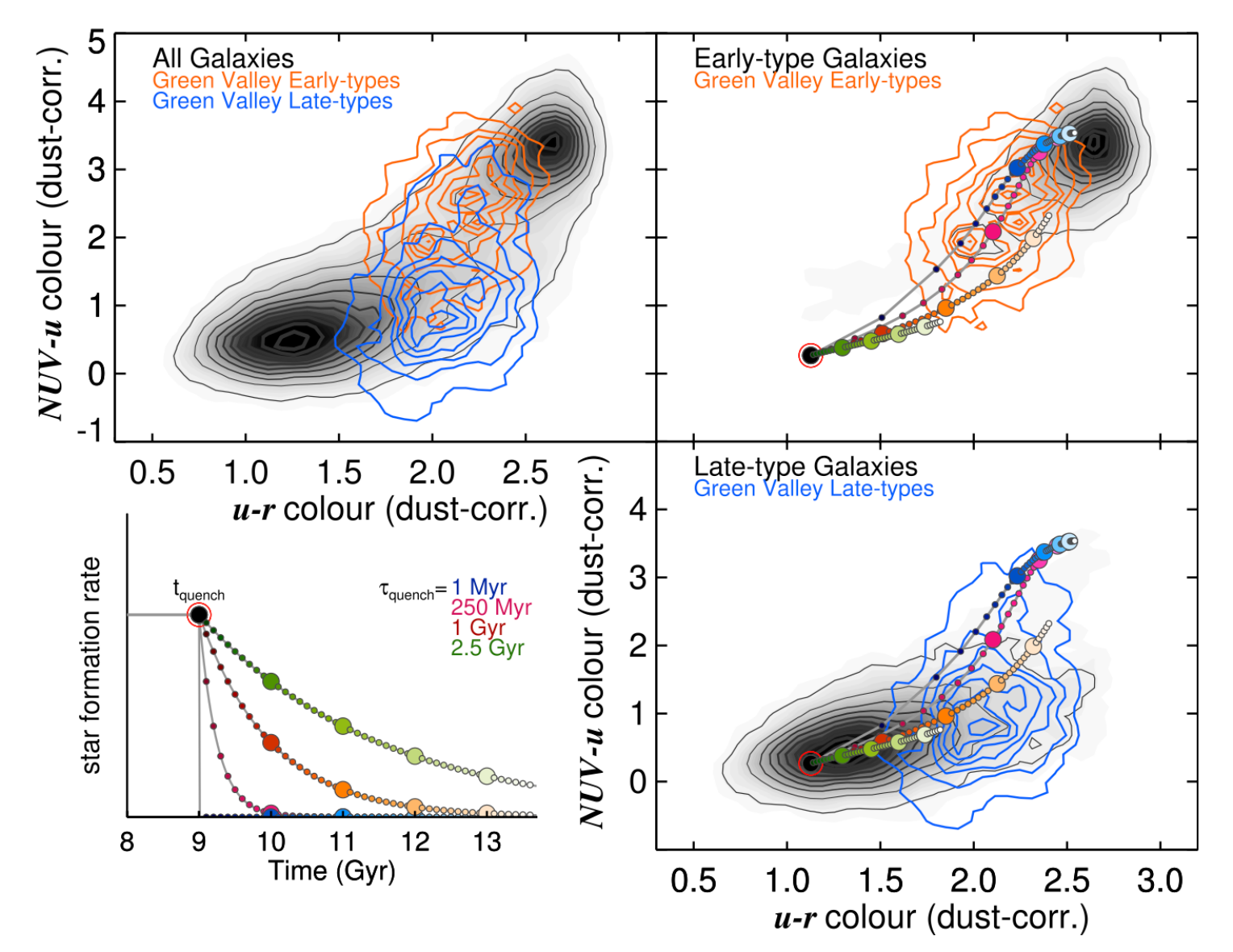}
\caption{The location of SDSS galaxies in (NUV - $u$) vs. ($u-r$) color space, reproduced from \cite{Schawinski2014}. The top-left panel shows the full galaxy population (grey contours), with early-type (red contours) and late-type (blue contours) galaxies within the green valley region overlaid. The right-hand panels show the results for early-type galaxies (top panel) and late-type panels (bottom panel), with evolutionary tracks from \cite{Bruzual2003} stellar synthesis models with different quenching timescales. The bottom-left panel shows exponentially declining SFR models for varying quenching timescales ($\tau_Q$). Assuming green valley systems are en route to quenching, the results in this figure clearly establish that early-type galaxies quench much faster ($\tau_Q \lesssim 0.5$\,Gyr) than late-type systems ($\tau_Q \gtrsim 1$\,Gyr).}\label{f36}
\end{figure}

To investigate this more quantitatively, \cite{Schawinski2014} construct stellar population synthesis models for quenching galaxies, under the assumption of an exponentially declining SFR. Explicitly, the decline in star formation rate ($\Psi$) is modeled as:

\begin{equation}
\Psi(t) \bigg|_{t \geq t_Q} = \Psi(t_Q) \, \cdot e^{-(t-t_Q)/\tau_Q} 
\end{equation}

\noindent where, $\Psi(t_Q)$ is the star formation rate at the time quenching begins ($t_Q$), and $\tau_Q$ is the quenching timescale. Higher values of $\tau_Q$ indicate slower quenching. On the right-hand panels of Fig.~\ref{f36}, evolutionary tracks through the color - color space are overlaid and compared to early-types (top panel) and late-types (bottom panel). These correspond to varying $\tau_Q$ values as depicted on the bottom-left panel. 

Early-type green valley systems are found to have much shorter quenching times than late-type systems, with the former having $\tau_Q < 0.5$\,Gyr and the latter having $\tau_Q > 1$\,Gyr. Similar results are found via CSP modeling in \cite{Smethurst2015}, confirming this general picture. Additionally, this later publication finds a general trend whereby more massive systems, and those in denser environments, tend to experience faster quenching than lower mass systems, and those in less dense environments. These trends are often attributed to the importance of mergers for fast quenching, which may lead to accelerated gas depletion in the ISM via a starburst phase, in addition to enhanced SN and radiatively efficient AGN feedback leading to outflows (e.g., \citealt{DiMatteo2005, Springel2005c, Hopkins2006, Hopkins2008}).

However, there are a number of caveats here. In both \cite{Schawinski2014} and \cite{Smethurst2015} green valley systems are assumed to be quenching. However, the late-type green valley galaxies could potentially be quenched only in their cores, i.e., having quenched bulges and star forming discs (e.g., \citealt{Mendel2013, Pan2016, Morselli2017, Belfiore2017}). In turn, this could lead to the observed curvature of the SFMS at high masses (e.g., \citealt{Abramson2014, Whitaker2014}). In this scenario, the discs may be forming stars normally, with only the bulge region suppressed in star formation, or fully quenched. Nonetheless, \cite{Bluck2022} show that bulge mass is the most effective parameter for predicting whether discs will be quenched, which strongly implies a close link between the quenching of these two structural components. It may be that the disc simply lags behind the bulge in quenching, when both structures are present. Hence, the essential idea that discs quench slower than spheroids is still valid. In any case, these works are very important for establishing that the green valley (as observed in optical colors) is not a single population, but instead splits into (at least) two separable populations, which are evolving in very different ways towards quiescence.  

Also leveraging the green valley population, morphological and structural transitions (especially computed via bulge mass, $M_B$, and the surface stellar mass density within 1\,kpc, $\Sigma_1$) are found to pre-date quenching at all redshifts up to cosmic noon (see, e.g., \citealt{Barro2013, Lang2014, Zolotov2015, Barro2017}). Explicitly, these studies find evidence for the central densities and bulge masses of green valley systems being higher than that of star forming systems, and closer to (and sometimes even higher than) the quenched population. Consequently, progenitor bias via size evolution cannot be the dominant cause of the difference in structures between star forming and quenched systems (as argued for in \citealt{Lilly2016}). These changes within galaxies occur {\it before} quenching and, hence, may potentially be the underlying cause thereof.

These results have led to the important idea of `compaction', whereby star forming galaxies experience dissipational collapse (via mergers or violent disc instabilities), leading to a rapid increase in central densities and, subsequently, quenching (see, \citealt{Cheung2012, Fang2013, Barro2013, Barro2014, Barro2017, Dekel2014, Zolotov2015, Tacchella2016a, Tacchella2016b, Tacchella2022}). Ultimately, these results are consistent with quenching via preventative AGN feedback, whereby systems with denser cores host more massive SMBHs, which are capable of stabilizing the CGM, shutting off cooling flows, and ultimately causing quenching (see Sects.~\ref{s2} and \ref{s4}; and \citealt{Bluck2014, Bluck2016, Terrazas2016, Terrazas2017, Terrazas2020, Piotrowska2022, Bluck2022, Bluck2023, Bluck2024}). 

Additionally, these results provide a potential explanation for the diversity of quenching timescales (see Sect.~\ref{s71}), as fast-quenching may arise from gas rich major mergers (e.g., \citealt{Schawinski2014, Zolotov2015, Tacchella2022}), with slow-quenching occurring later in the history of the Universe via secular evolution (e.g., \citealt{Cheung2013, Schawinski2014, Smethurst2015}). This narrative fits well with the observed increase in merger rates with increasing redshift (see, e.g., \citealt{Rawat2008, Conselice2009, Lopez2009, Bluck2009, Lotz2011, Bluck2012, Duncan2019}), leading to faster quenching timescales at earlier cosmic times (as observed). Finally, in order for SMBHs to reach the high masses needed to cause quenching they must grow via accretion and black hole mergers, both of which are enhanced during galaxy mergers, as confirmed by much observational evidence (see, e.g., \citealt{Ellison2008, Treister2011, Ellison2011, Koss2012, Goulding2018}).

\subsubsection{Post starburst galaxies}\label{s722}

Post starburst galaxies (PSBs) are classically defined spectroscopically as systems which exhibit no emission lines (indicating a dearth of O- and B-type stars), but present with deep Balmer absorption features (indicating the presence of many A-type stars). This implies that these systems must have recently, and rapidly, completed a star forming episode, hence the name (see, \citealt{Dressler1983, Couch1987, Zabludoff1996, Poggianti1999, Wild2009, French2015}). PSBs are also often referred to as `E+A' galaxies (elliptical-like spectra with A-type stars) and `K+A' systems (galaxies presenting with K- and A-type stars). 

Post starbursts are extremely rare in the local Universe, contributing $\lesssim1$\,\% of the galaxy population (see \citealt{Goto2005, Quintero2004, Wild2009, French2015, Rowlands2015}). However, there is evidence that this population becomes much more common at higher redshifts, reaching a few per cent of the total galaxy population at cosmic noon (see, \citealt{Whitaker2012, Wild2016, Forrest2020, Wild2020, Carnall2020}). This implies that PSBs are typically rarer than green valley systems, which (dependent upon method) make up $\sim 5 - 10$\,\% of the galaxy population (e.g., \citealt{Salim2007, Schawinski2014, Bluck2016}).

The most pressing question regarding PSBs from the perspective of this review is whether these systems are recently quenched galaxies, quenching galaxies, or the extremes of SFMS oscillations. Furthermore, if they are quenched or quenching systems, how common is the PSB phase in the general process of quenching? Put simply, are PSBs a critical part of quenching, or not? These questions have been investigated extensively in the literature (see especially, \citealt{Wild2009, Wild2010, Wild2016, Wild2020}). In the local Universe, it seems PSBs are not only rare in general but also can explain at most a few per cent of recent quenching (e.g., \citealt{Wild2009}). However, they are much more common in fast-quenching events, rising to $\sim$20 - 30\% prevalence (e.g., \citealt{Rowlands2018}). Alternatively, at high redshifts, PSBs may account for a significant fraction of all quenched systems, with estimates reaching up to 50\% or so (see, \citealt{Wild2016, Forrest2020, Carnall2023b}).

Since PSBs are rare (especially in the local Universe) and furthermore that wide-field spectroscopy is limited in the high-$z$ Universe, methods to identify PSBs in photometry have been developed. These techniques have proved instrumental in assessing the statistical significance of this population to the evolution of galaxies in general. The most successful technique is that of super-colors, derived via a principal component analysis (PCA; see \citealt{Wild2014, Wild2016}). This approach enables the formation of large statistical samples of PSB candidates, which form the basis of many contemporary analyses.

Many studies identify that morphological disturbances are prevalent in PSBs, qualitatively supporting a merger-driven, fast-quenching scenario (e.g., \citealt{Zabludoff1996, Yang2008, Wild2009, Pawlik2018}). Several other works identify an enhanced AGN fraction, or other evidence for AGN, in PSBs (see, \citealt{Yesuf2014, Alatalo2016, French2018}), although \citealt{Wild2010} provide evidence for a delay between the starburst and the AGN peak, suggesting that radiatively efficient AGN do not cause the departure from the SFMS. Alternatively, \citealt{Almaini2025} find no evidence for excess X-ray detected AGN in PSBs in the UDS field. 

Historically, a close link between PSBs and the cluster environment was found, with many local PSBs residing within cluster environments (see, e.g., \citealt{Poggianti1999, Poggianti2009, Dressler2013, Muzzin2014}). For these systems, it is highly likely that strong ram pressure stripping is the cause of the sudden reduction in star formation. However, PSBs are also prevalent in the field, especially so at earlier cosmic times (e.g., \citealt{Zabludoff1996, Blake2004, Wild2009, Rowlands2015, Pawlik2018, French2015, French2018}). This highlights that secular processes must also be capable of forming PSBs, with mergers (and possibly AGN) as the leading contenders for causing the required rapid departure from the SFMS.

In summary, post-starburst galaxies are widely believed to be recently quenched galaxies, rather than quenching systems or main sequence oscillations, which experienced fast-quenching of $\tau_Q \sim 0.1 - 1$\,Gyr (e.g., \citealt{Wild2009, Wild2010, Wild2014, French2018, Suess2022}). In the local Universe, they make up only a very small fraction of quenched systems, and are therefore most likely not a major aspect of late-time galaxy quenching (e.g., \citealt{Wild2009, French2015, Rowlands2015}). However, at higher redshifts, they become much more numerous, and may even dominate the fast-quenching pathway (e.g., \citealt{Wild2016, Belli2019, Tacchella2022, Suess2022, Carnall2023b}). These results are consistent with the work on quenching timescales (discussed in Sect.~\ref{s71}), whereby rapid quenching is generally found to be more prevalent at earlier cosmic times. Finally, evidence for mergers and AGN are often reported in the PSB population. However, contrary evidence also exists (especially in the case of AGN). Although some PSBs are clearly linked to dense environments, the majority appear in lower-density environments such as the field. Ultimately, in terms of contribution to the quenching literature, studies of PSBs have been useful for identifying fast-quenching systems, and investigating their triggers. However, a full consensus for their origin and the role they play within the broader framework of quenching remains elusive.


\section{Summary and conclusions}\label{s8}

In this review series we have sought to answer a simple question, \textit{why do galaxies cease forming stars?} Along the way, we have seen that this innocent enough seeming question is remarkably deep and connects with almost every aspect of cosmology, galaxy formation and evolution, and the physics within galaxies (from star formation and supernovae to black hole formation, growth, and feedback). In this final section we will do three things: (i)~briefly summarize the contents of Part~II of this review; (ii)~attempt an answer to the overarching question (separated into intrinsic and environmental routes); and (iii)~acknowledge what is still unknown to the field.

\subsection{Summary of Part~II}\label{s81}

In Sect.~\ref{s1}, we provide a brief overview of Part~I of this review series, recalling three fundamental problems with $\Lambda$CDM as a theory of galaxy formation and evolution, in lieu of strong baryonic feedback. Explicitly, we summarize: (i)~the cosmological problem (i.e., why star formation is so inefficient within dark matter haloes); (ii)~the clusters problem (i.e., why the hot gaseous halo surrounding high mass galaxies, groups, and clusters is thermodynamically stable); and (iii)~the bimodality problem (i.e., why galaxies separate out into two populations in terms of star formation rates, color, and the ages of their stellar populations). Additionally, we provide a summary of terminology, a brief review of key observational constraints on quenching, and an introduction to the simulation types interacted with in this part of the review.

In Sect.~\ref{s2}, we explore the quenching of centrals (and high-mass galaxies in general) from a theoretical point of view. We begin by presenting a summary of the need for strong baryonic feedback to account for the structure of the observed galaxy stellar mass function, within the framework of cosmological simulations. We present the theoretical case for quenching being fundamentally a maintenance problem, not a one time event. We discuss the role of supernovae (both Type Ia \& II) within contemporary models. We then discuss three important intrinsic quenching paradigms: (i)~halo mass quenching from virial shocks, (ii)~quasar-mode AGN feedback, and (iii)~radio-mode AGN feedback. We go on to review the connection between these theoretical ideas, and discuss at length their implementation within contemporary cosmological simulations. Finally, we discuss an alternative to feedback-driven quenching via the potential role of dynamical stabilization as a quenching aid.

In Sect.~\ref{s3}, we explore the quenching of satellite galaxies from a theoretical perspective. We begin by reviewing how galaxy mergers occur via dynamical friction, and discuss how the merger timescale increases with decreasing pair mass fraction. This is the fundamental origin of the group and cluster environment, in which satellite galaxies quench. We go on to review specific theoretical routes for quenching (especially low-mass) satellite galaxies, including: ram pressure stripping, dynamical stripping, galaxy - galaxy harassment, strangulation, pre-processing, and AGN-feedback modified environmental quenching. 

In Sect.~\ref{s4}, we turn our attention to testing the fundamental causes of galaxy quenching, as predicted via theory. We review direct evidence for SN and AGN feedback (in both the ejective and preventative modes). We explore where AGN reside on the star forming main sequence, noting many potential issues with this apparently logical approach. We then explore how supermassive black hole mass varies between star forming and quenched galaxies. Of particular note, we review strong evidence for supermassive black hole mass being the best predictor of cental (and high-mass satellite) quenching in both simulations and observations across cosmic time (from $z = 0 - 8$), clearly favoring the dominant role of AGN feedback. 

In Sect.~\ref{s5}, we explore methods to test the fundamental causes of satellite galaxy quenching, as predicted via various theoretical models. We explore the direct evidence for ram pressure and dynamical stripping operating in observations of satellite galaxies, within both the cluster and group environments. Additionally, we demonstrate that simulations and observations agree that low-mass satellites have their quenching ultimately dependent upon their environments, in stark contrast to high-mass satellites (which behave almost identically to centrals in terms of their quenching). This implies that the fundamental distinction between intrinsic and environmental quenching is not the central vs. satellite class, but rather the mass of the system.

In Sect.~\ref{s6}, we explore the observational evidence for the precise mechanisms of galaxy quenching. In particular, we explore whether reduction in gas fraction ($f_\mathrm{gas}$) or reduction in star formation efficiency (SFE) drives the reduction in sSFR away from the star forming main sequence towards quiescence. The consensus from the observational literature is that both $f_\mathrm{gas}$ and SFE reduce during quenching, at least in the local Universe. This requires a quenching cause which is capable of both lowering the gas mass of the ISM, as well as reducing the efficiency of star formation out of the remaining gas reservoir. Additionally, we present evidence utilizing stellar metallicity measurements and modeling that all quenching involves a significant phase of starvation (whereby stars continue to be formed out of highly metal enriched/ non diluted ISM gas). Crucially, high-mass quenching may be explained almost exclusively via starvation (e.g., preventative feedback) but low-mass quenching requires both starvation and significant ISM ejection (e.g., as a result of ram pressure or tidal stripping). 

In Sect.~\ref{s7}, we review various approach to timing the process of galaxy quenching, considering both when it occurs and how long it takes. We discuss strong evidence for cosmic downsizing, whereby more massive galaxies form their stars and quench earlier in the history of the Universe than less massive systems. We also discuss the diversity of quenching durations found in the observational literature, which hints at the need for multiple quenching triggers. In the second part of this section, we discuss various efforts to `catch quenching in action', including via the green valley and post starburst galaxies. We discuss the `compaction followed by quenching' scenario, providing evidence for this key route to quenching massive galaxies.


\subsection{Tentative answers to the big questions}\label{s82}

In this section we will state clearly what we believe are the answers to the big questions posed in this review article. It seems abundantly clear that quenching is governed in nature by two completely separate sets of physical processes: (i)~baryonic feedback in centrals and high-mass satellites; and (ii)~environmental processes in low-mass satellites. Any attempt to further elucidate the causes or mechanisms of quenching must first make this distinction, or else risk profound confusion. The clearest evidence of this separation presented within this review article is most probably to be found in Fig.~\ref{f2a}, with strong support from Figs.~\ref{f27} and \ref{f28} (for simulations and observations, respectively).

\subsubsection{Intrinsic quenching of high-mass galaxies}\label{s821}

In order to quench a massive galaxy one must prevent gas flows into the system. The formation of a hot static atmosphere around massive galaxies is an expected consequence of stable virial shocks forming (see Sect.~\ref{s231}), which may shield a high-mass galaxy from cosmological cold gas streams. However, cooling from the hot gas halo is still expected via free\,--\,free and metal line emission, especially as halo mass increases. Consequently, some form of heating is absolutely required to stabilize the hot gas halo around massive galaxies, groups, and clusters, solving the `clusters problem'. The only theoretically proposed mechanism that offers any help here is radio-mode AGN feedback (see Sects.~\ref{s233} and \ref{s25}). As a reminder, in this review radio-mode AGN feedback refers specifically to AGN feedback in the low Eddington ratio regime, likely via relativistic jets, but also potentially via other mechanisms.

Observational evidence for radio-mode AGN feedback is plentiful in clusters and high-mass groups (see Sect.~\ref{s43}). Moreover, the direct consequences of this mode of feedback on the galaxy population is found to be a strong dependence of quenching on SMBH mass (not accretion rate). Remarkably, this prediction is met in observations, via the use of various proxies, across $\sim$13\,Gyr of cosmic history (see Sect.~\ref{s46}). Furthermore, preventative feedback ultimately achieves quenching via starvation, which is consistent with the constraints via stellar metallicities (see Sect.~\ref{s62}). Finally, this mode of quenching naturally explains the observed reduction in gas content of high-mass quiescent systems (see Sect.~\ref{s61}). 

Therefore, radio-mode AGN feedback is highly likely to be the ultimate cause of massive galaxy quenching. Conversely, pure quasar-mode AGN feedback would theoretically lead to excessive rejuvenation of star formation in the late Universe (see Sects.~\ref{s232} and \ref{s24}). Moreover, the observational evidence for AGN-outflows rarely find cases of velocities high enough to fully escape the dark matter halo (see Sect.~\ref{s42}). Whilst this is not a requirement for quenching per se (as the quasar-mode feedback could in principle keep the CGM hot), it is a requirement in practice considering the rarity of full-blown quasars and the necessity to maintain quiescence over many Gyr. Furthermore, the expectations from quasar-mode AGN feedback are in direct contradiction with metallicity-based analyses, which demonstrate that high-mass galaxies quench primarily via starvation, not sudden ejection of their ISM (see Sect.~\ref{s62}). 

Alternatively, pure dynamical stabilization leads to the expectation of the ISM increasing in gas mass during quenching, which is the opposite of that observed (see Sect.~\ref{s61}). Supernova feedback is not effective in simulations at quenching high-mass haloes (see Sect.~\ref{s22}), and the lack of importance of stellar mass in population-based studies for predicting quenching effectively rules this cause out (e.g., Fig.~\ref{f20}). Equally, environmental parameters, including halo mass, are found to be of no value for predicting high-mass galaxy quenching when careful controls for nuisance parameters are applied (see, e.g., Fig.~\ref{f28}).

Nonetheless, there remain some partial inconsistencies with the radio-mode AGN quenching paradigm. For example, many studies find that, in addition to reducing the gas content of the ISM, the SFE is also reduced during quenching (see Sect.~\ref{s61}). Potentially, this may be explained by turbulence injection into the ISM via radio-mode feedback itself (see Fig.~\ref{f30}). However, reduction in SFE may also point towards dynamical stabilization operating within the broader framework of radio-mode starvation (see Sect.~\ref{s26}). Ultimately, these processes may cause star formation to halt earlier than expected with quenching regulated by pure radio-mode AGN feedback. 

Additionally, some cosmological simulations achieve massive galaxy quenching without an explicit radio-mode AGN prescription (notably, EAGLE). Nonetheless, all quenching of high-mass galaxies in EAGLE is achieved via AGN feedback (but here in a single mode, which aims to capture both high and low Eddington ratio feedback). Crucially, however, there is strong evidence for excessive rejuvenation in this simulation (exactly as expected without significant late-time halo heating). Furthermore, the quenched fraction - mass relationships in EAGLE are more discrepant with observations than in, e.g., IllustrisTNG (which does incorporate a mechanical feedback at low accretion rates akin to radio-mode feedback, which turns out to be the primary quenching mode of centrals). Finally, radio-mode AGN feedback by itself may struggle to account for the full diversity in quenching timescales (see Sects.~\ref{s71} and \ref{s72}). Hence, it remains possible that quasar-mode ejective feedback (possibly triggered in mergers) may on occasion rapidly quench massive galaxies. Nevertheless, it is of paramount importance to appreciate that this will not lead to full/ long-term quiescence without the radio-mode acting to stabilize the thermodynamics of the CGM.

Ultimately, it is our view that radio-mode AGN feedback is likely the fundamental cause of all massive galaxy quenching. In lieu of radio-mode feedback, the hot gas halo surrounding massive galaxies will cool and eventually replenish the ISM (no matter if it was previously depleted by a rival feedback mechanism). This will inevitably lead to extensive rejuvenation of star formation, which is simply not seen in local massive spheroids (which present with low levels of star formation for many Gyr). Other mechanisms may aid this quenching route, either by halting star formation in the ISM early (e.g., via dynamical stabilization) or speeding up the process of ISM depletion (e.g., via quasar-mode ejective feedback). Nevertheless, true long-term quenching requires the stabilization of cooling flows in the CGM, and \textit{only} radio-mode feedback offers a plausible theoretical route to achieve this. 

Fortunately, the observational evidence in favor of this underlying cause is now very strong within the high mass haloes of clusters and massive groups, where it is most needed theoretically (see especially Sects.~\ref{s4}, \ref{s5} and \ref{s6}). Nevertheless, it should be acknowledged that within lower mass haloes the evidence for relativistic jets is more limited and, hence, further work is needed to convincingly establish that this mode of AGN feedback can quench the vast majority of galaxies with $M_* \gtrsim 10^{11} \, M_\odot$, as required by observations at low redshifts. Nevertheless, that all mass quenching is ultimately related to SMBH mass, rather than stellar or halo mass, is firmly established by both the theoretical and observational literature reviewed here. At the very least, this points towards AGN in some feedback mode as the true origin of mass-quenching.

\subsubsection{Environmental quenching of low-mass satellites}\label{822}

Low-mass satellite galaxies are observed to quench in much greater numbers than low-mass centrals (see Sect.~\ref{s51}) and, moreover, exhibit no dependence on SMBH mass, or its best known proxies (see Sect.~\ref{s54}). Therefore, AGN feedback (in any mode) cannot be the underlying cause of low-mass satellite quenching. Conversely, low-mass satellite quenching is shown to be accurately predicted by environmental parameters, clearly implying an environmental origin here (see Sect.~\ref{s5}). Unlike massive galaxies, the quenching of low mass systems is found to involve a mix of \textit{both} partial ISM ejection and starvation (see Sect.~\ref{s62}). This suggests that we must look for environmental mechanisms capable of ejecting gas from the ISM, and simultaneously preventing gas re-accretion from the satellite's CGM or broader environment.

Ram pressure stripping is extremely effective at removing the CGM around high-mass satellite galaxies, which is only weakly bound to the system (see Sect.~\ref{s32}). With the CGM removed, cooling flows cannot replenish the ISM in these systems. Furthermore, given that satellites no longer reside at the center of their own dark matter haloes (but rather are in orbit of a group/ cluster central), replenishment of the ISM via cold gas streams is also halted for low-mass satellites. Taken together, essentially all satellite galaxies are expected to slow-quench via starvation. Nonetheless, some intrinsic feedback will still be necessary to offset gas return from stellar evolution (see Sect.~\ref{sec:stellar_quenching}.)

The quenching of satellites may be accelerated via removal of the ISM directly, potentially as a result of high-density ram pressure stripping, or else via dynamical stripping from satellite - central interactions, or satellite - satellite harassment (see Sects.~\ref{s32} and \ref{s33}). While there is abundant observational evidence for all of these environmental processes occurring in nature (see Sects.~\ref{s52} and \ref{s53}), the precise breakdown of the contribution of these processes to low-mass satellite quenching remains elusive. Nevertheless, the contemporary literature favors ram pressure stripping as the dominant mode of quenching in the cluster environment, with dynamical stripping as the principal quenching trigger in the group environment. However, in all cases of low-mass satellite quenching, the ultimate cause of long-term quiescence is a lack of cold gas inflows into the system.

Finally, we end this sub-section with comment on the fundamental link between all quenching types. It is important to appreciate that the CGM around centrals \textit{is} the local environment of satellites. Hence, the formation and stabilization of a hot static atmosphere around high-mass centrals is absolutely crucial for all types of quenching to occur. Initially, this environment is formed by shock heating (of cold gas inflows and via satellite galaxy accretion into the system). Thereafter, it is kept stable via preventative, radio-mode AGN feedback. The hot gas halo provides a shield against cold gas inflows from the IGM, decoupling the environment from direct interaction with the wider Universe. Provided CGM cooling is offset by radio-mode AGN heating, the central galaxy will quench via starvation (as discussed above). 

Furthermore, this hot dense medium provides the potential for ram pressure stripping, as satellites move at high relative velocities through it. Fascinatingly, the connection between radio-mode heating and environmental quenching has been explored in both simulations and observations, with a remarkable consistency found between them (see Sect.~\ref{s36}). Therefore, in the final analysis, whilst the mechanisms of quenching are different between centrals and satellites, the ultimate cause may be one and the same --- the formation and stabilization of the hot gaseous halo. Nonetheless, the key parameter for high-mass galaxy quenching is supermassive black hole mass (which reflects the prevention of cooling flows via historic AGN feedback) and the key parameter for low-mass satellite quenching is the mass of the halo itself, or a tracer of local galaxy density (which reflects the presence of a dense medium and the availability of other systems to interact with tidally).

\subsection{Looking forwards: What do we still not know?}\label{83}

Even if the above conclusions are correct, namely that high-mass quenching is driven by radio-mode AGN feedback and low-mass satellite quenching is driven via a mix of ram pressure and dynamical stripping, there remain many unanswered questions in the field. Here we provide a list of some prominent known unknowns. Of course, what may be even more troubling and fascinating are the unknown unknowns.

\begin{itemize}
    \item \textbf{Jet physics:} How are relativistic jets launched from the inner regions surrounding supermassive black holes? How best should we model radio-mode feedback in cosmological simulations? How exactly does the energy released in jets couple to the ISM and CGM within and surrounding galaxies? Why are jets more common at low Eddington ratios? Do relativistic jets stop forming at some point in the quiescent phase of galaxy evolution? If so, do we expect rejuvenation of massive spheroids eventually? If not, what is the ultimate fate of the vast quantity of baryons contained within the hot gaseous halo surrounding massive galaxies, groups, and clusters?

    \item \textbf{Quenching triggers:} Are mergers, starbursts, gas depletion, quasars, and ejective AGN feedback important for \textit{triggering} massive galaxy quenching? Can these processes explain the diversity of quenching timescales reported in the literature? What is the physics driving the transition from quasar-mode to radio-mode AGN feedback? How best should we model these processes in the next generation of cosmological simulations?

    \item \textbf{The origin of supermassive black holes:} What is the origin of supermassive black holes in the Universe? Do they form from mergers of Pop. III stars, or is direct collapse important? Are there primordial black holes formed at the end of inflation? If so, can these explain the early growth of SMBHs witnessed in JWST observations? Furthermore, if primordial black holes exist, can they explain (even partially) the dark matter?

    \item \textbf{The dark sector:} What is dark matter? What is dark energy? Do the fundamental nature of these components of $\Lambda$CDM have any bearing on structure formation, galaxy evolution, and, ultimately, star formation and quenching? Or, are the equations of state for each component all that is required to model these small-scale processes effectively? Is dark matter self-interacting? If so, what impact does this self interaction have on structure formation and galaxy evolution? For instance, is dark matter annihilation real, and, if so, how does this released energy affect the formation of stars in high mass galaxies? Is dark energy a cosmological constant? If not, how does it evolve, and does it spatially congregate? If so, how does this impact structure formation and ultimately the formation of stars within galaxies?

    \item  \textbf{The Hubble tension:} Is the Hubble tension real? If so, does it point towards new physics, new energy contributions to the Universe, or a more dynamic dark sector? Once these questions are resolved, how does the updated background cosmology impact the formation of structure, galaxies, and ultimately stars in the Universe?

    \item  \textbf{The origin of groups and clusters:} When do the first groups and clusters form in the history of the Universe? When is the transition from denser environments acting as an accelerator of galaxy evolution to a suppressor thereof? More precisely, when do group and cluster environments become hostile to future star formation within their galactic members? How are AGN and quasars linked to cosmic environment? Ultimately, does the environment in which a galaxy evolves determine its end state, or are there important contributions on smaller scales?

    \item \textbf{The details of environmental quenching:} What is the precise breakdown of environmental routes to low-mass satellite quenching? In what mass haloes, and in what locations within these haloes, is ram pressure and dynamical stripping most effective? What fraction of quenched satellites quench previously as centrals, or as satellites of lower mass structures? Overall, how prevalent is environmental quenching compared to intrinsic quenching, as a function of cosmic time?

    \item  \textbf{The future of our Universe:} Given that the vast majority of baryons are not in stars, is quenching the end of the story of galaxy evolution, or merely a phase? What is the ultimate fate of the ICM and CGM around massive galaxies? When will the last star form in the Universe? When will the last star go dark? Can we make reliable predictions about the future of our Universe?

\end{itemize}

Obviously, there is still a great deal for us to do as a scientific community! We hope this review has provided a helpful introduction to a vibrant field, which is striving to answer some of the biggest questions in the Universe. Even more importantly, we hope this review will inspire readers to challenge the conclusions reached here and push this field forwards.

\backmatter

\section*{Declarations}
\bmhead{Competing Interests} The author declares no competing interests.



\phantomsection
\addcontentsline{toc}{section}{References}
\bibliography{Quench_Review_submitted}

@article{Aird2012,
  author = {J. Aird and A. L. Coil and J. Moustakas and M. R. Blanton and others},
  title = {{PRIMUS: the dependence of AGN accretion on host stellar mass and color}},
  journal = {\apj},
  volume = {746},
  number = {1},
  pages = {90},
  year = {2012},
  doi = {10.1088/0004-637X/746/1/90}
}

@article{Baldry2006,
  author = {I. K. Baldry and M. L. Balogh and R. G. Bower and others},
  title = {{Galaxy bimodality versus stellar mass and environment}},
  journal = {\mnras},
  volume = {373},
  number = {2},
  pages = {469},
  year = {2006},
  doi = {10.1111/j.1365-2966.2006.11081.x}
}

@article{Behroozi2010,
  author = {P. S. Behroozi and C. Conroy and R. H. Wechsler},
  title = {{A comprehensive analysis of uncertainties affecting the stellar mass–halo mass relation for $0 < z < 4$}},
  journal = {\apj},
  volume = {717},
  number = {1},
  pages = {379},
  year = {2010},
  doi = {10.1088/0004-637X/717/1/379}
}

@article{Belfiore2017,
  author = {F. Belfiore and R. Maiolino and C. Maraston and others},
  title = {{SDSS-IV MaNGA–the spatially resolved transition from star formation to quiescence}},
  journal = {\mnras},
  volume = {466},
  number = {3},
  pages = {2570},
  year = {2017},
  doi = {10.1093/mnras/stw3211}
}

@article{Belfiore2018,
  author = {F. Belfiore and R. Maiolino and K. Bundy and others},
  title = {{SDSS IV MaNGA–sSFR profiles and the slow quenching of discs in green valley galaxies}},
  journal = {\mnras},
  volume = {477},
  number = {3},
  pages = {3014},
  year = {2018},
  doi = {10.1093/mnras/sty768}
}

@article{Bell2008,
  author = {E. F. Bell},
  title = {{Galaxy bulges and their black holes: a requirement for the quenching of star formation}},
  journal = {\apj},
  volume = {682},
  number = {1},
  pages = {355},
  year = {2008},
  doi = {10.1086/589551}
}

@article{Bell2012,
  author = {E. F. Bell and A. van der Wel and C. Papovich and others},
  title = {{What turns galaxies off? The different morphologies of star-forming and quiescent galaxies since $z \sim 2$ from CANDELS}},
  journal = {\apj},
  volume = {753},
  number = {2},
  pages = {167},
  year = {2012},
  doi = {10.1088/0004-637X/753/2/167}
}

@article{Bluck2014,
  author = {A. F. L. Bluck and J. T. Mendel and S. L. Ellison and others},
  title = {{Bulge mass is king: the dominant role of the bulge in determining the fraction of passive galaxies in the Sloan Digital Sky Survey}},
  journal = {\mnras},
  volume = {441},
  number = {1},
  pages = {599},
  year = {2014},
  doi = {10.1093/mnras/stu594}
}

@article{Bluck2016,
  author = {A. F. L. Bluck and J. T. Mendel and S. L. Ellison and others},
  title = {{The impact of galactic properties and environment on the quenching of central and satellite galaxies: a comparison between SDSS, Illustris and L-Galaxies}},
  journal = {\mnras},
  volume = {462},
  number = {3},
  pages = {2559},
  year = {2016},
  doi = {10.1093/mnras/stw1665}
}

@article{Bluck2020a,
  author = {A. F. L. Bluck and R. Maiolino and S. F. S{\'a}nchez and others},
  title = {{Are galactic star formation and quenching governed by local, global, or environmental phenomena?}},
  journal = {\mnras},
  volume = {492},
  number = {1},
  pages = {96},
  year = {2020},
  doi = {10.1093/mnras/stz3264}
}

@article{Bluck2020b,
  author = {A. F. L. Bluck and R. Maiolino and J. M. Piotrowska and others},
  title = {{How do central and satellite galaxies quench?–Insights from spatially resolved spectroscopy in the MaNGA survey}},
  journal = {\mnras},
  volume = {499},
  number = {1},
  pages = {230},
  year = {2020},
  doi = {10.1093/mnras/staa2806}
}

@article{Bluck2022,
  author = {A. F. L. Bluck and R. Maiolino and S. Brownson and others},
  title = {{The quenching of galaxies, bulges, and disks since cosmic noon–A machine learning approach for identifying causality in astronomical data}},
  journal = {\aap},
  volume = {659},
  pages = {160},
  year = {2022},
  doi = {10.1051/0004-6361/202142643}
}

@article{Bluck2023,
  author = {A. F. L. Bluck and J. M. Piotrowska and R. Maiolino},
  title = {{The Fundamental Signature of Star Formation Quenching from AGN Feedback: A Critical Dependence of Quiescence on Supermassive Black Hole Mass, Not Halo Mass or Environment}},
  journal = {\apj},
  volume = {944},
  number = {2},
  pages = {108},
  year = {2023},
  doi = {10.3847/1538-4357/acac7c}
}

@article{Bluck2024,
  author = {A. F. L. Bluck and C. J. Conselice and K. Ormerod and others},
  title = {{The impact of supermassive black holes on galaxy evolution from cosmic noon to the present day}},
  journal = {\apj},
  volume = {961},
  number = {1},
  pages = {163},
  year = {2024},
  doi = {10.1088/0004-637X/961/1/163}
}

@article{Bower2006,
  author = {R. G. Bower and A. J. Benson and R. Malbon and others},
  title = {{Breaking the hierarchy of galaxy formation}},
  journal = {\mnras},
  volume = {370},
  number = {2},
  pages = {645},
  year = {2006},
  doi = {10.1111/j.1365-2966.2006.10519.x}
}

@article{Bower2008,
  author = {R. G. Bower and I. G. McCarthy and A. J. Benson},
  title = {{The flip side of galaxy formation: a combined model of galaxy formation and cluster heating}},
  journal = {\mnras},
  volume = {390},
  number = {4},
  pages = {1399},
  year = {2008},
  doi = {10.1111/j.1365-2966.2008.13838.x}
}

@article{Brinchmann2004,
  author = {J. Brinchmann and S. Charlot and S. D. M. White and others},
  title = {{The physical properties of star-forming galaxies in the low-redshift Universe}},
  journal = {\mnras},
  volume = {351},
  number = {4},
  pages = {1151},
  year = {2004},
  doi = {10.1111/j.1365-2966.2004.07881.x}
}

@article{Brownson2020,
  author = {S. Brownson and F. Belfiore and R. Maiolino and L. Lin and S. Carniani},
  title = {{Molecular gas and star formation in green valley galaxies: A case study}},
  journal = {MNRAS Lett.},
  volume = {498},
  number = {1},
  pages = {L66},
  year = {2020},
  doi = {10.1093/mnrasl/slaa107}
}

@article{Brownson2022,
  author = {S. Brownson and A. F. L. Bluck and R. Maiolino and G. C. Jones},
  title = {{What drives galaxy quenching? A deep connection between galaxy kinematics and quenching in the local Universe}},
  journal = {\mnras},
  volume = {511},
  number = {2},
  pages = {1913},
  year = {2022},
  doi = {10.1093/mnras/stab3749}
}

@article{Cameron2009a,
  author = {E. Cameron and S. P. Driver},
  title = {{Galaxy evolution by color-log(n) type since redshift unity in the Hubble Ultra Deep Field}},
  journal = {\aap},
  volume = {493},
  number = {2},
  pages = {489},
  year = {2009},
  doi = {10.1051/0004-6361:20078558}
}

@article{Cameron2009b,
  author = {E. Cameron and S. P. Driver and A. W. Graham and J. Liske},
  title = {{The Millennium Galaxy Catalogue: Exploring the color-concentration bimodality via bulge-disk decomposition}},
  journal = {\apj},
  volume = {699},
  number = {1},
  pages = {105},
  year = {2009},
  doi = {10.1088/0004-637X/699/1/105}
}

@article{Cheung2012,
  author = {E. Cheung and S. M. Faber and D. C. Koo and A. A. Dutton and others},
  title = {{The dependence of quenching upon the inner structure of galaxies at $0.5 \leq z < 0.8$ in the DEEP2/AEGIS survey}},
  journal = {\apj},
  volume = {760},
  number = {2},
  pages = {131},
  year = {2012},
  doi = {10.1088/0004-637X/760/2/131}
}

@article{Cicone2012,
  author = {C. Cicone and C. Feruglio and R. Maiolino and F. Fiore and others},
  title = {{The physics and the structure of the quasar-driven outflow in Mrk 231}},
  journal = {\aap},
  volume = {543},
  pages = {A99},
  year = {2012},
  doi = {10.1051/0004-6361/201218793}
}

@article{Cicone2014,
  author = {C. Cicone and R. Maiolino and E. Sturm and J. Graci{\'a}-Carpio and others},
  title = {{Massive molecular outflows and evidence for AGN feedback from CO observations}},
  journal = {\aap},
  volume = {562},
  pages = {A21},
  year = {2014},
  doi = {10.1051/0004-6361/201322464}
}

@article{Cicone2015,
  author = {C. Cicone and R. Maiolino and S. Gallerani and R. Neri and others},
  title = {{Very extended cold gas, star formation and outflows in the halo of a bright quasar at $z > 6$}},
  journal = {\aap},
  volume = {574},
  pages = {A14},
  year = {2015},
  doi = {10.1051/0004-6361/201424980}
}

@article{Crain2015,
  author = {R. A. Crain and J. Schaye and R. G. Bower and others},
  title = {{The EAGLE simulations of galaxy formation: calibration of subgrid physics and model variations}},
  journal = {\mnras},
  volume = {450},
  number = {2},
  pages = {1937},
  year = {2015},
  doi = {10.1093/mnras/stv725}
}

@article{Croton2006,
  author = {D. J. Croton and V. Springel and S. D. M. White and G. De Lucia and others},
  title = {{The many lives of active galactic nuclei: cooling flows, black holes and the luminosities and colours of galaxies}},
  journal = {\mnras},
  volume = {365},
  number = {1},
  pages = {11},
  year = {2006},
  doi = {10.1111/j.1365-2966.2005.09675.x}
}

@article{Dave2019,
  author = {R. Dav{\'e} and D. Angl{\'e}s-Alc{\'a}zar and D. Narayanan and others},
  title = {{Simba: Cosmological simulations with black hole growth and feedback}},
  journal = {\mnras},
  volume = {486},
  number = {2},
  pages = {2827},
  year = {2019},
  doi = {10.1093/mnras/stz937}
}

@article{Dekel2006,
  author = {A. Dekel and Y. Birnboim},
  title = {{Galaxy bimodality due to cold flows and shock heating}},
  journal = {\mnras},
  volume = {368},
  number = {1},
  pages = {2},
  year = {2006},
  doi = {10.1111/j.1365-2966.2006.10145.x}
}

@article{Dekel2014,
  author = {A. Dekel and A. Burkert},
  title = {{Wet disc contraction to galactic blue nuggets and quenching to red nuggets}},
  journal = {\mnras},
  volume = {438},
  number = {2},
  pages = {1870},
  year = {2014},
  doi = {10.1093/mnras/stt2331}
}

@article{Dekel2009,
  author = {A. Dekel and Y. Birnboim and G. Engel and J. Freundlich and T. Goerdt and others},
  title = {{Cold streams in early massive hot haloes as the main mode of galaxy formation}},
  journal = {Nature},
  volume = {457},
  pages = {451},
  year = {2009},
  doi = {10.1038/nature07648}
}

@article{Dressler1980,
  author = {A. Dressler},
  title = {{Galaxy morphology in rich clusters: Implications for the formation and evolution of galaxies}},
  journal = {\apj},
  volume = {236},
  pages = {351},
  year = {1980},
  doi = {10.1086/157753}
}

@article{Driver2006,
  author = {S. P. Driver and P. D. Allen and A. W. Graham and others},
  title = {{The Millennium Galaxy Catalogue: Morphological classification and bimodality in the colour–concentration plane}},
  journal = {\mnras},
  volume = {368},
  number = {1},
  pages = {414},
  year = {2006},
  doi = {10.1111/j.1365-2966.2006.10114.x}
}

@article{Ebeling2014,
  author = {H. Ebeling and L. N. Stephenson and A. C. Edge},
  title = {{Jellyfish: evidence of extreme ram-pressure stripping in massive galaxy clusters}},
  journal = {\apjl},
  volume = {781},
  number = {2},
  pages = {L40},
  year = {2014},
  doi = {10.1088/2041-8205/781/2/L40}
}

@article{Efstathiou1990,
  author = {G. Efstathiou and W. J. Sutherland and S. J. Maddox},
  title = {{The cosmological constant and cold dark matter}},
  journal = {Nature},
  volume = {348},
  pages = {705},
  year = {1990},
  doi = {10.1038/348705a0}
}

@article{Ellison2018,
  author = {S. L. Ellison and S. F. S{\'a}nchez and H. Ibarra-Medel and others},
  title = {{Star formation is boosted (and quenched) from the inside-out: radial star formation profiles from MaNGA}},
  journal = {\mnras},
  volume = {474},
  number = {2},
  pages = {2039},
  year = {2018},
  doi = {10.1093/mnras/stx2882}
}

@article{Fabian1994,
  author = {A. C. Fabian},
  title = {{Cooling flows in clusters of galaxies}},
  journal = {\araa},
  volume = {32},
  pages = {277},
  year = {1994},
  doi = {10.1146/annurev.aa.32.090194.001425}
}

@article{Fabian1999,
  author = {A. C. Fabian},
  title = {{The obscured growth of massive black holes}},
  journal = {\mnras},
  volume = {308},
  number = {4},
  pages = {L39},
  year = {1999},
  doi = {10.1046/j.1365-8711.1999.02941.x}
}

@article{Fabian2012,
  author = {A. C. Fabian},
  title = {{Observational evidence of active galactic nuclei feedback}},
  journal = {\araa},
  volume = {50},
  pages = {455},
  year = {2012},
  doi = {10.1146/annurev-astro-081811-125521}
}

@article{Fabian2006,
  author = {A. C. Fabian and J. S. Sanders and G. B. Taylor and others},
  title = {{A very deep Chandra observation of the Perseus cluster: shocks, ripples and conduction}},
  journal = {\mnras},
  volume = {366},
  number = {2},
  pages = {417},
  year = {2006},
  doi = {10.1111/j.1365-2966.2005.09896.x}
}

@article{Fang2013,
  author = {J. J. Fang and S. M. Faber and D. C. Koo and A. Dekel},
  title = {{A link between star formation quenching and inner stellar mass density in Sloan Digital Sky Survey central galaxies}},
  journal = {\apj},
  volume = {776},
  number = {1},
  pages = {63},
  year = {2013},
  doi = {10.1088/0004-637X/776/1/63}
}

@article{Feruglio2010,
  author = {C. Feruglio and R. Maiolino and E. Piconcelli and N. Menci and others},
  title = {{Quasar feedback revealed by giant molecular outflows}},
  journal = {\aap},
  volume = {518},
  pages = {L155},
  year = {2010},
  doi = {10.1051/0004-6361/201015164}
}

@article{Frenk1988,
  author = {C. S. Frenk and S. D. M. White and M. Davis and G. Efstathiou},
  title = {{The formation of dark halos in a universe dominated by cold dark matter}},
  journal = {\apj},
  volume = {327},
  pages = {507},
  year = {1988},
  doi = {10.1086/166213}
}

@article{Fluetsch2019,
  author = {A. Fluetsch and R. Maiolino and S. Carniani and others},
  title = {{Cold molecular outflows in the local Universe and their feedback effect on galaxies}},
  journal = {\mnras},
  volume = {483},
  number = {4},
  pages = {4586},
  year = {2019},
  doi = {10.1093/mnras/sty3449}
}

@article{Fukugita2004,
  author = {M. Fukugita and P. J. E. Peebles},
  title = {{The cosmic energy inventory}},
  journal = {\apj},
  volume = {616},
  number = {2},
  pages = {643},
  year = {2004},
  doi = {10.1086/425155}
}

@article{GonzalezDelgado2014,
  author = {R. M. Gonzalez Delgado and E. Perez and R. C. Fernandes and others},
  title = {{The star formation history of CALIFA galaxies: Radial structures}},
  journal = {\aap},
  volume = {562},
  pages = {A47},
  year = {2014},
  doi = {10.1051/0004-6361/201322011}
}

@article{GonzalezDelgado2016,
  author = {R. M. Gonzalez Delgado and R. C. Fernandes and E. Perez and others},
  title = {{Star formation along the Hubble sequence: Radial structure of the star formation of CALIFA galaxies}},
  journal = {\aap},
  volume = {590},
  pages = {A44},
  year = {2016},
  doi = {10.1051/0004-6361/201628174}
}

@article{Henriques2013,
  author = {B. M. B. Henriques and S. D. M. White and P. A. Thomas and others},
  title = {{Simulations of the galaxy population constrained by observations from z = 3 to the present day: implications for galactic winds and the fate of their ejecta}},
  journal = {\mnras},
  volume = {431},
  number = {4},
  pages = {3373},
  year = {2013},
  doi = {10.1093/mnras/stt415}
}

@article{Henriques2015,
  author = {B. M. B. Henriques and S. D. M. White and P. A. Thomas and others},
  title = {{Galaxy formation in the Planck cosmology – I. Matching the observed evolution of star formation rates, colours and stellar masses}},
  journal = {\mnras},
  volume = {451},
  number = {3},
  pages = {2663},
  year = {2015},
  doi = {10.1093/mnras/stv705}
}

@article{Henriques2019,
  author = {B. M. B. Henriques and S. D. M. White and S. J. Lilly and others},
  title = {{The origin of the mass scales for maximal star formation efficiency and quenching: the critical role of supernovae}},
  journal = {\mnras},
  volume = {485},
  number = {3},
  pages = {3446},
  year = {2019},
  doi = {10.1093/mnras/stz668}
}

@article{Hickox2009,
  author = {R. C. Hickox and C. Jones and W. R. Forman and others},
  title = {{Host galaxies, clustering, Eddington ratios, and evolution of radio, X-ray, and infrared-selected AGNs}},
  journal = {\apj},
  volume = {696},
  number = {1},
  pages = {891},
  year = {2009},
  doi = {10.1088/0004-637X/696/1/891}
}

@article{Hickox2014,
  author = {R. C. Hickox and J. R. Mullaney and D. M. Alexander and others},
  title = {{Black hole variability and the star formation–active galactic nucleus connection: do all star-forming galaxies host an active galactic nucleus?}},
  journal = {\apj},
  volume = {782},
  number = {1},
  pages = {9},
  year = {2014},
  doi = {10.1088/0004-637X/782/1/9}
}

@article{HlavacekLarrondo2012,
  author = {J. Hlavacek-Larrondo and A. C. Fabian and A. C. Edge},
  title = {{Extreme AGN feedback in the MAssive Cluster Survey: a detailed study of X-ray cavities at $z > 0.3$}},
  journal = {\mnras},
  volume = {421},
  number = {2},
  pages = {1360},
  year = {2012},
  doi = {10.1111/j.1365-2966.2012.20403.x}
}

@article{HlavacekLarrondo2015,
  author = {J. Hlavacek-Larrondo and M. McDonald and B. A. Benson and others},
  title = {{X-Ray Cavities in a Sample of 83 SPT-selected Clusters of Galaxies: Tracing the Evolution of AGN Feedback in Clusters of Galaxies out to z=1.2}},
  journal = {\apj},
  volume = {805},
  number = {1},
  pages = {35},
  year = {2015},
  doi = {10.1088/0004-637X/805/1/35}
}

@article{HlavacekLarrondo2018,
  author = {J. Hlavacek-Larrondo and others},
  title = {{Mystery solved: discovery of extended radio emission in the merging galaxy cluster Abell 2146}},
  journal = {\mnras},
  volume = {475},
  number = {2},
  pages = {2743},
  year = {2018},
  doi = {10.1093/mnras/stx3160}
}

@article{Hopkins2006,
  author = {P. F. Hopkins and L. Hernquist and T. J. Cox and others},
  title = {{A unified, merger-driven model of the origin of starbursts, quasars, the cosmic X-ray background, supermassive black holes, and galaxy spheroids}},
  journal = {\apjs},
  volume = {163},
  number = {1},
  pages = {1},
  year = {2006},
  doi = {10.1086/499298}
}

@article{Hopkins2008,
  author = {P. F. Hopkins and L. Hernquist and T. J. Cox and others},
  title = {{A cosmological framework for the co-evolution of quasars, supermassive black holes, and elliptical galaxies. I. Galaxy mergers and quasar activity}},
  journal = {\apjs},
  volume = {175},
  number = {2},
  pages = {356},
  year = {2008},
  doi = {10.1086/524362}
}

@article{Kennicutt1998,
  author = {R. C. Kennicutt},
  title = {{Star formation in galaxies along the Hubble sequence}},
  journal = {\araa},
  volume = {36},
  pages = {189},
  year = {1998},
  doi = {10.1146/annurev.astro.36.1.189}
}

@article{Lang2014,
  author = {P. Lang and S. Wuyts and R. S. Somerville and others},
  title = {{Bulge Growth and Quenching since z = 2.5 in CANDELS/3D-HST}},
  journal = {\apj},
  volume = {788},
  number = {1},
  pages = {11},
  year = {2014},
  doi = {10.1088/0004-637X/788/1/11}
}

@article{Madau2014,
  author = {P. Madau and M. Dickinson},
  title = {{Cosmic star-formation history}},
  journal = {\araa},
  volume = {52},
  pages = {415},
  year = {2014},
  doi = {10.1146/annurev-astro-081811-125615}
}

@article{Maiolino2012,
  author = {R. Maiolino and S. Gallerani and R. Neri and others},
  title = {{Evidence of strong quasar feedback in the early Universe}},
  journal = {MNRAS Lett.},
  volume = {425},
  number = {1},
  pages = {L66},
  year = {2012},
  doi = {10.1111/j.1745-3933.2012.01306.x}
}

@article{Martig2009,
  author = {M. Martig and F. Bournaud and R. Teyssier and A. Dekel},
  title = {{Morphological quenching of star formation: making early-type galaxies red}},
  journal = {\apj},
  volume = {707},
  number = {1},
  pages = {250},
  year = {2009},
  doi = {10.1088/0004-637X/707/1/250}
}

@article{Moster2010,
  author = {B. P. Moster and R. S. Somerville and C. Maulbetsch and others},
  title = {{Constraints on the relationship between stellar mass and halo mass at low and high redshift}},
  journal = {\apj},
  volume = {710},
  number = {2},
  pages = {903},
  year = {2010},
  doi = {10.1088/0004-637X/710/2/903}
}

@article{Moster2013,
  author = {B. P. Moster and T. Naab and S. D. M. White},
  title = {{Galactic star formation and accretion histories from matching galaxies to dark matter haloes}},
  journal = {\mnras},
  volume = {428},
  number = {4},
  pages = {3121},
  year = {2013},
  doi = {10.1093/mnras/sts261}
}

@article{Nandra2007,
  author = {K. Nandra and others},
  title = {{The AEGIS Survey. X-ray Luminosity Functions from the Extended Groth Strip Deep X-ray Survey}},
  journal = {\apj},
  volume = {660},
  number = {2},
  pages = {L11},
  year = {2007},
  doi = {10.1086/517858}
}

@article{Nelson2018,
  author = {D. Nelson and V. Springel and R. Pakmor and others},
  title = {{The IllustrisTNG Simulations: Public Data Release}},
  journal = {\mnras},
  volume = {475},
  number = {1},
  pages = {624},
  year = {2018},
  doi = {10.1093/mnras/stx3040}
}

@article{Nelson2019,
  author = {D. Nelson and V. Springel and A. Pillepich and others},
  title = {{The Next Generation of Cosmological Hydrodynamical Simulations}},
  journal = {Computational Astrophysics and Cosmology},
  volume = {6},
  pages = {2},
  year = {2019},
  doi = {10.1186/s40668-019-0028-x}
}

@article{Peng2010,
  author = {Y.-J. Peng and S. J. Lilly and K. Kovac and others},
  title = {{Mass and Environment as Drivers of Galaxy Evolution in SDSS and zCOSMOS and the Origin of the Schechter Function}},
  journal = {\apj},
  volume = {721},
  number = {1},
  pages = {193},
  year = {2010},
  doi = {10.1088/0004-637X/721/1/193}
}

@article{Peng2012,
  author = {Y.-J. Peng and S. J. Lilly and A. Renzini and M. Carollo},
  title = {{A Mass-dependent Quenching Model for Star Formation Histories of Galaxies}},
  journal = {\apj},
  volume = {757},
  number = {1},
  pages = {4},
  year = {2012},
  doi = {10.1088/0004-637X/757/1/4}
}

@article{Pillepich2018,
  author = {A. Pillepich and D. Nelson and L. Hernquist and others},
  title = {{First results from the IllustrisTNG simulations: the stellar mass content of groups and clusters of galaxies}},
  journal = {\mnras},
  volume = {475},
  number = {1},
  pages = {648},
  year = {2018},
  doi = {10.1093/mnras/stx3112}
}

@article{Piotrowska2020,
  author = {J. M. Piotrowska and A. F. L. Bluck and R. Maiolino and A. Concas and Y. Peng},
  title = {{Towards a deeper understanding of the physics driving galaxy quenching–inferring trends in the gas content via extinction}},
  journal = {MNRAS Lett.},
  volume = {492},
  number = {1},
  pages = {L6},
  year = {2020},
  doi = {10.1093/mnrasl/slz167}
}

@article{Piotrowska2022,
  author = {J. M. Piotrowska and A. F. L. Bluck and R. Maiolino and Y. Peng},
  title = {{On the quenching of star formation in observed and simulated central galaxies: evidence for the role of integrated AGN feedback}},
  journal = {\mnras},
  volume = {512},
  number = {1},
  pages = {1052},
  year = {2022},
  doi = {10.1093/mnras/stab3673}
}

@article{Poggianti2017,
  author = {B. M. Poggianti and A. Moretti and M. Gullieuszik and others},
  title = {{GASP. I. Gas stripping phenomena in galaxies with MUSE}},
  journal = {\apj},
  volume = {844},
  number = {1},
  pages = {48},
  year = {2017},
  doi = {10.3847/1538-4357/aa78ed}
}

@article{Rosario2013,
  author = {D. J. Rosario and P. Santini and D. Lutz and H. Netzer and others},
  title = {{Nuclear activity is more prevalent in star-forming galaxies}},
  journal = {\apj},
  volume = {771},
  number = {1},
  pages = {63},
  year = {2013},
  doi = {10.1088/0004-637X/771/1/63}
}

@article{Saintonge2016,
  author = {A. Saintonge and B. Catinella and L. Cortese and others},
  title = {{Molecular and atomic gas along and across the main sequence of star-forming galaxies}},
  journal = {\mnras},
  volume = {462},
  number = {2},
  pages = {1749},
  year = {2016},
  doi = {10.1093/mnras/stw1715}
}

@article{Saintonge2017,
  author = {A. Saintonge and B. Catinella and L. J. Tacconi and others},
  title = {{xCOLD GASS: the complete IRAM 30 m legacy survey of molecular gas for galaxy evolution studies}},
  journal = {\apjs},
  volume = {233},
  number = {2},
  pages = {22},
  year = {2017},
  doi = {10.3847/1538-4365/aa97e0}
}

@article{Schaye2015,
  author = {J. Schaye and R. A. Crain and R. G. Bower and others},
  title = {{The EAGLE Project: Simulating the Evolution and Assembly of Galaxies and Their Environments}},
  journal = {\mnras},
  volume = {446},
  number = {1},
  pages = {521},
  year = {2015},
  doi = {10.1093/mnras/stu2058}
}

@article{Shull2012,
  author = {J. M. Shull and B. D. Smith and C. W. Danforth},
  title = {{The Baryon Census in a Multiphase Intergalactic Medium: 30\% of the Baryons May Still Be Missing}},
  journal = {\apj},
  volume = {759},
  number = {1},
  pages = {23},
  year = {2012},
  doi = {10.1088/0004-637X/759/1/23}
}

@article{Sijacki2007,
  author = {D. Sijacki and V. Springel and T. Di Matteo and L. Hernquist},
  title = {{A unified model for AGN feedback in cosmological simulations of structure formation}},
  journal = {\mnras},
  volume = {380},
  number = {3},
  pages = {877},
  year = {2007},
  doi = {10.1111/j.1365-2966.2007.12153.x}
}

@article{Somerville2015,
  author = {R. S. Somerville and R. Dav{\'e}},
  title = {{Physical Models of Galaxy Formation in a Cosmological Framework}},
  journal = {\araa},
  volume = {53},
  pages = {51},
  year = {2015},
  doi = {10.1146/annurev-astro-082812-140951}
}

@article{Strateva2001,
  author = {I. Strateva and {\v{Z}}. {Ivezi{\'c}} and G. R. Knapp and others},
  title = {{Color separation of galaxy types in the Sloan Digital Sky Survey imaging data}},
  journal = {\aj},
  volume = {122},
  number = {4},
  pages = {1861},
  year = {2001},
  doi = {10.1086/323301}
}

@article{Tacchella2015,
  author = {S. Tacchella and C. M. Carollo and A. Renzini and others},
  title = {{Evidence for mature bulges and an inside-out quenching phase 3 billion years after the Big Bang}},
  journal = {Science},
  volume = {348},
  number = {6232},
  pages = {314},
  year = {2015},
  doi = {10.1126/science.1261094}
}

@article{Tacchella2016a,
  author = {S. Tacchella and A. Dekel and C. M. Carollo and others},
  title = {{Evolution of density profiles in high-z galaxies: compaction and quenching inside-out}},
  journal = {\mnras},
  volume = {458},
  number = {1},
  pages = {242},
  year = {2016},
  doi = {10.1093/mnras/stw303}
}

@article{Tacchella2016b,
  author = {S. Tacchella and A. Dekel and C. M. Carollo and others},
  title = {{The confinement of star-forming galaxies into a main sequence through episodes of gas compaction, depletion and replenishment}},
  journal = {\mnras},
  volume = {457},
  number = {3},
  pages = {2790},
  year = {2016},
  doi = {10.1093/mnras/stw131}
}

@article{Terrazas2016,
  author = {B. A. Terrazas and E. F. Bell and B. M. B. Henriques and others},
  title = {{Quiescence correlates strongly with directly measured black hole mass in central galaxies}},
  journal = {\apjl},
  volume = {830},
  number = {1},
  pages = {L12},
  year = {2016},
  doi = {10.3847/2041-8205/830/1/L12}
}

@article{Terrazas2017,
  author = {B. A. Terrazas and E. F. Bell and J. Woo and B. M. B. Henriques},
  title = {{Supermassive black holes as the regulators of star formation in central galaxies}},
  journal = {\apj},
  volume = {844},
  number = {2},
  pages = {170},
  year = {2017},
  doi = {10.3847/1538-4357/aa7d07}
}

@article{Terrazas2020,
  author = {B. A. Terrazas and E. F. Bell and A. Pillepich and others},
  title = {{The relationship between black hole mass and galaxy properties: examining the black hole feedback model in IllustrisTNG}},
  journal = {\mnras},
  volume = {493},
  number = {2},
  pages = {1888},
  year = {2020},
  doi = {10.1093/mnras/staa374}
}

@article{vandenBosch2008,
  author = {F. C. van den Bosch and D. Aquino and X. Yang and others},
  title = {{The importance of satellite quenching for the build-up of the red sequence of present-day galaxies}},
  journal = {\mnras},
  volume = {387},
  number = {1},
  pages = {79},
  year = {2008},
  doi = {10.1111/j.1365-2966.2008.13230.x}
}

@article{Vogelsberger2014a,
  author = {M. Vogelsberger and S. Genel and V. Springel and P. Torrey and others},
  title = {{Properties of galaxies reproduced by a hydrodynamic simulation}},
  journal = {Nature},
  volume = {509},
  pages = {177},
  year = {2014},
  doi = {10.1038/nature13316}
}

@article{Vogelsberger2014b,
  author = {M. Vogelsberger and S. Genel and V. Springel and others},
  title = {{Introducing the Illustris Project: simulating the coevolution of dark and visible matter in the Universe}},
  journal = {\mnras},
  volume = {444},
  number = {2},
  pages = {1518},
  year = {2014},
  doi = {10.1093/mnras/stu1536}
}

@article{Ward2022,
  author = {S. R. Ward and C. M. Harrison and T. Costa and V. Mainieri},
  title = {{Cosmological simulations predict that AGN preferentially live in gas-rich, star-forming galaxies despite effective feedback}},
  journal = {\mnras},
  volume = {513},
  number = {2},
  pages = {2936},
  year = {2022},
  doi = {10.1093/mnras/stac901}
}

@article{Wake2012,
  author = {D. A. Wake and P. G. van Dokkum and M. Franx},
  title = {{Revealing Velocity Dispersion as the Best Indicator of a Galaxy's Color, Compared to Stellar Mass, Surface Mass Density, or Morphology}},
  journal = {\apjl},
  volume = {751},
  number = {2},
  pages = {L44},
  year = {2012},
  doi = {10.1088/2041-8205/751/2/L44}
}

@article{Weinberger2017,
  author = {R. Weinberger and V. Springel and L. Hernquist and others},
  title = {{The IllustrisTNG Simulations: AGN Feedback as a Driver of Galaxy Evolution}},
  journal = {\mnras},
  volume = {465},
  number = {1},
  pages = {329},
  year = {2017},
  doi = {10.1093/mnras/stx1187}
}

@article{Weinberger2018,
  author = {R. Weinberger and V. Springel and R. Pakmor and others},
  title = {{Supermassive black holes and their feedback effects in the IllustrisTNG simulation}},
  journal = {\mnras},
  volume = {479},
  number = {3},
  pages = {4056},
  year = {2018},
  doi = {10.1093/mnras/sty1733}
}

@article{Wild2009,
  author = {V. Wild and S. J. Charlot and R. Kauffmann},
  title = {{Understanding the physics of star formation quenching in the green valley}},
  journal = {\mnras},
  volume = {395},
  pages = {144},
  year = {2009},
  doi = {10.1111/j.1365-2966.2009.14668.x}
}

@article{Wild2010,
  author = {V. Wild and T. Heckman and S. Charlot},
  title = {{Timing the starburst–AGN connection}},
  journal = {\mnras},
  volume = {405},
  number = {2},
  pages = {933},
  year = {2010},
  doi = {10.1111/j.1365-2966.2010.16536.x}
}

@article{Wild2016,
  author = {V. Wild and O. Almaini and J. Dunlop and C. Simpson},
  title = {{The evolution of post-starburst galaxies from z = 2 to 0.5}},
  journal = {\mnras},
  volume = {463},
  number = {1},
  pages = {832},
  year = {2016},
  doi = {10.1093/mnras/stw1996}
}

@article{Woo2013,
  author = {J. Woo and A. Dekel and S. M. Faber and K. Noeske},
  title = {{Dependence of galaxy quenching on halo mass and distance from its centre}},
  journal = {\mnras},
  volume = {428},
  number = {4},
  pages = {3306},
  year = {2013},
  doi = {10.1093/mnras/sts274}
}

@article{Woo2015,
  author = {J. Woo and A. Dekel and S. M. Faber and D. C. Koo},
  title = {{Two conditions for galaxy quenching: compact centres and massive haloes}},
  journal = {\mnras},
  volume = {448},
  number = {1},
  pages = {237},
  year = {2015},
  doi = {10.1093/mnras/stu2755}
}

@article{Yang2009,
  author = {X. Yang and H. J. Mo and F. C. van den Bosch},
  title = {{Galaxy Groups in the SDSS DR4: III. The Luminosity and Stellar Mass Functions}},
  journal = {\apj},
  volume = {695},
  number = {2},
  pages = {900},
  year = {2009},
  doi = {10.1088/0004-637X/695/2/900}
}

@article{Zinger2020,
  author = {E. Zinger and A. Pillepich and D. Nelson and others},
  title = {{Ejective and preventative: the IllustrisTNG black hole feedback and its effects on the thermodynamics of the gas within and around galaxies}},
  journal = {\mnras},
  volume = {499},
  number = {1},
  pages = {768},
  year = {2020},
  doi = {10.1093/mnras/staa2607}
}

@BOOK{Mo2010,
       author = {{Mo}, Houjun and {van den Bosch}, Frank C. and {White}, Simon},
        title = "{Galaxy Formation and Evolution}",
      address = {Cambridge, UK},
    publisher = {Cambridge University Press},
         year = 2010,
       adsurl = {https://ui.adsabs.harvard.edu/abs/2010gfe..book.....M}
}

@ARTICLE{Springel2010,
       author = {{Springel}, Volker},
        title = "{E pur si muove: Galilean-invariant cosmological hydrodynamical simulations on a moving mesh}",
      journal = {\mnras},
         year = 2010,
        month = jan,
       volume = {401},
       number = {2},
        pages = {791-851},
          doi = {10.1111/j.1365-2966.2009.15715.x},
archivePrefix = {arXiv},
       eprint = {0901.4107},
 primaryClass = {astro-ph.CO},
       adsurl = {https://ui.adsabs.harvard.edu/abs/2010MNRAS.401..791S}
}

@ARTICLE{Springel2018,
       author = {{Springel}, Volker and {Pakmor}, R{\"u}diger and {Pillepich}, Annalisa and {Weinberger}, Rainer and {Nelson}, Dylan and {Hernquist}, Lars and {Vogelsberger}, Mark and {Genel}, Shy and {Torrey}, Paul and {Marinacci}, Federico and {Naiman}, Jill},
        title = "{First results from the IllustrisTNG simulations: matter and galaxy clustering}",
      journal = {\mnras},
         year = 2018,
        month = mar,
       volume = {475},
       number = {1},
        pages = {676-698},
          doi = {10.1093/mnras/stx3304},
archivePrefix = {arXiv},
       eprint = {1707.03397},
 primaryClass = {astro-ph.GA},
       adsurl = {https://ui.adsabs.harvard.edu/abs/2018MNRAS.475..676S}
}

@ARTICLE{Springel2005c,
       author = {{Springel}, Volker and {Di Matteo}, Tiziana and {Hernquist}, Lars},
        title = "{Black Holes in Galaxy Mergers: The Formation of Red Elliptical Galaxies}",
      journal = {\apjl},
         year = 2005,
        month = feb,
       volume = {620},
       number = {2},
        pages = {L79-L82},
          doi = {10.1086/428772},
archivePrefix = {arXiv},
       eprint = {astro-ph/0409436},
 primaryClass = {astro-ph},
       adsurl = {https://ui.adsabs.harvard.edu/abs/2005ApJ...620L..79S}
}

@BOOK{Hobson2006,
       author = {{Hobson}, M.~P. and {Efstathiou}, G.~P. and {Lasenby}, A.~N.},
        title = "{General Relativity}",
         year = 2006,
          doi = {10.2277/0521829518},
       adsurl = {https://ui.adsabs.harvard.edu/abs/2006gere.book.....H}
}

@BOOK{Longair2006,
       author = {{Longair}, Malcolm S.},
        title = "{Galaxy Formation}",
         year = 2008,
       adsurl = {https://ui.adsabs.harvard.edu/abs/2008gafo.book.....L}
}

@book{Peacock1999,
  author    = {John A. Peacock},
  title     = {Cosmological Physics},
  year      = {1999},
  publisher = {Cambridge University Press},
  address   = {Cambridge, UK},
  doi       = {10.1017/CBO9780511804533}
}

@ARTICLE{Bell2003,
       author = {{Bell}, Eric F. and {McIntosh}, Daniel H. and {Katz}, Neal and {Weinberg}, Martin D.},
        title = "{The Optical and Near-Infrared Properties of Galaxies. I. Luminosity and Stellar Mass Functions}",
      journal = {\apjs},
         year = 2003,
        month = dec,
       volume = {149},
       number = {2},
        pages = {289-312},
          doi = {10.1086/378847},
archivePrefix = {arXiv},
       eprint = {astro-ph/0302543},
 primaryClass = {astro-ph},
       adsurl = {https://ui.adsabs.harvard.edu/abs/2003ApJS..149..289B}
}

@ARTICLE{Muzzin2013,
       author = {{Muzzin}, Adam and {Marchesini}, Danilo and {Stefanon}, Mauro and {Franx}, Marijn and {McCracken}, Henry J. and {Milvang-Jensen}, Bo and {Dunlop}, James S. and {Fynbo}, J.~P.~U. and {Brammer}, Gabriel and {Labb{\'e}}, Ivo and {van Dokkum}, Pieter G.},
        title = "{The Evolution of the Stellar Mass Functions of Star-forming and Quiescent Galaxies to z = 4 from the COSMOS/UltraVISTA Survey}",
      journal = {\apj},
         year = 2013,
        month = nov,
       volume = {777},
       number = {1},
          eid = {18},
        pages = {18},
          doi = {10.1088/0004-637X/777/1/18},
archivePrefix = {arXiv},
       eprint = {1303.4409},
 primaryClass = {astro-ph.CO},
       adsurl = {https://ui.adsabs.harvard.edu/abs/2013ApJ...777...18M}
}

@ARTICLE{Baldry2012,
       author = {{Baldry}, I.~K. and {Driver}, S.~P. and {Loveday} and J. and {Taylor}, E.~N. and {Kelvin}, L.~S. and {Liske} and J. and {Norberg} and P. and {Robotham}, A.~S.~G. and {Brough} and S. and {Hopkins}, A.~M. and {Bamford}, S.~P. and {Peacock}, J.~A. and {Bland-Hawthorn} and J. and {Conselice}, C.~J. and {Croom}, S.~M. and {Jones}, D.~H. and {Parkinson}, H.~R. and {Popescu}, C.~C. and {Prescott} and M. and {Sharp}, R.~G. and {Tuffs}, R.~J.},
        title = "{Galaxy And Mass Assembly (GAMA): the galaxy stellar mass function at $z < 0.06$}",
      journal = {\mnras},
         year = 2012,
        month = mar,
       volume = {421},
       number = {1},
        pages = {621-634},
          doi = {10.1111/j.1365-2966.2012.20340.x},
archivePrefix = {arXiv},
       eprint = {1111.5707},
 primaryClass = {astro-ph.CO},
       adsurl = {https://ui.adsabs.harvard.edu/abs/2012MNRAS.421..621B}
}

@ARTICLE{Cole2000,
       author = {{Cole}, Shaun and {Lacey}, Cedric G. and {Baugh}, Carlton M. and {Frenk}, Carlos S.},
        title = "{Hierarchical galaxy formation}",
      journal = {\mnras},
         year = 2000,
        month = nov,
       volume = {319},
       number = {1},
        pages = {168-204},
          doi = {10.1046/j.1365-8711.2000.03879.x},
archivePrefix = {arXiv},
       eprint = {astro-ph/0007281},
 primaryClass = {astro-ph},
       adsurl = {https://ui.adsabs.harvard.edu/abs/2000MNRAS.319..168C}
}

@ARTICLE{Schawinski2014,
       author = {{Schawinski}, Kevin and {Urry} and C. Megan and {Simmons}, Brooke D. and {Fortson}, Lucy and {Kaviraj}, Sugata and {Keel}, William C. and {Lintott}, Chris J. and {Masters}, Karen L. and {Nichol}, Robert C. and {Sarzi}, Marc and {Skibba}, Ramin and {Treister}, Ezequiel and {Willett}, Kyle W. and {Wong} and O. Ivy and {Yi}, Sukyoung K.},
        title = "{The green valley is a red herring: Galaxy Zoo reveals two evolutionary pathways towards quenching of star formation in early- and late-type galaxies}",
      journal = {\mnras},
         year = 2014,
        month = may,
       volume = {440},
       number = {1},
        pages = {889-907},
          doi = {10.1093/mnras/stu327},
archivePrefix = {arXiv},
       eprint = {1402.4814},
 primaryClass = {astro-ph.GA},
       adsurl = {https://ui.adsabs.harvard.edu/abs/2014MNRAS.440..889S}
}

@ARTICLE{Bruzual2003,
       author = {{Bruzual} and G. and {Charlot}, S.},
        title = "{Stellar population synthesis at the resolution of 2003}",
      journal = {\mnras},
         year = 2003,
        month = oct,
       volume = {344},
       number = {4},
        pages = {1000-1028},
          doi = {10.1046/j.1365-8711.2003.06897.x},
archivePrefix = {arXiv},
       eprint = {astro-ph/0309134},
 primaryClass = {astro-ph},
       adsurl = {https://ui.adsabs.harvard.edu/abs/2003MNRAS.344.1000B}
}

@ARTICLE{Toomre1972,
       author = {{Toomre}, Alar and {Toomre}, Juri},
        title = "{Galactic Bridges and Tails}",
      journal = {\apj},
         year = 1972,
        month = dec,
       volume = {178},
        pages = {623-666},
          doi = {10.1086/151823},
       adsurl = {https://ui.adsabs.harvard.edu/abs/1972ApJ...178..623T}
}

@ARTICLE{Gensior2020,
       author = {{Gensior}, Jindra and {Kruijssen}, J.~M. Diederik and {Keller}, Benjamin W.},
        title = "{Heart of darkness: the influence of galactic dynamics on quenching star formation in galaxy spheroids}",
      journal = {\mnras},
         year = 2020,
        month = jun,
       volume = {495},
       number = {1},
        pages = {199-223},
          doi = {10.1093/mnras/staa1184},
archivePrefix = {arXiv},
       eprint = {2002.01484},
 primaryClass = {astro-ph.GA},
       adsurl = {https://ui.adsabs.harvard.edu/abs/2020MNRAS.495..199G}
}

@ARTICLE{York2000,
       author = {{York}, Donald G. and {Adelman} and J. and {Anderson}, Jr., John E. and {Anderson}, Scott F. and {Annis}, James and {Bahcall}, Neta A. and {Bakken}, J.~A. and {Barkhouser}, Robert and {Bastian}, Steven and {Berman}, Eileen and {Boroski}, William N. and {Bracker}, Steve and {Briegel}, Charlie and {Briggs}, John W. and {Brinkmann} and J. and {Brunner}, Robert and {Burles}, Scott and {Carey}, Larry and {Carr}, Michael A. and {Castander}, Francisco J. and {Chen}, Bing and {Colestock}, Patrick L. and {Connolly}, A.~J. and {Crocker}, J.~H. and {Csabai}, Istv{\'a}n and {Czarapata}, Paul C. and {Davis}, John Eric and {Doi}, Mamoru and {Dombeck}, Tom and {Eisenstein}, Daniel and {Ellman}, Nancy and {Elms}, Brian R. and {Evans}, Michael L. and {Fan}, Xiaohui and {Federwitz}, Glenn R. and {Fiscelli}, Larry and {Friedman}, Scott and {Frieman}, Joshua A. and {Fukugita}, Masataka and {Gillespie}, Bruce and {Gunn}, James E. and {Gurbani}, Vijay K. and {de Haas}, Ernst and {Haldeman}, Merle and {Harris}, Frederick H. and {Hayes} and J. and {Heckman}, Timothy M. and {Hennessy}, G.~S. and {Hindsley}, Robert B. and {Holm}, Scott and {Holmgren}, Donald J. and {Huang}, Chi-hao and {Hull}, Charles and {Husby}, Don and {Ichikawa}, Shin-Ichi and {Ichikawa}, Takashi and {Ivezi{\'c}}, {\v{Z}}eljko and {Kent}, Stephen and {Kim}, Rita S.~J. and {Kinney} and E. and {Klaene}, Mark and {Kleinman}, A.~N. and {Kleinman} and S. and {Knapp}, G.~R. and {Korienek}, John and {Kron}, Richard G. and {Kunszt}, Peter Z. and {Lamb}, D.~Q. and {Lee} and B. and {Leger} and R. French and {Limmongkol}, Siriluk and {Lindenmeyer}, Carl and {Long}, Daniel C. and {Loomis}, Craig and {Loveday}, Jon and {Lucinio}, Rich and {Lupton}, Robert H. and {MacKinnon}, Bryan and {Mannery}, Edward J. and {Mantsch}, P.~M. and {Margon}, Bruce and {McGehee}, Peregrine and {McKay}, Timothy A. and {Meiksin}, Avery and {Merelli}, Aronne and {Monet}, David G. and {Munn}, Jeffrey A. and {Narayanan}, Vijay K. and {Nash}, Thomas and {Neilsen}, Eric and {Neswold}, Rich and {Newberg}, Heidi Jo and {Nichol}, R.~C. and {Nicinski}, Tom and {Nonino}, Mario and {Okada}, Norio and {Okamura}, Sadanori and {Ostriker}, Jeremiah P. and {Owen}, Russell and {Pauls} and A. George and {Peoples}, John and {Peterson}, R.~L. and {Petravick}, Donald and {Pier}, Jeffrey R. and {Pope}, Adrian and {Pordes}, Ruth and {Prosapio}, Angela and {Rechenmacher}, Ron and {Quinn}, Thomas R. and {Richards}, Gordon T. and {Richmond}, Michael W. and {Rivetta}, Claudio H. and {Rockosi}, Constance M. and {Ruthmansdorfer}, Kurt and {Sandford}, Dale and {Schlegel}, David J. and {Schneider}, Donald P. and {Sekiguchi}, Maki and {Sergey}, Gary and {Shimasaku}, Kazuhiro and {Siegmund}, Walter A. and {Smee}, Stephen and {Smith} and J. Allyn and {Snedden} and S. and {Stone} and R. and {Stoughton}, Chris and {Strauss}, Michael A. and {Stubbs}, Christopher and {SubbaRao}, Mark and {Szalay}, Alexander S. and {Szapudi}, Istvan and {Szokoly}, Gyula P. and {Thakar}, Anirudda R. and {Tremonti}, Christy and {Tucker}, Douglas L. and {Uomoto}, Alan and {Vanden Berk}, Dan and {Vogeley}, Michael S. and {Waddell}, Patrick and {Wang}, Shu-i. and {Watanabe}, Masaru and {Weinberg}, David H. and {Yanny}, Brian and {Yasuda}, Naoki and {SDSS Collaboration}},
        title = "{The Sloan Digital Sky Survey: Technical Summary}",
      journal = {\aj},
         year = 2000,
        month = sep,
       volume = {120},
       number = {3},
        pages = {1579-1587},
          doi = {10.1086/301513},
archivePrefix = {arXiv},
       eprint = {astro-ph/0006396},
 primaryClass = {astro-ph},
       adsurl = {https://ui.adsabs.harvard.edu/abs/2000AJ....120.1579Y}
}

@ARTICLE{Abazajian2009,
       author = {{Abazajian}, Kevork N. and {Adelman-McCarthy}, Jennifer K. and {Ag{\"u}eros}, Marcel A. and {Allam}, Sahar S. and {Allende Prieto}, Carlos and {An}, Deokkeun and {Anderson}, Kurt S.~J. and {Anderson}, Scott F. and {Annis}, James and {Bahcall}, Neta A. and {Bailer-Jones}, C.~A.~L. and {Barentine}, J.~C. and {Bassett}, Bruce A. and {Becker}, Andrew C. and {Beers}, Timothy C. and {Bell}, Eric F. and {Belokurov}, Vasily and {Berlind}, Andreas A. and {Berman}, Eileen F. and {Bernardi}, Mariangela and {Bickerton}, Steven J. and {Bizyaev}, Dmitry and {Blakeslee}, John P. and {Blanton}, Michael R. and {Bochanski}, John J. and {Boroski}, William N. and {Brewington}, Howard J. and {Brinchmann}, Jarle and {Brinkmann} and J. and {Brunner}, Robert J. and {Budav{\'a}ri}, Tam{\'a}s and {Carey}, Larry N. and {Carliles}, Samuel and {Carr}, Michael A. and {Castander}, Francisco J. and {Cinabro}, David and {Connolly}, A.~J. and {Csabai}, Istv{\'a}n and {Cunha}, Carlos E. and {Czarapata}, Paul C. and {Davenport}, James R.~A. and {de Haas}, Ernst and {Dilday}, Ben and {Doi}, Mamoru and {Eisenstein}, Daniel J. and {Evans}, Michael L. and {Evans}, N.~W. and {Fan}, Xiaohui and {Friedman}, Scott D. and {Frieman}, Joshua A. and {Fukugita}, Masataka and {G{\"a}nsicke}, Boris T. and {Gates}, Evalyn and {Gillespie}, Bruce and {Gilmore} and G. and {Gonzalez}, Belinda and {Gonzalez}, Carlos F. and {Grebel}, Eva K. and {Gunn}, James E. and {Gy{\"o}ry}, Zsuzsanna and {Hall}, Patrick B. and {Harding}, Paul and {Harris}, Frederick H. and {Harvanek}, Michael and {Hawley}, Suzanne L. and {Hayes}, Jeffrey J.~E. and {Heckman}, Timothy M. and {Hendry}, John S. and {Hennessy}, Gregory S. and {Hindsley}, Robert B. and {Hoblitt} and J. and {Hogan}, Craig J. and {Hogg}, David W. and {Holtzman}, Jon A. and {Hyde}, Joseph B. and {Ichikawa}, Shin-ichi and {Ichikawa}, Takashi and {Im}, Myungshin and {Ivezi{\'c}}, {\v{Z}}eljko and {Jester}, Sebastian and {Jiang}, Linhua and {Johnson}, Jennifer A. and {Jorgensen}, Anders M. and {Juri{\'c}}, Mario and {Kent}, Stephen M. and {Kessler} and R. and {Kleinman}, S.~J. and {Knapp}, G.~R. and {Konishi}, Kohki and {Kron}, Richard G. and {Krzesinski}, Jurek and {Kuropatkin}, Nikolay and {Lampeitl}, Hubert and {Lebedeva}, Svetlana and {Lee}, Myung Gyoon and {Lee}, Young Sun and {French Leger} and R. and {L{\'e}pine}, S{\'e}bastien and {Li}, Nolan and {Lima}, Marcos and {Lin}, Huan and {Long}, Daniel C. and {Loomis}, Craig P. and {Loveday}, Jon and {Lupton}, Robert H. and {Magnier}, Eugene and {Malanushenko}, Olena and {Malanushenko}, Viktor and {Mandelbaum}, Rachel and {Margon}, Bruce and {Marriner}, John P. and {Mart{\'\i}nez-Delgado}, David and {Matsubara}, Takahiko and {McGehee}, Peregrine M. and {McKay}, Timothy A. and {Meiksin}, Avery and {Morrison}, Heather L. and {Mullally}, Fergal and {Munn}, Jeffrey A. and {Murphy}, Tara and {Nash}, Thomas and {Nebot}, Ada and {Neilsen}, Jr., Eric H. and {Newberg}, Heidi Jo and {Newman}, Peter R. and {Nichol}, Robert C. and {Nicinski}, Tom and {Nieto-Santisteban}, Maria and {Nitta}, Atsuko and {Okamura}, Sadanori and {Oravetz}, Daniel J. and {Ostriker}, Jeremiah P. and {Owen}, Russell and {Padmanabhan}, Nikhil and {Pan}, Kaike and {Park}, Changbom and {Pauls}, George and {Peoples}, Jr., John and {Percival}, Will J. and {Pier}, Jeffrey R. and {Pope}, Adrian C. and {Pourbaix}, Dimitri and {Price}, Paul A. and {Purger}, Norbert and {Quinn}, Thomas and {Raddick} and M. Jordan and {Re Fiorentin}, Paola and {Richards}, Gordon T. and {Richmond}, Michael W. and {Riess}, Adam G. and {Rix}, Hans-Walter and {Rockosi}, Constance M. and {Sako}, Masao and {Schlegel}, David J. and {Schneider}, Donald P. and {Scholz}, Ralf-Dieter and {Schreiber}, Matthias R. and {Schwope}, Axel D. and {Seljak}, Uro{\v{s}} and {Sesar}, Branimir and {Sheldon}, Erin and {Shimasaku}, Kazu and {Sibley}, Valena C. and {Simmons}, A.~E. and {Sivarani}, Thirupathi and {Allyn Smith} and J. and {Smith}, Martin C. and {Smol{\v{c}}i{\'c}}, Vernesa and {Snedden}, Stephanie A. and {Stebbins}, Albert and {Steinmetz}, Matthias and {Stoughton}, Chris and {Strauss}, Michael A. and {SubbaRao}, Mark and {Suto}, Yasushi and {Szalay}, Alexander S. and {Szapudi}, Istv{\'a}n and {Szkody}, Paula and {Tanaka}, Masayuki and {Tegmark}, Max and {Teodoro}, Luis F.~A. and {Thakar}, Aniruddha R. and {Tremonti}, Christy A. and {Tucker}, Douglas L. and {Uomoto}, Alan and {Vanden Berk}, Daniel E. and {Vandenberg}, Jan and {Vidrih} and S. and {Vogeley}, Michael S. and {Voges}, Wolfgang and {Vogt}, Nicole P. and {Wadadekar}, Yogesh and {Watters}, Shannon and {Weinberg}, David H. and {West}, Andrew A. and {White}, Simon D.~M. and {Wilhite}, Brian C. and {Wonders}, Alainna C. and {Yanny}, Brian and {Yocum}, D.~R.},
        title = "{The Seventh Data Release of the Sloan Digital Sky Survey}",
      journal = {\apjs},
         year = 2009,
        month = jun,
       volume = {182},
       number = {2},
        pages = {543-558},
          doi = {10.1088/0067-0049/182/2/543},
archivePrefix = {arXiv},
       eprint = {0812.0649},
 primaryClass = {astro-ph},
       adsurl = {https://ui.adsabs.harvard.edu/abs/2009ApJS..182..543A}
}

@ARTICLE{Bundy2015,
       author = {{Bundy}, Kevin and {Bershady}, Matthew A. and {Law}, David R. and {Yan}, Renbin and {Drory}, Niv and {MacDonald}, Nicholas and {Wake}, David A. and {Cherinka}, Brian and {S{\'a}nchez-Gallego}, Jos{\'e} R. and {Weijmans}, Anne-Marie and {Thomas}, Daniel and {Tremonti}, Christy and {Masters}, Karen and {Coccato}, Lodovico and {Diamond-Stanic}, Aleksandar M. and {Arag{\'o}n-Salamanca}, Alfonso and {Avila-Reese}, Vladimir and {Badenes}, Carles and {Falc{\'o}n-Barroso}, J{\'e}sus and {Belfiore}, Francesco and {Bizyaev}, Dmitry and {Blanc}, Guillermo A. and {Bland-Hawthorn}, Joss and {Blanton}, Michael R. and {Brownstein}, Joel R. and {Byler}, Nell and {Cappellari}, Michele and {Conroy}, Charlie and {Dutton}, Aaron A. and {Emsellem}, Eric and {Etherington}, James and {Frinchaboy}, Peter M. and {Fu}, Hai and {Gunn}, James E. and {Harding}, Paul and {Johnston}, Evelyn J. and {Kauffmann}, Guinevere and {Kinemuchi}, Karen and {Klaene}, Mark A. and {Knapen}, Johan H. and {Leauthaud}, Alexie and {Li}, Cheng and {Lin}, Lihwai and {Maiolino}, Roberto and {Malanushenko}, Viktor and {Malanushenko}, Elena and {Mao}, Shude and {Maraston}, Claudia and {McDermid}, Richard M. and {Merrifield}, Michael R. and {Nichol}, Robert C. and {Oravetz}, Daniel and {Pan}, Kaike and {Parejko}, John K. and {Sanchez}, Sebastian F. and {Schlegel}, David and {Simmons}, Audrey and {Steele}, Oliver and {Steinmetz}, Matthias and {Thanjavur}, Karun and {Thompson}, Benjamin A. and {Tinker}, Jeremy L. and {van den Bosch}, Remco C.~E. and {Westfall}, Kyle B. and {Wilkinson}, David and {Wright}, Shelley and {Xiao}, Ting and {Zhang}, Kai},
        title = "{Overview of the SDSS-IV MaNGA Survey: Mapping nearby Galaxies at Apache Point Observatory}",
      journal = {\apj},
         year = 2015,
        month = jan,
       volume = {798},
       number = {1},
          eid = {7},
        pages = {7},
          doi = {10.1088/0004-637X/798/1/7},
archivePrefix = {arXiv},
       eprint = {1412.1482},
 primaryClass = {astro-ph.GA},
       adsurl = {https://ui.adsabs.harvard.edu/abs/2015ApJ...798....7B}
}

@ARTICLE{Maiolino2020,
       author = {{Maiolino} and R. and {Cirasuolo} and M. and {Afonso} and J. and {Bauer}, F.~E. and {Bowler} and R. and {Cucciati} and O. and {Daddi} and E. and {De Lucia} and G. and {Evans} and C. and {Flores} and H. and {Gargiulo} and A. and {Garilli} and B. and {Jablonka} and P. and {Jarvis} and M. and {Kneib} and J. -P. and {Lilly} and S. and {Looser} and T. and {Magliocchetti} and M. and {Man} and Z. and {Mannucci} and F. and {Maurogordato} and S. and {McLure}, R.~J. and {Norberg} and P. and {Oesch} and P. and {Oliva} and E. and {Paltani} and S. and {Pappalardo} and C. and {Peng} and Y. and {Pentericci} and L. and {Pozzetti} and L. and {Renzini} and A. and {Rodrigues} and M. and {Royer} and F. and {Serjeant} and S. and {Vanzi} and L. and {Wild} and V. and {Zamorani}, G.},
        title = "{MOONRISE: The Main MOONS GTO Extragalactic Survey}",
      journal = {The Messenger},
         year = 2020,
        month = jun,
       volume = {180},
        pages = {24-29},
          doi = {10.18727/0722-6691/5197},
archivePrefix = {arXiv},
       eprint = {2009.00644},
 primaryClass = {astro-ph.GA},
       adsurl = {https://ui.adsabs.harvard.edu/abs/2020Msngr.180...24M}
}

@ARTICLE{Cirasuolo2020,
       author = {{Cirasuolo} and M. and {Fairley} and A. and {Rees} and P. and {Gonzalez}, O.~A. and {Taylor} and W. and {Maiolino} and R. and {Afonso} and J. and {Evans} and C. and {Flores} and H. and {Lilly} and S. and {Oliva} and E. and {Paltani} and S. and {Vanzi} and L. and {Abreu} and M. and {Accardo} and M. and {Adams} and N. and {{\'A}lvarez M{\'e}ndez} and D. and {Amans} and J. -P. and {Amarantidis} and S. and {Atek} and H. and {Atkinson} and D. and {Banerji} and M. and {Barrett} and J. and {Barrientos} and F. and {Bauer} and F. and {Beard} and S. and {B{\'e}chet} and C. and {Belfiore} and A. and {Bellazzini} and M. and {Benoist} and C. and {Best} and P. and {Biazzo} and K. and {Black} and M. and {Boettger} and D. and {Bonifacio} and P. and {Bowler} and R. and {Bragaglia} and A. and {Brierley} and S. and {Brinchmann} and J. and {Brinkmann} and M. and {Buat} and V. and {Buitrago} and F. and {Burgarella} and D. and {Burningham} and B. and {Buscher} and D. and {Cabral} and A. and {Caffau} and E. and {Cardoso} and L. and {Carnall} and A. and {Carollo} and M. and {Castillo} and R. and {Castignani} and G. and {Catelan} and M. and {Cicone} and C. and {Cimatti} and A. and {Cioni} and M. -R.~L. and {Clementini} and G. and {Cochrane} and W. and {Coelho} and J. and {Colling} and M. and {Contini} and T. and {Contreras} and R. and {Conzelmann} and R. and {Cresci} and G. and {Cropper} and M. and {Cucciati} and O. and {Cullen} and F. and {Cumani} and C. and {Curti} and M. and {Da Silva} and A. and {Daddi} and E. and {Dalessandro} and E. and {Dalessio} and F. and {Dauvin} and L. and {Davidson} and G. and {de Laverny} and P. and {Delplancke-Str{\"o}bele} and F. and {De Lucia} and G. and {Del Vecchio} and C. and {Dessauges-Zavadsky} and M. and {Di Matteo} and P. and {Dole} and H. and {Drass} and H. and {Dunlop} and J. and {D{\"u}nner} and R. and {Eales} and S. and {Ellis} and R. and {Enriques} and B. and {Fasola} and G. and {Ferguson} and A. and {Ferruzzi} and D. and {Fisher} and M. and {Flores} and M. and {Fontana} and A. and {Forchi} and V. and {Francois} and P. and {Franzetti} and P. and {Gargiulo} and A. and {Garilli} and B. and {Gaudemard} and J. and {Gieles} and M. and {Gilmore} and G. and {Ginolfi} and M. and {Gomes}, J.~M. and {Guinouard} and I. and {Gutierrez} and P. and {Haigron} and R. and {Hammer} and F. and {Hammersley} and P. and {Haniff} and C. and {Harrison} and C. and {Haywood} and M. and {Hill} and V. and {Hubin} and N. and {Humphrey} and A. and {Ibata} and R. and {Infante} and L. and {Ives} and D. and {Ivison} and R. and {Iwert} and O. and {Jablonka} and P. and {Jakob} and G. and {Jarvis} and M. and {King} and D. and {Kneib} and J. -P. and {Laporte} and P. and {Lawrence} and A. and {Lee} and D. and {Li Causi} and G. and {Lorenzoni} and S. and {Lucatello} and S. and {Luco} and Y. and {Macleod} and A. and {Magliocchetti} and M. and {Magrini} and L. and {Mainieri} and V. and {Maire} and C. and {Mannucci} and F. and {Martin} and N. and {Matute} and I. and {Maurogordato} and S. and {McGee} and S. and {Mcleod} and D. and {McLure} and R. and {McMahon} and R. and {Melse} and B. -T. and {Messias} and H. and {Mucciarelli} and A. and {Nisini} and B. and {Nix} and J. and {Norberg} and P. and {Oesch} and P. and {Oliveira} and A. and {Origlia} and L. and {Padilla} and N. and {Palsa} and R. and {Pancino} and E. and {Papaderos} and P. and {Pappalardo} and C. and {Parry} and I. and {Pasquini} and L. and {Peacock} and J. and {Pedichini} and F. and {Pello} and R. and {Peng} and Y. and {Pentericci} and L. and {Pfuhl} and O. and {Piazzesi} and R. and {Popovic} and D. and {Pozzetti} and L. and {Puech} and M. and {Puzia} and T. and {Raichoor} and A. and {Randich} and S. and {Recio-Blanco} and A. and {Reis} and S. and {Reix} and F. and {Renzini} and A. and {Rodrigues} and M. and {Rojas} and F. and {Rojas-Arriagada}, {\'A}. and {Rota} and S. and {Royer} and F. and {Sacco} and G. and {Sanchez-Janssen} and R. and {Sanna} and N. and {Santos} and P. and {Sarzi} and M. and {Schaerer} and D. and {Schiavon} and R. and {Schnell} and R. and {Schultheis} and M. and {Scodeggio} and M. and {Serjeant} and S. and {Shen} and T. -C. and {Simmonds} and C. and {Smoker} and J. and {Sobral} and D. and {Sordet} and M. and {Sp{\'e}rone}, D.},
        title = "{MOONS: The New Multi-Object Spectrograph for the VLT}",
      journal = {The Messenger},
         year = 2020,
        month = jun,
       volume = {180},
        pages = {10-17},
          doi = {10.18727/0722-6691/5195},
archivePrefix = {arXiv},
       eprint = {2009.00628},
 primaryClass = {astro-ph.IM},
       adsurl = {https://ui.adsabs.harvard.edu/abs/2020Msngr.180...10C}
}

@ARTICLE{Navarro1997,
       author = {{Navarro}, Julio F. and {Frenk}, Carlos S. and {White}, Simon D.~M.},
        title = "{A Universal Density Profile from Hierarchical Clustering}",
      journal = {\apj},
         year = 1997,
        month = dec,
       volume = {490},
       number = {2},
        pages = {493-508},
          doi = {10.1086/304888},
archivePrefix = {arXiv},
       eprint = {astro-ph/9611107},
 primaryClass = {astro-ph},
       adsurl = {https://ui.adsabs.harvard.edu/abs/1997ApJ...490..493N}
}

@ARTICLE{Navarro1996,
       author = {{Navarro}, Julio F. and {Frenk}, Carlos S. and {White}, Simon D.~M.},
        title = "{The Structure of Cold Dark Matter Halos}",
      journal = {\apj},
         year = 1996,
        month = may,
       volume = {462},
        pages = {563},
          doi = {10.1086/177173},
archivePrefix = {arXiv},
       eprint = {astro-ph/9508025},
 primaryClass = {astro-ph},
       adsurl = {https://ui.adsabs.harvard.edu/abs/1996ApJ...462..563N}
}

@ARTICLE{Navarro2004,
       author = {{Navarro}, J.~F. and {Hayashi} and E. and {Power} and C. and {Jenkins}, A.~R. and {Frenk}, C.~S. and {White}, S.~D.~M. and {Springel} and V. and {Stadel} and J. and {Quinn}, T.~R.},
        title = "{The inner structure of {\ensuremath{\Lambda}}CDM haloes - III. Universality and asymptotic slopes}",
      journal = {\mnras},
         year = 2004,
        month = apr,
       volume = {349},
       number = {3},
        pages = {1039-1051},
          doi = {10.1111/j.1365-2966.2004.07586.x},
archivePrefix = {arXiv},
       eprint = {astro-ph/0311231},
 primaryClass = {astro-ph},
       adsurl = {https://ui.adsabs.harvard.edu/abs/2004MNRAS.349.1039N}
}

@ARTICLE{Baugh2006,
       author = {{Baugh}, C.~M.},
        title = "{A primer on hierarchical galaxy formation: the semi-analytical approach}",
      journal = {Rep. Prog. Phys.},
         year = 2006,
        month = dec,
       volume = {69},
       number = {12},
        pages = {3101-3156},
          doi = {10.1088/0034-4885/69/12/R02},
archivePrefix = {arXiv},
       eprint = {astro-ph/0610031},
 primaryClass = {astro-ph},
       adsurl = {https://ui.adsabs.harvard.edu/abs/2006RPPh...69.3101B}
}

@ARTICLE{Springel2003,
       author = {{Springel}, Volker and {Hernquist}, Lars},
        title = "{Cosmological smoothed particle hydrodynamics simulations: a hybrid multiphase model for star formation}",
      journal = {\mnras},
         year = 2003,
        month = feb,
       volume = {339},
       number = {2},
        pages = {289-311},
          doi = {10.1046/j.1365-8711.2003.06206.x},
archivePrefix = {arXiv},
       eprint = {astro-ph/0206393},
 primaryClass = {astro-ph},
       adsurl = {https://ui.adsabs.harvard.edu/abs/2003MNRAS.339..289S}
}

@ARTICLE{Wuyts2011,
       author = {{Wuyts}, Stijn and {F{\"o}rster Schreiber}, Natascha M. and {van der Wel}, Arjen and {Magnelli}, Benjamin and {Guo}, Yicheng and {Genzel}, Reinhard and {Lutz}, Dieter and {Aussel}, Herv{\'e} and {Barro}, Guillermo and {Berta}, Stefano and {Cava}, Antonio and {Graci{\'a}-Carpio}, Javier and {Hathi}, Nimish P. and {Huang}, Kuang-Han and {Kocevski}, Dale D. and {Koekemoer}, Anton M. and {Lee}, Kyoung-Soo and {Le Floc'h}, Emeric and {McGrath}, Elizabeth J. and {Nordon}, Raanan and {Popesso}, Paola and {Pozzi}, Francesca and {Riguccini}, Laurie and {Rodighiero}, Giulia and {Saintonge}, Amelie and {Tacconi}, Linda},
        title = "{Galaxy Structure and Mode of Star Formation in the SFR-Mass Plane from $z \sim 2.5$ to $z \sim 0.1$}",
      journal = {\apj},
         year = 2011,
        month = dec,
       volume = {742},
       number = {2},
          eid = {96},
        pages = {96},
          doi = {10.1088/0004-637X/742/2/96},
archivePrefix = {arXiv},
       eprint = {1107.0317},
 primaryClass = {astro-ph.CO},
       adsurl = {https://ui.adsabs.harvard.edu/abs/2011ApJ...742...96W}
}

@ARTICLE{Behroozi2013,
       author = {{Behroozi}, Peter S. and {Wechsler}, Risa H. and {Conroy}, Charlie},
        title = "{The Average Star Formation Histories of Galaxies in Dark Matter Halos from z = 0-8}",
      journal = {\apj},
         year = 2013,
        month = jun,
       volume = {770},
       number = {1},
          eid = {57},
        pages = {57},
          doi = {10.1088/0004-637X/770/1/57},
archivePrefix = {arXiv},
       eprint = {1207.6105},
 primaryClass = {astro-ph.CO},
       adsurl = {https://ui.adsabs.harvard.edu/abs/2013ApJ...770...57B}
}

@ARTICLE{Fakhouri2010,
       author = {{Fakhouri}, Onsi and {Ma}, Chung-Pei and {Boylan-Kolchin}, Michael},
        title = "{The merger rates and mass assembly histories of dark matter haloes in the two Millennium simulations}",
      journal = {\mnras},
         year = 2010,
        month = aug,
       volume = {406},
       number = {4},
        pages = {2267-2278},
          doi = {10.1111/j.1365-2966.2010.16859.x},
archivePrefix = {arXiv},
       eprint = {1001.2304},
 primaryClass = {astro-ph.CO},
       adsurl = {https://ui.adsabs.harvard.edu/abs/2010MNRAS.406.2267F}
}

@ARTICLE{Press1974,
       author = {{Press}, William H. and {Schechter}, Paul},
        title = "{Formation of Galaxies and Clusters of Galaxies by Self-Similar Gravitational Condensation}",
      journal = {\apj},
         year = 1974,
        month = feb,
       volume = {187},
        pages = {425-438},
          doi = {10.1086/152650},
       adsurl = {https://ui.adsabs.harvard.edu/abs/1974ApJ...187..425P}
}

@ARTICLE{Butcher1978,
       author = {{Butcher} and H. and {Oemler}, Jr., A.},
        title = "{The evolution of galaxies in clusters. I. ISIT photometry of Cl 0024+1654 and 3C 295.}",
      journal = {\apj},
         year = 1978,
        month = jan,
       volume = {219},
        pages = {18-30},
          doi = {10.1086/155751},
       adsurl = {https://ui.adsabs.harvard.edu/abs/1978ApJ...219...18B}
}

@ARTICLE{Barro2017,
       author = {{Barro}, Guillermo and {Faber}, S.~M. and {Koo}, David C. and {Dekel}, Avishai and {Fang}, Jerome J. and {Trump}, Jonathan R. and {P{\'e}rez-Gonz{\'a}lez}, Pablo G. and {Pacifici}, Camilla and {Primack}, Joel R. and {Somerville}, Rachel S. and {Yan}, Haojing and {Guo}, Yicheng and {Liu}, Fengshan and {Ceverino}, Daniel and {Kocevski}, Dale D. and {McGrath}, Elizabeth},
        title = "{Structural and Star-forming Relations since $z \sim 3$: Connecting Compact Star-forming and Quiescent Galaxies}",
      journal = {\apj},
         year = 2017,
        month = may,
       volume = {840},
       number = {1},
          eid = {47},
        pages = {47},
          doi = {10.3847/1538-4357/aa6b05},
archivePrefix = {arXiv},
       eprint = {1509.00469},
 primaryClass = {astro-ph.GA},
       adsurl = {https://ui.adsabs.harvard.edu/abs/2017ApJ...840...47B}
}

@ARTICLE{Mendel2014,
       author = {{Mendel} and J. Trevor and {Simard}, Luc and {Palmer}, Michael and {Ellison}, Sara L. and {Patton}, David R.},
        title = "{A Catalog of Bulge, Disk, and Total Stellar Mass Estimates for the Sloan Digital Sky Survey}",
      journal = {\apjs},
         year = 2014,
        month = jan,
       volume = {210},
       number = {1},
          eid = {3},
        pages = {3},
          doi = {10.1088/0067-0049/210/1/3},
archivePrefix = {arXiv},
       eprint = {1310.8304},
 primaryClass = {astro-ph.CO},
       adsurl = {https://ui.adsabs.harvard.edu/abs/2014ApJS..210....3M}
}

@ARTICLE{Saglia2016,
       author = {{Saglia}, R.~P. and {Opitsch} and M. and {Erwin} and P. and {Thomas} and J. and {Beifiori} and A. and {Fabricius} and M. and {Mazzalay} and X. and {Nowak} and N. and {Rusli}, S.~P. and {Bender}, R.},
        title = "{The SINFONI Black Hole Survey: The Black Hole Fundamental Plane Revisited and the Paths of (Co)evolution of Supermassive Black Holes and Bulges}",
      journal = {\apj},
         year = 2016,
        month = feb,
       volume = {818},
       number = {1},
          eid = {47},
        pages = {47},
          doi = {10.3847/0004-637X/818/1/47},
archivePrefix = {arXiv},
       eprint = {1601.00974},
 primaryClass = {astro-ph.GA},
       adsurl = {https://ui.adsabs.harvard.edu/abs/2016ApJ...818...47S}
}

@ARTICLE{Ferrarese2000,
       author = {{Ferrarese}, Laura and {Merritt}, David},
        title = "{A Fundamental Relation between Supermassive Black Holes and Their Host Galaxies}",
      journal = {\apjl},
         year = 2000,
        month = aug,
       volume = {539},
       number = {1},
        pages = {L9-L12},
          doi = {10.1086/312838},
archivePrefix = {arXiv},
       eprint = {astro-ph/0006053},
 primaryClass = {astro-ph},
       adsurl = {https://ui.adsabs.harvard.edu/abs/2000ApJ...539L...9F}
}

@ARTICLE{Cappellari2013a,
       author = {{Cappellari}, Michele and {Scott}, Nicholas and {Alatalo}, Katherine and {Blitz}, Leo and {Bois}, Maxime and {Bournaud}, Fr{\'e}d{\'e}ric and {Bureau} and M. and {Crocker}, Alison F. and {Davies}, Roger L. and {Davis}, Timothy A. and {de Zeeuw}, P.~T. and {Duc}, Pierre-Alain and {Emsellem}, Eric and {Khochfar}, Sadegh and {Krajnovi{\'c}}, Davor and {Kuntschner}, Harald and {McDermid}, Richard M. and {Morganti}, Raffaella and {Naab}, Thorsten and {Oosterloo}, Tom and {Sarzi}, Marc and {Serra}, Paolo and {Weijmans}, Anne-Marie and {Young}, Lisa M.},
        title = "{The ATLAS$^{3D}$ project - XV. Benchmark for early-type galaxies scaling relations from 260 dynamical models: mass-to-light ratio, dark matter, Fundamental Plane and Mass Plane}",
      journal = {\mnras},
         year = 2013,
        month = jul,
       volume = {432},
       number = {3},
        pages = {1709-1741},
          doi = {10.1093/mnras/stt562},
archivePrefix = {arXiv},
       eprint = {1208.3522},
 primaryClass = {astro-ph.CO},
       adsurl = {https://ui.adsabs.harvard.edu/abs/2013MNRAS.432.1709C}
}

@ARTICLE{Forster2009,
       author = {{F{\"o}rster Schreiber}, N.~M. and {Genzel} and R. and {Bouch{\'e}} and N. and {Cresci} and G. and {Davies} and R. and {Buschkamp} and P. and {Shapiro} and K. and {Tacconi}, L.~J. and {Hicks}, E.~K.~S. and {Genel} and S. and {Shapley}, A.~E. and {Erb}, D.~K. and {Steidel}, C.~C. and {Lutz} and D. and {Eisenhauer} and F. and {Gillessen} and S. and {Sternberg} and A. and {Renzini} and A. and {Cimatti} and A. and {Daddi} and E. and {Kurk} and J. and {Lilly} and S. and {Kong} and X. and {Lehnert}, M.~D. and {Nesvadba} and N. and {Verma} and A. and {McCracken} and H. and {Arimoto} and N. and {Mignoli} and M. and {Onodera}, M.},
        title = "{The SINS Survey: SINFONI Integral Field Spectroscopy of $z \sim 2$ Star-forming Galaxies}",
      journal = {\apj},
         year = 2009,
        month = dec,
       volume = {706},
       number = {2},
        pages = {1364-1428},
          doi = {10.1088/0004-637X/706/2/1364},
archivePrefix = {arXiv},
       eprint = {0903.1872},
 primaryClass = {astro-ph.CO},
       adsurl = {https://ui.adsabs.harvard.edu/abs/2009ApJ...706.1364F}
}

@ARTICLE{Forster2006,
       author = {{F{\"o}rster Schreiber}, N.~M. and {Genzel} and R. and {Lehnert}, M.~D. and {Bouch{\'e}} and N. and {Verma} and A. and {Erb}, D.~K. and {Shapley}, A.~E. and {Steidel}, C.~C. and {Davies} and R. and {Lutz} and D. and {Nesvadba} and N. and {Tacconi}, L.~J. and {Eisenhauer} and F. and {Abuter} and R. and {Gilbert} and A. and {Gillessen} and S. and {Sternberg}, A.},
        title = "{SINFONI Integral Field Spectroscopy of $z \sim 2$ UV-selected Galaxies: Rotation Curves and Dynamical Evolution}",
      journal = {\apj},
         year = 2006,
        month = jul,
       volume = {645},
       number = {2},
        pages = {1062-1075},
          doi = {10.1086/504403},
archivePrefix = {arXiv},
       eprint = {astro-ph/0603559},
 primaryClass = {astro-ph},
       adsurl = {https://ui.adsabs.harvard.edu/abs/2006ApJ...645.1062F}
}

@ARTICLE{Lilly2016,
       author = {{Lilly}, Simon J. and {Carollo} and C. Marcella},
        title = "{Surface Density Effects in Quenching: Cause or Effect?}",
      journal = {\apj},
         year = 2016,
        month = dec,
       volume = {833},
       number = {1},
          eid = {1},
        pages = {1},
          doi = {10.3847/0004-637X/833/1/1},
archivePrefix = {arXiv},
       eprint = {1604.06459},
 primaryClass = {astro-ph.GA},
       adsurl = {https://ui.adsabs.harvard.edu/abs/2016ApJ...833....1L}
}

@ARTICLE{Goubert2024,
       author = {{Goubert}, Paul H. and {Bluck}, Asa F.~L. and {Piotrowska}, Joanna M. and {Maiolino}, Roberto},
        title = "{The role of environment and AGN feedback in quenching local galaxies: comparing cosmological hydrodynamical simulations to the SDSS}",
      journal = {\mnras},
         year = 2024,
        month = mar,
       volume = {528},
       number = {3},
        pages = {4891-4921},
          doi = {10.1093/mnras/stae269},
archivePrefix = {arXiv},
       eprint = {2401.12953},
 primaryClass = {astro-ph.GA},
       adsurl = {https://ui.adsabs.harvard.edu/abs/2024MNRAS.528.4891G}
}

@ARTICLE{Gunn1972,
       author = {{Gunn}, James E. and {Gott}, III and J. Richard},
        title = "{On the Infall of Matter Into Clusters of Galaxies and Some Effects on Their Evolution}",
      journal = {\apj},
         year = 1972,
        month = aug,
       volume = {176},
        pages = {1},
          doi = {10.1086/151605},
       adsurl = {https://ui.adsabs.harvard.edu/abs/1972ApJ...176....1G}
}

@ARTICLE{Moore1996,
       author = {{Moore}, Ben and {Katz}, Neal and {Lake}, George and {Dressler}, Alan and {Oemler}, Augustus},
        title = "{Galaxy harassment and the evolution of clusters of galaxies}",
      journal = {\nat},
         year = 1996,
        month = feb,
       volume = {379},
       number = {6566},
        pages = {613-616},
          doi = {10.1038/379613a0},
archivePrefix = {arXiv},
       eprint = {astro-ph/9510034},
 primaryClass = {astro-ph},
       adsurl = {https://ui.adsabs.harvard.edu/abs/1996Natur.379..613M}
}

@ARTICLE{Dekel2019,
       author = {{Dekel}, Avishai and {Lapiner}, Sharon and {Dubois}, Yohan},
        title = "{Origin of the Golden Mass of Galaxies and Black Holes}",
      journal = {arXiv e-prints},
         year = 2019,
        month = apr,
          eid = {arXiv:1904.08431},
        pages = {arXiv:1904.08431},
          doi = {10.48550/arXiv.1904.08431},
archivePrefix = {arXiv},
       eprint = {1904.08431},
 primaryClass = {astro-ph.GA},
       adsurl = {https://ui.adsabs.harvard.edu/abs/2019arXiv190408431D}
}

@ARTICLE{DiMatteo2005,
       author = {{Di Matteo}, Tiziana and {Springel}, Volker and {Hernquist}, Lars},
        title = "{Energy input from quasars regulates the growth and activity of black holes and their host galaxies}",
      journal = {\nat},
         year = 2005,
        month = feb,
       volume = {433},
       number = {7026},
        pages = {604-607},
          doi = {10.1038/nature03335},
archivePrefix = {arXiv},
       eprint = {astro-ph/0502199},
 primaryClass = {astro-ph},
       adsurl = {https://ui.adsabs.harvard.edu/abs/2005Natur.433..604D}
}

@ARTICLE{DiMatteo2008a,
       author = {{Di Matteo}, Tiziana and {Colberg}, J{\"o}rg and {Springel}, Volker and {Hernquist}, Lars and {Sijacki}, Debora},
        title = "{Direct Cosmological Simulations of the Growth of Black Holes and Galaxies}",
      journal = {\apj},
         year = 2008,
        month = mar,
       volume = {676},
       number = {1},
        pages = {33-53},
          doi = {10.1086/524921},
archivePrefix = {arXiv},
       eprint = {0705.2269},
 primaryClass = {astro-ph},
       adsurl = {https://ui.adsabs.harvard.edu/abs/2008ApJ...676...33D}
}

@ARTICLE{Guo2011,
       author = {{Guo}, Qi and {White}, Simon and {Boylan-Kolchin}, Michael and {De Lucia}, Gabriella and {Kauffmann}, Guinevere and {Lemson}, Gerard and {Li}, Cheng and {Springel}, Volker and {Weinmann}, Simone},
        title = "{From dwarf spheroidals to cD galaxies: simulating the galaxy population in a {\ensuremath{\Lambda}}CDM cosmology}",
      journal = {\mnras},
         year = 2011,
        month = may,
       volume = {413},
       number = {1},
        pages = {101-131},
          doi = {10.1111/j.1365-2966.2010.18114.x},
archivePrefix = {arXiv},
       eprint = {1006.0106},
 primaryClass = {astro-ph.CO},
       adsurl = {https://ui.adsabs.harvard.edu/abs/2011MNRAS.413..101G}
}

@ARTICLE{DeLucia2007a,
       author = {{De Lucia}, Gabriella and {Blaizot}, J{\'e}r{\'e}my},
        title = "{The hierarchical formation of the brightest cluster galaxies}",
      journal = {\mnras},
         year = 2007,
        month = feb,
       volume = {375},
       number = {1},
        pages = {2-14},
          doi = {10.1111/j.1365-2966.2006.11287.x},
archivePrefix = {arXiv},
       eprint = {astro-ph/0606519},
 primaryClass = {astro-ph},
       adsurl = {https://ui.adsabs.harvard.edu/abs/2007MNRAS.375....2D}
}

@ARTICLE{Vogelsberger2020,
       author = {{Vogelsberger}, Mark and {Marinacci}, Federico and {Torrey}, Paul and {Puchwein}, Ewald},
        title = "{Cosmological simulations of galaxy formation}",
      journal = {Nature Rev. Phys.},
         year = 2020,
        month = jan,
       volume = {2},
       number = {1},
        pages = {42-66},
          doi = {10.1038/s42254-019-0127-2},
archivePrefix = {arXiv},
       eprint = {1909.07976},
 primaryClass = {astro-ph.GA},
       adsurl = {https://ui.adsabs.harvard.edu/abs/2020NatRP...2...42V}
}

@ARTICLE{Soltan1982,
       author = {{Soltan}, A.},
        title = "{Masses of quasars}",
      journal = {\mnras},
         year = 1982,
        month = jul,
       volume = {200},
        pages = {115-122},
          doi = {10.1093/mnras/200.1.115},
       adsurl = {https://ui.adsabs.harvard.edu/abs/1982MNRAS.200..115S}
}

@ARTICLE{Silk1998,
       author = {{Silk}, Joseph and {Rees}, Martin J.},
        title = "{Quasars and galaxy formation}",
      journal = {\aap},
         year = 1998,
        month = mar,
       volume = {331},
        pages = {L1-L4},
          doi = {10.48550/arXiv.astro-ph/9801013},
archivePrefix = {arXiv},
       eprint = {astro-ph/9801013},
 primaryClass = {astro-ph},
       adsurl = {https://ui.adsabs.harvard.edu/abs/1998A&A...331L...1S}
}

@ARTICLE{Efstathiou1985,
       author = {{Efstathiou} and G. and {Davis} and M. and {White}, S.~D.~M. and {Frenk}, C.~S.},
        title = "{Numerical techniques for large cosmological N-body simulations}",
      journal = {\apjs},
         year = 1985,
        month = feb,
       volume = {57},
        pages = {241-260},
          doi = {10.1086/191003},
       adsurl = {https://ui.adsabs.harvard.edu/abs/1985ApJS...57..241E}
}

@ARTICLE{Zeldovich1970,
       author = {{Zel'dovich}, Ya. B.},
        title = "{Gravitational instability: An approximate theory for large density perturbations.}",
      journal = {\aap},
         year = 1970,
        month = mar,
       volume = {5},
        pages = {84-89},
       adsurl = {https://ui.adsabs.harvard.edu/abs/1970A&A.....5...84Z}
}

@ARTICLE{Springel2005a,
       author = {{Springel}, Volker and {White}, Simon D.~M. and {Jenkins}, Adrian and {Frenk}, Carlos S. and {Yoshida}, Naoki and {Gao}, Liang and {Navarro}, Julio and {Thacker}, Robert and {Croton}, Darren and {Helly}, John and {Peacock}, John A. and {Cole}, Shaun and {Thomas}, Peter and {Couchman}, Hugh and {Evrard}, August and {Colberg}, J{\"o}rg and {Pearce}, Frazer},
        title = "{Simulations of the formation, evolution and clustering of galaxies and quasars}",
      journal = {\nat},
         year = 2005,
        month = jun,
       volume = {435},
       number = {7042},
        pages = {629-636},
          doi = {10.1038/nature03597},
archivePrefix = {arXiv},
       eprint = {astro-ph/0504097},
 primaryClass = {astro-ph},
       adsurl = {https://ui.adsabs.harvard.edu/abs/2005Natur.435..629S}
}

@ARTICLE{Boylan2009,
       author = {{Boylan-Kolchin}, Michael and {Springel}, Volker and {White}, Simon D.~M. and {Jenkins}, Adrian and {Lemson}, Gerard},
        title = "{Resolving cosmic structure formation with the Millennium-II Simulation}",
      journal = {\mnras},
         year = 2009,
        month = sep,
       volume = {398},
       number = {3},
        pages = {1150-1164},
          doi = {10.1111/j.1365-2966.2009.15191.x},
archivePrefix = {arXiv},
       eprint = {0903.3041},
 primaryClass = {astro-ph.CO},
       adsurl = {https://ui.adsabs.harvard.edu/abs/2009MNRAS.398.1150B}
}

@ARTICLE{Moore1999a,
       author = {{Moore}, Ben and {Ghigna}, Sebastiano and {Governato}, Fabio and {Lake}, George and {Quinn}, Thomas and {Stadel}, Joachim and {Tozzi}, Paolo},
        title = "{Dark Matter Substructure within Galactic Halos}",
      journal = {\apjl},
         year = 1999,
        month = oct,
       volume = {524},
       number = {1},
        pages = {L19-L22},
          doi = {10.1086/312287},
archivePrefix = {arXiv},
       eprint = {astro-ph/9907411},
 primaryClass = {astro-ph},
       adsurl = {https://ui.adsabs.harvard.edu/abs/1999ApJ...524L..19M}
}

@ARTICLE{Kauffmann1993,
       author = {{Kauffmann} and G. and {White}, S.~D.~M. and {Guiderdoni}, B.},
        title = "{The formation and evolution of galaxies within merging dark matter haloes.}",
      journal = {\mnras},
         year = 1993,
        month = sep,
       volume = {264},
        pages = {201-218},
          doi = {10.1093/mnras/264.1.201},
       adsurl = {https://ui.adsabs.harvard.edu/abs/1993MNRAS.264..201K}
}

@ARTICLE{Springel2005b,
       author = {{Springel}, Volker},
        title = "{The cosmological simulation code GADGET-2}",
      journal = {\mnras},
         year = 2005,
        month = dec,
       volume = {364},
       number = {4},
        pages = {1105-1134},
          doi = {10.1111/j.1365-2966.2005.09655.x},
archivePrefix = {arXiv},
       eprint = {astro-ph/0505010},
 primaryClass = {astro-ph},
       adsurl = {https://ui.adsabs.harvard.edu/abs/2005MNRAS.364.1105S}
}

@ARTICLE{Hopkins2018,
       author = {{Hopkins}, Philip F. and {Wetzel}, Andrew and {Kere{\v{s}}}, Du{\v{s}}an and {Faucher-Gigu{\`e}re}, Claude-Andr{\'e} and {Quataert}, Eliot and {Boylan-Kolchin}, Michael and {Murray}, Norman and {Hayward}, Christopher C. and {Garrison-Kimmel}, Shea and {Hummels}, Cameron and {Feldmann}, Robert and {Torrey}, Paul and {Ma}, Xiangcheng and {Angl{\'e}s-Alc{\'a}zar}, Daniel and {Su}, Kung-Yi and {Orr}, Matthew and {Schmitz}, Denise and {Escala}, Ivanna and {Sanderson}, Robyn and {Grudi{\'c}}, Michael Y. and {Hafen}, Zachary and {Kim}, Ji-Hoon and {Fitts}, Alex and {Bullock}, James S. and {Wheeler}, Coral and {Chan}, T.~K. and {Elbert}, Oliver D. and {Narayanan}, Desika},
        title = "{FIRE-2 simulations: physics versus numerics in galaxy formation}",
      journal = {\mnras},
         year = 2018,
        month = oct,
       volume = {480},
       number = {1},
        pages = {800-863},
          doi = {10.1093/mnras/sty1690},
archivePrefix = {arXiv},
       eprint = {1702.06148},
 primaryClass = {astro-ph.GA},
       adsurl = {https://ui.adsabs.harvard.edu/abs/2018MNRAS.480..800H}
}

@ARTICLE{Teyssier2002,
       author = {{Teyssier}, R.},
        title = "{Cosmological hydrodynamics with adaptive mesh refinement. A new high resolution code called RAMSES}",
      journal = {\aap},
         year = 2002,
        month = apr,
       volume = {385},
        pages = {337-364},
          doi = {10.1051/0004-6361:20011817},
archivePrefix = {arXiv},
       eprint = {astro-ph/0111367},
 primaryClass = {astro-ph},
       adsurl = {https://ui.adsabs.harvard.edu/abs/2002A&A...385..337T}
}

@ARTICLE{Navarro1993,
       author = {{Navarro}, J.~F. and {White}, S.~D.~M.},
        title = "{Simulations of Dissipative Galaxy Formation in Hierarchically Clustering Universes - Part One - Tests of the Code}",
      journal = {\mnras},
         year = 1993,
        month = nov,
       volume = {265},
        pages = {271},
          doi = {10.1093/mnras/265.2.271},
       adsurl = {https://ui.adsabs.harvard.edu/abs/1993MNRAS.265..271N}
}

@ARTICLE{Moreno2015,
       author = {{Moreno}, Jorge and {Torrey}, Paul and {Ellison}, Sara L. and {Patton}, David R. and {Bluck}, Asa F.~L. and {Bansal}, Gunjan and {Hernquist}, Lars},
        title = "{Mapping galaxy encounters in numerical simulations: the spatial extent of induced star formation}",
      journal = {\mnras},
         year = 2015,
        month = apr,
       volume = {448},
       number = {2},
        pages = {1107-1117},
          doi = {10.1093/mnras/stv094},
archivePrefix = {arXiv},
       eprint = {1501.03573},
 primaryClass = {astro-ph.GA},
       adsurl = {https://ui.adsabs.harvard.edu/abs/2015MNRAS.448.1107M}
}

@ARTICLE{Moreno2019,
       author = {{Moreno}, Jorge and {Torrey}, Paul and {Ellison}, Sara L. and {Patton}, David R. and {Hopkins}, Philip F. and {Bueno}, Michael and {Hayward}, Christopher C. and {Narayanan}, Desika and {Kere{\v{s}}}, Du{\v{s}}an and {Bluck}, Asa F.~L. and {Hernquist}, Lars},
        title = "{Interacting galaxies on FIRE-2: the connection between enhanced star formation and interstellar gas content}",
      journal = {\mnras},
         year = 2019,
        month = may,
       volume = {485},
       number = {1},
        pages = {1320-1338},
          doi = {10.1093/mnras/stz417},
archivePrefix = {arXiv},
       eprint = {1902.02305},
 primaryClass = {astro-ph.GA},
       adsurl = {https://ui.adsabs.harvard.edu/abs/2019MNRAS.485.1320M}
}

@ARTICLE{Torrey2014,
       author = {{Torrey}, Paul and {Vogelsberger}, Mark and {Genel}, Shy and {Sijacki}, Debora and {Springel}, Volker and {Hernquist}, Lars},
        title = "{A model for cosmological simulations of galaxy formation physics: multi-epoch validation}",
      journal = {\mnras},
         year = 2014,
        month = mar,
       volume = {438},
       number = {3},
        pages = {1985-2004},
          doi = {10.1093/mnras/stt2295},
archivePrefix = {arXiv},
       eprint = {1305.4931},
 primaryClass = {astro-ph.CO},
       adsurl = {https://ui.adsabs.harvard.edu/abs/2014MNRAS.438.1985T}
}

@ARTICLE{Tremmel2019,
       author = {{Tremmel} and M. and {Quinn}, T.~R. and {Ricarte} and A. and {Babul} and A. and {Chadayammuri} and U. and {Natarajan} and P. and {Nagai} and D. and {Pontzen} and A. and {Volonteri}, M.},
        title = "{Introducing ROMULUSC: a cosmological simulation of a galaxy cluster with an unprecedented resolution}",
      journal = {\mnras},
         year = 2019,
        month = mar,
       volume = {483},
       number = {3},
        pages = {3336-3362},
          doi = {10.1093/mnras/sty3336},
archivePrefix = {arXiv},
       eprint = {1806.01282},
 primaryClass = {astro-ph.GA},
       adsurl = {https://ui.adsabs.harvard.edu/abs/2019MNRAS.483.3336T}
}

@ARTICLE{Pillepich2019,
       author = {{Pillepich}, Annalisa and {Nelson}, Dylan and {Springel}, Volker and {Pakmor}, R{\"u}diger and {Torrey}, Paul and {Weinberger}, Rainer and {Vogelsberger}, Mark and {Marinacci}, Federico and {Genel}, Shy and {van der Wel}, Arjen and {Hernquist}, Lars},
        title = "{First results from the TNG50 simulation: the evolution of stellar and gaseous discs across cosmic time}",
      journal = {\mnras},
         year = 2019,
        month = dec,
       volume = {490},
       number = {3},
        pages = {3196-3233},
          doi = {10.1093/mnras/stz2338},
archivePrefix = {arXiv},
       eprint = {1902.05553},
 primaryClass = {astro-ph.GA},
       adsurl = {https://ui.adsabs.harvard.edu/abs/2019MNRAS.490.3196P}
}

@ARTICLE{Jeans1902,
       author = {{Jeans}, J.~H.},
        title = "{The Stability of a Spherical Nebula}",
      journal = {Philosophical Transactions of the Royal Society of London Series A},
         year = 1902,
        month = jan,
       volume = {199},
        pages = {1-53},
          doi = {10.1098/rsta.1902.0012},
       adsurl = {https://ui.adsabs.harvard.edu/abs/1902RSPTA.199....1J}
}

@ARTICLE{Larson1981,
       author = {{Larson}, R.~B.},
        title = "{Turbulence and star formation in molecular clouds.}",
      journal = {\mnras},
         year = 1981,
        month = mar,
       volume = {194},
        pages = {809-826},
          doi = {10.1093/mnras/194.4.809},
       adsurl = {https://ui.adsabs.harvard.edu/abs/1981MNRAS.194..809L}
}

@ARTICLE{McKee2007,
       author = {{McKee}, Christopher F. and {Ostriker}, Eve C.},
        title = "{Theory of Star Formation}",
      journal = {\araa},
         year = 2007,
        month = sep,
       volume = {45},
       number = {1},
        pages = {565-687},
          doi = {10.1146/annurev.astro.45.051806.110602},
archivePrefix = {arXiv},
       eprint = {0707.3514},
 primaryClass = {astro-ph},
       adsurl = {https://ui.adsabs.harvard.edu/abs/2007ARA&A..45..565M}
}

@ARTICLE{Matteucci2006,
       author = {{Matteucci} and F. and {Panagia} and N. and {Pipino} and A. and {Mannucci} and F. and {Recchi} and S. and {Della Valle}, M.},
        title = "{A new formulation of the Type Ia supernova rate and its consequences on galactic chemical evolution}",
      journal = {\mnras},
         year = 2006,
        month = oct,
       volume = {372},
       number = {1},
        pages = {265-275},
          doi = {10.1111/j.1365-2966.2006.10848.x},
archivePrefix = {arXiv},
       eprint = {astro-ph/0607504},
 primaryClass = {astro-ph},
       adsurl = {https://ui.adsabs.harvard.edu/abs/2006MNRAS.372..265M}
}

@ARTICLE{Matteucci1986,
       author = {{Matteucci} and F. and {Greggio}, L.},
        title = "{Relative roles of type I and II supernovae in the chemical enrichment of the interstellar gas}",
      journal = {\aap},
         year = 1986,
        month = jan,
       volume = {154},
       number = {1-2},
        pages = {279-287},
       adsurl = {https://ui.adsabs.harvard.edu/abs/1986A&A...154..279M}
}

@ARTICLE{Marinacci2010,
       author = {{Marinacci}, Federico and {Binney}, James and {Fraternali}, Filippo and {Nipoti}, Carlo and {Ciotti}, Luca and {Londrillo}, Pasquale},
        title = "{The mode of gas accretion on to star-forming galaxies}",
      journal = {\mnras},
         year = 2010,
        month = may,
       volume = {404},
       number = {3},
        pages = {1464-1474},
          doi = {10.1111/j.1365-2966.2010.16352.x},
archivePrefix = {arXiv},
       eprint = {1001.2446},
 primaryClass = {astro-ph.GA},
       adsurl = {https://ui.adsabs.harvard.edu/abs/2010MNRAS.404.1464M}
}

@ARTICLE{Armillotta2016,
       author = {{Armillotta} and L. and {Fraternali} and F. and {Marinacci}, F.},
        title = "{Efficiency of gas cooling and accretion at the disc-corona interface}",
      journal = {\mnras},
         year = 2016,
        month = nov,
       volume = {462},
       number = {4},
        pages = {4157-4170},
          doi = {10.1093/mnras/stw1930},
archivePrefix = {arXiv},
       eprint = {1608.06290},
 primaryClass = {astro-ph.GA},
       adsurl = {https://ui.adsabs.harvard.edu/abs/2016MNRAS.462.4157A}
}

@ARTICLE{Katz1996,
       author = {{Katz}, Neal and {Weinberg}, David H. and {Hernquist}, Lars},
        title = "{Cosmological Simulations with TreeSPH}",
      journal = {\apjs},
         year = 1996,
        month = jul,
       volume = {105},
        pages = {19},
          doi = {10.1086/192305},
archivePrefix = {arXiv},
       eprint = {astro-ph/9509107},
 primaryClass = {astro-ph},
       adsurl = {https://ui.adsabs.harvard.edu/abs/1996ApJS..105...19K}
}

@ARTICLE{Della2008,
       author = {{Dalla Vecchia}, Claudio and {Schaye}, Joop},
        title = "{Simulating galactic outflows with kinetic supernova feedback}",
      journal = {\mnras},
         year = 2008,
        month = jul,
       volume = {387},
       number = {4},
        pages = {1431-1444},
          doi = {10.1111/j.1365-2966.2008.13322.x},
archivePrefix = {arXiv},
       eprint = {0801.2770},
 primaryClass = {astro-ph},
       adsurl = {https://ui.adsabs.harvard.edu/abs/2008MNRAS.387.1431D}
}

@ARTICLE{Krumholz2012,
       author = {{Krumholz}, Mark R. and {Dekel}, Avishai and {McKee}, Christopher F.},
        title = "{A Universal, Local Star Formation Law in Galactic Clouds, nearby Galaxies, High-redshift Disks, and Starbursts}",
      journal = {\apj},
         year = 2012,
        month = jan,
       volume = {745},
       number = {1},
          eid = {69},
        pages = {69},
          doi = {10.1088/0004-637X/745/1/69},
archivePrefix = {arXiv},
       eprint = {1109.4150},
 primaryClass = {astro-ph.CO},
       adsurl = {https://ui.adsabs.harvard.edu/abs/2012ApJ...745...69K}
}

@ARTICLE{Birnboim2003,
       author = {{Birnboim}, Yuval and {Dekel}, Avishai},
        title = "{Virial shocks in galactic haloes?}",
      journal = {\mnras},
         year = 2003,
        month = oct,
       volume = {345},
       number = {1},
        pages = {349-364},
          doi = {10.1046/j.1365-8711.2003.06955.x},
archivePrefix = {arXiv},
       eprint = {astro-ph/0302161},
 primaryClass = {astro-ph},
       adsurl = {https://ui.adsabs.harvard.edu/abs/2003MNRAS.345..349B}
}

@ARTICLE{Keres2005,
       author = {{Kere{\v{s}}}, Du{\v{s}}an and {Katz}, Neal and {Weinberg}, David H. and {Dav{\'e}}, Romeel},
        title = "{How do galaxies get their gas?}",
      journal = {\mnras},
         year = 2005,
        month = oct,
       volume = {363},
       number = {1},
        pages = {2-28},
          doi = {10.1111/j.1365-2966.2005.09451.x},
archivePrefix = {arXiv},
       eprint = {astro-ph/0407095},
 primaryClass = {astro-ph},
       adsurl = {https://ui.adsabs.harvard.edu/abs/2005MNRAS.363....2K}
}

@ARTICLE{Keres2009,
       author = {{Kere{\v{s}}}, Du{\v{s}}an and {Katz}, Neal and {Fardal}, Mark and {Dav{\'e}}, Romeel and {Weinberg}, David H.},
        title = "{Galaxies in a simulated {\ensuremath{\Lambda}}CDM Universe - I. Cold mode and hot cores}",
      journal = {\mnras},
         year = 2009,
        month = may,
       volume = {395},
       number = {1},
        pages = {160-179},
          doi = {10.1111/j.1365-2966.2009.14541.x},
archivePrefix = {arXiv},
       eprint = {0809.1430},
 primaryClass = {astro-ph},
       adsurl = {https://ui.adsabs.harvard.edu/abs/2009MNRAS.395..160K}
}

@ARTICLE{Rees1977,
       author = {{Rees}, M.~J. and {Ostriker}, J.~P.},
        title = "{Cooling, dynamics and fragmentation of massive gas clouds: clues to the masses and radii of galaxies and clusters.}",
      journal = {\mnras},
         year = 1977,
        month = jun,
       volume = {179},
        pages = {541-559},
          doi = {10.1093/mnras/179.4.541},
       adsurl = {https://ui.adsabs.harvard.edu/abs/1977MNRAS.179..541R}
}

@ARTICLE{Silk1977,
       author = {{Silk}, J.},
        title = "{On the fragmentation of cosmic gas clouds. I. The formation of galaxies and the first generation of stars.}",
      journal = {\apj},
         year = 1977,
        month = feb,
       volume = {211},
        pages = {638-648},
          doi = {10.1086/154972},
       adsurl = {https://ui.adsabs.harvard.edu/abs/1977ApJ...211..638S}
}

@ARTICLE{Mo1998,
       author = {{Mo}, H.~J. and {Mao}, Shude and {White}, Simon D.~M.},
        title = "{The formation of galactic discs}",
      journal = {\mnras},
         year = 1998,
        month = apr,
       volume = {295},
       number = {2},
        pages = {319-336},
          doi = {10.1046/j.1365-8711.1998.01227.x},
archivePrefix = {arXiv},
       eprint = {astro-ph/9707093},
 primaryClass = {astro-ph},
       adsurl = {https://ui.adsabs.harvard.edu/abs/1998MNRAS.295..319M}
}

@ARTICLE{Somerville2002,
       author = {{Somerville}, Rachel S.},
        title = "{Can Photoionization Squelching Resolve the Substructure Crisis?}",
      journal = {\apjl},
         year = 2002,
        month = jun,
       volume = {572},
       number = {1},
        pages = {L23-L26},
          doi = {10.1086/341444},
archivePrefix = {arXiv},
       eprint = {astro-ph/0107507},
 primaryClass = {astro-ph},
       adsurl = {https://ui.adsabs.harvard.edu/abs/2002ApJ...572L..23S}
}

@ARTICLE{Springel2005d,
       author = {{Springel}, Volker and {Di Matteo}, Tiziana and {Hernquist}, Lars},
        title = "{Modelling feedback from stars and black holes in galaxy mergers}",
      journal = {\mnras},
         year = 2005,
        month = aug,
       volume = {361},
       number = {3},
        pages = {776-794},
          doi = {10.1111/j.1365-2966.2005.09238.x},
archivePrefix = {arXiv},
       eprint = {astro-ph/0411108},
 primaryClass = {astro-ph},
       adsurl = {https://ui.adsabs.harvard.edu/abs/2005MNRAS.361..776S}
}

@article{Lynden-Bell1969,
  author = {Lynden-Bell, Donald},
  title = {Galactic Nuclei as Collapsed Old Quasars},
  journal = {Nature},
  volume = {223},
  pages = {690–694},
  year = {1969},
  doi = {10.1038/223690a0}
}

@article{Schmidt1969,
  author = {Schmidt, Maarten},
  title = {3C 273: A Star-Like Object with Large Red-Shift},
  journal = {Nature},
  volume = {197},
  pages = {1040–1040},
  year = {1963},
  doi = {10.1038/1971040a0}
}

@article{Greenstein1963,
  author = {Greenstein, Jesse L. and Matthews and T. A.},
  title = {Red-Shift of the Emission Lines in the Optical Spectrum of 3C 48},
  journal = {\apj},
  volume = {138},
  pages = {873},
  year = {1963},
  doi = {10.1086/147700}
}

@article{Schmidt1968,
  author = {Schmidt, Maarten},
  title = {Space Distribution and Luminosity Functions of Quasi-Stellar Radio Sources},
  journal = {\apj},
  volume = {151},
  pages = {393},
  year = {1968},
  doi = {10.1086/149446}
}

@article{Baldwin1977,
  author = {Baldwin, J. A.},
  title = {Luminosity Indicators in the Spectra of Quasi-Stellar Objects},
  journal = {\apj},
  volume = {214},
  pages = {679–684},
  year = {1977},
  doi = {10.1086/155294}
}

@article{Elvis1994,
  author = {Elvis, Martin and Wilkes, Belinda J. and McDowell, Jonathan C. and Green, Richard F. and Bechtold, Jill and Willner and S. P. and Oey and M. S. and Polomski and E. and Cutri, R.},
  title = {Atlas of Quasar Energy Distributions},
  journal = {\apjs},
  volume = {95},
  pages = {1–68},
  year = {1994},
  doi = {10.1086/192093}
}

@article{Rees1984,
  author = {Rees, M. J.},
  title = {Black Hole Models for Active Galactic Nuclei},
  journal = {\araa},
  volume = {22},
  pages = {471–506},
  year = {1984},
  doi = {10.1146/annurev.aa.22.090184.002351}
}

@article{Shakura1973,
  author = {Shakura and N. I. and Sunyaev and R. A.},
  title = {Black holes in binary systems. Observational appearance},
  journal = {Astronomy and Astrophysics},
  volume = {24},
  pages = {337–355},
  year = {1973}
}

@ARTICLE{Thorne1974,
       author = {{Thorne}, Kip S.},
        title = "{Disk-Accretion onto a Black Hole. II. Evolution of the Hole}",
      journal = {\apj},
         year = 1974,
        month = jul,
       volume = {191},
        pages = {507-520},
          doi = {10.1086/152991},
       adsurl = {https://ui.adsabs.harvard.edu/abs/1974ApJ...191..507T}
}

@ARTICLE{King2003a,
       author = {{King}, Andrew},
        title = "{Black Holes, Galaxy Formation, and the M$_{BH}$-{\ensuremath{\sigma}} Relation}",
      journal = {\apjl},
         year = 2003,
        month = oct,
       volume = {596},
       number = {1},
        pages = {L27-L29},
          doi = {10.1086/379143},
archivePrefix = {arXiv},
       eprint = {astro-ph/0308342},
 primaryClass = {astro-ph},
       adsurl = {https://ui.adsabs.harvard.edu/abs/2003ApJ...596L..27K}
}

@ARTICLE{King2003b,
       author = {{King}, A.~R. and {Pounds}, K.~A.},
        title = "{Black hole winds}",
      journal = {\mnras},
         year = 2003,
        month = oct,
       volume = {345},
       number = {2},
        pages = {657-659},
          doi = {10.1046/j.1365-8711.2003.06980.x},
archivePrefix = {arXiv},
       eprint = {astro-ph/0305541},
 primaryClass = {astro-ph},
       adsurl = {https://ui.adsabs.harvard.edu/abs/2003MNRAS.345..657K}
}

@ARTICLE{Murray2005,
       author = {{Murray}, Norman and {Quataert}, Eliot and {Thompson}, Todd A.},
        title = "{On the Maximum Luminosity of Galaxies and Their Central Black Holes: Feedback from Momentum-driven Winds}",
      journal = {\apj},
         year = 2005,
        month = jan,
       volume = {618},
       number = {2},
        pages = {569-585},
          doi = {10.1086/426067},
archivePrefix = {arXiv},
       eprint = {astro-ph/0406070},
 primaryClass = {astro-ph},
       adsurl = {https://ui.adsabs.harvard.edu/abs/2005ApJ...618..569M}
}

@book{Eddington1926,
  author = {Eddington, A. S.},
  title = {The Internal Constitution of the Stars},
  publisher = {Cambridge University Press},
  year = {1926}
}

@article{Hoyle1939,
  author = {Hoyle, F. and Lyttleton, R. A.},
  title = {The effect of interstellar matter on climatic variation},
  journal = {\mnras},
  volume = {99},
  pages = {180–182},
  year = {1939},
  adsurl = {https://ui.adsabs.harvard.edu/abs/1939MNRAS..99..180H}
}

@article{Bondi1944,
  author = {Bondi, H. and Hoyle, F.},
  title = {On the mechanism of accretion by stars},
  journal = {\mnras},
  volume = {104},
  pages = {273–282},
  year = {1944},
  adsurl = {https://ui.adsabs.harvard.edu/abs/1944MNRAS.104..273B}
}

@article{Bondi1952,
  author = {Bondi, H.},
  title = {On spherically symmetrical accretion},
  journal = {\mnras},
  volume = {112},
  pages = {195–204},
  year = {1952},
  adsurl = {https://ui.adsabs.harvard.edu/abs/1952MNRAS.112..195B}
}

@article{Narayan1994,
  author = {Narayan, Ramesh and Yi, Insu},
  title = {Advection-dominated Accretion: A Self-similar Solution},
  journal = {\apjl},
  volume = {428},
  pages = {L13–L16},
  year = {1994},
  doi = {10.1086/187381}
}

@article{Narayan1995,
  author = {Narayan, Ramesh and Yi, Insu},
  title = {Advection-dominated Accretion: Underfed Black Holes and Neutron Stars},
  journal = {\apj},
  volume = {452},
  pages = {710–735},
  year = {1995},
  doi = {10.1086/176343}
}

@ARTICLE{Heckman2014,
       author = {{Heckman}, Timothy M. and {Best}, Philip N.},
        title = "{The Coevolution of Galaxies and Supermassive Black Holes: Insights from Surveys of the Contemporary Universe}",
      journal = {\araa},
         year = 2014,
        month = aug,
       volume = {52},
        pages = {589-660},
          doi = {10.1146/annurev-astro-081913-035722},
archivePrefix = {arXiv},
       eprint = {1403.4620},
 primaryClass = {astro-ph.GA},
       adsurl = {https://ui.adsabs.harvard.edu/abs/2014ARA&A..52..589H}
}

@article{Costa2015,
  author = {Costa, Tiago and Sijacki, Debora and Haehnelt, Martin G.},
  title = {Feedback from active galactic nuclei in simulations of galaxy formation},
  journal = {\mnras},
  volume = {448},
  pages = {L30--L34},
  year = {2015},
  doi = {10.1093/mnrasl/slu193}
}

@article{Gabor2014,
  author = {Gabor, Jared M. and Bournaud, Fr{\'e}d{\'e}ric},
  title = {AGN feedback in high-redshift disk galaxies: the outflows do not stop star formation},
  journal = {\mnras},
  volume = {441},
  pages = {1615--1627},
  year = {2014},
  doi = {10.1093/mnras/stu687}
}

@article{Angles2017,
  author = {Angl{\'e}s-Alc{\'a}zar, Daniel and Faucher-Giguère, Claude-Andr{\'e} and Quataert, Eliot and Hopkins, Philip F. and Feldmann, Robert and Torrey, Paul and Wetzel, Andrew and Kereš, Dušan},
  title = {Quasar feedback and the inefficient quenching of massive galaxies at high redshift},
  journal = {MNRAS Lett.},
  volume = {472},
  pages = {L109--L114},
  year = {2017},
  doi = {10.1093/mnrasl/slx155}
}

@article{Noguchi1999,
  author = {Noguchi, Masafumi},
  title = {Early Evolution of Disk Galaxies: Formation of Bulges in Clumpy Young Galaxies},
  journal = {\apj},
  volume = {514},
  pages = {77--95},
  year = {1999},
  doi = {10.1086/306926}
}

@article{Dekel2009b,
  author = {Dekel, Avishai and Sari, Re'em and Ceverino, Daniel},
  title = {Formation of Massive Galaxies at High Redshift: Cold Streams, Clumpy Disks, and Compact Spheroids},
  journal = {\apj},
  volume = {703},
  pages = {785--801},
  year = {2009},
  doi = {10.1088/0004-637X/703/1/785}
}

@article{Ceverino2010,
  author = {Ceverino, Daniel and Dekel, Avishai and Bournaud, Fr{\'e}d{\'e}ric},
  title = {High-redshift clumpy disks and bulges in cosmological simulations},
  journal = {\mnras},
  volume = {404},
  pages = {2151--2169},
  year = {2010},
  doi = {10.1111/j.1365-2966.2010.16433.x}
}

@article{Ellison2011,
  author = {Ellison and S. L. and Patton and D. R. and Mendel and J. T. and Scudder and J. M.},
  title = {Galaxy pairs in the Sloan Digital Sky Survey - VII. The merger-AGN connection},
  journal = {\mnras},
  volume = {418},
  pages = {2043--2055},
  year = {2011},
  doi = {10.1111/j.1365-2966.2011.19562.x}
}

@article{Ellison2013,
  author = {Ellison and S. L. and Mendel and J. T. and Patton and D. R. and Scudder and J. M.},
  title = {Galaxy pairs in the Sloan Digital Sky Survey - VIII. The observational properties of post-merger galaxies},
  journal = {\mnras},
  volume = {435},
  pages = {3627--3647},
  year = {2013},
  doi = {10.1093/mnras/stt1562}
}

@article{Ellison2024,
  author = {Ellison and S. L. and Ferreira and L. and Bickley and R. and Grindlay and T. and Salim and S. and Byrne-Mamahit and S. and Satyapal and S. and Patton and D. R. and Scudder and J. M.},
  title = {Galaxy evolution in the post-merger regime. III -- The triggering of active galactic nuclei peaks immediately after coalescence},
  journal = {\mnras},
  volume = {518},
  pages = {1234--1248},
  year = {2024},
  doi = {10.1093/mnras/stad1234}
}

@ARTICLE{Haring2004,
       author = {{H{\"a}ring}, Nadine and {Rix}, Hans-Walter},
        title = "{On the Black Hole Mass-Bulge Mass Relation}",
      journal = {\apjl},
         year = 2004,
        month = apr,
       volume = {604},
       number = {2},
        pages = {L89-L92},
          doi = {10.1086/383567},
archivePrefix = {arXiv},
       eprint = {astro-ph/0402376},
 primaryClass = {astro-ph},
       adsurl = {https://ui.adsabs.harvard.edu/abs/2004ApJ...604L..89H}
}

@ARTICLE{Maggorian1998,
       author = {{Magorrian}, John and {Tremaine}, Scott and {Richstone}, Douglas and {Bender}, Ralf and {Bower}, Gary and {Dressler}, Alan and {Faber}, S.~M. and {Gebhardt}, Karl and {Green}, Richard and {Grillmair}, Carl and {Kormendy}, John and {Lauer}, Tod},
        title = "{The Demography of Massive Dark Objects in Galaxy Centers}",
      journal = {\aj},
         year = 1998,
        month = jun,
       volume = {115},
       number = {6},
        pages = {2285-2305},
          doi = {10.1086/300353},
archivePrefix = {arXiv},
       eprint = {astro-ph/9708072},
 primaryClass = {astro-ph},
       adsurl = {https://ui.adsabs.harvard.edu/abs/1998AJ....115.2285M}
}

@ARTICLE{Aird2015,
       author = {{Aird} and J. and {Coil}, A.~L. and {Georgakakis} and A. and {Nandra} and K. and {Barro} and G. and {P{\'e}rez-Gonz{\'a}lez}, P.~G.},
        title = "{The evolution of the X-ray luminosity functions of unabsorbed and absorbed AGNs out to $z \sim 5$}",
      journal = {\mnras},
         year = 2015,
        month = aug,
       volume = {451},
       number = {2},
        pages = {1892-1927},
          doi = {10.1093/mnras/stv1062},
archivePrefix = {arXiv},
       eprint = {1503.01120},
 primaryClass = {astro-ph.HE},
       adsurl = {https://ui.adsabs.harvard.edu/abs/2015MNRAS.451.1892A}
}

@ARTICLE{Aird2010,
       author = {{Aird} and J. and {Nandra} and K. and {Laird}, E.~S. and {Georgakakis} and A. and {Ashby}, M.~L.~N. and {Barmby} and P. and {Coil}, A.~L. and {Huang} and J. -S. and {Koekemoer}, A.~M. and {Steidel}, C.~C. and {Willmer}, C.~N.~A.},
        title = "{The evolution of the hard X-ray luminosity function of AGN}",
      journal = {\mnras},
         year = 2010,
        month = feb,
       volume = {401},
       number = {4},
        pages = {2531-2551},
          doi = {10.1111/j.1365-2966.2009.15829.x},
archivePrefix = {arXiv},
       eprint = {0910.1141},
 primaryClass = {astro-ph.CO},
       adsurl = {https://ui.adsabs.harvard.edu/abs/2010MNRAS.401.2531A}
}

@article{Ellison2019,
  author = {Ellison and S. L. and Viswanathan and A. and Patton and D. R. and Bottrell and C. and McConnachie and A. W. and Gwyn and S. and Cuillandre, J.-C.},
  title = {A definitive merger-AGN connection at $z \sim 0$ with CFIS: mergers have an excess of AGN and AGN hosts are more frequently disturbed},
  journal = {\mnras},
  volume = {487},
  pages = {2491--2504},
  year = {2019},
  doi = {10.1093/mnras/stz1453}
}

@article{Barnes1991,
  author = {Barnes and J. E. and Hernquist and L. E.},
  title = {Fueling starburst galaxies with gas-rich mergers},
  journal = {\apjl},
  volume = {370},
  pages = {L65--L68},
  year = {1991},
  doi = {10.1086/185978}
}

@article{Mihos1996,
  author = {Mihos and J. C. and Hernquist, L.},
  title = {Gasdynamics and Starbursts in Major Mergers},
  journal = {\apj},
  volume = {464},
  pages = {641--663},
  year = {1996},
  doi = {10.1086/177353}
}

@article{Barnes2002,
  author = {Barnes, Joshua E.},
  title = {Inflow and Starburst in Merging Galaxies},
  journal = {\mnras},
  volume = {333},
  pages = {481--494},
  year = {2002},
  doi = {10.1046/j.1365-8711.2002.05432.x}
}

@ARTICLE{Fluetsch2021,
       author = {{Fluetsch} and A. and {Maiolino} and R. and {Carniani} and S. and {Arribas} and S. and {Belfiore} and F. and {Bellocchi} and E. and {Cazzoli} and S. and {Cicone} and C. and {Cresci} and G. and {Fabian}, A.~C. and {Gallagher} and R. and {Ishibashi} and W. and {Mannucci} and F. and {Marconi} and A. and {Perna} and M. and {Sturm} and E. and {Venturi}, G.},
        title = "{Properties of the multiphase outflows in local (ultra)luminous infrared galaxies}",
      journal = {\mnras},
         year = 2021,
        month = aug,
       volume = {505},
       number = {4},
        pages = {5753-5783},
          doi = {10.1093/mnras/stab1666},
archivePrefix = {arXiv},
       eprint = {2006.13232},
 primaryClass = {astro-ph.GA},
       adsurl = {https://ui.adsabs.harvard.edu/abs/2021MNRAS.505.5753F}
}

@ARTICLE{Tacchella2016,
       author = {{Tacchella}, Sandro and {Dekel}, Avishai and {Carollo} and C. Marcella and {Ceverino}, Daniel and {DeGraf}, Colin and {Lapiner}, Sharon and {Mandelker}, Nir and {Primack Joel}, R.},
        title = "{The confinement of star-forming galaxies into a main sequence through episodes of gas compaction, depletion and replenishment}",
      journal = {\mnras},
         year = 2016,
        month = apr,
       volume = {457},
       number = {3},
        pages = {2790-2813},
          doi = {10.1093/mnras/stw131},
archivePrefix = {arXiv},
       eprint = {1509.02529},
 primaryClass = {astro-ph.GA},
       adsurl = {https://ui.adsabs.harvard.edu/abs/2016MNRAS.457.2790T}
}

@ARTICLE{Lilly2013,
       author = {{Lilly}, Simon J. and {Carollo} and C. Marcella and {Pipino}, Antonio and {Renzini}, Alvio and {Peng}, Yingjie},
        title = "{Gas Regulation of Galaxies: The Evolution of the Cosmic Specific Star Formation Rate, the Metallicity-Mass-Star-formation Rate Relation, and the Stellar Content of Halos}",
      journal = {\apj},
         year = 2013,
        month = aug,
       volume = {772},
       number = {2},
          eid = {119},
        pages = {119},
          doi = {10.1088/0004-637X/772/2/119},
archivePrefix = {arXiv},
       eprint = {1303.5059},
 primaryClass = {astro-ph.CO},
       adsurl = {https://ui.adsabs.harvard.edu/abs/2013ApJ...772..119L}
}

@ARTICLE{Omand2014,
       author = {{Omand}, Conor M.~B. and {Balogh}, Michael L. and {Poggianti}, Bianca M.},
        title = "{The connection between galaxy structure and quenching efficiency}",
      journal = {\mnras},
         year = 2014,
        month = may,
       volume = {440},
       number = {1},
        pages = {843-858},
          doi = {10.1093/mnras/stu331},
archivePrefix = {arXiv},
       eprint = {1402.3394},
 primaryClass = {astro-ph.GA},
       adsurl = {https://ui.adsabs.harvard.edu/abs/2014MNRAS.440..843O}
}

@ARTICLE{Genel2014,
       author = {{Genel}, Shy and {Vogelsberger}, Mark and {Springel}, Volker and {Sijacki}, Debora and {Nelson}, Dylan and {Snyder}, Greg and {Rodriguez-Gomez}, Vicente and {Torrey}, Paul and {Hernquist}, Lars},
        title = "{Introducing the Illustris project: the evolution of galaxy populations across cosmic time}",
      journal = {\mnras},
         year = 2014,
        month = nov,
       volume = {445},
       number = {1},
        pages = {175-200},
          doi = {10.1093/mnras/stu1654},
archivePrefix = {arXiv},
       eprint = {1405.3749},
 primaryClass = {astro-ph.CO},
       adsurl = {https://ui.adsabs.harvard.edu/abs/2014MNRAS.445..175G}
}

@ARTICLE{Suresh2015,
       author = {{Suresh}, Joshua and {Bird}, Simeon and {Vogelsberger}, Mark and {Genel}, Shy and {Torrey}, Paul and {Sijacki}, Debora and {Springel}, Volker and {Hernquist}, Lars},
        title = "{The impact of galactic feedback on the circumgalactic medium}",
      journal = {\mnras},
         year = 2015,
        month = mar,
       volume = {448},
       number = {1},
        pages = {895-909},
          doi = {10.1093/mnras/stu2762},
archivePrefix = {arXiv},
       eprint = {1501.02267},
 primaryClass = {astro-ph.GA},
       adsurl = {https://ui.adsabs.harvard.edu/abs/2015MNRAS.448..895S}
}

@ARTICLE{Nelson2015,
       author = {{Nelson}, Dylan and {Genel}, Shy and {Vogelsberger}, Mark and {Springel}, Volker and {Sijacki}, Debora and {Torrey}, Paul and {Hernquist}, Lars},
        title = "{The impact of feedback on cosmological gas accretion}",
      journal = {\mnras},
         year = 2015,
        month = mar,
       volume = {448},
       number = {1},
        pages = {59-74},
          doi = {10.1093/mnras/stv017},
archivePrefix = {arXiv},
       eprint = {1410.5425},
 primaryClass = {astro-ph.CO},
       adsurl = {https://ui.adsabs.harvard.edu/abs/2015MNRAS.448...59N}
}

@ARTICLE{Naiman2018,
       author = {{Naiman}, Jill P. and {Pillepich}, Annalisa and {Springel}, Volker and {Ramirez-Ruiz}, Enrico and {Torrey}, Paul and {Vogelsberger}, Mark and {Pakmor}, R{\"u}diger and {Nelson}, Dylan and {Marinacci}, Federico and {Hernquist}, Lars and {Weinberger}, Rainer and {Genel}, Shy},
        title = "{First results from the IllustrisTNG simulations: a tale of two elements - chemical evolution of magnesium and europium}",
      journal = {\mnras},
         year = 2018,
        month = jun,
       volume = {477},
       number = {1},
        pages = {1206-1224},
          doi = {10.1093/mnras/sty618},
archivePrefix = {arXiv},
       eprint = {1707.03401},
 primaryClass = {astro-ph.GA},
       adsurl = {https://ui.adsabs.harvard.edu/abs/2018MNRAS.477.1206N}
}

@ARTICLE{Vogelsberger2013,
       author = {{Vogelsberger}, Mark and {Genel}, Shy and {Sijacki}, Debora and {Torrey}, Paul and {Springel}, Volker and {Hernquist}, Lars},
        title = "{A model for cosmological simulations of galaxy formation physics}",
      journal = {\mnras},
         year = 2013,
        month = dec,
       volume = {436},
       number = {4},
        pages = {3031-3067},
          doi = {10.1093/mnras/stt1789},
archivePrefix = {arXiv},
       eprint = {1305.2913},
 primaryClass = {astro-ph.CO},
       adsurl = {https://ui.adsabs.harvard.edu/abs/2013MNRAS.436.3031V}
}

@ARTICLE{Scheuer1974,
       author = {{Scheuer}, P.~A.~G.},
        title = "{Models of extragalactic radio sources with a continuous energy supply from a central object}",
      journal = {\mnras},
         year = 1974,
        month = mar,
       volume = {166},
        pages = {513-528},
          doi = {10.1093/mnras/166.3.513},
       adsurl = {https://ui.adsabs.harvard.edu/abs/1974MNRAS.166..513S}
}

@ARTICLE{Dubois2014,
       author = {{Dubois} and Y. and {Pichon} and C. and {Welker} and C. and {Le Borgne} and D. and {Devriendt} and J. and {Laigle} and C. and {Codis} and S. and {Pogosyan} and D. and {Arnouts} and S. and {Benabed} and K. and {Bertin} and E. and {Blaizot} and J. and {Bouchet} and F. and {Cardoso} and J. -F. and {Colombi} and S. and {de Lapparent} and V. and {Desjacques} and V. and {Gavazzi} and R. and {Kassin} and S. and {Kimm} and T. and {McCracken} and H. and {Milliard} and B. and {Peirani} and S. and {Prunet} and S. and {Rouberol} and S. and {Silk} and J. and {Slyz} and A. and {Sousbie} and T. and {Teyssier} and R. and {Tresse} and L. and {Treyer} and M. and {Vibert} and D. and {Volonteri}, M.},
        title = "{Dancing in the dark: galactic properties trace spin swings along the cosmic web}",
      journal = {\mnras},
         year = 2014,
        month = oct,
       volume = {444},
       number = {2},
        pages = {1453-1468},
          doi = {10.1093/mnras/stu1227},
archivePrefix = {arXiv},
       eprint = {1402.1165},
 primaryClass = {astro-ph.CO},
       adsurl = {https://ui.adsabs.harvard.edu/abs/2014MNRAS.444.1453D}
}

@ARTICLE{Hirschmann2014,
       author = {{Hirschmann}, Michaela and {Dolag}, Klaus and {Saro}, Alexandro and {Bachmann}, Lisa and {Borgani}, Stefano and {Burkert}, Andreas},
        title = "{Cosmological simulations of black hole growth: AGN luminosities and downsizing}",
      journal = {\mnras},
         year = 2014,
        month = aug,
       volume = {442},
       number = {3},
        pages = {2304-2324},
          doi = {10.1093/mnras/stu1023},
archivePrefix = {arXiv},
       eprint = {1308.0333},
 primaryClass = {astro-ph.CO},
       adsurl = {https://ui.adsabs.harvard.edu/abs/2014MNRAS.442.2304H}
}

@ARTICLE{McCarthy2017,
       author = {{McCarthy}, Ian G. and {Schaye}, Joop and {Bird}, Simeon and {Le Brun}, Amandine M.~C.},
        title = "{The BAHAMAS project: calibrated hydrodynamical simulations for large-scale structure cosmology}",
      journal = {\mnras},
         year = 2017,
        month = mar,
       volume = {465},
       number = {3},
        pages = {2936-2965},
          doi = {10.1093/mnras/stw2792},
archivePrefix = {arXiv},
       eprint = {1603.02702},
 primaryClass = {astro-ph.CO},
       adsurl = {https://ui.adsabs.harvard.edu/abs/2017MNRAS.465.2936M}
}

@ARTICLE{Tremmel2017,
       author = {{Tremmel} and M. and {Karcher} and M. and {Governato} and F. and {Volonteri} and M. and {Quinn}, T.~R. and {Pontzen} and A. and {Anderson} and L. and {Bellovary}, J.},
        title = "{The Romulus cosmological simulations: a physical approach to the formation, dynamics and accretion models of SMBHs}",
      journal = {\mnras},
         year = 2017,
        month = sep,
       volume = {470},
       number = {1},
        pages = {1121-1139},
          doi = {10.1093/mnras/stx1160},
archivePrefix = {arXiv},
       eprint = {1607.02151},
 primaryClass = {astro-ph.GA},
       adsurl = {https://ui.adsabs.harvard.edu/abs/2017MNRAS.470.1121T}
}

@ARTICLE{Feldmann2023,
       author = {{Feldmann}, Robert and {Quataert}, Eliot and {Faucher-Gigu{\`e}re}, Claude-Andr{\'e} and {Hopkins}, Philip F. and {{\c{C}}atmabacak}, Onur and {Kere{\v{s}}}, Du{\v{s}}an and {Bassini}, Luigi and {Bernardini}, Mauro and {Bullock}, James S. and {Cenci}, Elia and {Gensior}, Jindra and {Liang}, Lichen and {Moreno}, Jorge and {Wetzel}, Andrew},
        title = "{FIREbox: simulating galaxies at high dynamic range in a cosmological volume}",
      journal = {\mnras},
         year = 2023,
        month = jul,
       volume = {522},
       number = {3},
        pages = {3831-3860},
          doi = {10.1093/mnras/stad1205},
archivePrefix = {arXiv},
       eprint = {2205.15325},
 primaryClass = {astro-ph.GA},
       adsurl = {https://ui.adsabs.harvard.edu/abs/2023MNRAS.522.3831F}
}

@ARTICLE{LeBrun2014,
       author = {{Le Brun}, Amandine M.~C. and {McCarthy}, Ian G. and {Schaye}, Joop and {Ponman}, Trevor J.},
        title = "{Towards a realistic population of simulated galaxy groups and clusters}",
      journal = {\mnras},
         year = 2014,
        month = jun,
       volume = {441},
       number = {2},
        pages = {1270-1290},
          doi = {10.1093/mnras/stu608},
archivePrefix = {arXiv},
       eprint = {1312.5462},
 primaryClass = {astro-ph.CO},
       adsurl = {https://ui.adsabs.harvard.edu/abs/2014MNRAS.441.1270L}
}

@article{Gnedin2000,
  author = {Gnedin, Nickolay Y.},
  title = {Cosmological Reionization by Stellar Sources},
  journal = {\apj},
  volume = {535},
  number = {2},
  pages = {530--554},
  year = {2000},
  doi = {10.1086/308868}
}

@article{Nagamine2004,
  author = {Nagamine, Kentaro and Springel, Volker and Hernquist, Lars and Machacek, Michael},
  title = {Star Formation History and the Metal Enrichment of the Intergalactic Medium},
  journal = {\mnras},
  volume = {348},
  number = {2},
  pages = {421--442},
  year = {2004},
  doi = {10.1111/j.1365-2966.2004.07383.x}
}

@article{Crain2009,
  author = {Crain, Robert A. and Theuns, Tom and Dalla Vecchia, Claudio and Eke, Vincent R. and Frenk, Carlos S. and Jenkins, Adrian and Kay, Scott T. and Peacock, John A. and Pearce, Frazer R. and Schaye, Joop and Springel, Volker and Thomas, Peter A. and White, Simon D. M.},
  title = {Galaxies-Intergalactic Medium Interaction Calculation - I. Galaxy formation as a function of large-scale environment},
  journal = {\mnras},
  volume = {399},
  number = {4},
  pages = {1773--1794},
  year = {2009},
  doi = {10.1111/j.1365-2966.2009.15398.x}
}

@article{Schaye2010,
  author = {Schaye, Joop and Dalla Vecchia, Claudio and Booth and C. M. and Wiersma and R. P. C. and Theuns, Tom and Haas and M. R. and Bertone, Serena and Duffy and A. R. and McCarthy and I. G. and van de Voort, Freeke},
  title = {The physics driving the cosmic star formation history},
  journal = {\mnras},
  volume = {402},
  number = {3},
  pages = {1536--1560},
  year = {2010},
  doi = {10.1111/j.1365-2966.2009.16029.x}
}

@article{DiMatteo2012,
  author = {Di Matteo, Tiziana and Khandai, Nishikanta and DeGraf, Colin and Feng, Yu and Croft, Rupert A. C. and Lopez, Javier and Springel, Volker},
  title = {Cold Flows and the First Quasars},
  journal = {\apjl},
  volume = {745},
  number = {1},
  pages = {L29},
  year = {2012},
  doi = {10.1088/2041-8205/745/1/L29}
}

@ARTICLE{Hunter2001,
       author = {{Hunter}, Deidre A. and {Elmegreen}, Bruce G. and {van Woerden}, Hugo},
        title = "{Neutral Hydrogen and Star Formation in the Irregular Galaxy NGC 2366}",
      journal = {\apj},
         year = 2001,
        month = aug,
       volume = {556},
       number = {2},
        pages = {773-800},
          doi = {10.1086/321611},
archivePrefix = {arXiv},
       eprint = {astro-ph/0104091},
 primaryClass = {astro-ph},
       adsurl = {https://ui.adsabs.harvard.edu/abs/2001ApJ...556..773H}
}

@ARTICLE{Elmegreen2002,
       author = {{Elmegreen}, Bruce G.},
        title = "{Star Formation from Galaxies to Globules}",
      journal = {\apj},
         year = 2002,
        month = sep,
       volume = {577},
       number = {1},
        pages = {206-220},
          doi = {10.1086/342177},
archivePrefix = {arXiv},
       eprint = {astro-ph/0207114},
 primaryClass = {astro-ph},
       adsurl = {https://ui.adsabs.harvard.edu/abs/2002ApJ...577..206E}
}

@ARTICLE{Kawata2007,
       author = {{Kawata}, Daisuke and {Cen}, Renyue and {Ho}, Luis C.},
        title = "{Gravitational Stability of Circumnuclear Disks in Elliptical Galaxies}",
      journal = {\apj},
         year = 2007,
        month = nov,
       volume = {669},
       number = {1},
        pages = {232-240},
          doi = {10.1086/521299},
archivePrefix = {arXiv},
       eprint = {0706.0005},
 primaryClass = {astro-ph},
       adsurl = {https://ui.adsabs.harvard.edu/abs/2007ApJ...669..232K}
}

@ARTICLE{Gensior2021,
       author = {{Gensior}, Jindra and {Kruijssen}, J.~M. Diederik},
        title = "{The elephant in the bathtub: when the physics of star formation regulate the baryon cycle of galaxies}",
      journal = {\mnras},
         year = 2021,
        month = jan,
       volume = {500},
       number = {2},
        pages = {2000-2011},
          doi = {10.1093/mnras/staa3453},
archivePrefix = {arXiv},
       eprint = {2011.01235},
 primaryClass = {astro-ph.GA},
       adsurl = {https://ui.adsabs.harvard.edu/abs/2021MNRAS.500.2000G}
}

@ARTICLE{Toomre1964,
       author = {{Toomre}, A.},
        title = "{On the gravitational stability of a disk of stars.}",
      journal = {\apj},
         year = 1964,
        month = may,
       volume = {139},
        pages = {1217-1238},
          doi = {10.1086/147861},
       adsurl = {https://ui.adsabs.harvard.edu/abs/1964ApJ...139.1217T}
}

@ARTICLE{Lin2019,
       author = {{Lin}, Lihwai and {Pan}, Hsi-An and {Ellison}, Sara L. and {Belfiore}, Francesco and {Shi}, Yong and {S{\'a}nchez}, Sebasti{\'a}n F. and {Hsieh}, Bau-Ching and {Rowlands}, Kate and {Ramya} and S. and {Thorp}, Mallory D. and {Li}, Cheng and {Maiolino}, Roberto},
        title = "{The ALMaQUEST Survey: The Molecular Gas Main Sequence and the Origin of the Star-forming Main Sequence}",
      journal = {\apjl},
         year = 2019,
        month = oct,
       volume = {884},
       number = {2},
          eid = {L33},
        pages = {L33},
          doi = {10.3847/2041-8213/ab4815},
archivePrefix = {arXiv},
       eprint = {1909.11243},
 primaryClass = {astro-ph.GA},
       adsurl = {https://ui.adsabs.harvard.edu/abs/2019ApJ...884L..33L}
}

@ARTICLE{Boylan2008,
       author = {{Boylan-Kolchin}, Michael and {Ma}, Chung-Pei and {Quataert}, Eliot},
        title = "{Dynamical friction and galaxy merging time-scales}",
      journal = {\mnras},
         year = 2008,
        month = jan,
       volume = {383},
       number = {1},
        pages = {93-101},
          doi = {10.1111/j.1365-2966.2007.12530.x},
archivePrefix = {arXiv},
       eprint = {0707.2960},
 primaryClass = {astro-ph},
       adsurl = {https://ui.adsabs.harvard.edu/abs/2008MNRAS.383...93B}
}

@BOOK{Binney2008,
       author = {{Binney}, James and {Tremaine}, Scott},
        title = {Galactic Dynamics},
      edition = {2nd},
      address = {Princeton, NJ},
    publisher = {Princeton University Press},
         year = 2008,
       adsurl = {https://ui.adsabs.harvard.edu/abs/2008gady.book.....B}
}

@article{Chandrasekhar1943,
  author       = {Chandrasekhar, S.},
  title        = {Dynamical Friction. I. General Considerations: the Coefficient of Dynamical Friction},
  journal      = {\apj},
  year         = {1943},
  volume       = {97},
  pages        = {255--262},
  doi          = {10.1086/144517}
}

@ARTICLE{Abadi1999,
       author = {{Abadi}, Mario G. and {Moore}, Ben and {Bower}, Richard G.},
        title = "{Ram pressure stripping of spiral galaxies in clusters}",
      journal = {\mnras},
         year = 1999,
        month = oct,
       volume = {308},
       number = {4},
        pages = {947-954},
          doi = {10.1046/j.1365-8711.1999.02715.x},
archivePrefix = {arXiv},
       eprint = {astro-ph/9903436},
 primaryClass = {astro-ph},
       adsurl = {https://ui.adsabs.harvard.edu/abs/1999MNRAS.308..947A}
}

@ARTICLE{Roediger2005,
       author = {{Roediger} and E. and {Hensler}, G.},
        title = "{Ram pressure stripping of disk galaxies. From high to low density environments}",
      journal = {\aap},
         year = 2005,
        month = apr,
       volume = {433},
       number = {3},
        pages = {875-895},
          doi = {10.1051/0004-6361:20042131},
       adsurl = {https://ui.adsabs.harvard.edu/abs/2005A&A...433..875R}
}

@ARTICLE{Tonnesen2009,
       author = {{Tonnesen}, Stephanie and {Bryan}, Greg L.},
        title = "{Gas Stripping in Simulated Galaxies with a Multiphase Interstellar Medium}",
      journal = {\apj},
         year = 2009,
        month = apr,
       volume = {694},
       number = {2},
        pages = {789-804},
          doi = {10.1088/0004-637X/694/2/789},
archivePrefix = {arXiv},
       eprint = {0901.2115},
 primaryClass = {astro-ph.GA},
       adsurl = {https://ui.adsabs.harvard.edu/abs/2009ApJ...694..789T}
}

@ARTICLE{Jachym2007,
       author = {{J{\'a}chym} and P. and {Palou{\v{s}}} and J. and {K{\"o}ppen} and J. and {Combes}, F.},
        title = "{Gas stripping in galaxy clusters: a new SPH simulation approach}",
      journal = {\aap},
         year = 2007,
        month = sep,
       volume = {472},
       number = {1},
        pages = {5-20},
          doi = {10.1051/0004-6361:20066442},
archivePrefix = {arXiv},
       eprint = {0706.3631},
 primaryClass = {astro-ph},
       adsurl = {https://ui.adsabs.harvard.edu/abs/2007A&A...472....5J}
}

@ARTICLE{Jachym2009,
       author = {{J{\'a}chym} and P. and {K{\"o}ppen} and J. and {Palou{\v{s}}} and J. and {Combes}, F.},
        title = "{Ram pressure stripping of tilted galaxies}",
      journal = {\aap},
         year = 2009,
        month = jun,
       volume = {500},
       number = {2},
        pages = {693-703},
          doi = {10.1051/0004-6361/200811469},
archivePrefix = {arXiv},
       eprint = {0904.3886},
 primaryClass = {astro-ph.CO},
       adsurl = {https://ui.adsabs.harvard.edu/abs/2009A&A...500..693J}
}

@ARTICLE{Yun2019,
       author = {{Yun}, Kiyun and {Pillepich}, Annalisa and {Zinger}, Elad and {Nelson}, Dylan and {Donnari}, Martina and {Joshi}, Gandhali and {Rodriguez-Gomez}, Vicente and {Genel}, Shy and {Weinberger}, Rainer and {Vogelsberger}, Mark and {Hernquist}, Lars},
        title = "{Jellyfish galaxies with the IllustrisTNG simulations - I. Gas-stripping phenomena in the full cosmological context}",
      journal = {\mnras},
         year = 2019,
        month = feb,
       volume = {483},
       number = {1},
        pages = {1042-1066},
          doi = {10.1093/mnras/sty3156},
archivePrefix = {arXiv},
       eprint = {1810.00005},
 primaryClass = {astro-ph.GA},
       adsurl = {https://ui.adsabs.harvard.edu/abs/2019MNRAS.483.1042Y}
}

@ARTICLE{Rohr2023,
       author = {{Rohr}, Eric and {Pillepich}, Annalisa and {Nelson}, Dylan and {Zinger}, Elad and {Joshi}, Gandhali D. and {Ayromlou}, Mohammadreza},
        title = "{Jellyfish galaxies with the IllustrisTNG simulations - when, where, and for how long does ram pressure stripping of cold gas occur?}",
      journal = {\mnras},
         year = 2023,
        month = sep,
       volume = {524},
       number = {3},
        pages = {3502-3525},
          doi = {10.1093/mnras/stad2101},
archivePrefix = {arXiv},
       eprint = {2304.09196},
 primaryClass = {astro-ph.GA},
       adsurl = {https://ui.adsabs.harvard.edu/abs/2023MNRAS.524.3502R}
}

@ARTICLE{Goller2023,
       author = {{G{\"o}ller}, Junia and {Joshi}, Gandhali D. and {Rohr}, Eric and {Zinger}, Elad and {Pillepich}, Annalisa},
        title = "{Jellyfish galaxies with the IllustrisTNG simulations - No enhanced population-wide star formation according to TNG50}",
      journal = {\mnras},
         year = 2023,
        month = nov,
       volume = {525},
       number = {3},
        pages = {3551-3570},
          doi = {10.1093/mnras/stad2551},
archivePrefix = {arXiv},
       eprint = {2304.09199},
 primaryClass = {astro-ph.GA},
       adsurl = {https://ui.adsabs.harvard.edu/abs/2023MNRAS.525.3551G}
}

@ARTICLE{Zinger2024,
       author = {{Zinger}, Elad and {Joshi}, Gandhali D. and {Pillepich}, Annalisa and {Rohr}, Eric and {Nelson}, Dylan},
        title = "{Jellyfish galaxies with the IllustrisTNG simulations - citizen-science results towards large distances, low-mass hosts, and high redshifts}",
      journal = {\mnras},
         year = 2024,
        month = jan,
       volume = {527},
       number = {3},
        pages = {8257-8289},
          doi = {10.1093/mnras/stad3716},
archivePrefix = {arXiv},
       eprint = {2304.09202},
 primaryClass = {astro-ph.GA},
       adsurl = {https://ui.adsabs.harvard.edu/abs/2024MNRAS.527.8257Z}
}

@ARTICLE{Lee2022,
       author = {{Lee}, Jaehyun and {Kimm}, Taysun and {Blaizot}, J{\'e}r{\'e}my and {Katz}, Harley and {Lee}, Wonki and {Sheen}, Yun-Kyeong and {Devriendt}, Julien and {Slyz}, Adrianne},
        title = "{Simulating Jellyfish Galaxies: A Case Study for a Gas-rich Dwarf Galaxy}",
      journal = {\apj},
         year = 2022,
        month = apr,
       volume = {928},
       number = {2},
          eid = {144},
        pages = {144},
          doi = {10.3847/1538-4357/ac5595},
archivePrefix = {arXiv},
       eprint = {2201.01316},
 primaryClass = {astro-ph.GA},
       adsurl = {https://ui.adsabs.harvard.edu/abs/2022ApJ...928..144L}
}

@ARTICLE{King1962,
       author = {{King}, Ivan},
        title = "{The structure of star clusters. I. an empirical density law}",
      journal = {\aj},
         year = 1962,
        month = oct,
       volume = {67},
        pages = {471},
          doi = {10.1086/108756},
       adsurl = {https://ui.adsabs.harvard.edu/abs/1962AJ.....67..471K}
}

@ARTICLE{Hoemer1957,
       author = {{von Hoerner}, S.},
        title = "{Internal structure of globular clusters}",
      journal = {\apj},
         year = 1957,
        month = mar,
       volume = {125},
        pages = {451},
          doi = {10.1086/146321},
       adsurl = {https://ui.adsabs.harvard.edu/abs/1957ApJ...125..451V}
}

@ARTICLE{Tormen1998,
       author = {{Tormen}, Giuseppe and {Diaferio}, Antonaldo and {Syer}, D.},
        title = "{Survival of substructure within dark matter haloes}",
      journal = {\mnras},
         year = 1998,
        month = sep,
       volume = {299},
       number = {3},
        pages = {728-742},
          doi = {10.1046/j.1365-8711.1998.01775.x},
archivePrefix = {arXiv},
       eprint = {astro-ph/9712222},
 primaryClass = {astro-ph},
       adsurl = {https://ui.adsabs.harvard.edu/abs/1998MNRAS.299..728T}
}

@ARTICLE{Johnston1995,
       author = {{Johnston}, Kathryn V. and {Spergel}, David N. and {Hernquist}, Lars},
        title = "{The Disruption of the Sagittarius Dwarf Galaxy}",
      journal = {\apj},
         year = 1995,
        month = oct,
       volume = {451},
        pages = {598},
          doi = {10.1086/176247},
archivePrefix = {arXiv},
       eprint = {astro-ph/9502005},
 primaryClass = {astro-ph},
       adsurl = {https://ui.adsabs.harvard.edu/abs/1995ApJ...451..598J}
}

@ARTICLE{Hayashi2003,
       author = {{Hayashi}, Eric and {Navarro}, Julio F. and {Taylor}, James E. and {Stadel}, Joachim and {Quinn}, Thomas},
        title = "{The Structural Evolution of Substructure}",
      journal = {\apj},
         year = 2003,
        month = feb,
       volume = {584},
       number = {2},
        pages = {541-558},
          doi = {10.1086/345788},
archivePrefix = {arXiv},
       eprint = {astro-ph/0203004},
 primaryClass = {astro-ph},
       adsurl = {https://ui.adsabs.harvard.edu/abs/2003ApJ...584..541H}
}

@ARTICLE{Read2006,
       author = {{Read}, J.~I. and {Wilkinson}, M.~I. and {Evans}, N.~W. and {Gilmore} and G. and {Kleyna}, Jan T.},
        title = "{The tidal stripping of satellites}",
      journal = {\mnras},
         year = 2006,
        month = feb,
       volume = {366},
       number = {2},
        pages = {429-437},
          doi = {10.1111/j.1365-2966.2005.09861.x},
archivePrefix = {arXiv},
       eprint = {astro-ph/0506687},
 primaryClass = {astro-ph},
       adsurl = {https://ui.adsabs.harvard.edu/abs/2006MNRAS.366..429R}
}

@ARTICLE{Penarrubia2008,
       author = {{Pe{\~n}arrubia}, Jorge and {Navarro}, Julio F. and {McConnachie}, Alan W.},
        title = "{The Tidal Evolution of Local Group Dwarf Spheroidals}",
      journal = {\apj},
         year = 2008,
        month = jan,
       volume = {673},
       number = {1},
        pages = {226-240},
          doi = {10.1086/523686},
archivePrefix = {arXiv},
       eprint = {0708.3087},
 primaryClass = {astro-ph},
       adsurl = {https://ui.adsabs.harvard.edu/abs/2008ApJ...673..226P}
}

@ARTICLE{Cooper2010,
       author = {{Cooper}, A.~P. and {Cole} and S. and {Frenk}, C.~S. and {White}, S.~D.~M. and {Helly} and J. and {Benson}, A.~J. and {De Lucia} and G. and {Helmi} and A. and {Jenkins} and A. and {Navarro}, J.~F. and {Springel} and V. and {Wang}, J.},
        title = "{Galactic stellar haloes in the CDM model}",
      journal = {\mnras},
         year = 2010,
        month = aug,
       volume = {406},
       number = {2},
        pages = {744-766},
          doi = {10.1111/j.1365-2966.2010.16740.x},
archivePrefix = {arXiv},
       eprint = {0910.3211},
 primaryClass = {astro-ph.GA},
       adsurl = {https://ui.adsabs.harvard.edu/abs/2010MNRAS.406..744C}
}

@ARTICLE{Fattahi2020,
       author = {{Fattahi}, Azadeh and {Deason}, Alis J. and {Frenk}, Carlos S. and {Simpson}, Christine M. and {G{\'o}mez}, Facundo A. and {Grand}, Robert J.~J. and {Monachesi}, Antonela and {Marinacci}, Federico and {Pakmor}, R{\"u}diger},
        title = "{A tale of two populations: surviving and destroyed dwarf galaxies and the build-up of the Milky Way's stellar halo}",
      journal = {\mnras},
         year = 2020,
        month = oct,
       volume = {497},
       number = {4},
        pages = {4459-4471},
          doi = {10.1093/mnras/staa2221},
archivePrefix = {arXiv},
       eprint = {2002.12043},
 primaryClass = {astro-ph.GA},
       adsurl = {https://ui.adsabs.harvard.edu/abs/2020MNRAS.497.4459F}
}

@ARTICLE{Grand2017,
       author = {{Grand}, Robert J.~J. and {G{\'o}mez}, Facundo A. and {Marinacci}, Federico and {Pakmor}, R{\"u}diger and {Springel}, Volker and {Campbell}, David J.~R. and {Frenk}, Carlos S. and {Jenkins}, Adrian and {White}, Simon D.~M.},
        title = "{The Auriga Project: the properties and formation mechanisms of disc galaxies across cosmic time}",
      journal = {\mnras},
         year = 2017,
        month = may,
       volume = {467},
       number = {1},
        pages = {179-207},
          doi = {10.1093/mnras/stx071},
archivePrefix = {arXiv},
       eprint = {1610.01159},
 primaryClass = {astro-ph.GA},
       adsurl = {https://ui.adsabs.harvard.edu/abs/2017MNRAS.467..179G}
}

@ARTICLE{Joshi2020,
       author = {{Joshi}, Gandhali D. and {Pillepich}, Annalisa and {Nelson}, Dylan and {Marinacci}, Federico and {Springel}, Volker and {Rodriguez-Gomez}, Vicente and {Vogelsberger}, Mark and {Hernquist}, Lars},
        title = "{The fate of disc galaxies in IllustrisTNG clusters}",
      journal = {\mnras},
         year = 2020,
        month = aug,
       volume = {496},
       number = {3},
        pages = {2673-2703},
          doi = {10.1093/mnras/staa1668},
archivePrefix = {arXiv},
       eprint = {2004.01191},
 primaryClass = {astro-ph.GA},
       adsurl = {https://ui.adsabs.harvard.edu/abs/2020MNRAS.496.2673J}
}

@ARTICLE{Peng2015,
       author = {{Peng} and Y. and {Maiolino} and R. and {Cochrane}, R.},
        title = "{Strangulation as the primary mechanism for shutting down star formation in galaxies}",
      journal = {\nat},
         year = 2015,
        month = may,
       volume = {521},
       number = {7551},
        pages = {192-195},
          doi = {10.1038/nature14439},
archivePrefix = {arXiv},
       eprint = {1505.03143},
 primaryClass = {astro-ph.GA},
       adsurl = {https://ui.adsabs.harvard.edu/abs/2015Natur.521..192P}
}

@ARTICLE{Maier2019a,
       author = {{Maier} and C. and {Ziegler}, B.~L. and {Haines}, C.~P. and {Smith}, G.~P.},
        title = "{Slow-then-rapid quenching as traced by tentative evidence for enhanced metallicities of cluster galaxies at $z \sim 0.2$ in the slow quenching phase}",
      journal = {\aap},
         year = 2019,
        month = jan,
       volume = {621},
          eid = {A131},
        pages = {A131},
          doi = {10.1051/0004-6361/201834290},
archivePrefix = {arXiv},
       eprint = {1809.07675},
 primaryClass = {astro-ph.GA},
       adsurl = {https://ui.adsabs.harvard.edu/abs/2019A&A...621A.131M}
}

@ARTICLE{Maier2019b,
       author = {{Maier} and C. and {Hayashi} and M. and {Ziegler}, B.~L. and {Kodama}, T.},
        title = "{Cluster induced quenching of galaxies in the massive cluster XMMXCS J2215.9-1738 at $z \sim 1.5$ traced by enhanced metallicities inside half R$_{200}$}",
      journal = {\aap},
         year = 2019,
        month = jun,
       volume = {626},
          eid = {A14},
        pages = {A14},
          doi = {10.1051/0004-6361/201935522},
archivePrefix = {arXiv},
       eprint = {1903.09591},
 primaryClass = {astro-ph.GA},
       adsurl = {https://ui.adsabs.harvard.edu/abs/2019A&A...626A..14M}
}

@ARTICLE{Kawata2008,
       author = {{Kawata}, Daisuke and {Mulchaey}, John S.},
        title = "{Strangulation in Galaxy Groups}",
      journal = {\apjl},
         year = 2008,
        month = jan,
       volume = {672},
       number = {2},
        pages = {L103},
          doi = {10.1086/526544},
archivePrefix = {arXiv},
       eprint = {0707.3814},
 primaryClass = {astro-ph},
       adsurl = {https://ui.adsabs.harvard.edu/abs/2008ApJ...672L.103K}
}

@ARTICLE{Bahe2015,
       author = {{Bah{\'e}}, Yannick M. and {McCarthy}, Ian G.},
        title = "{Star formation quenching in simulated group and cluster galaxies: when, how, and why?}",
      journal = {\mnras},
         year = 2015,
        month = feb,
       volume = {447},
       number = {1},
        pages = {969-992},
          doi = {10.1093/mnras/stu2293},
archivePrefix = {arXiv},
       eprint = {1410.8161},
 primaryClass = {astro-ph.GA},
       adsurl = {https://ui.adsabs.harvard.edu/abs/2015MNRAS.447..969B}
}

@ARTICLE{Trussler2020,
       author = {{Trussler}, James and {Maiolino}, Roberto and {Maraston}, Claudia and {Peng}, Yingjie and {Thomas}, Daniel and {Goddard}, Daniel and {Lian}, Jianhui},
        title = "{Both starvation and outflows drive galaxy quenching}",
      journal = {\mnras},
         year = 2020,
        month = feb,
       volume = {491},
       number = {4},
        pages = {5406-5434},
          doi = {10.1093/mnras/stz3286},
archivePrefix = {arXiv},
       eprint = {1811.09283},
 primaryClass = {astro-ph.GA},
       adsurl = {https://ui.adsabs.harvard.edu/abs/2020MNRAS.491.5406T}
}

@ARTICLE{Martin2021,
       author = {{Mart{\'\i}n-Navarro}, Ignacio and {Pillepich}, Annalisa and {Nelson}, Dylan and {Rodriguez-Gomez}, Vicente and {Donnari}, Martina and {Hernquist}, Lars and {Springel}, Volker},
        title = "{Anisotropic satellite galaxy quenching modulated by black hole activity}",
      journal = {\nat},
         year = 2021,
        month = jun,
       volume = {594},
       number = {7862},
        pages = {187-190},
          doi = {10.1038/s41586-021-03545-9},
archivePrefix = {arXiv},
       eprint = {2106.04587},
 primaryClass = {astro-ph.GA},
       adsurl = {https://ui.adsabs.harvard.edu/abs/2021Natur.594..187M}
}

@ARTICLE{Stott2022,
       author = {{Stott}, John P.},
        title = "{Evidence for anisotropic quenching in massive galaxy clusters at $z \approx 0.5$}",
      journal = {\mnras},
         year = 2022,
        month = apr,
       volume = {511},
       number = {2},
        pages = {2659-2664},
          doi = {10.1093/mnras/stac089},
archivePrefix = {arXiv},
       eprint = {2112.03937},
 primaryClass = {astro-ph.GA},
       adsurl = {https://ui.adsabs.harvard.edu/abs/2022MNRAS.511.2659S}
}

@ARTICLE{Ando2023,
       author = {{Ando}, Makoto and {Shimasaku}, Kazuhiro and {Ito}, Kei},
        title = "{Detection of anisotropic satellite quenching in galaxy clusters up to z   1}",
      journal = {\mnras},
         year = 2023,
        month = feb,
       volume = {519},
       number = {1},
        pages = {13-25},
          doi = {10.1093/mnras/stac3251},
archivePrefix = {arXiv},
       eprint = {2209.00015},
 primaryClass = {astro-ph.GA},
       adsurl = {https://ui.adsabs.harvard.edu/abs/2023MNRAS.519...13A}
}

@ARTICLE{Zabludoff1998,
       author = {{Zabludoff}, Ann I. and {Mulchaey}, John S.},
        title = "{The Properties of Poor Groups of Galaxies. I. Spectroscopic Survey and Results}",
      journal = {\apj},
         year = 1998,
        month = mar,
       volume = {496},
       number = {1},
        pages = {39-72},
          doi = {10.1086/305355},
archivePrefix = {arXiv},
       eprint = {astro-ph/9708132},
 primaryClass = {astro-ph},
       adsurl = {https://ui.adsabs.harvard.edu/abs/1998ApJ...496...39Z}
}

@ARTICLE{Hou2014,
       author = {{Hou}, Annie and {Parker}, Laura C. and {Harris}, William E.},
        title = "{The pre-processing of subhaloes in SDSS groups and clusters}",
      journal = {\mnras},
         year = 2014,
        month = jul,
       volume = {442},
       number = {1},
        pages = {406-418},
          doi = {10.1093/mnras/stu829},
archivePrefix = {arXiv},
       eprint = {1404.7504},
 primaryClass = {astro-ph.GA},
       adsurl = {https://ui.adsabs.harvard.edu/abs/2014MNRAS.442..406H}
}

@ARTICLE{Lopes2024,
       author = {{Lopes}, Paulo A.~A. and {Ribeiro}, Andr{\'e} L.~B. and {Brambila}, Douglas},
        title = "{The role of groups in galaxy evolution: compelling evidence of pre-processing out to the turnaround radius of clusters}",
      journal = {\mnras},
         year = 2024,
        month = jan,
       volume = {527},
       number = {1},
        pages = {L19-L25},
          doi = {10.1093/mnrasl/slad134},
archivePrefix = {arXiv},
       eprint = {2309.11578},
 primaryClass = {astro-ph.CO},
       adsurl = {https://ui.adsabs.harvard.edu/abs/2024MNRAS.527L..19L}
}

@ARTICLE{Fujita2004,
       author = {{Fujita}, Yutaka},
        title = "{Pre-Processing of Galaxies before Entering a Cluster}",
      journal = {\pasj},
         year = 2004,
        month = feb,
       volume = {56},
        pages = {29-43},
          doi = {10.1093/pasj/56.1.29},
archivePrefix = {arXiv},
       eprint = {astro-ph/0311193},
 primaryClass = {astro-ph},
       adsurl = {https://ui.adsabs.harvard.edu/abs/2004PASJ...56...29F}
}

@ARTICLE{Merritt1984,
       author = {{Merritt}, D.},
        title = "{Relaxation and tidal stripping in rich clusters of galaxies. II. Evolution of the luminosity distribution.}",
      journal = {\apj},
         year = 1984,
        month = jan,
       volume = {276},
        pages = {26-37},
          doi = {10.1086/161590},
       adsurl = {https://ui.adsabs.harvard.edu/abs/1984ApJ...276...26M}
}

@ARTICLE{Richstone1983,
       author = {{Richstone}, D.~O. and {Malumuth}, E.~M.},
        title = "{The evolution of clusters of galaxies. I - Very rich clusters.}",
      journal = {\apj},
         year = 1983,
        month = may,
       volume = {268},
        pages = {30-46},
          doi = {10.1086/160926},
       adsurl = {https://ui.adsabs.harvard.edu/abs/1983ApJ...268...30R}
}

@ARTICLE{Murante2007,
       author = {{Murante}, Giuseppe and {Giovalli}, Martina and {Gerhard}, Ortwin and {Arnaboldi}, Magda and {Borgani}, Stefano and {Dolag}, Klaus},
        title = "{The importance of mergers for the origin of intracluster stars in cosmological simulations of galaxy clusters}",
      journal = {\mnras},
         year = 2007,
        month = may,
       volume = {377},
       number = {1},
        pages = {2-16},
          doi = {10.1111/j.1365-2966.2007.11568.x},
archivePrefix = {arXiv},
       eprint = {astro-ph/0701925},
 primaryClass = {astro-ph},
       adsurl = {https://ui.adsabs.harvard.edu/abs/2007MNRAS.377....2M}
}

@ARTICLE{Murante2004,
       author = {{Murante} and G. and {Arnaboldi} and M. and {Gerhard} and O. and {Borgani} and S. and {Cheng}, L.~M. and {Diaferio} and A. and {Dolag} and K. and {Moscardini} and L. and {Tormen} and G. and {Tornatore} and L. and {Tozzi}, P.},
        title = "{The Diffuse Light in Simulations of Galaxy Clusters}",
      journal = {\apjl},
         year = 2004,
        month = jun,
       volume = {607},
       number = {2},
        pages = {L83-L86},
          doi = {10.1086/421348},
archivePrefix = {arXiv},
       eprint = {astro-ph/0404025},
 primaryClass = {astro-ph},
       adsurl = {https://ui.adsabs.harvard.edu/abs/2004ApJ...607L..83M}
}

@ARTICLE{Moster2011,
       author = {{Moster}, Benjamin P. and {Macci{\`o}}, Andrea V. and {Somerville}, Rachel S. and {Naab}, Thorsten and {Cox}, T.~J.},
        title = "{The effects of a hot gaseous halo in galaxy major mergers}",
      journal = {\mnras},
         year = 2011,
        month = aug,
       volume = {415},
       number = {4},
        pages = {3750-3770},
          doi = {10.1111/j.1365-2966.2011.18984.x},
archivePrefix = {arXiv},
       eprint = {1104.0246},
 primaryClass = {astro-ph.GA},
       adsurl = {https://ui.adsabs.harvard.edu/abs/2011MNRAS.415.3750M}
}

@ARTICLE{Cox2006,
       author = {{Cox}, T.~J. and {Jonsson}, Patrik and {Primack}, Joel R. and {Somerville}, Rachel S.},
        title = "{Feedback in simulations of disc-galaxy major mergers}",
      journal = {\mnras},
         year = 2006,
        month = dec,
       volume = {373},
       number = {3},
        pages = {1013-1038},
          doi = {10.1111/j.1365-2966.2006.11107.x},
archivePrefix = {arXiv},
       eprint = {astro-ph/0503201},
 primaryClass = {astro-ph},
       adsurl = {https://ui.adsabs.harvard.edu/abs/2006MNRAS.373.1013C}
}

@ARTICLE{Torrey2012,
       author = {{Torrey}, Paul and {Cox}, T.~J. and {Kewley}, Lisa and {Hernquist}, Lars},
        title = "{The Metallicity Evolution of Interacting Galaxies}",
      journal = {\apj},
         year = 2012,
        month = feb,
       volume = {746},
       number = {1},
          eid = {108},
        pages = {108},
          doi = {10.1088/0004-637X/746/1/108},
archivePrefix = {arXiv},
       eprint = {1107.0001},
 primaryClass = {astro-ph.GA},
       adsurl = {https://ui.adsabs.harvard.edu/abs/2012ApJ...746..108T}
}

@ARTICLE{Kormendy2004,
       author = {{Kormendy}, John and {Kennicutt}, Jr., Robert C.},
        title = "{Secular Evolution and the Formation of Pseudobulges in Disk Galaxies}",
      journal = {\araa},
         year = 2004,
        month = sep,
       volume = {42},
       number = {1},
        pages = {603-683},
          doi = {10.1146/annurev.astro.42.053102.134024},
archivePrefix = {arXiv},
       eprint = {astro-ph/0407343},
 primaryClass = {astro-ph},
       adsurl = {https://ui.adsabs.harvard.edu/abs/2004ARA&A..42..603K}
}

@ARTICLE{Conselice2003a,
       author = {{Conselice}, Christopher J.},
        title = "{The Relationship between Stellar Light Distributions of Galaxies and Their Formation Histories}",
      journal = {\apjs},
         year = 2003,
        month = jul,
       volume = {147},
       number = {1},
        pages = {1-28},
          doi = {10.1086/375001},
archivePrefix = {arXiv},
       eprint = {astro-ph/0303065},
 primaryClass = {astro-ph},
       adsurl = {https://ui.adsabs.harvard.edu/abs/2003ApJS..147....1C}
}

@ARTICLE{Conselice2003b,
       author = {{Conselice}, Christopher J. and {Bershady}, Matthew A. and {Dickinson}, Mark and {Papovich}, Casey},
        title = "{A Direct Measurement of Major Galaxy Mergers at z<\raisebox{-0.5ex}\textasciitilde3}",
      journal = {\aj},
         year = 2003,
        month = sep,
       volume = {126},
       number = {3},
        pages = {1183-1207},
          doi = {10.1086/377318},
archivePrefix = {arXiv},
       eprint = {astro-ph/0306106},
 primaryClass = {astro-ph},
       adsurl = {https://ui.adsabs.harvard.edu/abs/2003AJ....126.1183C}
}

@ARTICLE{Conselice2014,
       author = {{Conselice}, Christopher J.},
        title = "{The Evolution of Galaxy Structure Over Cosmic Time}",
      journal = {\araa},
         year = 2014,
        month = aug,
       volume = {52},
        pages = {291-337},
          doi = {10.1146/annurev-astro-081913-040037},
archivePrefix = {arXiv},
       eprint = {1403.2783},
 primaryClass = {astro-ph.GA},
       adsurl = {https://ui.adsabs.harvard.edu/abs/2014ARA&A..52..291C}
}

@ARTICLE{Moreno2013,
       author = {{Moreno}, Jorge and {Bluck}, Asa F.~L. and {Ellison}, Sara L. and {Patton}, David R. and {Torrey}, Paul and {Moster}, Benjamin P.},
        title = "{The dynamics of galaxy pairs in a cosmological setting}",
      journal = {\mnras},
         year = 2013,
        month = dec,
       volume = {436},
       number = {2},
        pages = {1765-1786},
          doi = {10.1093/mnras/stt1694},
archivePrefix = {arXiv},
       eprint = {1309.1191},
 primaryClass = {astro-ph.CO},
       adsurl = {https://ui.adsabs.harvard.edu/abs/2013MNRAS.436.1765M}
}

@ARTICLE{McIntosh2008,
       author = {{McIntosh}, Daniel H. and {Guo}, Yicheng and {Hertzberg}, Jen and {Katz}, Neal and {Mo}, H.~J. and {van den Bosch}, Frank C. and {Yang}, Xiaohu},
        title = "{Ongoing assembly of massive galaxies by major merging in large groups and clusters from the SDSS}",
      journal = {\mnras},
         year = 2008,
        month = aug,
       volume = {388},
       number = {4},
        pages = {1537-1556},
          doi = {10.1111/j.1365-2966.2008.13531.x},
archivePrefix = {arXiv},
       eprint = {0710.2157},
 primaryClass = {astro-ph},
       adsurl = {https://ui.adsabs.harvard.edu/abs/2008MNRAS.388.1537M}
}

@ARTICLE{Bundy2009,
       author = {{Bundy}, Kevin and {Fukugita}, Masataka and {Ellis}, Richard S. and {Targett}, Thomas A. and {Belli}, Sirio and {Kodama}, Tadayuki},
        title = "{The Greater Impact of Mergers on the Growth of Massive Galaxies: Implications for Mass Assembly and Evolution since z sime 1}",
      journal = {\apj},
         year = 2009,
        month = jun,
       volume = {697},
       number = {2},
        pages = {1369-1383},
          doi = {10.1088/0004-637X/697/2/1369},
archivePrefix = {arXiv},
       eprint = {0902.1188},
 primaryClass = {astro-ph.CO},
       adsurl = {https://ui.adsabs.harvard.edu/abs/2009ApJ...697.1369B}
}

@ARTICLE{Ellison2010,
       author = {{Ellison}, Sara L. and {Patton}, David R. and {Simard}, Luc and {McConnachie}, Alan W. and {Baldry}, Ivan K. and {Mendel} and J. Trevor},
        title = "{Galaxy pairs in the Sloan Digital Sky Survey - II. The effect of environment on interactions}",
      journal = {\mnras},
         year = 2010,
        month = sep,
       volume = {407},
       number = {3},
        pages = {1514-1528},
          doi = {10.1111/j.1365-2966.2010.17076.x},
archivePrefix = {arXiv},
       eprint = {1002.4418},
 primaryClass = {astro-ph.CO},
       adsurl = {https://ui.adsabs.harvard.edu/abs/2010MNRAS.407.1514E}
}

@ARTICLE{Lin2010,
       author = {{Lin}, Lihwai and {Cooper}, Michael C. and {Jian}, Hung-Yu and {Koo}, David C. and {Patton}, David R. and {Yan}, Renbin and {Willmer}, Christopher N.~A. and {Coil}, Alison L. and {Chiueh}, Tzihong and {Croton}, Darren J. and {Gerke}, Brian F. and {Lotz}, Jennifer and {Guhathakurta}, Puragra and {Newman}, Jeffrey A.},
        title = "{Where do Wet, Dry, and Mixed Galaxy Mergers Occur? A Study of the Environments of Close Galaxy Pairs in the DEEP2 Galaxy Redshift Survey}",
      journal = {\apj},
         year = 2010,
        month = aug,
       volume = {718},
       number = {2},
        pages = {1158-1170},
          doi = {10.1088/0004-637X/718/2/1158},
archivePrefix = {arXiv},
       eprint = {1001.4560},
 primaryClass = {astro-ph.CO},
       adsurl = {https://ui.adsabs.harvard.edu/abs/2010ApJ...718.1158L}
}

@ARTICLE{Conselice2006,
       author = {{Conselice}, Christopher J.},
        title = "{Early and Rapid Merging as a Formation Mechanism of Massive Galaxies: Empirical Constraints}",
      journal = {\apj},
         year = 2006,
        month = feb,
       volume = {638},
       number = {2},
        pages = {686-702},
          doi = {10.1086/499067},
archivePrefix = {arXiv},
       eprint = {astro-ph/0507146},
 primaryClass = {astro-ph},
       adsurl = {https://ui.adsabs.harvard.edu/abs/2006ApJ...638..686C}
}

\end{document}